\newcommand{\punkto}[9]{\put(#3
   ,0){\setlength{\unitlength}{2.5cm}\begin{picture}(0,0)(0,-0.7)
   \put(0,#4){\circle*{   0.030}}
   \put(0,#4){\line(0,1){#5}}
   \put(0,#4){\line(0,-1){#5}}
   \end{picture}}}
\newcommand{\punktoo}[9]{\put(#3
   ,0){\setlength{\unitlength}{2.5cm}\begin{picture}(0,0)(0,-0.7)
   \put(0,#4){\circle{   0.040}}
   \end{picture}}}
\newcommand{\punktn}[9]{\put(#3
   ,0){\setlength{\unitlength}{2.5cm}\begin{picture}(0,0)(0,-0.7)
   \put(0,#4){\circle*{   0.030}}
   \put(0,#4){\line(0,1){#5}}
   \put(0,#4){\line(0,-1){#5}}
   \end{picture}}}
\newcommand{\punktnm}[9]{\put(#3
   ,0){\setlength{\unitlength}{2.5cm}\begin{picture}(0,0)(0,-0.7)
   \put(0,#4){\circle{   0.040}}
   \end{picture}}}
\newcommand{\punktm}[9]{\put(#3
   ,0){\setlength{\unitlength}{2.5cm}\begin{picture}(0,0)(0,-0.6)
   \put(0,#4){\circle*{   0.030}}
   \put(0,#4){\line(0,1){#5}}
   \put(0,#4){\line(0,-1){#5}}
   \end{picture}}}
\newcommand{\punktmm}[9]{\put(#3
   ,0){\setlength{\unitlength}{2.5cm}\begin{picture}(0,0)(0,-0.6)
   \put(0,#4){\circle{   0.040}}
   \end{picture}}}
\newcommand{\punktl}[9]{\put(#3
   ,0){\setlength{\unitlength}{2.5cm}\begin{picture}(0,0)(0,-0.4)
   \put(0,#4){\circle*{   0.030}}
   \put(0,#4){\line(0,1){#5}}
   \put(0,#4){\line(0,-1){#5}}
   \end{picture}}}
\newcommand{\punktll}[9]{\put(#3
   ,0){\setlength{\unitlength}{2.5cm}\begin{picture}(0,0)(0,-0.4)
   \put(0,#4){\circle{   0.040}}
   \end{picture}}}
\newcommand{\punktk}[9]{\put(#3
   ,0){\setlength{\unitlength}{2.5cm}\begin{picture}(0,0)(0,-0.5)
   \put(0,#4){\circle*{   0.030}}
   \put(0,#4){\line(0,1){#5}}
   \put(0,#4){\line(0,-1){#5}}
   \end{picture}}}
\newcommand{\punktkk}[9]{\put(#3
   ,0){\setlength{\unitlength}{2.5cm}\begin{picture}(0,0)(0,-0.5)
   \put(0,#4){\circle{   0.040}}
   \end{picture}}}
\newcommand{\punktj}[9]{\put(#3
   ,0){\setlength{\unitlength}{2.5cm}\begin{picture}(0,0)(0,-0.4)
   \put(0,#4){\circle*{   0.030}}
   \put(0,#4){\line(0,1){#5}}
   \put(0,#4){\line(0,-1){#5}}
   \end{picture}}}
\newcommand{\punktjj}[9]{\put(#3
   ,0){\setlength{\unitlength}{2.5cm}\begin{picture}(0,0)(0,-0.4)
   \put(0,#4){\circle{   0.040}}
   \end{picture}}}
\newcommand{\punkti}[9]{\put(#3
   ,0){\setlength{\unitlength}{2.5cm}\begin{picture}(0,0)(0,-1.4)
   \put(0,#4){\circle*{   0.030}}
   \put(0,#4){\line(0,1){#5}}
   \put(0,#4){\line(0,-1){#5}}
   \end{picture}}}
\newcommand{\punktii}[9]{\put(#3
   ,0){\setlength{\unitlength}{2.5cm}\begin{picture}(0,0)(0,-1.4)
   \put(0,#4){\circle{   0.040}}
   \end{picture}}}
\newcommand{\punkth}[9]{\put(#3
   ,0){\setlength{\unitlength}{2.5cm}\begin{picture}(0,0)(0,-1.2)
   \put(0,#4){\circle*{   0.030}}
   \put(0,#4){\line(0,1){#5}}
   \put(0,#4){\line(0,-1){#5}}
   \end{picture}}}
\newcommand{\punkthh}[9]{\put(#3
   ,0){\setlength{\unitlength}{2.5cm}\begin{picture}(0,0)(0,-1.2)
   \put(0,#4){\circle{   0.040}}
   \end{picture}}}
\newcommand{\punktg}[9]{\put(#3
   ,0){\setlength{\unitlength}{2.5cm}\begin{picture}(0,0)(0,-1)
   \put(0,#4){\circle*{   0.030}}
   \put(0,#4){\line(0,1){#5}}
   \put(0,#4){\line(0,-1){#5}}
   \end{picture}}}
\newcommand{\punktgg}[9]{\put(#3
   ,0){\setlength{\unitlength}{2.5cm}\begin{picture}(0,0)(0,-1)
   \put(0,#4){\circle{   0.040}}
   \end{picture}}}
\newcommand{\punktf}[9]{\put(#3
   ,0){\setlength{\unitlength}{2.5cm}\begin{picture}(0,0)(0,-0.6)
   \put(0,#4){\circle*{   0.030}}
   \put(0,#4){\line(0,1){#5}}
   \put(0,#4){\line(0,-1){#5}}
   \end{picture}}}
\newcommand{\punktff}[9]{\put(#3
   ,0){\setlength{\unitlength}{2.5cm}\begin{picture}(0,0)(0,-0.6)
   \put(0,#4){\circle{   0.040}}
   \end{picture}}}
\newcommand{\punkte}[9]{\put(#3
   ,0){\setlength{\unitlength}{2.5cm}\begin{picture}(0,0)(0,-0.6)
   \put(0,#4){\circle*{   0.030}}
   \put(0,#4){\line(0,1){#5}}
   \put(0,#4){\line(0,-1){#5}}
   \end{picture}}}
\newcommand{\punktee}[9]{\put(#3
   ,0){\setlength{\unitlength}{2.5cm}\begin{picture}(0,0)(0,-0.6)
   \put(0,#4){\circle{   0.040}}
   \end{picture}}}
\newcommand{\punktd}[9]{\put(#3
   ,0){\setlength{\unitlength}{2.5cm}\begin{picture}(0,0)(0,-1.4)
   \put(0,#4){\circle*{   0.030}}
   \put(0,#4){\line(0,1){#5}}
   \put(0,#4){\line(0,-1){#5}}
   \end{picture}}}
\newcommand{\punktdd}[9]{\put(#3
   ,0){\setlength{\unitlength}{2.5cm}\begin{picture}(0,0)(0,-1.4)
   \put(0,#4){\circle{   0.040}}
   \end{picture}}}
\newcommand{\punktc}[9]{\put(#3
   ,0){\setlength{\unitlength}{2.5cm}\begin{picture}(0,0)(0,-2)
   \put(0,#4){\circle*{   0.030}}
   \put(0,#4){\line(0,1){#5}}
   \put(0,#4){\line(0,-1){#5}}
   \end{picture}}}
\newcommand{\punktcc}[9]{\put(#3
   ,0){\setlength{\unitlength}{2.5cm}\begin{picture}(0,0)(0,-2)
   \put(0,#4){\circle{   0.040}}
   \end{picture}}}
\newcommand{\punktb}[9]{\put(#3
   ,0){\setlength{\unitlength}{2.5cm}\begin{picture}(0,0)(0,-0.8)
   \put(0,#4){\circle*{   0.030}}
   \put(0,#4){\line(0,1){#5}}
   \put(0,#4){\line(0,-1){#5}}
   \end{picture}}}
\newcommand{\punktbb}[9]{\put(#3
   ,0){\setlength{\unitlength}{2.5cm}\begin{picture}(0,0)(0,-0.8)
   \put(0,#4){\circle{   0.040}}
   \end{picture}}}
\newcommand{\punkta}[9]{\put(#3
   ,0){\setlength{\unitlength}{2.5cm}\begin{picture}(0,0)(0,-0.5)
   \put(0,#4){\circle*{   0.030}}
   \put(0,#4){\line(0,1){#5}}
   \put(0,#4){\line(0,-1){#5}}
   \end{picture}}}

\newcommand{\T}[1]{\rule[1.5ex]{0cm}{#1cm}}
\newcommand{\U}[1]{\rule[-#1cm]{0cm}{#1cm}}

\documentstyle{laa}
\begin{document}

\thesaurus{03(11.17.3)}
\title{The Hamburg Quasar Monitoring Program (HQM) at Calar Alto\thanks{Based
    on observations collected at the
    German-Spanish Astronomical Centre, Calar Alto,
    operated by the Max-Planck-Institut f\"ur Astronomie (MPIA), Heidelberg,
    jointly with the Spanish National Commission for Astronomy}:\\
    I$\!$I$\!$I. Lightcurves of optically violent variable sources}
\author{K.-J.~Schramm
     \and
        U.~Borgeest
     \and
        D.~K\"uhl
     \and
        J.~von~Linde
     \and
        M.D.~Linnert
     \and
        T.~Schramm}
\institute{Hamburger Sternwarte, Gojenbergsweg 112,
           D-W\,2050 Hamburg 80, Germany }
\date{Received date; accepted date}
\maketitle

\begin{abstract}
HQM is an optical broad-band photometric monitoring program carried out
since September 1988. We use a CCD camera at the MPIA 1.2$\,$m
telescope. Fully automatic photometric reduction relative to stars in the
frames is done within a few minutes after each exposure, thus interesting
brightness changes can be followed in detail. The typical photometric error
is 1--2\,\% for a 17.5\,mag quasar. We here present lightcurves of 14
known violently variable sources and compare them with literature data.
For two BL\,Lac objects, 1E\,1229+645 and 4C\,56.27 (1823+568), this paper
is the first variability study. We have also carried out POSS photometry to
obtain indications for variability on a longer timescale.
\end{abstract}

\section{Introduction}

In this paper, being the third one in a series, we show some lightcurves
for quasars which were known to be optically violent variables (OVVs) prior
to our program or which were identified as BL\,Lac objects due to their
optical spectra so that vi\-o\-lent
variability can be assumed. We {\em define\/} an object as an OVV if it
shows variations $\ga$\,0.5\,mag with gradients $\ga$\,5\,mag\,yr$^{-1}$ in
the quasar restframe. We show only those lightcurves for which measurements
exist sufficiently spread over a time-span $\ga$\,2\,yr. The OVV class which is
possibly
equivalent to the blazar class has sometimes (e.g.\ Kayser et al.\
\cite{KRS86}, Schneider \& Weiss \cite{SW87}) been proposed to contain good
microlensing candidates. Our intention was to test this hypothesis by
comparing the high quality lightcurves of our program with simulated
microlensing lightcurves. A detailed physical discussion of the flaring
characteristics of the OVVs in our sample will appear in a subsequent
publication (see also Borgeest \& Schramm \cite{BS92}).

For HQM, we use a CCD camera which was equipped with different chips, in
1988 with an RCA 15$ \mu $ chip ($640 \! \times \! 1024$, pixel size
0.315$''$) and later various, but similar, coated GEC 22$ \mu $ chips ($410
\! \times \! 580$, pixel size 0.462$''$). We measure the quasar fluxes
through a standard Johnson $R\/$ broad-band filter relative to stars
included in the frames. The data reduction is carried out automatically,
immediately after the observation, on a $\mu$VAX\,3200 workstation. The
software package ``HQM'' has been developed in Hamburg (for a short
description see Borgeest \& Schramm 1993, hereafter Paper~I); it is much
faster than standard image processing software. A 0.01\,mag accuracy (in
relative photometry) in the lightcurve of a $\sim 17.5\,$mag quasar can be
reached in this way also for ``non-photometric'' conditions with a typical
exposure time of 500 sec. One of the most interesting OVVs in our sample is
3C\,345; a detailed discussion of possible
variability mechanisms can be found in a separate publication (Schramm et
al. \cite{SBCW93}; cf.\ also Borgeest \& Schramm \cite{BS92} and Schramm \&
Borgeest \cite{SB92}). For another three objects, we already published
or are preparing separate papers:
0836+710 (von~Linde et~al.\ \cite{LBSG93}) and
PKS\,0420-014 (Wagner et al.\ \cite{WBHS93}) which have both been found to
be highly luminous in the $\gamma$-ray waveband as well as PKS\,1510-089
(Valtaoja et al.\ \cite{VBPS93}) for which we have obtained simultaneous
mm-radio data. In Paper~I and in Schramm et al.\ (\cite{SBKL93},
Paper~I$\!$I) we have plotted lightcurves of weakly variable objects.

Extensive monitoring programs which contain violently variable quasars have
been carried out by several investigators:
Lightcurves of many bright quasars have been obtained from the Harvard
historical plate collection (e.g.\ Angione \cite{Ang73})
spanning periods of up to 100 years. Monitoring data obtained until
1973 are reviewed and critically discussed by Grandi \& Tifft
(\cite{GT74}). Lloyd (\cite{Llo84}, hereafter L84) reports on lightcurves
of 36 radio sources from the Herstmonceux Optical Monitoring program for
the period 1966-1980. At the Rosemary Hill
Observatory more than 200, mostly radio-selected quasars have been monitored
since 1968, although not all objects over the total period (Pica et
al.~\cite{PPSL80}, hereafter PPSL; Pica \& Smith \cite{PS83}, hereafter
PS83; Smith et al.\ \cite{SNLC93}, hereafter SNLC, and Refs.\ therein);
more recent data on OVVs can also be found in Webb et al.\ (\cite{WSLF88}).
Another program was carried out at the Asagio Observatory (e.g.\ Barbieri
et al.~\cite{BRZ79}, hereafter BRZ) over the period 1967 to 1977.
Moore \& Stockman (\cite{MS84}) have
collected a catalog of the observational properties of
239 quasars, including variability data. A more recent monitoring program
of BL\,Lac objects and quasars, starting in the early '80s, is carried out
at the Tuorla Observatory, Finland (Sillanp\"a\"a et al.\
\cite{SHK88,SMV91}).

\begin{table*}
\caption[ ]{Properties of the HQM-quasars discussed in this article. Following
            the name, we have listed the emission redshift, the absolute
            visual magnitude and the radio flux at 6\,cm (all from V\'eron
            \& V\'eron-Cetty \cite{VV91}). In the following columns
            redshifts and restframe equivalent widths (together for both
            lines) of MgII-absorption line systems are given. The next
            columns list properties of foreground galaxies, $\theta_{\rm
            gal}$ is the angular separation and $r_{\rm gal}$ the physical
            ($H_0=50$, $q_0=0$) distance of the galaxy from the sight-line,
            $\theta_{\rm a}$ is the approximate radius of a galaxy on a
            POSS blue print. The next column lists the variability
            classification. In the last column we give the results of our
            POSS photometry; a ``+'' indicates that the quasar became
            brighter in recent years. Following abbreviations are used:
            med: medium, nd: not detected}
\begin{flushleft}
\scriptsize
\begin{tabular}{lllcc|ccr|lccclr|cr|l}
\hline
\hline
           &                    & &&
           &                    &                     &
           &                    &&
           &                    &                     &
           &                    &
           &                    \\

Object     & Name               & $z_{\rm em}$ &$-M_V$&$S_6$
           & $z_{\rm abs}$ & \multicolumn{1}{c}{$W_{\rm r}$}
           & \multicolumn{1}{c|}{Ref}
           & $z_{\rm gal}$
           & \multicolumn{2}{c}{$\theta_{\rm gal}$}
           & $r_{\rm gal}$      & \multicolumn{1}{c}{$V_{\rm gal}$}
           & \multicolumn{1}{c|}{Ref}
           & \multicolumn{1}{c}{Var} & \multicolumn{1}{c|}{Ref}
           & \multicolumn{1}{c}{$\Delta R_{40}$}\\

           &                    &&&[Jy]
           &                    & \multicolumn{1}{c}{[\AA]}         &
           &                    & \multicolumn{1}{c}{[\arcsec]}
           & $\!\!\!$[$\theta_{\rm a}$]$\!\!\!$
           & $\!\!\!\!$[kpc]$\!\!\!\!$              &               &
           &                    &
\U{0.2}    & \multicolumn{1}{c}{$[$mag$]$}    \\
\hline
\hline
0007+106   & IIIZw2             & 0.089 & 23.3& 0.42\T{0.2}
           & --                 &                     &
           & --                 &&
           &                    &                     &
           & OVV                & \ref{c}
           & $-0.15\pm0.50^{\triangleright}$                   \\
0215+015   & PKS                & 1.715  &30.0&0.36
           & 1.345              & 3.2                 &  \ref{6b}
           & --                 &&
           &                    &                     &
           & OVV                & \ref{6c}
           & \multicolumn{1}{c}{--}     \\
0219+428   & 3C$\,$66A          & 0.444     &27.2& 1.04
           & --                 &                     &
           & 0.021              & 145&4.9
           & 90                 & 15.4                & \ref{6a}
           & OVV                & \ref{2}
           & $+0.56\pm0.20$          \\
0235+164   & AO                 & 0.934  &28.7&1.94
           & 0.524              & 4.8                 & \ref{7a}
           & 0.524              & 0.5&
           & 3.6                & 20.1                & \ref{8}
           & OVV                & \ref{6c}
           & $+2.36\pm0.25$             \\
           &                    & &&
           &                    &                     &
           &                    & 2&
           &                    & 20.9                &
           &                    &
           &                   \\
           &                    & &&
           &                    &                     &
           & 0.525              & 6&
           & 9.0                & 21.9                &
           &                    &
           &                       \\
           &                    & &&
           & 0.852              & 0.9                 &
           &                    &&
           &                    &                     &
           &                    &
           &                    \\
0735+178   & PKS                &$>$.424    &$>$26.0&1.99
           & 0.424              & 2.8                 & \ref{9f}
           & --                 &&
           &                    &                     &
           & OVV                & \ref{6c}\U{0.2}
           & $-0.34\pm0.44$   \\
\hline
0836+710   & 4C$\,$71.07        & 2.17 &30.1&2.57\T{0.2}
           & 0.914              &                     & \ref{13aa}
           & --                 &&
           &                    &                     &
           & nd                 & \ref{12a}
           & $-0.09\pm0.10$                \\
0851+202   & OJ$\,$287          & 0.306  &26.0&2.61
           & --                 &                     &
           &                    & 3&
           &                    & 21                  & \ref{32}
           & OVV                &  \ref{6c}
           & $-1.14\pm0.24$ \\
           &                    & &&
           &                    &                     &
           & )$^{\star}$        &&
           &                    &                     & \ref{32}
           &                    &
           &                      \\
1219+285   & ON$\,$231          & 0.102&22.7&0.72
           & --                 &                     &
           & --                 &&
           &                    &                     &
           & OVV                & \ref{6c}
           & $-0.44\pm0.14$ \\
1229+645   & 1E                 & 0.164     &23.0&0.042
           & --                 &                     &
           & 0.009              & 81&2.0
           & 21                 & 14.2                & \ref{10}
           & --                 &
           & $-1.00\pm0.31^{\triangleright}$          \\
           &                    & &&
           &                    &                     &
           & )$^{\sharp}$        &&
           &                    &                     & \ref{32}
           &                    &
           &   \\
1253$-$055 & 3C$\,$279          & 0.538  &25.1&15.34
           & --                 &                     &
           & --                 &&
           &                    &                     &
           & OVV                & \ref{21a}\U{0.2}
           & $+1.67\pm0.50^{\triangleleft}$\\
\hline
1308+326   & B2                 & 0.996  &29.1&1.59\T{0.2}
           & 0.879              & 0.4                 & \ref{3a}
           &                    & 5.4&
           &                    & 21                  & \ref{32}
           & OVV                & \ref{6c}
           & $-1.17\pm0.36^{\triangleleft}$     \\
1638+398   & NRAO$\,$512        & 1.66  &29.4&1.16
           & --                 &                     &
           & )$^{\sharp}$        &&
           &                    &                     & \ref{32}
           & OVV                & \ref{6c}
           & $-0.87\pm0.27$   \\
1641+399   & 3C$\,$345          & 0.595         &27.1&5.65
           & --                 &                     &
           &                    & 4.5&
           &                    &                     & \ref{32}
           & OVV                & \ref{26c}
           & \multicolumn{1}{c}{--} \\
           &                    & &&
           &                    &                     &
           &                    & 6.6&
           &                    &                     &
           &                    &
           &                       \\
1749+701   & W1                 & 0.76?  &26.7&1.09
           & --                 &                     &
           & --                 &&
           &                    &                     &
           & med                & \ref{27b}
           & $-0.59\pm0.21$\\
1823+568   & 4C\,56.27          & 0.66?&25.0&1.67
           & --                 &                     &
           & --                 &&
           &                    &                     &
           & --                 &\U{0.2}
           & \multicolumn{1}{c}{--}     \\
\hline
2223$-$052 & 3C$\,$446          & 1.404  &27.1&4.07\T{0.2}
           & 0.847              &                     & \ref{28ee}
           & --                 &&
           &                    &                     &
           & OVV                & \ref{6c}\U{0.2}
           & $+1.40\pm0.17$  \\
\hline
\multicolumn{16}{l}{\ \ \ } \\
\multicolumn{16}{l}{$^{\sharp}$\ Faint galaxy/galaxies within few arcsec} \\
\multicolumn{16}{l}{$^{\star}$\ There is a number excess of
             faint galaxies inside $30\arcsec$}  \\
\multicolumn{16}{l}{$^{\triangleleft}$\ There are only less than five stars
             in the CCD frames which are useful for POSS photometry}\\
\multicolumn{16}{l}{$^{\triangleright}$\ All stars useful for POSS
             photometry are fainter than the quasar}\\
\multicolumn{16}{l}{\ }\\
\hline
\hline
\end{tabular}
\end{flushleft}
\normalsize
\end{table*}

\begin{table}
\begin{flushleft}

\vspace*{-0.3cm}
\caption{References for Table~1}
\begin{tabular}{l}
\hline
\parbox{8cm}{%
\begin{enumerate}
\item \label{32} This work
\item \label{c} Lloyd \cite{Llo84}
\item \label{6b} Gaskell \cite{Gas82}
\item \label{6c} Webb et al.\ \cite{WSLF88}
\item \label{6a} Monk et al.\ \cite{MPPB86}
\item \label{2} Pica et al.\ \cite{PPSL80}
\item \label{7a} Wolfe \& Wills \cite{WW77}
\item \label{8} Stickel et al.\ \cite{SFK88}
\item \label{9f} Peterson et. al.\ \cite{PCSW77}
\item \label{13aa}  Stickel \& K\"uhr \cite{SK92}
\item \label{12a} Wagner et al.\ \cite{WSQW90}
\item \label{10} Stocke et al.\ \cite{SSMG87}
\item \label{21a} Eachus \& Liller \cite{EL75}
\item \label{3a} Bergeron \& Boiss\'e \cite{BB84}
\item \label{26c} Schramm et al.\ \cite{SBCW93}
\item \label{27b} Arp et al.\ \cite{ASWO79}
\item \label{28ee} Perry et al.\ \cite{PBB78}
\end{enumerate}} \\
\hline
\end{tabular}
\end{flushleft}
\end{table}

\section{Discussion of the lightcurves}

Some interesting properties of the quasars under consideration are given in
Table~1; we also give there the results of a POSS photometry for which we used
our CCD frames to calibrate red POSS prints (for a short description see
Paper~I). The given values are variations with respect to the reference
magnitude $R_0$ indicated in each lightcurve.
All lightcurves are shown with the same time and magnitude scale
in Fig.~1. Those parts of the lightcurves which are better sampled
are plotted with higher time resolution in Figs.~2\,--\,8. In this
section we also discuss literature data; however, we concentrate only on a
few publications for each object. A more complete list of references can be
found in Hewitt \& Burbidge (\cite{HB87,HB89}).

\noindent
{\bf I$\!$I$\!$I\,Zw\,2 (0007+106).} This type-1 Seyfert galaxy was
included in the early Rosemary Hill sample; PPSL found only marginal
variability, comparable to our HQM data. Violent variability with a total
amplitude of 1.7\,mag was recorded by L84.

\noindent
{\bf PKS\,0215+015} is a BL\,Lac object with a very high redshift.
WSLF have monitored this object since 1981 and detected variations
over a 3.5\,mag range. Our data indicate violent variability, too; however,
both observed time-span and amplitude are smaller.

\begin{figure*}

\vspace*{0.5cm}

\begin{picture}(17 ,2.5 )(-1,0)
\put(0,0){\setlength{\unitlength}{0.0085cm}%
\begin{picture}(2000, 294.117)(7250,0)
\put(7250,0){\framebox(2000, 294.117)[tl]{\begin{picture}(0,0)(0,0)
        \put(0,0){\makebox(0,0)[tr]{$\Delta R$\hspace*{0.2cm}}}
        \put(2000,0){\makebox(0,0)[tr]{\large{0007+106}\T{0.4}
                                 \hspace*{0.5cm}}}
        \put(2000,-
294.117){\setlength{\unitlength}{1cm}\begin{picture}(0,0)(0,0)
        \end{picture}}
    \end{picture}}}

\thicklines
\put(7250,0){\setlength{\unitlength}{2.5cm}\begin{picture}(0,0)(0,-0.5)
   \put(0,0){\setlength{\unitlength}{1cm}\begin{picture}(0,0)(0,0)
        \put(0,0){\line(1,0){0.3}}
        \end{picture}}
   \end{picture}}

\put(9250,0){\setlength{\unitlength}{2.5cm}\begin{picture}(0,0)(0,-0.5)
   \put(0,0){\setlength{\unitlength}{1cm}\begin{picture}(0,0)(0,0)
        \put(0,0){\line(-1,0){0.3}}
        \end{picture}}
   \end{picture}}

\thinlines
\put(7250,0){\setlength{\unitlength}{2.5cm}\begin{picture}(0,0)(0,-0.5)
   \multiput(0,0)(0,0.1){5}{\setlength{\unitlength}{1cm}%
\begin{picture}(0,0)(0,0)
        \put(0,0){\line(1,0){0.12}}
        \end{picture}}
   \end{picture}}

\put(7250,0){\setlength{\unitlength}{2.5cm}\begin{picture}(0,0)(0,-0.5)
   \multiput(0,0)(0,-0.1){5}{\setlength{\unitlength}{1cm}%
\begin{picture}(0,0)(0,0)
        \put(0,0){\line(1,0){0.12}}
        \end{picture}}
   \end{picture}}

\put(9250,0){\setlength{\unitlength}{2.5cm}\begin{picture}(0,0)(0,-0.5)
   \multiput(0,0)(0,0.1){5}{\setlength{\unitlength}{1cm}%
\begin{picture}(0,0)(0,0)
        \put(0,0){\line(-1,0){0.12}}
        \end{picture}}
   \end{picture}}

\put(9250,0){\setlength{\unitlength}{2.5cm}\begin{picture}(0,0)(0,-0.5)
   \multiput(0,0)(0,-0.1){5}{\setlength{\unitlength}{1cm}%
\begin{picture}(0,0)(0,0)
        \put(0,0){\line(-1,0){0.12}}
        \end{picture}}
   \end{picture}}

\put(7250,0){\setlength{\unitlength}{2.5cm}\begin{picture}(0,0)(0,-0.5)
   \put(0,0.2){\setlength{\unitlength}{1cm}\begin{picture}(0,0)(0,0)
        \put(0,0){\line(1,0){0.12}}
        \put(-0.2,0){\makebox(0,0)[r]{\bf 0.2}}
        \end{picture}}
   \put(0,0.0){\setlength{\unitlength}{1cm}\begin{picture}(0,0)(0,0)
        \put(0,0){\line(1,0){0.12}}
        \put(-0.2,0){\makebox(0,0)[r]{\bf 0.0}}
        \end{picture}}
   \put(0,-0.4){\setlength{\unitlength}{1cm}\begin{picture}(0,0)(0,0)
        \put(0,0){\line(1,0){0.12}}
        \put(-0.2,0){\makebox(0,0)[r]{\bf -0.4}}
        \end{picture}}
   \put(0,-0.2){\setlength{\unitlength}{1cm}\begin{picture}(0,0)(0,0)
        \put(0,0){\line(1,0){0.12}}
        \put(-0.2,0){\makebox(0,0)[r]{\bf -0.2}}
        \end{picture}}
   \put(0,0.0){\setlength{\unitlength}{1cm}\begin{picture}(0,0)(0,0)
        \put(0,0){\line(1,0){0.12}}
        \put(-0.2,0){\makebox(0,0)[r]{\bf 0.0}}
        \end{picture}}
   \put(0,0.0){\setlength{\unitlength}{1cm}\begin{picture}(0,0)(0,0)
        \put(0,0){\line(1,0){0.12}}
        \put(-0.2,0){\makebox(0,0)[r]{\bf 0.0}}
        \end{picture}}
   \put(0,0.0){\setlength{\unitlength}{1cm}\begin{picture}(0,0)(0,0)
        \put(0,0){\line(1,0){0.12}}
        \put(-0.2,0){\makebox(0,0)[r]{\bf 0.0}}
        \end{picture}}
   \put(0,0.0){\setlength{\unitlength}{1cm}\begin{picture}(0,0)(0,0)
        \put(0,0){\line(1,0){0.12}}
        \put(-0.2,0){\makebox(0,0)[r]{\bf 0.0}}
        \end{picture}}
   \put(0,0.4){\setlength{\unitlength}{1cm}\begin{picture}(0,0)(0,0)
        \put(0,0){\line(1,0){0.12}}
        \end{picture}}
   \end{picture}}

   \put(7527.5, 294.117){\setlength{\unitlength}{1cm}\begin{picture}(0,0)(0,0)
        \put(0,0){\line(0,-1){0.2}}
        \put(0,0.2){\makebox(0,0)[b]{\bf 1989}}
   \end{picture}}
   \put(7892.5, 294.117){\setlength{\unitlength}{1cm}\begin{picture}(0,0)(0,0)
        \put(0,0){\line(0,-1){0.2}}
        \put(0,0.2){\makebox(0,0)[b]{\bf 1990}}
   \end{picture}}
   \put(8257.5, 294.117){\setlength{\unitlength}{1cm}\begin{picture}(0,0)(0,0)
        \put(0,0){\line(0,-1){0.2}}
        \put(0,0.2){\makebox(0,0)[b]{\bf 1991}}
  \end{picture}}
   \put(8622.5, 294.117){\setlength{\unitlength}{1cm}\begin{picture}(0,0)(0,0)
        \put(0,0){\line(0,-1){0.2}}
        \put(0,0.2){\makebox(0,0)[b]{\bf 1992}}
   \end{picture}}
   \put(8987.5, 294.117){\setlength{\unitlength}{1cm}\begin{picture}(0,0)(0,0)
        \put(0,0){\line(0,-1){0.2}}
        \put(0,0.2){\makebox(0,0)[b]{\bf 1993}}
   \end{picture}}

    \multiput(7250,0)(50,0){40}%
        {\setlength{\unitlength}{1cm}\begin{picture}(0,0)(0,0)
        \put(0,0){\line(0,1){0.12}}
    \end{picture}}
    \put(7500,0){\setlength{\unitlength}{1cm}\begin{picture}(0,0)(0,0)
        \put(0,0){\line(0,1){0.2}}
    \end{picture}}
    \put(8000,0){\setlength{\unitlength}{1cm}\begin{picture}(0,0)(0,0)
        \put(0,0){\line(0,1){0.2}}
    \end{picture}}
   \put(7750,0){\setlength{\unitlength}{1cm}\begin{picture}(0,0)(0,0)
        \put(0,0){\line(0,1){0.2}}
    \end{picture}}
   \put(8250,0){\setlength{\unitlength}{1cm}\begin{picture}(0,0)(0,0)
        \put(0,0){\line(0,1){0.2}}
    \end{picture}}
    \put(8500,0){\setlength{\unitlength}{1cm}\begin{picture}(0,0)(0,0)
        \put(0,0){\line(0,1){0.2}}
    \end{picture}}
    \put(8750,0){\setlength{\unitlength}{1cm}\begin{picture}(0,0)(0,0)
       \put(0,0){\line(0,1){0.2}}
    \end{picture}}
    \put(9000,0){\setlength{\unitlength}{1cm}\begin{picture}(0,0)(0,0)
       \put(0,0){\line(0,1){0.2}}
    \end{picture}}

\punkta{02/11/89}{21:56}{7833.414}{  0.118}{0.015}{ 500}{1.90}{
536}{1.2CA}
\punkta{14/12/89}{19:36}{7875.317}{  0.094}{0.014}{ 500}{2.19}{
967}{1.2CA}
\punkta{19/12/89}{18:41}{7880.279}{  0.040}{0.031}{ 100}{1.48}{
484}{1.2CA}
\punkta{19/12/89}{18:51}{7880.286}{  0.061}{0.014}{ 500}{1.97}{
635}{1.2CA}
\punkta{29/09/90}{22:27}{8164.436}{  0.176}{0.021}{ 100}{1.56}{
771}{1.2CA}
\punkta{22/08/91}{03:55}{8490.664}{ -0.074}{0.024}{ 100}{1.62}{
356}{1.2CA}
\punkta{22/08/91}{04:07}{8490.672}{ -0.058}{0.024}{ 500}{1.65}{
808}{1.2CA}
\punkta{23/08/91}{00:19}{8491.514}{ -0.058}{0.027}{ 100}{1.61}{
815}{1.2CA}
\punkta{23/08/91}{00:27}{8491.519}{ -0.051}{0.025}{ 500}{1.64}{
3020}{1.2CA}
\punkta{22/09/92}{00:35}{8887.525}{ -0.150}{0.037}{ 100}{1.64}{
395}{1.2CA}
\punkta{22/09/92}{00:38}{8887.527}{ -0.161}{0.041}{ 100}{1.33}{
400}{1.2CA}

\end{picture}}

\end{picture}

\vspace*{-0.02cm}

\begin{picture}(17 ,4 )(-1,0)
\put(0,0){\setlength{\unitlength}{0.0085cm}%
\begin{picture}(2000, 470.588)(7250,0)
\put(7250,0){\framebox(2000, 470.588)[tl]{\begin{picture}(0,0)(0,0)
        \put(2000,0){\makebox(0,0)[tr]{\large{0215+015}\T{0.4}
                                 \hspace*{0.5cm}}}
        \put(2000,-
470.588){\setlength{\unitlength}{1cm}\begin{picture}(0,0)(0,0)
        \end{picture}}
    \end{picture}}}

\thicklines
\put(7250,0){\setlength{\unitlength}{2.5cm}\begin{picture}(0,0)(0,-0.8)
   \put(0,0){\setlength{\unitlength}{1cm}\begin{picture}(0,0)(0,0)
        \put(0,0){\line(1,0){0.3}}
        \end{picture}}
   \end{picture}}

\put(9250,0){\setlength{\unitlength}{2.5cm}\begin{picture}(0,0)(0,-0.8)
   \put(0,0){\setlength{\unitlength}{1cm}\begin{picture}(0,0)(0,0)
        \put(0,0){\line(-1,0){0.3}}
        \end{picture}}
   \end{picture}}

\thinlines
\put(7250,0){\setlength{\unitlength}{2.5cm}\begin{picture}(0,0)(0,-0.8)
   \multiput(0,0)(0,0.1){8}{\setlength{\unitlength}{1cm}%
\begin{picture}(0,0)(0,0)
        \put(0,0){\line(1,0){0.12}}
        \end{picture}}
   \end{picture}}

\put(7250,0){\setlength{\unitlength}{2.5cm}\begin{picture}(0,0)(0,-0.8)
   \multiput(0,0)(0,-0.1){8}{\setlength{\unitlength}{1cm}%
\begin{picture}(0,0)(0,0)
        \put(0,0){\line(1,0){0.12}}
        \end{picture}}
   \end{picture}}

\put(9250,0){\setlength{\unitlength}{2.5cm}\begin{picture}(0,0)(0,-0.8)
   \multiput(0,0)(0,0.1){8}{\setlength{\unitlength}{1cm}%
\begin{picture}(0,0)(0,0)
        \put(0,0){\line(-1,0){0.12}}
        \end{picture}}
   \end{picture}}

\put(9250,0){\setlength{\unitlength}{2.5cm}\begin{picture}(0,0)(0,-0.8)
   \multiput(0,0)(0,-0.1){8}{\setlength{\unitlength}{1cm}%
\begin{picture}(0,0)(0,0)
        \put(0,0){\line(-1,0){0.12}}
        \end{picture}}
   \end{picture}}

\put(7250,0){\setlength{\unitlength}{2.5cm}\begin{picture}(0,0)(0,-0.8)
   \put(0,0.2){\setlength{\unitlength}{1cm}\begin{picture}(0,0)(0,0)
        \put(0,0){\line(1,0){0.12}}
        \put(-0.2,0){\makebox(0,0)[r]{\bf 0.2}}
        \end{picture}}
   \put(0,0.0){\setlength{\unitlength}{1cm}\begin{picture}(0,0)(0,0)
        \put(0,0){\line(1,0){0.12}}
        \put(-0.2,0){\makebox(0,0)[r]{\bf 0.0}}
        \end{picture}}
   \put(0,-0.4){\setlength{\unitlength}{1cm}\begin{picture}(0,0)(0,0)
        \put(0,0){\line(1,0){0.12}}
        \put(-0.2,0){\makebox(0,0)[r]{\bf -0.4}}
        \end{picture}}
   \put(0,-0.2){\setlength{\unitlength}{1cm}\begin{picture}(0,0)(0,0)
        \put(0,0){\line(1,0){0.12}}
        \put(-0.2,0){\makebox(0,0)[r]{\bf -0.2}}
        \end{picture}}
   \put(0,-0.6){\setlength{\unitlength}{1cm}\begin{picture}(0,0)(0,0)
        \put(0,0){\line(1,0){0.12}}
        \put(-0.2,0){\makebox(0,0)[r]{\bf -0.6}}
        \end{picture}}
   \put(0,0.4){\setlength{\unitlength}{1cm}\begin{picture}(0,0)(0,0)
        \put(0,0){\line(1,0){0.12}}
        \put(-0.2,0){\makebox(0,0)[r]{\bf 0.4}}
        \end{picture}}
   \put(0,0.6){\setlength{\unitlength}{1cm}\begin{picture}(0,0)(0,0)
        \put(0,0){\line(1,0){0.12}}
        \put(-0.2,0){\makebox(0,0)[r]{\bf 0.6}}
        \end{picture}}
   \put(0,0.0){\setlength{\unitlength}{1cm}\begin{picture}(0,0)(0,0)
        \put(0,0){\line(1,0){0.12}}
        \put(-0.2,0){\makebox(0,0)[r]{\bf 0.0}}
        \end{picture}}
   \put(0,0.6){\setlength{\unitlength}{1cm}\begin{picture}(0,0)(0,0)
        \put(0,0){\line(1,0){0.12}}
        \end{picture}}
   \end{picture}}

   \put(7527.5, 470.588){\setlength{\unitlength}{1cm}\begin{picture}(0,0)(0,0)
        \put(0,0){\line(0,-1){0.2}}
   \end{picture}}
   \put(7892.5, 470.588){\setlength{\unitlength}{1cm}\begin{picture}(0,0)(0,0)
        \put(0,0){\line(0,-1){0.2}}
   \end{picture}}
   \put(8257.5, 470.588){\setlength{\unitlength}{1cm}\begin{picture}(0,0)(0,0)
        \put(0,0){\line(0,-1){0.2}}
  \end{picture}}
   \put(8622.5, 470.588){\setlength{\unitlength}{1cm}\begin{picture}(0,0)(0,0)
        \put(0,0){\line(0,-1){0.2}}
   \end{picture}}
   \put(8987.5, 470.588){\setlength{\unitlength}{1cm}\begin{picture}(0,0)(0,0)
        \put(0,0){\line(0,-1){0.2}}
   \end{picture}}

    \multiput(7250,0)(50,0){40}%
        {\setlength{\unitlength}{1cm}\begin{picture}(0,0)(0,0)
        \put(0,0){\line(0,1){0.12}}
    \end{picture}}
    \put(7500,0){\setlength{\unitlength}{1cm}\begin{picture}(0,0)(0,0)
        \put(0,0){\line(0,1){0.2}}
    \end{picture}}
    \put(8000,0){\setlength{\unitlength}{1cm}\begin{picture}(0,0)(0,0)
        \put(0,0){\line(0,1){0.2}}
    \end{picture}}
   \put(7750,0){\setlength{\unitlength}{1cm}\begin{picture}(0,0)(0,0)
        \put(0,0){\line(0,1){0.2}}
    \end{picture}}
   \put(8250,0){\setlength{\unitlength}{1cm}\begin{picture}(0,0)(0,0)
        \put(0,0){\line(0,1){0.2}}
    \end{picture}}
    \put(8500,0){\setlength{\unitlength}{1cm}\begin{picture}(0,0)(0,0)
        \put(0,0){\line(0,1){0.2}}
    \end{picture}}
    \put(8750,0){\setlength{\unitlength}{1cm}\begin{picture}(0,0)(0,0)
       \put(0,0){\line(0,1){0.2}}
    \end{picture}}
    \put(9000,0){\setlength{\unitlength}{1cm}\begin{picture}(0,0)(0,0)
       \put(0,0){\line(0,1){0.2}}
    \end{picture}}

\punktb{24/07/89}{03:30}{7731.646}{  0.140}{0.017}{ 500}{1.61}{
1620}{1.2CA}
\punktb{02/09/89}{03:27}{7771.644}{ -0.157}{0.018}{ 500}{1.79}{
881}{1.2CA}
\punktb{01/10/89}{12:52}{7801.036}{  0.074}{0.077}{ 500}{1.76}{
545}{1.2CA}
\punktb{01/11/89}{22:54}{7832.454}{ -0.204}{0.039}{ 500}{1.42}{
547}{1.2CA}
\punktbb{19/12/89}{23:33}{7880.481}{  0.066}{0.099}{ 100}{2.38}{
572}{1.2CA}
\punktb{19/12/89}{23:44}{7880.489}{  0.091}{0.035}{ 500}{2.53}{
640}{1.2CA}
\punktb{25/09/90}{03:16}{8159.637}{  0.030}{0.016}{ 500}{1.68}{
905}{1.2CA}
\punktb{26/09/90}{03:16}{8160.637}{ -0.013}{0.065}{ 500}{1.66}{
910}{1.2CA}
\punktb{18/10/90}{01:26}{8182.560}{  0.047}{0.049}{ 100}{2.06}{
547}{1.2CA}
\punktb{20/10/90}{01:07}{8184.547}{ -0.226}{0.038}{ 100}{1.17}{
549}{1.2CA}
\punktb{20/10/90}{01:17}{8184.554}{ -0.246}{0.033}{ 500}{1.51}{
847}{1.2CA}
\punktb{20/10/90}{01:28}{8184.562}{ -0.282}{0.028}{ 500}{1.12}{
862}{1.2CA}
\punktbb{15/02/91}{19:33}{8303.315}{  0.124}{0.057}{ 100}{1.80}{
361}{1.2CA}
\punktb{15/02/91}{19:41}{8303.321}{  0.138}{0.024}{ 305}{1.84}{
562}{1.2CA}
\punktb{22/02/91}{19:22}{8310.307}{  0.138}{0.059}{ 500}{1.91}{
3063}{1.2CA}
\punktbb{23/02/91}{18:56}{8311.289}{  0.170}{0.081}{ 100}{1.41}{
804}{1.2CA}
\punktb{23/02/91}{19:04}{8311.295}{  0.160}{0.042}{ 210}{1.39}{
1085}{1.2CA}
\punktb{24/02/91}{23:10}{8312.466}{  0.020}{0.006}{ 100}{1.52}{
974}{1.2CA}
\punktb{24/02/91}{23:30}{8312.480}{  0.059}{0.011}{ 500}{1.55}{
2278}{1.2CA}
\punktb{24/08/91}{02:57}{8492.623}{  0.532}{0.076}{ 100}{1.52}{
605}{1.2CA}
\punktbb{18/10/91}{00:59}{8547.541}{ -0.029}{0.086}{ 100}{1.42}{
330}{1.2CA}
\punktb{18/10/91}{01:09}{8547.548}{ -0.076}{0.044}{ 500}{1.41}{
566}{1.2CA}
\punktb{21/09/92}{02:58}{8886.624}{ -0.671}{0.017}{ 100}{0.92}{
472}{1.2CA}
\punktb{21/09/92}{03:01}{8886.626}{ -0.643}{0.020}{ 100}{0.97}{
472}{1.2CA}
\punktb{21/09/92}{03:01}{8886.626}{ -0.643}{0.020}{ 100}{0.97}{
472}{1.2CA}

\end{picture}}

\end{picture}

\vspace*{-0.02cm}

\begin{picture}(17 ,4 )(-1,0)
\put(0,0){\setlength{\unitlength}{0.0085cm}%
\begin{picture}(2000, 470.588)(7250,0)
\put(7250,0){\framebox(2000, 470.588)[tl]{\begin{picture}(0,0)(0,0)
        \put(2000,0){\makebox(0,0)[tr]{\large{0219+428}\T{0.4}
                                 \hspace*{0.5cm}}}
        \put(2000,-
470.588){\setlength{\unitlength}{1cm}\begin{picture}(0,0)(0,0)
        \end{picture}}
    \end{picture}}}

\thicklines
\put(7250,0){\setlength{\unitlength}{2.5cm}\begin{picture}(0,0)(0,-0.8)
   \put(0,0){\setlength{\unitlength}{1cm}\begin{picture}(0,0)(0,0)
        \put(0,0){\line(1,0){0.3}}
        \end{picture}}
   \end{picture}}

\put(9250,0){\setlength{\unitlength}{2.5cm}\begin{picture}(0,0)(0,-0.8)
   \put(0,0){\setlength{\unitlength}{1cm}\begin{picture}(0,0)(0,0)
        \put(0,0){\line(-1,0){0.3}}
        \end{picture}}
   \end{picture}}

\thinlines
\put(7250,0){\setlength{\unitlength}{2.5cm}\begin{picture}(0,0)(0,-0.8)
   \multiput(0,0)(0,0.1){8}{\setlength{\unitlength}{1cm}%
\begin{picture}(0,0)(0,0)
        \put(0,0){\line(1,0){0.12}}
        \end{picture}}
   \end{picture}}

\put(7250,0){\setlength{\unitlength}{2.5cm}\begin{picture}(0,0)(0,-0.8)
   \multiput(0,0)(0,-0.1){8}{\setlength{\unitlength}{1cm}%
\begin{picture}(0,0)(0,0)
        \put(0,0){\line(1,0){0.12}}
        \end{picture}}
   \end{picture}}

\put(9250,0){\setlength{\unitlength}{2.5cm}\begin{picture}(0,0)(0,-0.8)
   \multiput(0,0)(0,0.1){8}{\setlength{\unitlength}{1cm}%
\begin{picture}(0,0)(0,0)
        \put(0,0){\line(-1,0){0.12}}
        \end{picture}}
   \end{picture}}

\put(9250,0){\setlength{\unitlength}{2.5cm}\begin{picture}(0,0)(0,-0.8)
   \multiput(0,0)(0,-0.1){8}{\setlength{\unitlength}{1cm}%
\begin{picture}(0,0)(0,0)
        \put(0,0){\line(-1,0){0.12}}
        \end{picture}}
   \end{picture}}

\put(7250,0){\setlength{\unitlength}{2.5cm}\begin{picture}(0,0)(0,-0.8)
   \put(0,0.2){\setlength{\unitlength}{1cm}\begin{picture}(0,0)(0,0)
        \put(0,0){\line(1,0){0.12}}
        \put(-0.2,0){\makebox(0,0)[r]{\bf 0.2}}
        \end{picture}}
   \put(0,0.0){\setlength{\unitlength}{1cm}\begin{picture}(0,0)(0,0)
        \put(0,0){\line(1,0){0.12}}
        \put(-0.2,0){\makebox(0,0)[r]{\bf 0.0}}
        \end{picture}}
   \put(0,-0.4){\setlength{\unitlength}{1cm}\begin{picture}(0,0)(0,0)
        \put(0,0){\line(1,0){0.12}}
        \put(-0.2,0){\makebox(0,0)[r]{\bf -0.4}}
        \end{picture}}
   \put(0,-0.2){\setlength{\unitlength}{1cm}\begin{picture}(0,0)(0,0)
        \put(0,0){\line(1,0){0.12}}
        \put(-0.2,0){\makebox(0,0)[r]{\bf -0.2}}
        \end{picture}}
   \put(0,-0.6){\setlength{\unitlength}{1cm}\begin{picture}(0,0)(0,0)
        \put(0,0){\line(1,0){0.12}}
        \put(-0.2,0){\makebox(0,0)[r]{\bf -0.6}}
        \end{picture}}
   \put(0,0.4){\setlength{\unitlength}{1cm}\begin{picture}(0,0)(0,0)
        \put(0,0){\line(1,0){0.12}}
        \put(-0.2,0){\makebox(0,0)[r]{\bf 0.4}}
        \end{picture}}
   \put(0,0.0){\setlength{\unitlength}{1cm}\begin{picture}(0,0)(0,0)
        \put(0,0){\line(1,0){0.12}}
        \put(-0.2,0){\makebox(0,0)[r]{\bf 0.0}}
        \end{picture}}
   \put(0,0.6){\setlength{\unitlength}{1cm}\begin{picture}(0,0)(0,0)
        \put(0,0){\line(1,0){0.12}}
        \put(-0.2,0){\makebox(0,0)[r]{\bf 0.6}}
        \end{picture}}
   \put(0,0.6){\setlength{\unitlength}{1cm}\begin{picture}(0,0)(0,0)
        \put(0,0){\line(1,0){0.12}}
        \end{picture}}
   \end{picture}}

   \put(7527.5, 470.588){\setlength{\unitlength}{1cm}\begin{picture}(0,0)(0,0)
        \put(0,0){\line(0,-1){0.2}}
   \end{picture}}
   \put(7892.5, 470.588){\setlength{\unitlength}{1cm}\begin{picture}(0,0)(0,0)
        \put(0,0){\line(0,-1){0.2}}
   \end{picture}}
   \put(8257.5, 470.588){\setlength{\unitlength}{1cm}\begin{picture}(0,0)(0,0)
        \put(0,0){\line(0,-1){0.2}}
  \end{picture}}
   \put(8622.5, 470.588){\setlength{\unitlength}{1cm}\begin{picture}(0,0)(0,0)
        \put(0,0){\line(0,-1){0.2}}
   \end{picture}}
   \put(8987.5, 470.588){\setlength{\unitlength}{1cm}\begin{picture}(0,0)(0,0)
        \put(0,0){\line(0,-1){0.2}}
   \end{picture}}

    \multiput(7250,0)(50,0){40}%
        {\setlength{\unitlength}{1cm}\begin{picture}(0,0)(0,0)
        \put(0,0){\line(0,1){0.12}}
    \end{picture}}
    \put(7500,0){\setlength{\unitlength}{1cm}\begin{picture}(0,0)(0,0)
        \put(0,0){\line(0,1){0.2}}
    \end{picture}}
    \put(8000,0){\setlength{\unitlength}{1cm}\begin{picture}(0,0)(0,0)
        \put(0,0){\line(0,1){0.2}}
    \end{picture}}
   \put(7750,0){\setlength{\unitlength}{1cm}\begin{picture}(0,0)(0,0)
        \put(0,0){\line(0,1){0.2}}
    \end{picture}}
   \put(8250,0){\setlength{\unitlength}{1cm}\begin{picture}(0,0)(0,0)
        \put(0,0){\line(0,1){0.2}}
    \end{picture}}
    \put(8500,0){\setlength{\unitlength}{1cm}\begin{picture}(0,0)(0,0)
        \put(0,0){\line(0,1){0.2}}
    \end{picture}}
    \put(8750,0){\setlength{\unitlength}{1cm}\begin{picture}(0,0)(0,0)
       \put(0,0){\line(0,1){0.2}}
    \end{picture}}
    \put(9000,0){\setlength{\unitlength}{1cm}\begin{picture}(0,0)(0,0)
       \put(0,0){\line(0,1){0.2}}
    \end{picture}}

\punktb{08/10/88}{23:31}{7443.480}{ -0.333}{0.041}{ 600}{1.54}{
730}{1.2CA}
\punktb{10/10/88}{00:26}{7444.518}{ -0.343}{0.025}{ 500}{2.42}{
941}{1.2CA}
\punktb{17/08/89}{00:35}{7755.525}{  0.227}{0.013}{ 500}{1.67}{
2564}{1.2CA}
\punktb{17/08/89}{00:48}{7755.534}{  0.229}{0.017}{ 500}{1.73}{
2450}{1.2CA}
\punktb{17/08/89}{01:00}{7755.542}{  0.230}{0.023}{ 500}{1.63}{
2186}{1.2CA}
\punktb{17/08/89}{01:11}{7755.550}{  0.231}{0.013}{ 500}{1.90}{
1960}{1.2CA}
\punktb{17/08/89}{01:21}{7755.557}{  0.236}{0.010}{ 500}{1.83}{
1652}{1.2CA}
\punktb{17/08/89}{01:31}{7755.564}{  0.233}{0.010}{ 500}{1.79}{
1352}{1.2CA}
\punktb{17/08/89}{01:41}{7755.571}{  0.227}{0.009}{ 500}{1.56}{
1117}{1.2CA}
\punktb{17/08/89}{01:51}{7755.578}{  0.236}{0.010}{ 500}{1.73}{
949}{1.2CA}
\punktb{01/09/89}{15:38}{7771.152}{  0.140}{0.023}{****}{1.77}{
729}{1.2CA}
\punktb{03/11/89}{00:22}{7833.515}{  0.027}{0.024}{ 500}{2.00}{
626}{1.2CA}
\punktb{13/12/89}{20:55}{7874.372}{  0.244}{0.026}{ 500}{2.05}{
3973}{1.2CA}
\punktbb{14/12/89}{00:42}{7874.530}{  0.245}{0.044}{ 100}{1.56}{
1105}{1.2CA}
\punktb{14/12/89}{20:21}{7875.348}{  0.239}{0.025}{ 150}{1.89}{
1072}{1.2CA}
\punktb{14/12/89}{23:50}{7875.493}{  0.282}{0.007}{ 500}{2.51}{
1453}{1.2CA}
\punktbb{15/12/89}{21:30}{7876.396}{  0.256}{0.035}{ 100}{2.50}{
569}{1.2CA}
\punktb{15/12/89}{21:38}{7876.401}{  0.272}{0.009}{ 500}{2.35}{
831}{1.2CA}
\punktb{19/12/89}{20:43}{7880.364}{  0.267}{0.027}{ 100}{1.13}{
731}{1.2CA}
\punktb{19/12/89}{20:52}{7880.370}{  0.271}{0.010}{ 500}{1.37}{
607}{1.2CA}
\punktb{20/12/89}{20:05}{7881.337}{  0.341}{0.022}{ 100}{1.80}{
479}{1.2CA}
\punktb{20/12/89}{20:13}{7881.342}{  0.353}{0.007}{ 500}{1.68}{
594}{1.2CA}
\punktbb{22/12/89}{20:12}{7883.342}{  0.561}{0.036}{ 106}{2.92}{
530}{1.2CA}
\punktb{22/12/89}{20:22}{7883.349}{  0.570}{0.015}{ 500}{2.81}{
599}{1.2CA}
\punktb{27/01/90}{19:43}{7919.322}{  0.260}{0.039}{ 532}{2.92}{
640}{1.2CA}
\punktb{28/02/90}{18:55}{7951.289}{  0.008}{0.022}{  50}{1.22}{
642}{1.2CA}
\punktb{28/02/90}{19:00}{7951.292}{ -0.008}{0.019}{  50}{1.31}{
620}{1.2CA}
\punktb{28/02/90}{19:06}{7951.296}{  0.002}{0.019}{ 500}{1.58}{
839}{1.2CA}
\punktb{25/09/90}{22:40}{8160.445}{ -0.085}{0.010}{ 100}{1.72}{
558}{1.2CA}
\punktb{25/09/90}{22:44}{8160.448}{ -0.081}{0.010}{ 180}{1.62}{
607}{1.2CA}
\punktb{28/09/90}{02:27}{8162.602}{ -0.167}{0.013}{ 100}{1.43}{
550}{1.2CA}
\punktb{30/09/90}{00:41}{8164.529}{ -0.141}{0.019}{ 100}{2.55}{
676}{1.2CA}
\punktb{30/09/90}{00:49}{8164.535}{ -0.162}{0.011}{ 500}{2.45}{
1212}{1.2CA}
\punktb{03/10/90}{01:35}{8167.566}{ -0.233}{0.027}{ 100}{2.58}{
928}{1.2CA}
\punktb{03/10/90}{01:44}{8167.573}{ -0.227}{0.013}{ 500}{2.49}{
2750}{1.2CA}
\punktb{20/10/90}{02:13}{8184.593}{ -0.130}{0.015}{ 100}{1.18}{
546}{1.2CA}
\punktb{20/10/90}{02:22}{8184.599}{ -0.149}{0.014}{ 500}{1.11}{
796}{1.2CA}
\punktb{21/10/90}{01:38}{8185.568}{ -0.112}{0.010}{ 100}{2.18}{
544}{1.2CA}
\punktb{21/10/90}{02:08}{8185.589}{ -0.109}{0.020}{ 500}{2.89}{
787}{1.2CA}
\punktb{15/02/91}{19:52}{8303.328}{ -0.381}{0.010}{ 116}{1.49}{
386}{1.2CA}
\punktb{15/02/91}{19:58}{8303.332}{ -0.357}{0.017}{ 300}{1.54}{
515}{1.2CA}
\punktb{23/02/91}{20:39}{8311.361}{ -0.309}{0.022}{ 100}{1.38}{
833}{1.2CA}
\punktb{23/02/91}{20:47}{8311.366}{ -0.321}{0.015}{ 300}{1.56}{
1962}{1.2CA}
\punktbb{24/02/91}{23:50}{8312.494}{ -0.261}{0.031}{ 100}{2.16}{
820}{1.2CA}
\punktb{24/02/91}{23:56}{8312.498}{ -0.288}{0.026}{ 300}{2.00}{
1746}{1.2CA}
\punktb{27/02/91}{19:14}{8315.302}{ -0.258}{0.045}{ 100}{2.10}{
820}{1.2CA}
\punktb{27/02/91}{19:14}{8315.302}{ -0.350}{0.026}{ 500}{4.68}{
2342}{1.2CA}
\punktbb{25/07/91}{03:13}{8462.634}{ -0.164}{0.034}{ 100}{1.52}{
497}{1.2CA}
\punktb{25/07/91}{03:19}{8462.638}{ -0.181}{0.013}{ 306}{1.39}{
667}{1.2CA}
\punktb{29/07/91}{03:44}{8466.656}{ -0.165}{0.019}{ 150}{2.32}{
534}{1.2CA}
\punktb{29/07/91}{03:53}{8466.662}{ -0.171}{0.013}{ 500}{2.37}{
1193}{1.2CA}
\punktb{30/07/91}{02:22}{8467.599}{ -0.169}{0.023}{ 100}{3.34}{
450}{1.2CA}
\punktb{30/07/91}{02:30}{8467.604}{ -0.160}{0.020}{ 500}{3.77}{
1205}{1.2CA}
\punktbb{31/07/91}{02:05}{8468.587}{ -0.110}{0.043}{ 100}{1.87}{
615}{1.2CA}
\punktb{31/07/91}{02:18}{8468.596}{ -0.133}{0.032}{ 500}{2.20}{
2525}{1.2CA}
\punktb{01/08/91}{03:28}{8469.645}{ -0.106}{0.015}{ 100}{1.31}{
430}{1.2CA}
\punktb{01/08/91}{03:34}{8469.649}{ -0.114}{0.021}{ 300}{1.17}{
780}{1.2CA}
\punktb{02/08/91}{03:01}{8470.626}{ -0.079}{0.015}{ 100}{1.04}{
433}{1.2CA}
\punktb{02/08/91}{03:04}{8470.628}{ -0.087}{0.015}{ 100}{0.95}{
432}{1.2CA}
\punktb{06/08/91}{02:57}{8474.623}{ -0.213}{0.030}{ 100}{1.16}{
350}{1.2CA}
\punktb{06/08/91}{03:05}{8474.629}{ -0.219}{0.040}{ 500}{0.99}{
652}{1.2CA}
\punktb{07/08/91}{03:56}{8475.664}{ -0.228}{0.012}{ 100}{1.07}{
309}{1.2CA}
\punktb{07/08/91}{04:04}{8475.670}{ -0.219}{0.020}{ 500}{1.13}{
598}{1.2CA}
\punktb{08/08/91}{03:11}{8476.633}{ -0.227}{0.017}{ 100}{1.39}{
345}{1.2CA}
\punktb{08/08/91}{03:19}{8476.638}{ -0.238}{0.010}{ 500}{1.41}{
715}{1.2CA}
\punktb{24/08/91}{03:16}{8492.637}{  0.027}{0.017}{ 100}{1.12}{
469}{1.2CA}
\punktb{24/08/91}{03:27}{8492.644}{  0.019}{0.015}{ 500}{1.33}{
1242}{1.2CA}
\punktbb{25/08/91}{00:12}{8493.509}{  0.020}{0.040}{ 100}{2.13}{
881}{1.2CA}
\punktb{25/08/91}{00:20}{8493.514}{  0.022}{0.027}{ 500}{2.05}{
3273}{1.2CA}
\punktb{27/08/91}{03:33}{8495.649}{  0.201}{0.025}{ 100}{1.98}{
669}{1.2CA}
\punktb{27/08/91}{03:42}{8495.654}{  0.205}{0.025}{ 500}{2.23}{
2627}{1.2CA}
\punktb{23/09/91}{03:25}{8522.642}{  0.108}{0.028}{ 500}{2.19}{
2369}{1.2CA}
\punktb{19/10/91}{03:08}{8548.631}{ -0.034}{0.008}{ 100}{1.81}{
313}{1.2CA}
\punktb{19/10/91}{03:18}{8548.638}{ -0.042}{0.018}{ 500}{1.78}{
556}{1.2CA}
\punktb{21/10/91}{00:17}{8550.512}{  0.069}{0.013}{ 100}{0.97}{
566}{1.2CA}
\punktb{21/10/91}{00:24}{8550.517}{  0.079}{0.016}{ 300}{1.03}{
1199}{1.2CA}
\punktb{29/10/91}{02:04}{8558.586}{ -0.064}{0.019}{ 100}{0.95}{
368}{1.2CA}
\punktb{29/10/91}{02:12}{8558.592}{ -0.090}{0.018}{ 500}{1.00}{
812}{1.2CA}
\punktb{22/09/92}{01:53}{8887.579}{ -0.328}{0.025}{ 150}{2.15}{
518}{1.2CA}
\punktb{22/09/92}{01:57}{8887.582}{ -0.337}{0.026}{ 208}{2.20}{
632}{1.2CA}
\punktb{25/09/92}{05:03}{8890.711}{ -0.323}{0.019}{ 100}{0.92}{
649}{1.2CA}
\punktb{25/09/92}{05:03}{8890.711}{ -0.323}{0.019}{ 100}{0.92}{
649}{1.2CA}

\end{picture}}

\end{picture}

\vspace*{-0.02cm}

\begin{picture}(17 ,10 )(-1,0)
\put(0,0){\setlength{\unitlength}{0.0085cm}%
\begin{picture}(2000,1176.470)(7250,0)
\put(7250,0){\framebox(2000,1176.470)[tl]{\begin{picture}(0,0)(0,0)
        \put(2000,0){\makebox(0,0)[tr]{\large{0235+164}\T{0.4}
                                 \hspace*{0.5cm}}}

\put(2000,-1176.470){\setlength{\unitlength}{1cm}\begin{picture}(0,0)(0,0)
            \put(0,-1){\makebox(0,0)[br]{\bf J.D.\,2,440,000\,+}}
        \end{picture}}
    \end{picture}}}

\thicklines
\put(7250,0){\setlength{\unitlength}{2.5cm}\begin{picture}(0,0)(0,-2)
   \put(0,0){\setlength{\unitlength}{1cm}\begin{picture}(0,0)(0,0)
        \put(0,0){\line(1,0){0.3}}
        \end{picture}}
   \end{picture}}

\put(9250,0){\setlength{\unitlength}{2.5cm}\begin{picture}(0,0)(0,-2)
   \put(0,0){\setlength{\unitlength}{1cm}\begin{picture}(0,0)(0,0)
        \put(0,0){\line(-1,0){0.3}}
        \end{picture}}
   \end{picture}}

\thinlines
\put(7250,0){\setlength{\unitlength}{2.5cm}\begin{picture}(0,0)(0,-2)
   \multiput(0,0)(0,0.1){20}{\setlength{\unitlength}{1cm}%
\begin{picture}(0,0)(0,0)
        \put(0,0){\line(1,0){0.12}}
        \end{picture}}
   \end{picture}}

\put(7250,0){\setlength{\unitlength}{2.5cm}\begin{picture}(0,0)(0,-2)
   \multiput(0,0)(0,-0.1){20}{\setlength{\unitlength}{1cm}%
\begin{picture}(0,0)(0,0)
        \put(0,0){\line(1,0){0.12}}
        \end{picture}}
   \end{picture}}

\put(9250,0){\setlength{\unitlength}{2.5cm}\begin{picture}(0,0)(0,-2)
   \multiput(0,0)(0,0.1){20}{\setlength{\unitlength}{1cm}%
\begin{picture}(0,0)(0,0)
        \put(0,0){\line(-1,0){0.12}}
        \end{picture}}
   \end{picture}}

\put(9250,0){\setlength{\unitlength}{2.5cm}\begin{picture}(0,0)(0,-2)
   \multiput(0,0)(0,-0.1){20}{\setlength{\unitlength}{1cm}%
\begin{picture}(0,0)(0,0)
        \put(0,0){\line(-1,0){0.12}}
        \end{picture}}
   \end{picture}}

\put(7250,0){\setlength{\unitlength}{2.5cm}\begin{picture}(0,0)(0,-2)
   \put(0,0.5){\setlength{\unitlength}{1cm}\begin{picture}(0,0)(0,0)
        \put(0,0){\line(1,0){0.12}}
        \put(-0.2,0){\makebox(0,0)[r]{\bf 0.5}}
        \end{picture}}
   \put(0,1.0){\setlength{\unitlength}{1cm}\begin{picture}(0,0)(0,0)
        \put(0,0){\line(1,0){0.12}}
        \put(-0.2,0){\makebox(0,0)[r]{\bf 1.0}}
        \end{picture}}
   \put(0,-0.5){\setlength{\unitlength}{1cm}\begin{picture}(0,0)(0,0)
        \put(0,0){\line(1,0){0.12}}
        \put(-0.2,0){\makebox(0,0)[r]{\bf -0.5}}
        \end{picture}}
   \put(0,-1.0){\setlength{\unitlength}{1cm}\begin{picture}(0,0)(0,0)
        \put(0,0){\line(1,0){0.12}}
        \put(-0.2,0){\makebox(0,0)[r]{\bf -1.0}}
        \end{picture}}
   \put(0,1.5){\setlength{\unitlength}{1cm}\begin{picture}(0,0)(0,0)
        \put(0,0){\line(1,0){0.12}}
        \put(-0.2,0){\makebox(0,0)[r]{\bf 1.5}}
        \end{picture}}
   \put(0,-1.5){\setlength{\unitlength}{1cm}\begin{picture}(0,0)(0,0)
        \put(0,0){\line(1,0){0.12}}
        \put(-0.2,0){\makebox(0,0)[r]{\bf -1.5}}
        \end{picture}}
   \put(0,0.0){\setlength{\unitlength}{1cm}\begin{picture}(0,0)(0,0)
        \put(0,0){\line(1,0){0.12}}
        \put(-0.2,0){\makebox(0,0)[r]{\bf 0.0}}
        \end{picture}}
   \put(0,0.0){\setlength{\unitlength}{1cm}\begin{picture}(0,0)(0,0)
        \put(0,0){\line(1,0){0.12}}
        \put(-0.2,0){\makebox(0,0)[r]{\bf 0.0}}
        \end{picture}}
   \put(0,0.0){\setlength{\unitlength}{1cm}\begin{picture}(0,0)(0,0)
        \put(0,0){\line(1,0){0.12}}
        \end{picture}}
   \end{picture}}

   \put(7527.5,1176.470){\setlength{\unitlength}{1cm}\begin{picture}(0,0)(0,0)
        \put(0,0){\line(0,-1){0.2}}
   \end{picture}}
   \put(7892.5,1176.470){\setlength{\unitlength}{1cm}\begin{picture}(0,0)(0,0)
        \put(0,0){\line(0,-1){0.2}}
   \end{picture}}
   \put(8257.5,1176.470){\setlength{\unitlength}{1cm}\begin{picture}(0,0)(0,0)
        \put(0,0){\line(0,-1){0.2}}
  \end{picture}}
   \put(8622.5,1176.470){\setlength{\unitlength}{1cm}\begin{picture}(0,0)(0,0)
        \put(0,0){\line(0,-1){0.2}}
   \end{picture}}
   \put(8987.5,1176.470){\setlength{\unitlength}{1cm}\begin{picture}(0,0)(0,0)
        \put(0,0){\line(0,-1){0.2}}
   \end{picture}}

    \multiput(7250,0)(50,0){40}%
        {\setlength{\unitlength}{1cm}\begin{picture}(0,0)(0,0)
        \put(0,0){\line(0,1){0.12}}
    \end{picture}}
    \put(7500,0){\setlength{\unitlength}{1cm}\begin{picture}(0,0)(0,0)
        \put(0,0){\line(0,1){0.2}}
        \put(0,-0.2){\makebox(0,0)[t]{\bf 7500}}
    \end{picture}}
    \put(8000,0){\setlength{\unitlength}{1cm}\begin{picture}(0,0)(0,0)
        \put(0,0){\line(0,1){0.2}}
        \put(0,-0.2){\makebox(0,0)[t]{\bf 8000}}
    \end{picture}}
   \put(7750,0){\setlength{\unitlength}{1cm}\begin{picture}(0,0)(0,0)
        \put(0,0){\line(0,1){0.2}}
        \put(0,-0.2){\makebox(0,0)[t]{\bf 7750}}
    \end{picture}}
   \put(8250,0){\setlength{\unitlength}{1cm}\begin{picture}(0,0)(0,0)
        \put(0,0){\line(0,1){0.2}}
        \put(0,-0.2){\makebox(0,0)[t]{\bf 8250}}
    \end{picture}}
    \put(8500,0){\setlength{\unitlength}{1cm}\begin{picture}(0,0)(0,0)
        \put(0,0){\line(0,1){0.2}}
        \put(0,-0.2){\makebox(0,0)[t]{\bf 8500}}
    \end{picture}}
    \put(8750,0){\setlength{\unitlength}{1cm}\begin{picture}(0,0)(0,0)
       \put(0,0){\line(0,1){0.2}}
        \put(0,-0.2){\makebox(0,0)[t]{\bf 8750}}
    \end{picture}}
    \put(9000,0){\setlength{\unitlength}{1cm}\begin{picture}(0,0)(0,0)
       \put(0,0){\line(0,1){0.2}}
        \put(0,-0.2){\makebox(0,0)[t]{\bf 9000}}
    \end{picture}}

\punktc{04/09/88}{04:09}{7408.673}{ -1.876}{0.036}{1000}{1.62}{
1236}{1.2CA}
\punktc{05/09/88}{03:21}{7409.640}{ -1.623}{0.055}{ 500}{1.97}{
689}{1.2CA}
\punktc{05/10/88}{02:49}{7439.618}{ -0.975}{0.020}{ 500}{1.20}{
750}{1.2CA}
\punktc{08/10/88}{01:46}{7442.574}{ -1.087}{0.049}{ 500}{1.20}{
707}{1.2CA}
\punktc{24/07/89}{03:43}{7731.655}{ -0.555}{0.021}{ 500}{1.56}{
1135}{1.2CA}
\punktc{16/08/89}{03:17}{7754.637}{ -1.660}{0.014}{1000}{1.48}{
1595}{1.2CA}
\punktc{01/09/89}{03:27}{7770.644}{ -1.755}{0.026}{ 500}{2.09}{
1157}{1.2CA}
\punktc{02/09/89}{03:13}{7771.634}{ -1.636}{0.025}{ 500}{1.50}{
851}{1.2CA}
\punktc{01/10/89}{13:04}{7801.045}{ -1.168}{0.033}{ 500}{1.58}{
578}{1.2CA}
\punktc{31/10/89}{22:56}{7831.456}{ -1.569}{0.096}{ 500}{1.41}{
553}{1.2CA}
\punktc{01/11/89}{23:07}{7832.464}{ -1.430}{0.087}{ 500}{1.34}{
532}{1.2CA}
\punktc{03/11/89}{00:06}{7833.504}{ -1.233}{0.052}{ 500}{2.02}{
580}{1.2CA}
\punktc{13/12/89}{19:54}{7874.330}{ -0.195}{0.035}{ 500}{2.20}{
2396}{1.2CA}
\punktc{13/12/89}{23:33}{7874.481}{  0.041}{0.047}{ 500}{2.02}{
4169}{1.2CA}
\punktc{14/12/89}{01:02}{7874.543}{ -0.095}{0.036}{ 100}{1.89}{
1069}{1.2CA}
\punktc{14/12/89}{18:35}{7875.274}{  0.413}{0.012}{ 500}{3.09}{
660}{1.2CA}
\punktc{15/12/89}{00:52}{7875.536}{  0.356}{0.026}{ 500}{3.37}{
1648}{1.2CA}
\punktc{15/12/89}{19:37}{7876.318}{ -0.324}{0.020}{ 100}{1.94}{
502}{1.2CA}
\punktc{15/12/89}{19:44}{7876.323}{ -0.337}{0.047}{ 500}{2.08}{
606}{1.2CA}
\punktc{15/12/89}{20:23}{7876.350}{ -0.392}{0.017}{ 500}{2.55}{
674}{1.2CA}
\punktc{15/12/89}{20:51}{7876.369}{ -0.469}{0.013}{ 500}{2.35}{
964}{1.2CA}
\punktc{15/12/89}{20:51}{7876.369}{ -0.475}{0.005}{ 500}{2.52}{
807}{1.2CA}
\punktc{19/12/89}{19:03}{7880.294}{ -0.611}{0.019}{ 500}{1.55}{
629}{1.2CA}
\punktc{19/12/89}{21:04}{7880.378}{ -0.606}{0.022}{ 250}{1.74}{
534}{1.2CA}
\punktc{20/12/89}{18:55}{7881.289}{ -0.854}{0.026}{ 500}{2.05}{
630}{1.2CA}
\punktc{20/12/89}{20:24}{7881.350}{ -1.097}{0.014}{ 500}{1.87}{
629}{1.2CA}
\punktc{21/12/89}{17:58}{7882.249}{ -0.299}{0.030}{ 500}{3.97}{
646}{1.2CA}
\punktc{22/12/89}{19:45}{7883.323}{  0.500}{0.034}{ 500}{3.11}{
611}{1.2CA}
\punktc{22/12/89}{21:15}{7883.386}{  0.376}{0.012}{ 500}{2.75}{
616}{1.2CA}
\punktc{27/01/90}{18:43}{7919.280}{ -0.084}{0.034}{ 500}{2.72}{
655}{1.2CA}
\punktc{29/01/90}{19:54}{7921.330}{  0.480}{0.009}{ 500}{2.68}{
629}{1.2CA}
\punktc{08/02/90}{19:29}{7931.312}{  0.487}{0.010}{ 500}{2.06}{
1393}{1.2CA}
\punktc{26/02/90}{20:43}{7949.363}{  0.129}{0.012}{ 500}{2.51}{
925}{1.2CA}
\punktc{27/02/90}{19:12}{7950.300}{  0.116}{0.012}{ 500}{1.67}{
852}{1.2CA}
\punktc{26/09/90}{02:35}{8160.608}{  0.614}{0.012}{ 100}{1.98}{
601}{1.2CA}
\punktc{26/09/90}{02:41}{8160.612}{  0.625}{0.013}{ 500}{1.95}{
877}{1.2CA}
\punktc{26/09/90}{23:01}{8161.459}{  0.160}{0.012}{ 100}{1.79}{
585}{1.2CA}
\punktc{26/09/90}{23:07}{8161.464}{  0.163}{0.009}{ 500}{1.62}{
840}{1.2CA}
\punktc{27/09/90}{23:38}{8162.485}{  0.344}{0.014}{ 100}{1.80}{
637}{1.2CA}
\punktc{27/09/90}{23:49}{8162.493}{  0.344}{0.026}{ 500}{1.94}{
1118}{1.2CA}
\punktc{27/09/90}{23:56}{8162.498}{  0.334}{0.012}{ 250}{1.50}{
764}{1.2CA}
\punktc{28/09/90}{03:26}{8162.644}{  0.430}{0.007}{ 300}{1.60}{
720}{1.2CA}
\punktc{29/09/90}{23:21}{8164.473}{  0.235}{0.026}{  50}{2.37}{
634}{1.2CA}
\punktc{29/09/90}{23:24}{8164.475}{  0.218}{0.023}{ 100}{2.39}{
753}{1.2CA}
\punktc{29/09/90}{23:31}{8164.480}{  0.231}{0.015}{ 500}{2.60}{
1629}{1.2CA}
\punktc{30/09/90}{02:09}{8164.590}{  0.203}{0.011}{ 330}{1.51}{
1139}{1.2CA}
\punktc{30/09/90}{05:04}{8164.712}{  0.217}{0.043}{ 150}{2.91}{
685}{1.2CA}
\punktc{30/09/90}{23:21}{8165.473}{  0.215}{0.026}{  50}{2.35}{
638}{1.2CA}
\punktc{04/10/90}{00:31}{8168.522}{ -0.100}{0.057}{ 500}{3.08}{
2877}{1.2CA}
\punktc{17/10/90}{01:42}{8181.571}{  0.811}{0.005}{ 500}{2.03}{
814}{1.2CA}
\punktc{17/10/90}{02:40}{8181.611}{  0.838}{0.019}{ 500}{1.95}{
802}{1.2CA}
\punktc{18/10/90}{01:33}{8182.565}{  0.793}{0.011}{ 100}{1.98}{
553}{1.2CA}
\punktc{18/10/90}{01:42}{8182.571}{  0.809}{0.019}{ 500}{2.02}{
809}{1.2CA}
\punktc{18/10/90}{02:14}{8182.593}{  0.812}{0.014}{ 500}{4.37}{
970}{1.2CA}
\punktc{18/10/90}{02:31}{8182.605}{  0.835}{0.008}{ 100}{1.76}{
585}{1.2CA}
\punktc{18/10/90}{02:40}{8182.611}{  0.839}{0.019}{ 500}{1.94}{
786}{1.2CA}
\punktc{19/10/90}{02:01}{8183.585}{  0.813}{0.024}{ 100}{4.43}{
575}{1.2CA}
\punktc{19/10/90}{02:04}{8183.586}{  0.858}{0.019}{ 500}{1.26}{
869}{1.2CA}
\punktc{19/10/90}{02:14}{8183.593}{  0.811}{0.013}{ 500}{4.35}{
911}{1.2CA}
\punktc{20/10/90}{01:55}{8184.580}{  0.857}{0.019}{ 100}{1.37}{
563}{1.2CA}
\punktc{20/10/90}{02:04}{8184.586}{  0.858}{0.008}{ 500}{1.24}{
864}{1.2CA}
\punktc{20/10/90}{02:30}{8184.604}{  0.905}{0.011}{ 500}{3.52}{
1044}{1.2CA}
\punktc{21/10/90}{02:20}{8185.598}{  0.910}{0.021}{ 100}{3.99}{
578}{1.2CA}
\punktc{21/10/90}{02:30}{8185.604}{  0.904}{0.011}{ 500}{3.50}{
950}{1.2CA}
\punktc{22/12/90}{09:00}{8247.875}{  0.658}{0.057}{ 100}{3.38}{
492}{1.2CA}
\punktc{31/01/91}{20:07}{8288.339}{ -0.355}{0.035}{ 500}{3.70}{
1563}{1.2CA}
\punktcc{01/02/91}{20:18}{8289.346}{ -0.238}{0.063}{1801}{3.19}{
2798}{1.2CA}
\punktcc{02/02/91}{19:04}{8290.295}{ -0.239}{0.074}{ 100}{4.81}{
345}{1.2CA}
\punktc{02/02/91}{19:12}{8290.300}{ -0.125}{0.026}{ 500}{4.58}{
791}{1.2CA}
\punktcc{02/02/91}{19:41}{8290.321}{ -0.164}{0.040}{ 100}{3.90}{
378}{1.2CA}
\punktc{02/02/91}{19:46}{8290.324}{ -0.126}{0.027}{ 270}{4.07}{
545}{1.2CA}
\punktc{02/02/91}{19:52}{8290.328}{ -0.134}{0.024}{ 300}{3.79}{
584}{1.2CA}
\punktc{02/02/91}{21:37}{8290.401}{ -0.112}{0.038}{ 300}{3.05}{
679}{1.2CA}
\punktc{04/02/91}{18:05}{8292.254}{  0.115}{0.049}{1800}{3.69}{
1797}{1.2CA}
\punktcc{07/02/91}{18:34}{8295.274}{  0.271}{0.056}{ 100}{4.02}{
647}{1.2CA}
\punktc{07/02/91}{18:47}{8295.283}{  0.267}{0.023}{ 300}{4.02}{
730}{1.2CA}
\punktc{07/02/91}{20:03}{8295.336}{  0.264}{0.034}{ 150}{3.90}{
593}{1.2CA}
\punktc{07/02/91}{20:11}{8295.341}{  0.236}{0.017}{ 500}{3.92}{
806}{1.2CA}
\punktc{09/02/91}{19:18}{8297.304}{ -0.002}{0.027}{ 250}{1.65}{
475}{1.2CA}
\punktc{09/02/91}{20:10}{8297.341}{ -0.008}{0.017}{ 150}{1.95}{
408}{1.2CA}
\punktc{09/02/91}{20:15}{8297.344}{ -0.040}{0.026}{ 100}{1.97}{
379}{1.2CA}
\punktc{09/02/91}{20:20}{8297.347}{ -0.055}{0.022}{ 300}{2.01}{
520}{1.2CA}
\punktc{10/02/91}{18:49}{8298.284}{  0.053}{0.059}{ 150}{2.33}{
547}{1.2CA}
\punktc{11/02/91}{19:48}{8299.325}{ -0.764}{0.039}{ 500}{4.14}{
816}{1.2CA}
\punktc{14/02/91}{18:49}{8302.284}{ -0.577}{0.059}{ 500}{4.00}{
1676}{1.2CA}
\punktc{14/02/91}{19:31}{8302.313}{ -0.554}{0.045}{ 100}{3.07}{
395}{1.2CA}
\punktc{14/02/91}{19:38}{8302.318}{ -0.590}{0.024}{ 500}{3.25}{
902}{1.2CA}
\punktc{14/02/91}{21:07}{8302.380}{ -0.708}{0.045}{ 300}{3.31}{
735}{1.2CA}
\punktcc{15/02/91}{18:50}{8303.285}{ -0.952}{0.091}{ 100}{3.88}{
403}{1.2CA}
\punktc{15/02/91}{19:04}{8303.295}{ -1.037}{0.020}{ 100}{1.81}{
391}{1.2CA}
\punktc{15/02/91}{19:11}{8303.300}{ -1.072}{0.011}{ 344}{1.74}{
656}{1.2CA}
\punktcc{15/02/91}{20:57}{8303.374}{ -1.031}{0.042}{ 100}{2.10}{
621}{1.2CA}
\punktc{15/02/91}{21:03}{8303.378}{ -1.033}{0.017}{ 300}{2.14}{
645}{1.2CA}
\punktc{22/02/91}{18:11}{8310.258}{ -1.074}{0.072}{1800}{3.05}{
5205}{1.2CA}
\punktcc{22/02/91}{18:54}{8310.288}{ -1.047}{0.089}{ 100}{1.84}{
1507}{1.2CA}
\punktc{23/02/91}{19:11}{8311.299}{ -0.720}{0.025}{ 100}{1.37}{
759}{1.2CA}
\punktc{23/02/91}{19:17}{8311.304}{ -0.746}{0.016}{ 300}{1.24}{
1716}{1.2CA}
\punktcc{24/02/91}{18:32}{8312.272}{ -0.123}{0.082}{1800}{2.89}{
3299}{1.2CA}
\punktc{24/02/91}{23:38}{8312.485}{ -0.098}{0.023}{ 100}{1.53}{
732}{1.2CA}
\punktc{24/02/91}{23:44}{8312.489}{ -0.052}{0.019}{ 300}{1.48}{
1531}{1.2CA}
\punktcc{26/02/91}{18:44}{8314.281}{ -0.269}{0.054}{1801}{4.03}{
4921}{1.2CA}
\punktc{06/08/91}{03:30}{8474.646}{ -0.721}{0.049}{ 100}{1.29}{
347}{1.2CA}
\punktc{06/08/91}{03:41}{8474.654}{ -0.803}{0.043}{ 500}{1.27}{
669}{1.2CA}
\punktc{07/08/91}{04:13}{8475.676}{ -1.341}{0.017}{ 100}{1.29}{
391}{1.2CA}
\punktc{07/08/91}{04:21}{8475.682}{ -1.366}{0.017}{ 500}{1.35}{
1914}{1.2CA}
\punktcc{08/08/91}{04:01}{8476.667}{ -1.618}{0.053}{ 100}{2.13}{
363}{1.2CA}
\punktc{08/08/91}{04:11}{8476.675}{ -1.632}{0.029}{ 500}{2.16}{
949}{1.2CA}
\punktc{10/08/91}{03:02}{8478.627}{  0.030}{0.027}{ 100}{3.35}{
334}{1.2CA}
\punktc{10/08/91}{03:11}{8478.633}{ -0.012}{0.050}{ 500}{3.77}{
886}{1.2CA}
\punktc{24/08/91}{03:35}{8492.650}{ -0.122}{0.023}{ 100}{1.39}{
491}{1.2CA}
\punktc{24/08/91}{03:43}{8492.655}{ -0.150}{0.017}{ 500}{1.77}{
1204}{1.2CA}
\punktc{27/08/91}{03:57}{8495.665}{ -0.723}{0.062}{ 500}{2.24}{
2914}{1.2CA}
\punktc{19/09/91}{04:14}{8518.677}{ -0.842}{0.045}{ 500}{1.52}{
509}{1.2CA}
\punktc{20/09/91}{03:27}{8519.644}{ -0.806}{0.045}{ 500}{1.21}{
465}{1.2CA}
\punktc{18/10/91}{01:36}{8547.567}{ -1.325}{0.025}{ 500}{1.50}{
569}{1.2CA}
\punktc{19/10/91}{03:01}{8548.626}{ -1.643}{0.070}{ 500}{2.23}{
594}{1.2CA}
\punktc{21/10/91}{00:08}{8550.506}{ -1.853}{0.068}{ 500}{1.39}{
1801}{1.2CA}
\punktc{29/10/91}{01:26}{8558.560}{ -1.495}{0.051}{ 100}{1.24}{
380}{1.2CA}
\punktc{29/10/91}{01:36}{8558.567}{ -1.484}{0.025}{ 500}{1.11}{
865}{1.2CA}
\punktc{01/02/92}{18:51}{8654.286}{ -0.352}{0.017}{ 500}{1.72}{
749}{1.2CA}
\punktc{02/02/92}{18:55}{8655.288}{ -0.210}{0.016}{ 500}{2.13}{
678}{1.2CA}
\punktc{07/02/92}{19:18}{8660.305}{ -0.121}{0.013}{ 100}{1.36}{
345}{1.2CA}
\punktc{07/02/92}{19:30}{8660.313}{ -0.131}{0.036}{ 500}{1.42}{
661}{1.2CA}
\punktc{08/02/92}{19:03}{8661.294}{  0.347}{0.014}{ 100}{1.41}{
442}{1.2CA}
\punktc{08/02/92}{19:13}{8661.301}{  0.324}{0.030}{ 500}{1.53}{
1124}{1.2CA}
\punktc{08/02/92}{20:01}{8661.334}{  0.318}{0.009}{ 100}{1.40}{
445}{1.2CA}
\punktc{08/02/92}{20:09}{8661.340}{  0.328}{0.013}{ 500}{1.47}{
1098}{1.2CA}
\punktcc{13/02/92}{18:29}{8666.271}{ -0.449}{0.061}{ 100}{2.10}{
1156}{1.2CA}
\punktc{13/02/92}{19:32}{8666.314}{ -0.311}{0.036}{ 500}{2.31}{
4721}{1.2CA}
\punktc{14/02/92}{18:54}{8667.288}{ -0.600}{0.018}{ 500}{1.48}{
3252}{1.2CA}
\punktc{23/09/92}{05:12}{8888.717}{  1.212}{0.044}{ 100}{1.38}{
834}{1.2CA}
\punktc{23/09/92}{05:15}{8888.719}{  1.219}{0.043}{ 100}{1.51}{
1055}{1.2CA}
\punktc{24/09/92}{05:03}{8889.711}{  1.120}{0.034}{ 100}{2.05}{
834}{1.2CA}
\punktc{24/09/92}{05:06}{8889.713}{  1.118}{0.037}{ 100}{2.21}{
883}{1.2CA}
\punktc{25/09/92}{05:11}{8890.716}{  0.927}{0.050}{ 100}{0.91}{
1684}{1.2CA}
\punktc{25/09/92}{05:11}{8890.716}{  0.927}{0.050}{ 100}{0.91}{
1684}{1.2CA}
\punktc{17/02/93}{19:33}{9036.315}{ -0.601}{0.058}{ 120}{1.65}{
590}{1.2CA}

\end{picture}}

\end{picture}

\vspace*{1cm}
\caption{HQM-lightcurves in the Johnson-$R\/$ band. Plotted are variations
$\Delta R$=$R_0$$-$$R$; the reference magnitude $R_0$ is indicated by thick
dashes; it was not possible to determine $R_0$ for all quasars, see text.
Measurements obtained with too short exposure time or under bad
atmospheric conditions or those with only one reference star are shown by
open circles; reliable error bars can in these cases not be given}
\end{figure*}
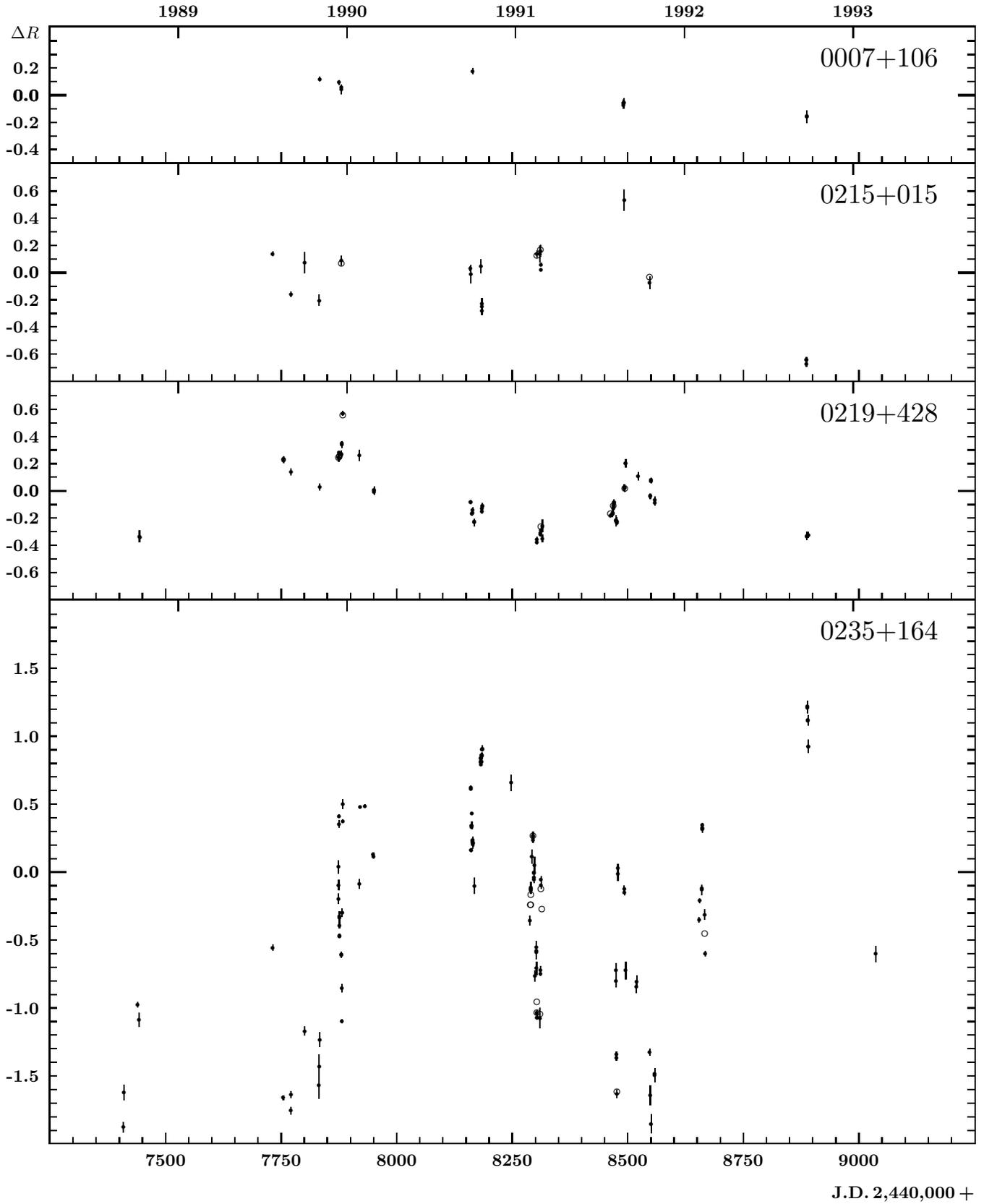

\begin{figure*}

\vspace*{0.5cm}

\begin{picture}(17 ,7 )(-1,0)
\put(0,0){\setlength{\unitlength}{0.0085cm}%
\begin{picture}(2000, 823.529)(7250,0)
\put(7250,0){\framebox(2000, 823.529)[tl]{\begin{picture}(0,0)(0,0)
        \put(0,0){\makebox(0,0)[tr]{$\Delta R$\hspace*{0.2cm}}}
        \put(2000,0){\makebox(0,0)[tr]{\large{0735+178}\T{0.4}
                                 \hspace*{0.5cm}}}
        \put(2000,-
823.529){\setlength{\unitlength}{1cm}\begin{picture}(0,0)(0,0)
        \end{picture}}
    \end{picture}}}

\thicklines
\put(7250,0){\setlength{\unitlength}{2.5cm}\begin{picture}(0,0)(0,-1.4)
   \put(0,0){\setlength{\unitlength}{1cm}\begin{picture}(0,0)(0,0)
        \put(0,0){\line(1,0){0.3}}
        \end{picture}}
   \end{picture}}

\put(9250,0){\setlength{\unitlength}{2.5cm}\begin{picture}(0,0)(0,-1.4)
   \put(0,0){\setlength{\unitlength}{1cm}\begin{picture}(0,0)(0,0)
        \put(0,0){\line(-1,0){0.3}}
        \end{picture}}
   \end{picture}}

\thinlines
\put(7250,0){\setlength{\unitlength}{2.5cm}\begin{picture}(0,0)(0,-1.4)
   \multiput(0,0)(0,0.1){14}{\setlength{\unitlength}{1cm}%
\begin{picture}(0,0)(0,0)
        \put(0,0){\line(1,0){0.12}}
        \end{picture}}
   \end{picture}}

\put(7250,0){\setlength{\unitlength}{2.5cm}\begin{picture}(0,0)(0,-1.4)
   \multiput(0,0)(0,-0.1){14}{\setlength{\unitlength}{1cm}%
\begin{picture}(0,0)(0,0)
        \put(0,0){\line(1,0){0.12}}
        \end{picture}}
   \end{picture}}

\put(9250,0){\setlength{\unitlength}{2.5cm}\begin{picture}(0,0)(0,-1.4)
   \multiput(0,0)(0,0.1){14}{\setlength{\unitlength}{1cm}%
\begin{picture}(0,0)(0,0)
        \put(0,0){\line(-1,0){0.12}}
        \end{picture}}
   \end{picture}}

\put(9250,0){\setlength{\unitlength}{2.5cm}\begin{picture}(0,0)(0,-1.4)
   \multiput(0,0)(0,-0.1){14}{\setlength{\unitlength}{1cm}%
\begin{picture}(0,0)(0,0)
        \put(0,0){\line(-1,0){0.12}}
        \end{picture}}
   \end{picture}}

\put(7250,0){\setlength{\unitlength}{2.5cm}\begin{picture}(0,0)(0,-1.4)
   \put(0,0.5){\setlength{\unitlength}{1cm}\begin{picture}(0,0)(0,0)
        \put(0,0){\line(1,0){0.12}}
        \put(-0.2,0){\makebox(0,0)[r]{\bf 0.5}}
        \end{picture}}
   \put(0,0.0){\setlength{\unitlength}{1cm}\begin{picture}(0,0)(0,0)
        \put(0,0){\line(1,0){0.12}}
        \put(-0.2,0){\makebox(0,0)[r]{\bf 0.0}}
        \end{picture}}
   \put(0,-1.0){\setlength{\unitlength}{1cm}\begin{picture}(0,0)(0,0)
        \put(0,0){\line(1,0){0.12}}
        \put(-0.2,0){\makebox(0,0)[r]{\bf -1.0}}
        \end{picture}}
   \put(0,-0.5){\setlength{\unitlength}{1cm}\begin{picture}(0,0)(0,0)
        \put(0,0){\line(1,0){0.12}}
        \put(-0.2,0){\makebox(0,0)[r]{\bf -0.5}}
        \end{picture}}
   \put(0,1.0){\setlength{\unitlength}{1cm}\begin{picture}(0,0)(0,0)
        \put(0,0){\line(1,0){0.12}}
        \put(-0.2,0){\makebox(0,0)[r]{\bf 1.0}}
        \end{picture}}
   \put(0,0.0){\setlength{\unitlength}{1cm}\begin{picture}(0,0)(0,0)
        \put(0,0){\line(1,0){0.12}}
        \put(-0.2,0){\makebox(0,0)[r]{\bf 0.0}}
        \end{picture}}
   \put(0,0.0){\setlength{\unitlength}{1cm}\begin{picture}(0,0)(0,0)
        \put(0,0){\line(1,0){0.12}}
        \put(-0.2,0){\makebox(0,0)[r]{\bf 0.0}}
        \end{picture}}
   \put(0,0.0){\setlength{\unitlength}{1cm}\begin{picture}(0,0)(0,0)
        \put(0,0){\line(1,0){0.12}}
        \put(-0.2,0){\makebox(0,0)[r]{\bf 0.0}}
        \end{picture}}
   \put(0,0.0){\setlength{\unitlength}{1cm}\begin{picture}(0,0)(0,0)
        \put(0,0){\line(1,0){0.12}}
        \end{picture}}
   \end{picture}}

   \put(7527.5, 823.529){\setlength{\unitlength}{1cm}\begin{picture}(0,0)(0,0)
        \put(0,0){\line(0,-1){0.2}}
        \put(0,0.2){\makebox(0,0)[b]{\bf 1989}}
   \end{picture}}
   \put(7892.5, 823.529){\setlength{\unitlength}{1cm}\begin{picture}(0,0)(0,0)
        \put(0,0){\line(0,-1){0.2}}
        \put(0,0.2){\makebox(0,0)[b]{\bf 1990}}
   \end{picture}}
   \put(8257.5, 823.529){\setlength{\unitlength}{1cm}\begin{picture}(0,0)(0,0)
        \put(0,0){\line(0,-1){0.2}}
        \put(0,0.2){\makebox(0,0)[b]{\bf 1991}}
  \end{picture}}
   \put(8622.5, 823.529){\setlength{\unitlength}{1cm}\begin{picture}(0,0)(0,0)
        \put(0,0){\line(0,-1){0.2}}
        \put(0,0.2){\makebox(0,0)[b]{\bf 1992}}
   \end{picture}}
   \put(8987.5, 823.529){\setlength{\unitlength}{1cm}\begin{picture}(0,0)(0,0)
        \put(0,0){\line(0,-1){0.2}}
        \put(0,0.2){\makebox(0,0)[b]{\bf 1993}}
   \end{picture}}

    \multiput(7250,0)(50,0){40}%
        {\setlength{\unitlength}{1cm}\begin{picture}(0,0)(0,0)
        \put(0,0){\line(0,1){0.12}}
    \end{picture}}
    \put(7500,0){\setlength{\unitlength}{1cm}\begin{picture}(0,0)(0,0)
        \put(0,0){\line(0,1){0.2}}
    \end{picture}}
    \put(8000,0){\setlength{\unitlength}{1cm}\begin{picture}(0,0)(0,0)
        \put(0,0){\line(0,1){0.2}}
    \end{picture}}
   \put(7750,0){\setlength{\unitlength}{1cm}\begin{picture}(0,0)(0,0)
        \put(0,0){\line(0,1){0.2}}
    \end{picture}}
   \put(8250,0){\setlength{\unitlength}{1cm}\begin{picture}(0,0)(0,0)
        \put(0,0){\line(0,1){0.2}}
    \end{picture}}
    \put(8500,0){\setlength{\unitlength}{1cm}\begin{picture}(0,0)(0,0)
        \put(0,0){\line(0,1){0.2}}
    \end{picture}}
    \put(8750,0){\setlength{\unitlength}{1cm}\begin{picture}(0,0)(0,0)
       \put(0,0){\line(0,1){0.2}}
    \end{picture}}
    \put(9000,0){\setlength{\unitlength}{1cm}\begin{picture}(0,0)(0,0)
       \put(0,0){\line(0,1){0.2}}
    \end{picture}}

\punktd{17/10/88}{04:44}{7451.698}{ -1.191}{0.031}{ 500}{1.85}{
618}{3.5CA}
\punktd{17/10/89}{02:35}{7816.608}{ -0.075}{0.023}{1000}{2.16}{
1845}{3.5CA}
\punktd{03/11/89}{01:58}{7833.582}{ -0.302}{0.021}{ 500}{2.58}{
553}{3.5CA}
\punktdd{20/12/89}{01:58}{7880.582}{  0.297}{0.052}{ 100}{1.92}{
538}{3.5CA}
\punktd{20/12/89}{02:02}{7880.585}{  0.309}{0.016}{ 250}{2.36}{
727}{3.5CA}
\punktd{20/12/89}{04:01}{7880.668}{  0.278}{0.008}{ 500}{1.95}{
718}{3.5CA}
\punktd{20/12/89}{13:56}{7881.081}{  0.152}{0.025}{ 500}{2.33}{
639}{3.5CA}
\punktdd{21/12/89}{01:49}{7881.576}{  0.144}{0.063}{ 100}{2.61}{
545}{3.5CA}
\punktd{21/12/89}{01:56}{7881.581}{  0.153}{0.025}{ 500}{2.33}{
639}{3.5CA}
\punktd{06/01/90}{00:47}{7897.533}{  0.309}{0.027}{ 500}{2.58}{
876}{3.5CA}
\punktdd{27/01/90}{20:59}{7919.375}{  0.386}{0.076}{ 797}{3.46}{
609}{3.5CA}
\punktd{09/02/90}{19:54}{7932.329}{  0.105}{0.011}{ 500}{1.52}{
659}{3.5CA}
\punktd{10/02/90}{21:35}{7933.400}{  0.091}{0.030}{ 500}{2.99}{
1420}{3.5CA}
\punktd{10/02/90}{21:46}{7933.407}{  0.108}{0.027}{ 500}{2.99}{
1361}{3.5CA}
\punktdd{26/02/90}{21:11}{7949.383}{ -0.145}{0.090}{  50}{1.41}{
467}{3.5CA}
\punktd{26/02/90}{21:19}{7949.388}{ -0.134}{0.031}{ 500}{2.31}{
731}{3.5CA}
\punktd{26/02/90}{21:44}{7949.406}{ -0.134}{0.013}{ 500}{2.43}{
618}{3.5CA}
\punktdd{27/02/90}{20:24}{7950.350}{ -0.131}{0.090}{  50}{1.46}{
492}{3.5CA}
\punktdd{27/02/90}{20:27}{7950.352}{ -0.101}{0.088}{  50}{1.41}{
482}{3.5CA}
\punktd{27/02/90}{20:35}{7950.358}{ -0.110}{0.031}{ 500}{1.58}{
704}{3.5CA}
\punktd{28/02/90}{20:38}{7951.360}{ -0.097}{0.019}{  50}{1.29}{
461}{3.5CA}
\punktd{28/02/90}{20:46}{7951.366}{ -0.095}{0.015}{ 500}{1.40}{
601}{3.5CA}
\punktdd{30/09/90}{03:43}{8164.655}{  0.236}{0.061}{ 100}{2.48}{
560}{3.5CA}
\punktd{30/09/90}{03:52}{8164.661}{  0.241}{0.029}{ 500}{2.38}{
875}{3.5CA}
\punktd{04/10/90}{03:43}{8168.655}{  0.054}{0.021}{ 500}{3.35}{
2119}{3.5CA}
\punktd{17/10/90}{03:33}{8181.648}{  0.257}{0.014}{ 500}{2.29}{
847}{3.5CA}
\punktd{18/10/90}{03:24}{8182.642}{  0.257}{0.039}{ 100}{2.82}{
542}{3.5CA}
\punktd{18/10/90}{03:33}{8182.648}{  0.253}{0.018}{ 500}{2.29}{
848}{3.5CA}
\punktd{19/10/90}{04:41}{8183.695}{  0.454}{0.012}{ 500}{1.56}{
817}{3.5CA}
\punktd{20/10/90}{04:14}{8184.677}{  0.476}{0.027}{ 500}{4.27}{
772}{3.5CA}
\punktd{20/10/90}{04:32}{8184.689}{  0.450}{0.032}{ 100}{1.63}{
556}{3.5CA}
\punktd{20/10/90}{04:41}{8184.695}{  0.450}{0.016}{ 500}{1.55}{
817}{3.5CA}
\punktd{21/10/90}{04:04}{8185.670}{  0.511}{0.030}{ 100}{3.77}{
543}{3.5CA}
\punktd{21/10/90}{04:14}{8185.677}{  0.473}{0.027}{ 500}{4.25}{
773}{3.5CA}
\punktd{27/11/90}{04:00}{8222.667}{  0.143}{0.009}{ 500}{2.14}{
731}{3.5CA}
\punktdd{18/12/90}{14:03}{8244.085}{  0.036}{0.090}{ 100}{5.11}{
437}{3.5CA}
\punktd{22/12/90}{11:28}{8247.978}{ -0.066}{0.020}{ 500}{2.70}{
616}{3.5CA}
\punktd{22/12/90}{13:48}{8248.076}{ -0.077}{0.025}{ 500}{3.24}{
642}{3.5CA}
\punktdd{31/01/91}{20:21}{8288.348}{  0.095}{0.048}{ 100}{3.76}{
845}{3.5CA}
\punktd{31/01/91}{20:32}{8288.356}{ -0.034}{0.023}{ 500}{4.06}{
3833}{3.5CA}
\punktdd{01/02/91}{00:16}{8288.512}{ -0.050}{0.045}{ 100}{3.13}{
957}{3.5CA}
\punktd{01/02/91}{00:24}{8288.517}{ -0.045}{0.019}{ 500}{3.10}{
2035}{3.5CA}
\punktdd{01/02/91}{01:29}{8288.562}{ -0.066}{0.041}{ 100}{2.26}{
681}{3.5CA}
\punktd{01/02/91}{01:36}{8288.567}{ -0.050}{0.020}{ 500}{2.35}{
2180}{3.5CA}
\punktd{01/02/91}{21:24}{8289.392}{ -0.197}{0.035}{1203}{3.13}{
2278}{3.5CA}
\punktd{02/02/91}{01:22}{8289.557}{ -0.246}{0.038}{1200}{4.51}{
3686}{3.5CA}
\punktd{02/02/91}{19:30}{8290.313}{ -0.188}{0.024}{ 500}{5.16}{
614}{3.5CA}
\punktd{02/02/91}{20:36}{8290.359}{ -0.185}{0.022}{ 500}{4.90}{
562}{3.5CA}
\punktd{02/02/91}{22:15}{8290.427}{ -0.161}{0.043}{ 100}{3.44}{
389}{3.5CA}
\punktd{02/02/91}{22:32}{8290.440}{ -0.169}{0.026}{ 300}{3.55}{
584}{3.5CA}
\punktd{03/02/91}{00:25}{8290.518}{ -0.157}{0.043}{ 100}{2.40}{
595}{3.5CA}
\punktd{03/02/91}{00:33}{8290.523}{ -0.132}{0.021}{ 500}{2.52}{
1068}{3.5CA}
\punktd{04/02/91}{18:56}{8292.289}{ -0.024}{0.047}{1201}{3.45}{
1089}{3.5CA}
\punktdd{04/02/91}{22:05}{8292.420}{ -0.030}{0.103}{1200}{5.22}{
875}{3.5CA}
\punktdd{06/02/91}{21:33}{8294.398}{  0.122}{0.278}{1200}{4.25}{
1274}{3.5CA}
\punktdd{06/02/91}{21:42}{8294.404}{ -0.001}{0.039}{1212}{4.41}{
1261}{3.5CA}
\punktdd{06/02/91}{22:59}{8294.458}{  0.071}{0.045}{ 110}{3.57}{
344}{3.5CA}
\punktd{06/02/91}{23:04}{8294.461}{  0.095}{0.031}{ 229}{3.48}{
427}{3.5CA}
\punktd{06/02/91}{23:37}{8294.484}{  0.079}{0.023}{ 300}{3.17}{
497}{3.5CA}
\punktd{07/02/91}{00:47}{8294.533}{  0.107}{0.041}{ 100}{2.26}{
341}{3.5CA}
\punktd{07/02/91}{19:27}{8295.311}{ -0.039}{0.030}{ 100}{4.67}{
335}{3.5CA}
\punktd{07/02/91}{19:34}{8295.316}{ -0.047}{0.021}{ 500}{4.74}{
625}{3.5CA}
\punktdd{07/02/91}{20:33}{8295.356}{ -0.131}{0.073}{ 200}{4.22}{
422}{3.5CA}
\punktdd{09/02/91}{00:12}{8296.509}{  0.119}{0.084}{ 100}{4.30}{
407}{3.5CA}
\punktd{09/02/91}{00:19}{8296.513}{  0.028}{0.051}{ 200}{3.92}{
492}{3.5CA}
\punktd{09/02/91}{00:48}{8296.534}{  0.024}{0.016}{ 100}{2.90}{
339}{3.5CA}
\punktd{09/02/91}{00:55}{8296.539}{  0.025}{0.030}{ 500}{2.55}{
628}{3.5CA}
\punktdd{09/02/91}{02:24}{8296.600}{  0.025}{0.108}{ 150}{1.94}{
381}{3.5CA}
\punktdd{09/02/91}{19:42}{8297.321}{  0.249}{0.147}{ 100}{1.69}{
328}{3.5CA}
\punktd{09/02/91}{20:48}{8297.367}{  0.233}{0.059}{ 100}{1.32}{
324}{3.5CA}
\punktd{09/02/91}{20:55}{8297.372}{  0.218}{0.008}{ 500}{1.42}{
594}{3.5CA}
\punktd{09/02/91}{22:20}{8297.431}{  0.238}{0.010}{ 100}{1.63}{
328}{3.5CA}
\punktd{09/02/91}{22:31}{8297.439}{  0.243}{0.021}{ 300}{1.73}{
448}{3.5CA}
\punktd{10/02/91}{00:15}{8297.511}{  0.243}{0.043}{ 150}{1.94}{
386}{3.5CA}
\punktdd{10/02/91}{00:22}{8297.516}{  0.213}{0.059}{ 150}{2.14}{
379}{3.5CA}
\punktdd{10/02/91}{19:29}{8298.312}{  0.116}{0.055}{ 100}{3.20}{
323}{3.5CA}
\punktd{10/02/91}{19:36}{8298.317}{  0.150}{0.029}{ 500}{3.12}{
575}{3.5CA}
\punktd{10/02/91}{19:47}{8298.325}{  0.183}{0.025}{ 500}{3.59}{
579}{3.5CA}
\punktd{11/02/91}{19:55}{8299.330}{  0.120}{0.034}{ 100}{4.82}{
323}{3.5CA}
\punktd{11/02/91}{20:02}{8299.335}{  0.135}{0.021}{ 500}{4.74}{
546}{3.5CA}
\punktdd{11/02/91}{20:31}{8299.355}{  0.113}{0.063}{ 100}{3.97}{
317}{3.5CA}
\punktdd{11/02/91}{20:36}{8299.359}{  0.115}{0.053}{ 206}{4.43}{
392}{3.5CA}
\punktd{11/02/91}{22:25}{8299.435}{  0.154}{0.037}{ 150}{4.50}{
340}{3.5CA}
\punktd{11/02/91}{22:32}{8299.439}{  0.127}{0.024}{ 300}{4.32}{
425}{3.5CA}
\punktd{14/02/91}{18:58}{8302.291}{ -0.047}{0.045}{ 100}{3.96}{
343}{3.5CA}
\punktd{14/02/91}{19:06}{8302.296}{ -0.023}{0.024}{ 500}{4.24}{
609}{3.5CA}
\punktd{14/02/91}{20:26}{8302.351}{ -0.036}{0.022}{ 100}{3.69}{
324}{3.5CA}
\punktd{14/02/91}{20:29}{8302.354}{ -0.013}{0.041}{ 110}{3.67}{
338}{3.5CA}
\punktdd{14/02/91}{20:33}{8302.356}{ -0.046}{0.048}{  96}{3.70}{
410}{3.5CA}
\punktd{14/02/91}{20:37}{8302.359}{ -0.022}{0.018}{ 300}{3.66}{
470}{3.5CA}
\punktdd{14/02/91}{21:36}{8302.400}{ -0.029}{0.048}{ 100}{3.53}{
365}{3.5CA}
\punktd{14/02/91}{21:47}{8302.408}{ -0.034}{0.016}{ 500}{3.62}{
589}{3.5CA}
\punktdd{14/02/91}{23:13}{8302.467}{ -0.021}{0.043}{ 100}{2.82}{
344}{3.5CA}
\punktd{14/02/91}{23:19}{8302.472}{ -0.019}{0.016}{ 300}{3.01}{
463}{3.5CA}
\punktd{15/02/91}{01:49}{8302.576}{ -0.020}{0.011}{ 100}{2.05}{
340}{3.5CA}
\punktd{15/02/91}{01:57}{8302.582}{ -0.020}{0.010}{ 500}{2.17}{
655}{3.5CA}
\punktd{15/02/91}{20:07}{8303.339}{ -0.002}{0.009}{ 100}{1.60}{
331}{3.5CA}
\punktd{15/02/91}{20:14}{8303.343}{ -0.003}{0.022}{ 300}{1.94}{
470}{3.5CA}
\punktd{15/02/91}{21:45}{8303.407}{  0.001}{0.008}{ 100}{1.59}{
384}{3.5CA}
\punktd{15/02/91}{21:53}{8303.412}{  0.006}{0.010}{ 300}{1.66}{
408}{3.5CA}
\punktd{16/02/91}{00:02}{8303.502}{ -0.001}{0.039}{ 100}{1.61}{
322}{3.5CA}
\punktd{16/02/91}{00:11}{8303.508}{  0.010}{0.011}{ 300}{1.63}{
406}{3.5CA}
\punktd{16/02/91}{01:39}{8303.569}{  0.010}{0.039}{ 100}{1.53}{
326}{3.5CA}
\punktd{16/02/91}{01:45}{8303.573}{  0.013}{0.009}{ 300}{1.46}{
450}{3.5CA}
\punktd{19/02/91}{19:35}{8307.317}{  0.082}{0.034}{1200}{2.62}{
1150}{3.5CA}
\punktd{19/02/91}{20:50}{8307.368}{  0.086}{0.046}{1391}{2.69}{
1036}{3.5CA}
\punktd{19/02/91}{22:31}{8307.438}{  0.090}{0.057}{1200}{2.61}{
922}{3.5CA}
\punktd{20/02/91}{00:32}{8307.522}{  0.088}{0.055}{1201}{2.93}{
1034}{3.5CA}
\punktd{20/02/91}{19:29}{8308.312}{  0.089}{0.072}{1200}{2.98}{
1173}{3.5CA}
\punktd{20/02/91}{20:57}{8308.373}{  0.092}{0.033}{1200}{2.73}{
1222}{3.5CA}
\punktd{20/02/91}{22:29}{8308.437}{  0.122}{0.077}{1426}{2.68}{
1490}{3.5CA}
\punktd{21/02/91}{19:04}{8309.295}{  0.114}{0.035}{1200}{3.37}{
1920}{3.5CA}
\punktd{21/02/91}{20:43}{8309.363}{  0.091}{0.053}{1320}{3.15}{
2032}{3.5CA}
\punktd{21/02/91}{22:32}{8309.439}{  0.083}{0.056}{1200}{3.23}{
1697}{3.5CA}
\punktd{22/02/91}{19:16}{8310.303}{  0.149}{0.056}{2189}{3.07}{
1660}{3.5CA}
\punktd{22/02/91}{20:47}{8310.366}{  0.158}{0.032}{1200}{2.99}{
1739}{3.5CA}
\punktd{22/02/91}{22:37}{8310.442}{  0.144}{0.031}{1200}{3.16}{
1745}{3.5CA}
\punktd{22/02/91}{22:42}{8310.446}{  0.182}{0.013}{ 100}{1.54}{
408}{3.5CA}
\punktd{22/02/91}{22:48}{8310.450}{  0.179}{0.020}{ 300}{1.61}{
704}{3.5CA}
\punktd{23/02/91}{21:06}{8311.380}{  0.294}{0.011}{ 100}{1.01}{
476}{3.5CA}
\punktd{23/02/91}{21:15}{8311.386}{  0.297}{0.010}{ 300}{1.09}{
904}{3.5CA}
\punktd{25/02/91}{01:10}{8312.549}{  0.315}{0.039}{ 100}{1.94}{
1691}{3.5CA}
\punktd{25/02/91}{01:17}{8312.554}{  0.291}{0.026}{ 300}{1.93}{
3960}{3.5CA}
\punktd{16/03/91}{21:21}{8332.390}{  0.985}{0.012}{ 500}{3.90}{
613}{3.5CA}
\punktd{18/03/91}{19:48}{8334.326}{  0.764}{0.031}{ 100}{2.57}{
341}{3.5CA}
\punktd{18/03/91}{21:06}{8334.380}{  0.777}{0.027}{ 100}{2.55}{
324}{3.5CA}
\punktd{18/03/91}{21:16}{8334.387}{  0.771}{0.013}{ 500}{2.53}{
555}{3.5CA}
\punktd{19/03/91}{20:29}{8335.354}{  0.677}{0.029}{ 100}{1.67}{
326}{3.5CA}
\punktd{19/03/91}{20:48}{8335.367}{  0.652}{0.014}{ 500}{1.61}{
592}{3.5CA}
\punktd{20/09/91}{04:28}{8519.686}{ -0.360}{0.008}{ 500}{1.40}{
599}{3.5CA}
\punktd{17/10/91}{05:34}{8546.732}{  0.009}{0.025}{ 500}{2.45}{
3323}{3.5CA}
\punktd{18/10/91}{03:55}{8547.664}{ -0.138}{0.018}{ 100}{1.55}{
335}{3.5CA}
\punktd{18/10/91}{04:06}{8547.671}{ -0.128}{0.019}{ 500}{1.73}{
628}{3.5CA}
\punktd{19/10/91}{05:07}{8548.713}{ -0.227}{0.013}{ 100}{1.61}{
335}{3.5CA}
\punktd{19/10/91}{05:14}{8548.719}{ -0.230}{0.014}{ 500}{2.10}{
644}{3.5CA}
\punktd{21/10/91}{03:26}{8550.643}{ -0.310}{0.050}{ 100}{1.72}{
409}{3.5CA}
\punktd{21/10/91}{03:34}{8550.649}{ -0.310}{0.023}{ 500}{1.74}{
928}{3.5CA}
\punktd{27/10/91}{05:31}{8556.730}{ -0.229}{0.049}{ 100}{2.12}{
901}{3.5CA}
\punktd{27/10/91}{05:36}{8556.734}{ -0.223}{0.036}{ 250}{2.19}{
2401}{3.5CA}
\punktd{01/02/92}{22:55}{8654.455}{ -0.270}{0.054}{ 100}{1.46}{
304}{3.5CA}
\punktd{03/02/92}{00:41}{8655.529}{ -0.210}{0.025}{ 500}{1.67}{
545}{3.5CA}
\punktd{06/02/92}{22:11}{8659.425}{ -0.183}{0.022}{ 500}{1.62}{
482}{3.5CA}
\punktd{07/02/92}{22:11}{8660.425}{ -0.182}{0.005}{ 500}{1.62}{
482}{3.5CA}
\punktd{07/02/92}{23:42}{8660.488}{ -0.245}{0.006}{ 500}{1.28}{
509}{3.5CA}
\punktd{08/02/92}{23:24}{8661.476}{ -0.313}{0.018}{ 500}{1.03}{
483}{3.5CA}
\punktd{09/02/92}{23:01}{8662.459}{ -0.303}{0.053}{ 100}{1.00}{
314}{3.5CA}
\punktd{09/02/92}{23:10}{8662.466}{ -0.308}{0.009}{ 500}{1.13}{
505}{3.5CA}
\punktdd{10/02/92}{20:48}{8663.367}{ -0.337}{0.054}{ 100}{1.30}{
341}{3.5CA}
\punktd{10/02/92}{20:58}{8663.374}{ -0.326}{0.025}{ 500}{0.96}{
668}{3.5CA}
\punktdd{11/02/92}{20:25}{8664.351}{ -0.322}{0.055}{ 100}{0.98}{
377}{3.5CA}
\punktd{11/02/92}{20:33}{8664.357}{ -0.339}{0.014}{ 500}{1.08}{
851}{3.5CA}
\punktdd{07/03/92}{01:05}{8688.546}{ -0.511}{0.046}{ 150}{1.40}{
263}{3.5CA}
\punktd{07/03/92}{01:12}{8688.550}{ -0.503}{0.021}{ 300}{1.33}{
343}{3.5CA}
\punktdd{09/03/92}{01:17}{8690.554}{ -0.508}{0.041}{ 150}{1.75}{
247}{3.5CA}
\punktd{09/03/92}{01:23}{8690.558}{ -0.493}{0.014}{ 300}{1.55}{
303}{3.5CA}
\punktdd{10/03/92}{01:20}{8691.556}{ -0.507}{0.058}{ 150}{1.87}{
258}{3.5CA}
\punktd{10/03/92}{01:25}{8691.560}{ -0.517}{0.030}{ 300}{1.85}{
328}{3.5CA}
\punktdd{11/03/92}{00:55}{8692.539}{ -0.556}{0.048}{ 150}{1.51}{
292}{3.5CA}
\punktd{11/03/92}{01:01}{8692.543}{ -0.552}{0.038}{ 300}{1.51}{
398}{3.5CA}
\punktd{13/03/92}{01:26}{8694.560}{ -0.546}{0.028}{ 150}{1.50}{
943}{3.5CA}
\punktd{13/03/92}{01:30}{8694.563}{ -0.576}{0.023}{ 150}{1.43}{
902}{3.5CA}
\punktdd{16/03/92}{00:54}{8697.538}{ -0.476}{0.054}{ 150}{1.64}{
1176}{3.5CA}
\punktd{16/03/92}{01:01}{8697.542}{ -0.506}{0.028}{ 500}{1.57}{
3482}{3.5CA}
\punktd{17/03/92}{01:08}{8698.548}{ -0.500}{0.043}{ 150}{1.21}{
787}{3.5CA}
\punktd{17/03/92}{01:14}{8698.552}{ -0.494}{0.030}{ 300}{1.20}{
1389}{3.5CA}
\punktd{18/03/92}{01:09}{8699.548}{ -0.484}{0.037}{ 200}{1.89}{
813}{3.5CA}
\punktd{18/03/92}{01:14}{8699.552}{ -0.444}{0.034}{ 300}{1.88}{
1136}{3.5CA}
\punktd{19/03/92}{01:29}{8700.562}{ -0.491}{0.045}{ 300}{1.94}{
1772}{3.5CA}
\punktdd{20/03/92}{00:58}{8701.540}{ -0.499}{0.058}{ 150}{1.41}{
410}{3.5CA}
\punktd{20/03/92}{01:03}{8701.544}{ -0.539}{0.011}{ 300}{1.47}{
654}{3.5CA}
\punktd{21/03/92}{02:06}{8702.588}{ -0.491}{0.042}{ 200}{1.98}{
460}{3.5CA}
\punktd{21/03/92}{02:13}{8702.593}{ -0.536}{0.016}{ 300}{2.23}{
629}{3.5CA}
\punktdd{23/03/92}{00:25}{8704.517}{ -0.504}{0.047}{ 150}{1.73}{
261}{3.5CA}
\punktd{23/03/92}{00:30}{8704.521}{ -0.503}{0.025}{ 300}{1.72}{
332}{3.5CA}
\punktdd{24/03/92}{00:49}{8705.534}{ -0.521}{0.046}{ 150}{1.90}{
245}{3.5CA}
\punktd{24/03/92}{00:54}{8705.538}{ -0.519}{0.039}{ 300}{1.73}{
313}{3.5CA}
\punktd{25/03/92}{00:40}{8706.528}{ -0.501}{0.030}{ 150}{1.78}{
251}{3.5CA}
\punktd{25/03/92}{00:44}{8706.531}{ -0.497}{0.045}{ 150}{1.64}{
250}{3.5CA}
\punktd{23/09/92}{04:43}{8888.697}{ -1.048}{0.026}{ 100}{1.48}{
624}{3.5CA}
\punktd{23/09/92}{04:46}{8888.699}{ -1.033}{0.024}{ 100}{1.49}{
609}{3.5CA}
\punktd{25/09/92}{04:52}{8890.703}{ -0.952}{0.011}{ 100}{1.13}{
557}{3.5CA}
\punktd{09/02/93}{21:04}{9028.378}{ -0.357}{0.016}{ 200}{1.30}{
448}{3.5CA}
\punktd{15/02/93}{22:37}{9034.443}{ -0.619}{0.020}{ 200}{1.70}{
116}{3.5CA}
\punktd{15/02/93}{22:42}{9034.446}{ -0.616}{0.019}{ 200}{1.72}{
-195}{3.5CA}
\punktd{17/02/93}{19:46}{9036.324}{ -0.593}{0.022}{ 180}{1.55}{
546}{3.5CA}
\punktd{17/02/93}{19:46}{9036.324}{ -0.593}{0.022}{ 180}{1.55}{
546}{3.5CA}

\end{picture}}

\end{picture}

\vspace*{-0.02cm}

\begin{picture}(17 ,3 )(-1,0)
\put(0,0){\setlength{\unitlength}{0.0085cm}%
\begin{picture}(2000, 352.941)(7250,0)
\put(7250,0){\framebox(2000, 352.941)[tl]{\begin{picture}(0,0)(0,0)
        \put(2000,0){\makebox(0,0)[tr]{\large{0836+710}\T{0.4}
                                 \hspace*{0.5cm}}}
        \put(2000,-
352.941){\setlength{\unitlength}{1cm}\begin{picture}(0,0)(0,0)
        \end{picture}}
    \end{picture}}}

\thicklines
\put(7250,0){\setlength{\unitlength}{2.5cm}\begin{picture}(0,0)(0,-0.6)
   \put(0,0){\setlength{\unitlength}{1cm}\begin{picture}(0,0)(0,0)
        \put(0,0){\line(1,0){0.3}}
        \end{picture}}
   \end{picture}}

\put(9250,0){\setlength{\unitlength}{2.5cm}\begin{picture}(0,0)(0,-0.6)
   \put(0,0){\setlength{\unitlength}{1cm}\begin{picture}(0,0)(0,0)
        \put(0,0){\line(-1,0){0.3}}
        \end{picture}}
   \end{picture}}

\thinlines
\put(7250,0){\setlength{\unitlength}{2.5cm}\begin{picture}(0,0)(0,-0.6)
   \multiput(0,0)(0,0.1){6}{\setlength{\unitlength}{1cm}%
\begin{picture}(0,0)(0,0)
        \put(0,0){\line(1,0){0.12}}
        \end{picture}}
   \end{picture}}

\put(7250,0){\setlength{\unitlength}{2.5cm}\begin{picture}(0,0)(0,-0.6)
   \multiput(0,0)(0,-0.1){6}{\setlength{\unitlength}{1cm}%
\begin{picture}(0,0)(0,0)
        \put(0,0){\line(1,0){0.12}}
        \end{picture}}
   \end{picture}}

\put(9250,0){\setlength{\unitlength}{2.5cm}\begin{picture}(0,0)(0,-0.6)
   \multiput(0,0)(0,0.1){6}{\setlength{\unitlength}{1cm}%
\begin{picture}(0,0)(0,0)
        \put(0,0){\line(-1,0){0.12}}
        \end{picture}}
   \end{picture}}

\put(9250,0){\setlength{\unitlength}{2.5cm}\begin{picture}(0,0)(0,-0.6)
   \multiput(0,0)(0,-0.1){6}{\setlength{\unitlength}{1cm}%
\begin{picture}(0,0)(0,0)
        \put(0,0){\line(-1,0){0.12}}
        \end{picture}}
   \end{picture}}

\put(7250,0){\setlength{\unitlength}{2.5cm}\begin{picture}(0,0)(0,-0.6)
   \put(0,0.4){\setlength{\unitlength}{1cm}\begin{picture}(0,0)(0,0)
        \put(0,0){\line(1,0){0.12}}
        \put(-0.2,0){\makebox(0,0)[r]{\bf 0.4}}
        \end{picture}}
   \put(0,0.2){\setlength{\unitlength}{1cm}\begin{picture}(0,0)(0,0)
        \put(0,0){\line(1,0){0.12}}
        \put(-0.2,0){\makebox(0,0)[r]{\bf 0.2}}
        \end{picture}}
   \put(0,-0.2){\setlength{\unitlength}{1cm}\begin{picture}(0,0)(0,0)
        \put(0,0){\line(1,0){0.12}}
        \put(-0.2,0){\makebox(0,0)[r]{\bf -0.2}}
        \end{picture}}
   \put(0,-0.4){\setlength{\unitlength}{1cm}\begin{picture}(0,0)(0,0)
        \put(0,0){\line(1,0){0.12}}
        \put(-0.2,0){\makebox(0,0)[r]{\bf -0.4}}
        \end{picture}}
   \put(0,0.0){\setlength{\unitlength}{1cm}\begin{picture}(0,0)(0,0)
        \put(0,0){\line(1,0){0.12}}
        \put(-0.2,0){\makebox(0,0)[r]{\bf 0.0}}
        \end{picture}}
   \put(0,0.0){\setlength{\unitlength}{1cm}\begin{picture}(0,0)(0,0)
        \put(0,0){\line(1,0){0.12}}
        \put(-0.2,0){\makebox(0,0)[r]{\bf 0.0}}
        \end{picture}}
   \put(0,0.0){\setlength{\unitlength}{1cm}\begin{picture}(0,0)(0,0)
        \put(0,0){\line(1,0){0.12}}
        \put(-0.2,0){\makebox(0,0)[r]{\bf 0.0}}
        \end{picture}}
   \put(0,0.0){\setlength{\unitlength}{1cm}\begin{picture}(0,0)(0,0)
        \put(0,0){\line(1,0){0.12}}
        \put(-0.2,0){\makebox(0,0)[r]{\bf 0.0}}
        \end{picture}}
   \put(0,0.4){\setlength{\unitlength}{1cm}\begin{picture}(0,0)(0,0)
        \put(0,0){\line(1,0){0.12}}
        \end{picture}}
   \end{picture}}

   \put(7527.5, 352.941){\setlength{\unitlength}{1cm}\begin{picture}(0,0)(0,0)
        \put(0,0){\line(0,-1){0.2}}
   \end{picture}}
   \put(7892.5, 352.941){\setlength{\unitlength}{1cm}\begin{picture}(0,0)(0,0)
        \put(0,0){\line(0,-1){0.2}}
   \end{picture}}
   \put(8257.5, 352.941){\setlength{\unitlength}{1cm}\begin{picture}(0,0)(0,0)
        \put(0,0){\line(0,-1){0.2}}
  \end{picture}}
   \put(8622.5, 352.941){\setlength{\unitlength}{1cm}\begin{picture}(0,0)(0,0)
        \put(0,0){\line(0,-1){0.2}}
   \end{picture}}
   \put(8987.5, 352.941){\setlength{\unitlength}{1cm}\begin{picture}(0,0)(0,0)
        \put(0,0){\line(0,-1){0.2}}
   \end{picture}}

    \multiput(7250,0)(50,0){40}%
        {\setlength{\unitlength}{1cm}\begin{picture}(0,0)(0,0)
        \put(0,0){\line(0,1){0.12}}
    \end{picture}}
    \put(7500,0){\setlength{\unitlength}{1cm}\begin{picture}(0,0)(0,0)
        \put(0,0){\line(0,1){0.2}}
    \end{picture}}
    \put(8000,0){\setlength{\unitlength}{1cm}\begin{picture}(0,0)(0,0)
        \put(0,0){\line(0,1){0.2}}
    \end{picture}}
   \put(7750,0){\setlength{\unitlength}{1cm}\begin{picture}(0,0)(0,0)
        \put(0,0){\line(0,1){0.2}}
    \end{picture}}
   \put(8250,0){\setlength{\unitlength}{1cm}\begin{picture}(0,0)(0,0)
        \put(0,0){\line(0,1){0.2}}
    \end{picture}}
    \put(8500,0){\setlength{\unitlength}{1cm}\begin{picture}(0,0)(0,0)
        \put(0,0){\line(0,1){0.2}}
    \end{picture}}
    \put(8750,0){\setlength{\unitlength}{1cm}\begin{picture}(0,0)(0,0)
       \put(0,0){\line(0,1){0.2}}
    \end{picture}}
    \put(9000,0){\setlength{\unitlength}{1cm}\begin{picture}(0,0)(0,0)
       \put(0,0){\line(0,1){0.2}}
    \end{picture}}

\punkte{01/10/89}{17:03}{7801.210}{ -0.204}{0.044}{ 500}{1.40}{
605}{1.2CA}
\punkte{01/11/89}{02:55}{7831.622}{ -0.222}{0.032}{ 500}{1.23}{
640}{1.2CA}
\punkte{02/11/89}{02:53}{7832.620}{ -0.236}{0.014}{ 500}{1.73}{
515}{1.2CA}
\punkte{06/01/90}{23:12}{7898.467}{ -0.279}{0.076}{ 500}{2.21}{
937}{1.2CA}
\punkte{21/05/90}{20:47}{8033.366}{ -0.230}{0.021}{ 400}{2.09}{
2181}{1.2CA}
\punkte{22/05/90}{21:24}{8034.392}{ -0.233}{0.010}{ 400}{2.79}{
1787}{1.2CA}
\punkte{22/05/90}{21:32}{8034.398}{ -0.250}{0.024}{ 400}{3.41}{
1754}{1.2CA}
\punkte{23/02/91}{21:53}{8311.412}{ -0.342}{0.025}{ 100}{1.09}{
428}{1.2CA}
\punkte{23/02/91}{22:02}{8311.418}{ -0.324}{0.019}{ 300}{1.26}{
782}{1.2CA}
\punktee{27/05/91}{21:26}{8404.394}{ -0.187}{0.155}{ 150}{1.70}{
1132}{1.2CA}
\punkte{27/10/91}{04:13}{8556.676}{ -0.209}{0.040}{ 500}{1.92}{
1786}{1.2CA}
\punkte{28/10/91}{04:24}{8557.683}{ -0.200}{0.022}{ 250}{1.86}{
837}{1.2CA}
\punkte{04/01/92}{00:19}{8625.514}{ -0.059}{0.027}{ 600}{1.23}{
1565}{1.2CA}
\punkte{04/01/92}{00:27}{8625.519}{ -0.057}{0.029}{ 600}{1.36}{
1632}{1.2CA}
\punkte{04/01/92}{00:33}{8625.523}{ -0.054}{0.027}{ 600}{1.54}{
1601}{1.2CA}
\punkte{04/01/92}{00:40}{8625.528}{ -0.063}{0.027}{ 600}{1.21}{
1611}{1.2CA}
\punkte{28/01/92}{12:19}{8650.013}{  0.036}{0.011}{ 700}{1.02}{
13141}{1.2CA}
\punkte{06/02/92}{02:16}{8658.595}{ -0.003}{0.014}{2000}{1.12}{
1622}{1.2CA}
\punkte{06/02/92}{03:06}{8658.630}{ -0.040}{0.034}{1000}{1.44}{
1575}{1.2CA}
\punkte{06/02/92}{23:11}{8659.467}{ -0.094}{0.049}{ 500}{1.49}{
477}{1.2CA}
\punkte{11/02/92}{21:38}{8664.402}{  0.124}{0.018}{ 100}{1.12}{
376}{1.2CA}
\punkte{11/02/92}{21:55}{8664.413}{  0.139}{0.022}{ 500}{1.27}{
839}{1.2CA}
\punkte{12/02/92}{00:40}{8664.528}{  0.181}{0.078}{ 100}{1.18}{
350}{1.2CA}
\punkte{12/02/92}{00:49}{8664.535}{  0.171}{0.015}{ 500}{1.24}{
697}{1.2CA}
\punkte{12/02/92}{01:54}{8664.580}{  0.162}{0.017}{ 500}{1.26}{
504}{1.2CA}
\punkte{13/02/92}{21:18}{8666.388}{  0.046}{0.081}{ 100}{2.01}{
513}{1.2CA}
\punkte{14/02/92}{09:15}{8666.886}{  0.091}{0.056}{ 600}{1.75}{
1544}{1.2CA}
\punktee{14/02/92}{21:33}{8667.399}{  0.203}{0.111}{ 600}{1.43}{
2766}{1.2CA}
\punkte{15/02/92}{13:20}{8668.056}{  0.325}{0.077}{ 600}{1.40}{
1889}{1.2CA}
\punkte{16/02/92}{02:01}{8668.585}{  0.440}{0.037}{1000}{1.36}{
2538}{1.2CA}
\punkte{16/02/92}{21:27}{8669.394}{  0.486}{0.022}{1000}{1.50}{
3454}{1.2CA}
\punkte{16/02/92}{22:09}{8669.423}{  0.443}{0.054}{1000}{1.53}{
4534}{1.2CA}
\punktee{16/02/92}{23:37}{8669.484}{  0.411}{0.102}{ 600}{1.48}{
846}{1.2CA}
\punkte{17/02/92}{00:48}{8669.533}{  0.394}{0.050}{1000}{1.59}{
3010}{1.2CA}
\punktee{17/02/92}{02:25}{8669.601}{  0.327}{0.108}{ 600}{1.54}{
888}{1.2CA}
\punkte{17/02/92}{03:12}{8669.634}{  0.296}{0.084}{1000}{1.74}{
3665}{1.2CA}
\punkte{18/08/92}{22:16}{8853.428}{ -0.121}{0.025}{ 700}{1.89}{
747}{1.2CA}
\punkte{21/09/92}{03:20}{8886.640}{ -0.229}{0.010}{ 100}{1.13}{
615}{1.2CA}
\punkte{22/09/92}{04:00}{8887.667}{ -0.305}{0.048}{ 200}{3.14}{
696}{1.2CA}
\punktee{21/01/93}{21:50}{9009.410}{ -0.189}{0.211}{ 120}{1.93}{
1530}{1.2CA}
\punktee{21/01/93}{21:54}{9009.413}{ -0.266}{0.185}{ 120}{2.03}{
1583}{1.2CA}
\punkte{08/02/93}{00:27}{9026.519}{ -0.114}{0.020}{ 200}{1.89}{
3320}{1.2CA}
\punkte{09/02/93}{01:47}{9027.575}{ -0.137}{0.003}{ 300}{1.63}{
2160}{1.2CA}
\punkte{16/02/93}{23:30}{9035.480}{ -0.155}{0.005}{ 200}{2.28}{
473}{1.2CA}
\punkte{16/02/93}{23:35}{9035.483}{ -0.148}{0.012}{ 200}{2.23}{
485}{1.2CA}
\punkte{16/02/93}{23:35}{9035.483}{ -0.148}{0.012}{ 200}{2.23}{
485}{1.2CA}
\punkte{}{}{9129.361}{-0.194}{0.042}{}{}{}{1.2CA}%
\punktee{}{}{9130.360}{-0.265}{0.0}{}{}{}{1.2CA}%
\punkte{}{}{9130.354}{-0.205}{0.035}{}{}{}{1.2CA}%
\punkte{}{}{9195.489}{-0.145}{0.025}{}{}{}{1.2CA}%

\end{picture}}

\end{picture}

\vspace*{-0.02cm}

\begin{picture}(17 ,5 )(-1,0)
\put(0,0){\setlength{\unitlength}{0.0085cm}%
\begin{picture}(2000, 588.235)(7250,0)
\put(7250,0){\framebox(2000, 588.235)[tl]{\begin{picture}(0,0)(0,0)
        \put(2000,0){\makebox(0,0)[tr]{\large{0851+202}\T{0.4}
                                 \hspace*{0.5cm}}}
        \put(2000,-
588.235){\setlength{\unitlength}{1cm}\begin{picture}(0,0)(0,0)
        \end{picture}}
    \end{picture}}}

\thicklines
\put(7250,0){\setlength{\unitlength}{2.5cm}\begin{picture}(0,0)(0,-1)
   \put(0,0){\setlength{\unitlength}{1cm}\begin{picture}(0,0)(0,0)
        \put(0,0){\line(1,0){0.3}}
        \end{picture}}
   \end{picture}}

\put(9250,0){\setlength{\unitlength}{2.5cm}\begin{picture}(0,0)(0,-1)
   \put(0,0){\setlength{\unitlength}{1cm}\begin{picture}(0,0)(0,0)
        \put(0,0){\line(-1,0){0.3}}
        \end{picture}}
   \end{picture}}

\thinlines
\put(7250,0){\setlength{\unitlength}{2.5cm}\begin{picture}(0,0)(0,-1)
   \multiput(0,0)(0,0.1){10}{\setlength{\unitlength}{1cm}%
\begin{picture}(0,0)(0,0)
        \put(0,0){\line(1,0){0.12}}
        \end{picture}}
   \end{picture}}

\put(7250,0){\setlength{\unitlength}{2.5cm}\begin{picture}(0,0)(0,-1)
   \multiput(0,0)(0,-0.1){10}{\setlength{\unitlength}{1cm}%
\begin{picture}(0,0)(0,0)
        \put(0,0){\line(1,0){0.12}}
        \end{picture}}
   \end{picture}}

\put(9250,0){\setlength{\unitlength}{2.5cm}\begin{picture}(0,0)(0,-1)
   \multiput(0,0)(0,0.1){10}{\setlength{\unitlength}{1cm}%
\begin{picture}(0,0)(0,0)
        \put(0,0){\line(-1,0){0.12}}
        \end{picture}}
   \end{picture}}

\put(9250,0){\setlength{\unitlength}{2.5cm}\begin{picture}(0,0)(0,-1)
   \multiput(0,0)(0,-0.1){10}{\setlength{\unitlength}{1cm}%
\begin{picture}(0,0)(0,0)
        \put(0,0){\line(-1,0){0.12}}
        \end{picture}}
   \end{picture}}

\put(7250,0){\setlength{\unitlength}{2.5cm}\begin{picture}(0,0)(0,-1)
   \put(0,0.4){\setlength{\unitlength}{1cm}\begin{picture}(0,0)(0,0)
        \put(0,0){\line(1,0){0.12}}
        \put(-0.2,0){\makebox(0,0)[r]{\bf 0.4}}
        \end{picture}}
   \put(0,0.2){\setlength{\unitlength}{1cm}\begin{picture}(0,0)(0,0)
        \put(0,0){\line(1,0){0.12}}
        \put(-0.2,0){\makebox(0,0)[r]{\bf 0.2}}
        \end{picture}}
   \put(0,-0.2){\setlength{\unitlength}{1cm}\begin{picture}(0,0)(0,0)
        \put(0,0){\line(1,0){0.12}}
        \put(-0.2,0){\makebox(0,0)[r]{\bf -0.2}}
        \end{picture}}
   \put(0,-0.4){\setlength{\unitlength}{1cm}\begin{picture}(0,0)(0,0)
        \put(0,0){\line(1,0){0.12}}
        \put(-0.2,0){\makebox(0,0)[r]{\bf -0.4}}
        \end{picture}}
   \put(0,0.6){\setlength{\unitlength}{1cm}\begin{picture}(0,0)(0,0)
        \put(0,0){\line(1,0){0.12}}
        \put(-0.2,0){\makebox(0,0)[r]{\bf 0.6}}
        \end{picture}}
   \put(0,-0.6){\setlength{\unitlength}{1cm}\begin{picture}(0,0)(0,0)
        \put(0,0){\line(1,0){0.12}}
        \put(-0.2,0){\makebox(0,0)[r]{\bf -0.6}}
        \end{picture}}
   \put(0,-0.8){\setlength{\unitlength}{1cm}\begin{picture}(0,0)(0,0)
        \put(0,0){\line(1,0){0.12}}
        \put(-0.2,0){\makebox(0,0)[r]{\bf -0.8}}
        \end{picture}}
   \put(0,0.0){\setlength{\unitlength}{1cm}\begin{picture}(0,0)(0,0)
        \put(0,0){\line(1,0){0.12}}
        \put(-0.2,0){\makebox(0,0)[r]{\bf 0.0}}
        \end{picture}}
   \put(0,0.8){\setlength{\unitlength}{1cm}\begin{picture}(0,0)(0,0)
        \put(0,0){\line(1,0){0.12}}
        \put(-0.2,0){\makebox(0,0)[r]{\bf 0.8}}
        \end{picture}}
   \end{picture}}

   \put(7527.5, 588.235){\setlength{\unitlength}{1cm}\begin{picture}(0,0)(0,0)
        \put(0,0){\line(0,-1){0.2}}
   \end{picture}}
   \put(7892.5, 588.235){\setlength{\unitlength}{1cm}\begin{picture}(0,0)(0,0)
        \put(0,0){\line(0,-1){0.2}}
   \end{picture}}
   \put(8257.5, 588.235){\setlength{\unitlength}{1cm}\begin{picture}(0,0)(0,0)
        \put(0,0){\line(0,-1){0.2}}
  \end{picture}}
   \put(8622.5, 588.235){\setlength{\unitlength}{1cm}\begin{picture}(0,0)(0,0)
        \put(0,0){\line(0,-1){0.2}}
   \end{picture}}
   \put(8987.5, 588.235){\setlength{\unitlength}{1cm}\begin{picture}(0,0)(0,0)
        \put(0,0){\line(0,-1){0.2}}
   \end{picture}}

    \multiput(7250,0)(50,0){40}%
        {\setlength{\unitlength}{1cm}\begin{picture}(0,0)(0,0)
        \put(0,0){\line(0,1){0.12}}
    \end{picture}}
    \put(7500,0){\setlength{\unitlength}{1cm}\begin{picture}(0,0)(0,0)
        \put(0,0){\line(0,1){0.2}}
    \end{picture}}
    \put(8000,0){\setlength{\unitlength}{1cm}\begin{picture}(0,0)(0,0)
        \put(0,0){\line(0,1){0.2}}
    \end{picture}}
   \put(7750,0){\setlength{\unitlength}{1cm}\begin{picture}(0,0)(0,0)
        \put(0,0){\line(0,1){0.2}}
    \end{picture}}
   \put(8250,0){\setlength{\unitlength}{1cm}\begin{picture}(0,0)(0,0)
        \put(0,0){\line(0,1){0.2}}
    \end{picture}}
    \put(8500,0){\setlength{\unitlength}{1cm}\begin{picture}(0,0)(0,0)
        \put(0,0){\line(0,1){0.2}}
    \end{picture}}
    \put(8750,0){\setlength{\unitlength}{1cm}\begin{picture}(0,0)(0,0)
       \put(0,0){\line(0,1){0.2}}
    \end{picture}}
    \put(9000,0){\setlength{\unitlength}{1cm}\begin{picture}(0,0)(0,0)
       \put(0,0){\line(0,1){0.2}}
    \end{picture}}

%
\punktg{17/10/89}{03:45}{7816.657}{  0.041}{0.011}{1052}{2.21}{
1727}{1.2CA}
\punktg{01/11/89}{04:39}{7831.694}{ -0.180}{0.011}{ 500}{1.37}{
553}{1.2CA}
\punktg{02/11/89}{04:10}{7832.674}{ -0.173}{0.015}{ 500}{1.72}{
578}{1.2CA}
\punktg{03/11/89}{04:17}{7833.679}{ -0.228}{0.021}{ 500}{2.47}{
589}{1.2CA}
\punktg{20/12/89}{02:09}{7880.590}{ -0.390}{0.023}{ 100}{2.11}{
521}{1.2CA}
\punktg{20/12/89}{02:13}{7880.593}{ -0.395}{0.022}{ 250}{2.30}{
595}{1.2CA}
\punktg{20/12/89}{04:12}{7880.675}{ -0.393}{0.011}{ 500}{2.09}{
727}{1.2CA}
\punktg{21/12/89}{00:42}{7881.530}{ -0.377}{0.021}{ 110}{2.13}{
480}{1.2CA}
\punktg{21/12/89}{00:50}{7881.535}{ -0.408}{0.017}{ 500}{2.89}{
597}{1.2CA}
\punktg{21/12/89}{03:13}{7881.635}{ -0.411}{0.010}{ 500}{1.94}{
627}{1.2CA}
\punktg{27/02/90}{22:01}{7950.418}{  0.063}{0.032}{  50}{1.45}{
500}{1.2CA}
\punktg{27/02/90}{22:13}{7950.426}{  0.071}{0.013}{ 500}{1.48}{
661}{1.2CA}
\punktg{24/05/90}{20:46}{8036.366}{ -0.558}{0.025}{ 400}{1.67}{
2778}{1.2CA}
\punktg{24/05/90}{21:13}{8036.384}{ -0.570}{0.023}{ 256}{1.60}{
1888}{1.2CA}
\punktg{24/05/90}{21:23}{8036.391}{ -0.541}{0.022}{ 400}{1.60}{
2777}{1.2CA}
\punktg{22/12/90}{13:28}{8248.061}{ -0.579}{0.037}{ 500}{2.70}{
638}{1.2CA}
\punktg{09/02/91}{21:14}{8297.385}{  0.225}{0.013}{ 100}{1.62}{
329}{1.2CA}
\punktg{09/02/91}{21:19}{8297.389}{  0.229}{0.025}{ 100}{1.72}{
338}{1.2CA}
\punktg{09/02/91}{21:24}{8297.392}{  0.204}{0.017}{ 300}{1.84}{
460}{1.2CA}
\punktg{16/02/91}{00:17}{8303.512}{  0.183}{0.020}{ 100}{1.89}{
309}{1.2CA}
\punktg{16/02/91}{00:25}{8303.518}{  0.197}{0.020}{ 300}{1.84}{
413}{1.2CA}
\punktg{22/02/91}{23:29}{8310.479}{  0.104}{0.010}{ 100}{1.32}{
664}{1.2CA}
\punktg{22/02/91}{23:35}{8310.483}{  0.106}{0.015}{ 300}{1.35}{
636}{1.2CA}
\punktg{23/02/91}{23:27}{8311.478}{ -0.053}{0.015}{ 100}{1.22}{
627}{1.2CA}
\punktg{23/02/91}{23:36}{8311.484}{ -0.046}{0.014}{ 300}{1.20}{
776}{1.2CA}
\punktg{25/02/91}{02:27}{8312.602}{ -0.016}{0.019}{ 100}{2.06}{
513}{1.2CA}
\punktg{25/02/91}{02:34}{8312.607}{  0.010}{0.011}{ 300}{2.44}{
1013}{1.2CA}
\punktg{24/05/91}{20:38}{8401.360}{  0.400}{0.024}{ 100}{2.09}{
535}{1.2CA}
\punktg{24/05/91}{20:43}{8401.364}{  0.382}{0.030}{ 300}{2.03}{
1042}{1.2CA}
\punktgg{25/05/91}{20:47}{8402.366}{  0.242}{0.057}{ 100}{3.74}{
541}{1.2CA}
\punktgg{25/05/91}{20:54}{8402.371}{  0.251}{0.065}{ 300}{4.36}{
1110}{1.2CA}
\punktg{19/10/91}{04:15}{8548.677}{  0.611}{0.018}{ 100}{1.99}{
338}{1.2CA}
\punktg{19/10/91}{04:22}{8548.683}{  0.607}{0.013}{ 500}{2.16}{
649}{1.2CA}
\punktg{21/10/91}{04:15}{8550.678}{  0.677}{0.018}{ 100}{2.05}{
350}{1.2CA}
\punktg{21/10/91}{04:23}{8550.683}{  0.683}{0.010}{ 500}{1.69}{
669}{1.2CA}
\punktg{27/10/91}{05:16}{8556.720}{  0.650}{0.023}{ 100}{2.46}{
560}{1.2CA}
\punktg{27/10/91}{05:24}{8556.725}{  0.647}{0.024}{ 500}{2.21}{
2093}{1.2CA}
\punktg{10/01/92}{01:27}{8631.561}{ -0.422}{0.009}{ 240}{2.48}{
1433}{1.2CA}
\punktg{10/01/92}{01:32}{8631.564}{ -0.345}{0.010}{ 240}{2.40}{
1442}{1.2CA}
\punktg{10/01/92}{01:32}{8631.564}{ -0.437}{0.021}{ 240}{2.35}{
1437}{1.2CA}
\punktg{10/01/92}{01:36}{8631.567}{ -0.425}{0.017}{ 240}{2.42}{
1472}{1.2CA}
\punktg{10/01/92}{01:37}{8631.567}{ -0.374}{0.019}{ 240}{2.43}{
1470}{1.2CA}
\punktg{10/01/92}{01:41}{8631.571}{ -0.391}{0.017}{ 240}{2.46}{
1478}{1.2CA}
\punktg{10/01/92}{01:41}{8631.571}{ -0.375}{0.017}{ 240}{2.48}{
1479}{1.2CA}
\punktg{10/01/92}{05:26}{8631.727}{ -0.402}{0.017}{ 240}{3.22}{
2001}{1.2CA}
\punktg{10/01/92}{05:26}{8631.727}{ -0.414}{0.016}{ 240}{3.21}{
2000}{1.2CA}
\punktg{10/01/92}{05:32}{8631.731}{ -0.426}{0.018}{ 300}{3.45}{
2442}{1.2CA}
\punktg{10/01/92}{05:32}{8631.731}{ -0.439}{0.011}{ 300}{3.62}{
2425}{1.2CA}
\punktg{10/01/92}{05:38}{8631.735}{ -0.388}{0.030}{ 300}{3.70}{
2444}{1.2CA}
\punktg{10/01/92}{05:38}{8631.735}{ -0.402}{0.013}{ 300}{3.84}{
2415}{1.2CA}
\punktgg{02/02/92}{01:53}{8654.579}{ -0.095}{0.041}{ 100}{1.65}{
303}{1.2CA}
\punktg{02/02/92}{02:01}{8654.584}{ -0.116}{0.035}{ 500}{1.95}{
518}{1.2CA}
\punktg{07/02/92}{00:23}{8659.516}{ -0.046}{0.031}{ 100}{1.35}{
302}{1.2CA}
\punktgg{08/02/92}{00:53}{8660.537}{ -0.046}{0.029}{ 100}{1.28}{
307}{1.2CA}
\punktg{08/02/92}{01:04}{8660.545}{  0.020}{0.034}{ 500}{1.47}{
504}{1.2CA}
\punktg{09/02/92}{01:42}{8661.571}{  0.042}{0.026}{ 500}{1.49}{
504}{1.2CA}
\punktgg{11/02/92}{01:21}{8663.557}{  0.034}{0.037}{ 100}{0.98}{
307}{1.2CA}
\punktg{11/02/92}{01:29}{8663.562}{  0.086}{0.010}{ 500}{0.98}{
512}{1.2CA}
\punktgg{11/02/92}{22:04}{8664.420}{  0.036}{0.039}{ 100}{1.06}{
354}{1.2CA}
\punktg{11/02/92}{22:58}{8664.457}{  0.062}{0.032}{ 500}{1.17}{
679}{1.2CA}
\punktgg{13/02/92}{22:13}{8666.426}{  0.081}{0.037}{ 100}{1.61}{
544}{1.2CA}
\punktg{13/02/92}{22:20}{8666.431}{  0.176}{0.036}{ 500}{2.05}{
1629}{1.2CA}
\punktgg{14/02/92}{23:24}{8667.475}{  0.180}{0.039}{ 100}{0.96}{
852}{1.2CA}
\punktg{14/02/92}{23:32}{8667.481}{  0.154}{0.047}{ 500}{1.22}{
3042}{1.2CA}
\punktgg{06/03/92}{02:40}{8687.611}{  0.166}{0.048}{ 150}{1.51}{
245}{1.2CA}
\punktg{06/03/92}{02:48}{8687.617}{  0.049}{0.047}{ 500}{1.45}{
383}{1.2CA}
\punktg{09/03/92}{02:32}{8690.606}{  0.051}{0.050}{ 150}{1.68}{
241}{1.2CA}
\punktg{09/03/92}{02:36}{8690.609}{  0.054}{0.043}{ 150}{1.35}{
242}{1.2CA}
\punktg{11/03/92}{02:29}{8692.604}{  0.058}{0.039}{ 150}{2.04}{
262}{1.2CA}
\punktg{09/02/93}{22:32}{9028.439}{ -0.623}{0.025}{ 200}{1.81}{
1651}{1.2CA}
\punktg{09/02/93}{22:37}{9028.443}{ -0.618}{0.027}{ 200}{1.89}{
1501}{1.2CA}
\punktg{09/02/93}{22:37}{9028.443}{ -0.618}{0.027}{ 200}{1.89}{
1501}{1.2CA}

\end{picture}}

\end{picture}

\vspace*{-0.02cm}

\begin{picture}(17 ,5 )(-1,0)
\put(0,0){\setlength{\unitlength}{0.0085cm}%
\begin{picture}(2000, 588.235)(7250,0)
\put(7250,0){\framebox(2000, 588.235)[tl]{\begin{picture}(0,0)(0,0)
        \put(2000,0){\makebox(0,0)[tr]{\large{1219+285}\T{0.4}
                                 \hspace*{0.5cm}}}
        \put(2000,-
588.235){\setlength{\unitlength}{1cm}\begin{picture}(0,0)(0,0)
            \put(0,-1){\makebox(0,0)[br]{\bf J.D.\,2,440,000\,+}}
        \end{picture}}
    \end{picture}}}

\thicklines
\put(7250,0){\setlength{\unitlength}{2.5cm}\begin{picture}(0,0)(0,-1)
   \put(0,0){\setlength{\unitlength}{1cm}\begin{picture}(0,0)(0,0)
        \put(0,0){\line(1,0){0.3}}
        \end{picture}}
   \end{picture}}

\put(9250,0){\setlength{\unitlength}{2.5cm}\begin{picture}(0,0)(0,-1)
   \put(0,0){\setlength{\unitlength}{1cm}\begin{picture}(0,0)(0,0)
        \put(0,0){\line(-1,0){0.3}}
        \end{picture}}
   \end{picture}}

\thinlines
\put(7250,0){\setlength{\unitlength}{2.5cm}\begin{picture}(0,0)(0,-1)
   \multiput(0,0)(0,0.1){10}{\setlength{\unitlength}{1cm}%
\begin{picture}(0,0)(0,0)
        \put(0,0){\line(1,0){0.12}}
        \end{picture}}
   \end{picture}}

\put(7250,0){\setlength{\unitlength}{2.5cm}\begin{picture}(0,0)(0,-1)
   \multiput(0,0)(0,-0.1){10}{\setlength{\unitlength}{1cm}%
\begin{picture}(0,0)(0,0)
        \put(0,0){\line(1,0){0.12}}
        \end{picture}}
   \end{picture}}

\put(9250,0){\setlength{\unitlength}{2.5cm}\begin{picture}(0,0)(0,-1)
   \multiput(0,0)(0,0.1){10}{\setlength{\unitlength}{1cm}%
\begin{picture}(0,0)(0,0)
        \put(0,0){\line(-1,0){0.12}}
        \end{picture}}
   \end{picture}}

\put(9250,0){\setlength{\unitlength}{2.5cm}\begin{picture}(0,0)(0,-1)
   \multiput(0,0)(0,-0.1){10}{\setlength{\unitlength}{1cm}%
\begin{picture}(0,0)(0,0)
        \put(0,0){\line(-1,0){0.12}}
        \end{picture}}
   \end{picture}}

\put(7250,0){\setlength{\unitlength}{2.5cm}\begin{picture}(0,0)(0,-1)
   \put(0,0.2){\setlength{\unitlength}{1cm}\begin{picture}(0,0)(0,0)
        \put(0,0){\line(1,0){0.12}}
        \put(-0.2,0){\makebox(0,0)[r]{\bf 0.2}}
        \end{picture}}
   \put(0,0.4){\setlength{\unitlength}{1cm}\begin{picture}(0,0)(0,0)
        \put(0,0){\line(1,0){0.12}}
        \put(-0.2,0){\makebox(0,0)[r]{\bf 0.4}}
        \end{picture}}
   \put(0,-0.2){\setlength{\unitlength}{1cm}\begin{picture}(0,0)(0,0)
        \put(0,0){\line(1,0){0.12}}
        \put(-0.2,0){\makebox(0,0)[r]{\bf -0.2}}
        \end{picture}}
   \put(0,-0.4){\setlength{\unitlength}{1cm}\begin{picture}(0,0)(0,0)
        \put(0,0){\line(1,0){0.12}}
        \put(-0.2,0){\makebox(0,0)[r]{\bf -0.4}}
        \end{picture}}
   \put(0,0.6){\setlength{\unitlength}{1cm}\begin{picture}(0,0)(0,0)
        \put(0,0){\line(1,0){0.12}}
        \put(-0.2,0){\makebox(0,0)[r]{\bf 0.6}}
        \end{picture}}
   \put(0,-0.6){\setlength{\unitlength}{1cm}\begin{picture}(0,0)(0,0)
        \put(0,0){\line(1,0){0.12}}
        \put(-0.2,0){\makebox(0,0)[r]{\bf -0.6}}
        \end{picture}}
   \put(0,-0.8){\setlength{\unitlength}{1cm}\begin{picture}(0,0)(0,0)
        \put(0,0){\line(1,0){0.12}}
        \put(-0.2,0){\makebox(0,0)[r]{\bf -0.8}}
        \end{picture}}
   \put(0,0.0){\setlength{\unitlength}{1cm}\begin{picture}(0,0)(0,0)
        \put(0,0){\line(1,0){0.12}}
        \put(-0.2,0){\makebox(0,0)[r]{\bf 0.0}}
        \end{picture}}
   \put(0,0.8){\setlength{\unitlength}{1cm}\begin{picture}(0,0)(0,0)
        \put(0,0){\line(1,0){0.12}}
        \put(-0.2,0){\makebox(0,0)[r]{\bf 0.8}}
        \end{picture}}
   \end{picture}}

   \put(7527.5, 588.235){\setlength{\unitlength}{1cm}\begin{picture}(0,0)(0,0)
        \put(0,0){\line(0,-1){0.2}}
   \end{picture}}
   \put(7892.5, 588.235){\setlength{\unitlength}{1cm}\begin{picture}(0,0)(0,0)
        \put(0,0){\line(0,-1){0.2}}
   \end{picture}}
   \put(8257.5, 588.235){\setlength{\unitlength}{1cm}\begin{picture}(0,0)(0,0)
        \put(0,0){\line(0,-1){0.2}}
  \end{picture}}
   \put(8622.5, 588.235){\setlength{\unitlength}{1cm}\begin{picture}(0,0)(0,0)
        \put(0,0){\line(0,-1){0.2}}
   \end{picture}}
   \put(8987.5, 588.235){\setlength{\unitlength}{1cm}\begin{picture}(0,0)(0,0)
        \put(0,0){\line(0,-1){0.2}}
   \end{picture}}

    \multiput(7250,0)(50,0){40}%
        {\setlength{\unitlength}{1cm}\begin{picture}(0,0)(0,0)
        \put(0,0){\line(0,1){0.12}}
    \end{picture}}
    \put(7500,0){\setlength{\unitlength}{1cm}\begin{picture}(0,0)(0,0)
        \put(0,0){\line(0,1){0.2}}
        \put(0,-0.2){\makebox(0,0)[t]{\bf 7500}}
    \end{picture}}
    \put(8000,0){\setlength{\unitlength}{1cm}\begin{picture}(0,0)(0,0)
        \put(0,0){\line(0,1){0.2}}
        \put(0,-0.2){\makebox(0,0)[t]{\bf 8000}}
    \end{picture}}
   \put(7750,0){\setlength{\unitlength}{1cm}\begin{picture}(0,0)(0,0)
        \put(0,0){\line(0,1){0.2}}
        \put(0,-0.2){\makebox(0,0)[t]{\bf 7750}}
    \end{picture}}
   \put(8250,0){\setlength{\unitlength}{1cm}\begin{picture}(0,0)(0,0)
        \put(0,0){\line(0,1){0.2}}
        \put(0,-0.2){\makebox(0,0)[t]{\bf 8250}}
    \end{picture}}
    \put(8500,0){\setlength{\unitlength}{1cm}\begin{picture}(0,0)(0,0)
        \put(0,0){\line(0,1){0.2}}
        \put(0,-0.2){\makebox(0,0)[t]{\bf 8500}}
    \end{picture}}
    \put(8750,0){\setlength{\unitlength}{1cm}\begin{picture}(0,0)(0,0)
       \put(0,0){\line(0,1){0.2}}
        \put(0,-0.2){\makebox(0,0)[t]{\bf 8750}}
    \end{picture}}
    \put(9000,0){\setlength{\unitlength}{1cm}\begin{picture}(0,0)(0,0)
       \put(0,0){\line(0,1){0.2}}
        \put(0,-0.2){\makebox(0,0)[t]{\bf 9000}}
    \end{picture}}

\punktg{11/06/89}{21:09}{7689.381}{ -0.667}{0.009}{ 500}{1.32}{
1338}{1.2CA}
\punktg{15/12/89}{04:48}{7875.700}{  0.016}{0.019}{ 100}{2.95}{
573}{1.2CA}
\punktg{15/12/89}{04:56}{7875.706}{  0.040}{0.011}{ 500}{2.52}{
1064}{1.2CA}
\punktg{20/12/89}{05:11}{7880.716}{  0.277}{0.007}{ 300}{1.80}{
1051}{1.2CA}
\punktgg{28/02/90}{01:54}{7950.579}{ -0.427}{0.043}{  50}{1.46}{
505}{1.2CA}
\punktg{28/02/90}{02:02}{7950.585}{ -0.474}{0.038}{ 500}{1.61}{
600}{1.2CA}
\punktg{01/02/91}{01:44}{8288.573}{ -0.159}{0.037}{ 100}{2.51}{
715}{1.2CA}
\punktg{01/02/91}{01:50}{8288.577}{ -0.259}{0.020}{ 100}{2.60}{
658}{1.2CA}
\punktg{01/02/91}{01:59}{8288.583}{ -0.304}{0.014}{ 500}{2.61}{
2108}{1.2CA}
\punktgg{01/02/91}{23:47}{8289.491}{ -0.360}{0.100}{ 600}{3.76}{
2850}{1.2CA}
\punktgg{02/02/91}{05:07}{8289.713}{ -0.173}{0.014}{ 600}{3.83}{
4277}{1.2CA}
\punktgg{02/02/91}{05:36}{8289.733}{ -0.187}{0.115}{ 600}{3.73}{
3796}{1.2CA}
\punktg{05/02/91}{01:15}{8292.553}{ -0.005}{0.044}{ 600}{3.60}{
1031}{1.2CA}
\punktg{07/02/91}{00:28}{8294.520}{ -0.206}{0.011}{ 100}{2.26}{
326}{1.2CA}
\punktg{07/02/91}{00:32}{8294.523}{ -0.214}{0.029}{  64}{2.26}{
306}{1.2CA}
\punktgg{07/02/91}{00:39}{8294.527}{ -0.107}{0.094}{ 500}{2.40}{
492}{1.2CA}
\punktg{09/02/91}{01:05}{8296.545}{ -0.187}{0.036}{ 100}{2.50}{
316}{1.2CA}
\punktgg{09/02/91}{01:11}{8296.550}{ -0.084}{0.067}{ 200}{2.67}{
404}{1.2CA}
\punktg{09/02/91}{01:17}{8296.554}{  0.020}{0.024}{ 300}{2.48}{
481}{1.2CA}
\punktgg{09/02/91}{23:41}{8297.487}{ -0.128}{0.053}{ 100}{2.34}{
328}{1.2CA}
\punktg{09/02/91}{23:50}{8297.494}{ -0.188}{0.013}{ 500}{2.83}{
658}{1.2CA}
\punktg{15/02/91}{00:31}{8302.522}{ -0.045}{0.026}{ 100}{4.36}{
310}{1.2CA}
\punktg{15/02/91}{00:57}{8302.540}{ -0.054}{0.019}{ 500}{3.56}{
662}{1.2CA}
\punktg{15/02/91}{03:01}{8302.626}{ -0.090}{0.011}{ 100}{2.02}{
451}{1.2CA}
\punktg{15/02/91}{03:13}{8302.634}{ -0.080}{0.082}{ 500}{2.14}{
526}{1.2CA}
\punktg{15/02/91}{22:02}{8303.419}{ -0.107}{0.028}{ 100}{1.97}{
341}{1.2CA}
\punktg{15/02/91}{22:11}{8303.425}{ -0.254}{0.076}{ 300}{1.93}{
471}{1.2CA}
\punktg{16/02/91}{00:39}{8303.528}{ -0.311}{0.054}{ 300}{1.90}{
385}{1.2CA}
\punktgg{16/02/91}{01:53}{8303.579}{ -0.343}{0.057}{ 100}{1.69}{
330}{1.2CA}
\punktg{16/02/91}{02:02}{8303.585}{ -0.304}{0.056}{ 200}{1.50}{
343}{1.2CA}
\punktg{16/02/91}{04:45}{8303.698}{ -0.197}{0.013}{ 100}{1.64}{
329}{1.2CA}
\punktg{16/02/91}{04:55}{8303.705}{ -0.224}{0.026}{ 300}{1.55}{
385}{1.2CA}
\punktg{19/02/91}{23:57}{8307.498}{ -0.159}{0.060}{ 601}{2.34}{
554}{1.2CA}
\punktg{20/02/91}{01:53}{8307.579}{ -0.106}{0.046}{ 600}{2.65}{
546}{1.2CA}
\punktg{20/02/91}{04:19}{8307.680}{ -0.154}{0.048}{ 600}{2.73}{
623}{1.2CA}
\punktgg{20/02/91}{23:29}{8308.479}{ -0.289}{0.108}{ 600}{3.56}{
775}{1.2CA}
\punktg{21/02/91}{04:05}{8308.670}{ -0.308}{0.046}{ 600}{2.40}{
688}{1.2CA}
\punktg{21/02/91}{21:39}{8309.402}{ -0.133}{0.060}{ 600}{3.04}{
896}{1.2CA}
\punktg{21/02/91}{23:25}{8309.476}{ -0.125}{0.063}{ 600}{3.73}{
704}{1.2CA}
\punktg{22/02/91}{21:52}{8310.412}{  0.221}{0.048}{ 600}{3.20}{
950}{1.2CA}
\punktg{22/02/91}{23:29}{8310.479}{  0.217}{0.043}{ 601}{3.05}{
793}{1.2CA}
\punktgg{23/02/91}{03:18}{8310.638}{  0.046}{0.079}{ 100}{1.37}{
306}{1.2CA}
\punktg{23/02/91}{03:28}{8310.645}{  0.038}{0.076}{ 200}{1.51}{
364}{1.2CA}
\punktg{24/02/91}{21:35}{8312.400}{ -0.064}{0.048}{ 600}{2.40}{
1094}{1.2CA}
\punktg{24/02/91}{22:31}{8312.438}{ -0.071}{0.048}{ 600}{2.71}{
977}{1.2CA}
\punktg{24/02/91}{23:29}{8312.479}{ -0.045}{0.054}{ 601}{3.42}{
901}{1.2CA}
\punktg{25/02/91}{00:53}{8312.537}{ -0.066}{0.044}{ 601}{3.11}{
911}{1.2CA}
\punktg{25/02/91}{03:33}{8312.649}{ -0.100}{0.047}{ 600}{2.81}{
837}{1.2CA}
\punktg{25/02/91}{22:01}{8313.417}{ -0.145}{0.083}{ 600}{2.96}{
1716}{1.2CA}
\punktg{25/02/91}{22:57}{8313.457}{ -0.145}{0.081}{ 600}{3.04}{
1608}{1.2CA}
\punktg{25/02/91}{23:42}{8313.488}{ -0.191}{0.072}{ 600}{2.82}{
1444}{1.2CA}
\punktg{26/02/91}{01:30}{8313.563}{ -0.227}{0.063}{ 600}{2.86}{
1351}{1.2CA}
\punktg{26/02/91}{02:56}{8313.623}{ -0.221}{0.050}{ 600}{3.25}{
1282}{1.2CA}
\punktg{26/02/91}{04:50}{8313.702}{ -0.175}{0.072}{ 600}{3.65}{
875}{1.2CA}
\punktg{26/02/91}{22:05}{8314.421}{  0.162}{0.049}{ 600}{3.32}{
3184}{1.2CA}
\punktg{26/02/91}{23:10}{8314.466}{  0.080}{0.046}{ 600}{3.58}{
2528}{1.2CA}
\punktg{27/02/91}{00:22}{8314.515}{  0.120}{0.045}{ 600}{3.48}{
2279}{1.2CA}
\punktg{27/02/91}{01:34}{8314.566}{  0.110}{0.044}{ 600}{2.97}{
2110}{1.2CA}
\punktg{27/02/91}{02:56}{8314.623}{  0.130}{0.121}{ 600}{3.20}{
2271}{1.2CA}
\punktg{27/02/91}{04:14}{8314.676}{  0.166}{0.044}{ 600}{3.83}{
2068}{1.2CA}
\punktg{27/02/91}{05:02}{8314.710}{  0.188}{0.142}{ 600}{3.97}{
1539}{1.2CA}
\punktg{19/03/91}{04:40}{8334.695}{  0.179}{0.012}{ 100}{2.51}{
360}{1.2CA}
\punktg{20/03/91}{00:23}{8335.516}{  0.131}{0.090}{ 100}{1.38}{
300}{1.2CA}
\punktg{20/03/91}{00:41}{8335.529}{  0.115}{0.104}{ 500}{1.36}{
450}{1.2CA}
\punktg{22/05/91}{00:01}{8398.501}{  0.156}{0.084}{ 200}{1.99}{
567}{1.2CA}
\punktg{02/02/92}{05:12}{8654.717}{  0.720}{0.018}{ 500}{1.76}{
493}{1.2CA}
\punktg{03/02/92}{04:59}{8655.708}{  0.608}{0.040}{ 500}{1.74}{
497}{1.2CA}
\punktg{09/02/92}{04:57}{8661.706}{  0.203}{0.052}{ 500}{1.04}{
456}{1.2CA}
\punktg{12/02/92}{04:53}{8664.704}{  0.391}{0.011}{ 100}{1.29}{
323}{1.2CA}
\punktg{12/02/92}{04:57}{8664.707}{  0.366}{0.055}{ 100}{1.27}{
314}{1.2CA}
\punktg{12/02/92}{05:09}{8664.715}{  0.273}{0.055}{ 500}{1.29}{
504}{1.2CA}
\punktgg{14/02/92}{02:10}{8666.591}{  0.597}{0.062}{ 100}{1.97}{
350}{1.2CA}
\punktg{14/02/92}{02:23}{8666.600}{  0.450}{0.033}{ 500}{2.51}{
708}{1.2CA}
\punktg{15/02/92}{03:43}{8667.655}{  0.434}{0.025}{ 100}{1.37}{
360}{1.2CA}
\punktg{15/02/92}{03:54}{8667.663}{  0.383}{0.054}{ 500}{1.23}{
738}{1.2CA}
\punktg{16/02/92}{03:22}{8668.640}{  0.328}{0.035}{ 100}{1.31}{
506}{1.2CA}
\punktg{16/02/92}{03:32}{8668.647}{  0.266}{0.052}{ 300}{1.45}{
1043}{1.2CA}
\punktg{16/02/92}{03:32}{8668.647}{  0.266}{0.052}{ 300}{1.45}{
1043}{1.2CA}

\end{picture}}

\end{picture}

\addtocounter{figure}{-1}

\vspace*{1cm}
\caption{(continued)}
\end{figure*}

\begin{figure*}

\vspace*{0.5cm}

\begin{picture}(17 ,3 )(-1,0)
\put(0,0){\setlength{\unitlength}{0.0085cm}%
\begin{picture}(2000, 352.941)(7250,0)
\put(7250,0){\framebox(2000, 352.941)[tl]{\begin{picture}(0,0)(0,0)
        \put(0,0){\makebox(0,0)[tr]{$\Delta R$\hspace*{0.2cm}}}
        \put(2000,0){\makebox(0,0)[tr]{\large{1229+645}\T{2.2}
                                 \hspace*{0.5cm}}}
        \put(2000,-
352.941){\setlength{\unitlength}{1cm}\begin{picture}(0,0)(0,0)
        \end{picture}}
    \end{picture}}}

\thicklines
\put(7250,0){\setlength{\unitlength}{2.5cm}\begin{picture}(0,0)(0,-0.6)
   \put(0,0){\setlength{\unitlength}{1cm}\begin{picture}(0,0)(0,0)
        \put(0,0){\line(1,0){0.3}}
        \end{picture}}
   \end{picture}}

\put(9250,0){\setlength{\unitlength}{2.5cm}\begin{picture}(0,0)(0,-0.6)
   \put(0,0){\setlength{\unitlength}{1cm}\begin{picture}(0,0)(0,0)
        \put(0,0){\line(-1,0){0.3}}
        \end{picture}}
   \end{picture}}

\thinlines
\put(7250,0){\setlength{\unitlength}{2.5cm}\begin{picture}(0,0)(0,-0.6)
   \multiput(0,0)(0,0.1){6}{\setlength{\unitlength}{1cm}%
\begin{picture}(0,0)(0,0)
        \put(0,0){\line(1,0){0.12}}
        \end{picture}}
   \end{picture}}

\put(7250,0){\setlength{\unitlength}{2.5cm}\begin{picture}(0,0)(0,-0.6)
   \multiput(0,0)(0,-0.1){6}{\setlength{\unitlength}{1cm}%
\begin{picture}(0,0)(0,0)
        \put(0,0){\line(1,0){0.12}}
        \end{picture}}
   \end{picture}}

\put(9250,0){\setlength{\unitlength}{2.5cm}\begin{picture}(0,0)(0,-0.6)
   \multiput(0,0)(0,0.1){6}{\setlength{\unitlength}{1cm}%
\begin{picture}(0,0)(0,0)
        \put(0,0){\line(-1,0){0.12}}
        \end{picture}}
   \end{picture}}

\put(9250,0){\setlength{\unitlength}{2.5cm}\begin{picture}(0,0)(0,-0.6)
   \multiput(0,0)(0,-0.1){6}{\setlength{\unitlength}{1cm}%
\begin{picture}(0,0)(0,0)
        \put(0,0){\line(-1,0){0.12}}
        \end{picture}}
   \end{picture}}

\put(7250,0){\setlength{\unitlength}{2.5cm}\begin{picture}(0,0)(0,-0.6)
   \put(0,0.2){\setlength{\unitlength}{1cm}\begin{picture}(0,0)(0,0)
        \put(0,0){\line(1,0){0.12}}
        \put(-0.2,0){\makebox(0,0)[r]{\bf 0.2}}
        \end{picture}}
   \put(0,0.2){\setlength{\unitlength}{1cm}\begin{picture}(0,0)(0,0)
        \put(0,0){\line(1,0){0.12}}
        \put(-0.2,0){\makebox(0,0)[r]{\bf 0.2}}
        \end{picture}}
   \put(0,-0.2){\setlength{\unitlength}{1cm}\begin{picture}(0,0)(0,0)
        \put(0,0){\line(1,0){0.12}}
        \put(-0.2,0){\makebox(0,0)[r]{\bf -0.2}}
        \end{picture}}
   \put(0,-0.4){\setlength{\unitlength}{1cm}\begin{picture}(0,0)(0,0)
        \put(0,0){\line(1,0){0.12}}
        \put(-0.2,0){\makebox(0,0)[r]{\bf -0.4}}
        \end{picture}}
   \put(0,0.0){\setlength{\unitlength}{1cm}\begin{picture}(0,0)(0,0)
        \put(0,0){\line(1,0){0.12}}
        \put(-0.2,0){\makebox(0,0)[r]{\bf 0.0}}
        \end{picture}}
   \put(0,0.0){\setlength{\unitlength}{1cm}\begin{picture}(0,0)(0,0)
        \put(0,0){\line(1,0){0.12}}
        \put(-0.2,0){\makebox(0,0)[r]{\bf 0.0}}
        \end{picture}}
   \put(0,0.0){\setlength{\unitlength}{1cm}\begin{picture}(0,0)(0,0)
        \put(0,0){\line(1,0){0.12}}
        \put(-0.2,0){\makebox(0,0)[r]{\bf 0.0}}
        \end{picture}}
   \put(0,0.0){\setlength{\unitlength}{1cm}\begin{picture}(0,0)(0,0)
        \put(0,0){\line(1,0){0.12}}
        \put(-0.2,0){\makebox(0,0)[r]{\bf 0.0}}
        \end{picture}}
   \put(0,0.4){\setlength{\unitlength}{1cm}\begin{picture}(0,0)(0,0)
        \put(0,0){\line(1,0){0.12}}
        \end{picture}}
   \end{picture}}

   \put(7527.5, 352.941){\setlength{\unitlength}{1cm}\begin{picture}(0,0)(0,0)
        \put(0,0){\line(0,-1){0.2}}
        \put(0,0.2){\makebox(0,0)[b]{\bf 1989}}
   \end{picture}}
   \put(7892.5, 352.941){\setlength{\unitlength}{1cm}\begin{picture}(0,0)(0,0)
        \put(0,0){\line(0,-1){0.2}}
        \put(0,0.2){\makebox(0,0)[b]{\bf 1990}}
   \end{picture}}
   \put(8257.5, 352.941){\setlength{\unitlength}{1cm}\begin{picture}(0,0)(0,0)
        \put(0,0){\line(0,-1){0.2}}
        \put(0,0.2){\makebox(0,0)[b]{\bf 1991}}
  \end{picture}}
   \put(8622.5, 352.941){\setlength{\unitlength}{1cm}\begin{picture}(0,0)(0,0)
        \put(0,0){\line(0,-1){0.2}}
        \put(0,0.2){\makebox(0,0)[b]{\bf 1992}}
   \end{picture}}
   \put(8987.5, 352.941){\setlength{\unitlength}{1cm}\begin{picture}(0,0)(0,0)
        \put(0,0){\line(0,-1){0.2}}
        \put(0,0.2){\makebox(0,0)[b]{\bf 1993}}
   \end{picture}}

    \multiput(7250,0)(50,0){40}%
        {\setlength{\unitlength}{1cm}\begin{picture}(0,0)(0,0)
        \put(0,0){\line(0,1){0.12}}
    \end{picture}}
    \put(7500,0){\setlength{\unitlength}{1cm}\begin{picture}(0,0)(0,0)
        \put(0,0){\line(0,1){0.2}}
    \end{picture}}
    \put(8000,0){\setlength{\unitlength}{1cm}\begin{picture}(0,0)(0,0)
        \put(0,0){\line(0,1){0.2}}
    \end{picture}}
   \put(7750,0){\setlength{\unitlength}{1cm}\begin{picture}(0,0)(0,0)
        \put(0,0){\line(0,1){0.2}}
    \end{picture}}
   \put(8250,0){\setlength{\unitlength}{1cm}\begin{picture}(0,0)(0,0)
        \put(0,0){\line(0,1){0.2}}
    \end{picture}}
    \put(8500,0){\setlength{\unitlength}{1cm}\begin{picture}(0,0)(0,0)
        \put(0,0){\line(0,1){0.2}}
    \end{picture}}
    \put(8750,0){\setlength{\unitlength}{1cm}\begin{picture}(0,0)(0,0)
       \put(0,0){\line(0,1){0.2}}
    \end{picture}}
    \put(9000,0){\setlength{\unitlength}{1cm}\begin{picture}(0,0)(0,0)
       \put(0,0){\line(0,1){0.2}}
    \end{picture}}

\punktf{12/06/89}{22:33}{7690.440}{ -0.161}{0.017}{ 500}{1.45}{
1145}{1.2CA}
\punktf{30/06/89}{20:50}{7708.368}{ -0.254}{0.020}{ 300}{1.19}{
3440}{1.2CA}
\punktf{14/08/89}{20:33}{7753.356}{  0.049}{0.024}{ 500}{1.91}{
1808}{1.2CA}
\punktf{21/05/90}{22:41}{8033.445}{ -0.205}{0.011}{ 500}{1.94}{
2093}{1.2CA}
\punktf{01/08/90}{20:54}{8105.371}{  0.107}{0.036}{ 500}{2.28}{
1118}{1.2CA}
\punktff{15/02/91}{21:16}{8303.387}{  0.074}{0.056}{ 100}{1.91}{
387}{1.2CA}
\punktf{15/02/91}{21:24}{8303.392}{  0.202}{0.038}{ 300}{2.18}{
432}{1.2CA}
\punktf{23/02/91}{01:55}{8310.580}{ -0.158}{0.041}{ 100}{0.98}{
479}{1.2CA}
\punktf{23/02/91}{02:03}{8310.586}{ -0.088}{0.030}{ 500}{1.21}{
643}{1.2CA}
\punktff{24/02/91}{01:44}{8311.573}{  0.024}{0.073}{ 100}{1.39}{
733}{1.2CA}
\punktf{23/05/91}{22:27}{8400.435}{ -0.140}{0.055}{ 100}{1.08}{
396}{1.2CA}
\punktf{23/05/91}{22:31}{8400.438}{ -0.140}{0.052}{ 100}{1.14}{
398}{1.2CA}
\punktf{23/05/91}{22:35}{8400.442}{ -0.131}{0.049}{ 300}{1.14}{
668}{1.2CA}
\punktf{02/02/92}{05:49}{8654.743}{  0.036}{0.025}{ 500}{1.73}{
512}{1.2CA}
\punktf{09/02/93}{03:04}{9027.628}{  0.260}{0.005}{ 200}{1.68}{
1406}{1.2CA}
\punktf{09/02/93}{03:12}{9027.634}{  0.297}{0.005}{ 399}{1.82}{
2622}{1.2CA}
\punktf{09/02/93}{03:12}{9027.634}{  0.297}{0.005}{ 399}{1.82}{
2622}{1.2CA}

\end{picture}}

\end{picture}

\vspace*{-0.02cm}

\begin{picture}(17 ,10 )(-1,0)
\put(0,0){\setlength{\unitlength}{0.0085cm}%
\begin{picture}(2000,1176.470)(7250,0)
\put(7250,0){\framebox(2000,1176.470)[tl]{\begin{picture}(0,0)(0,0)
        \put(2000,0){\makebox(0,0)[tr]{\large{1253$-$055}\T{0.4}
                                 \hspace*{0.5cm}}}

\put(2000,-1176.470){\setlength{\unitlength}{1cm}\begin{picture}(0,0)(0,0)
            \put(0,-1){\makebox(0,0)[br]{\bf J.D.\,2,440,000\,+}}
        \end{picture}}
    \end{picture}}}

\thicklines
\put(7250,0){\setlength{\unitlength}{2.5cm}\begin{picture}(0,0)(0,-2)
   \put(0,0){\setlength{\unitlength}{1cm}\begin{picture}(0,0)(0,0)
        \put(0,0){\line(1,0){0.3}}
        \end{picture}}
   \end{picture}}

\put(9250,0){\setlength{\unitlength}{2.5cm}\begin{picture}(0,0)(0,-2)
   \put(0,0){\setlength{\unitlength}{1cm}\begin{picture}(0,0)(0,0)
        \put(0,0){\line(-1,0){0.3}}
        \end{picture}}
   \end{picture}}

\thinlines
\put(7250,0){\setlength{\unitlength}{2.5cm}\begin{picture}(0,0)(0,-2)
   \multiput(0,0)(0,0.1){20}{\setlength{\unitlength}{1cm}%
\begin{picture}(0,0)(0,0)
        \put(0,0){\line(1,0){0.12}}
        \end{picture}}
   \end{picture}}

\put(7250,0){\setlength{\unitlength}{2.5cm}\begin{picture}(0,0)(0,-2)
   \multiput(0,0)(0,-0.1){20}{\setlength{\unitlength}{1cm}%
\begin{picture}(0,0)(0,0)
        \put(0,0){\line(1,0){0.12}}
        \end{picture}}
   \end{picture}}

\put(9250,0){\setlength{\unitlength}{2.5cm}\begin{picture}(0,0)(0,-2)
   \multiput(0,0)(0,0.1){20}{\setlength{\unitlength}{1cm}%
\begin{picture}(0,0)(0,0)
        \put(0,0){\line(-1,0){0.12}}
        \end{picture}}
   \end{picture}}

\put(9250,0){\setlength{\unitlength}{2.5cm}\begin{picture}(0,0)(0,-2)
   \multiput(0,0)(0,-0.1){20}{\setlength{\unitlength}{1cm}%
\begin{picture}(0,0)(0,0)
        \put(0,0){\line(-1,0){0.12}}
        \end{picture}}
   \end{picture}}

\put(7250,0){\setlength{\unitlength}{2.5cm}\begin{picture}(0,0)(0,-2)
   \put(0,0.5){\setlength{\unitlength}{1cm}\begin{picture}(0,0)(0,0)
        \put(0,0){\line(1,0){0.12}}
        \put(-0.2,0){\makebox(0,0)[r]{\bf 0.5}}
        \end{picture}}
   \put(0,1.0){\setlength{\unitlength}{1cm}\begin{picture}(0,0)(0,0)
        \put(0,0){\line(1,0){0.12}}
        \put(-0.2,0){\makebox(0,0)[r]{\bf 1.0}}
        \end{picture}}
   \put(0,-0.5){\setlength{\unitlength}{1cm}\begin{picture}(0,0)(0,0)
        \put(0,0){\line(1,0){0.12}}
        \put(-0.2,0){\makebox(0,0)[r]{\bf -0.5}}
        \end{picture}}
   \put(0,-1.0){\setlength{\unitlength}{1cm}\begin{picture}(0,0)(0,0)
        \put(0,0){\line(1,0){0.12}}
        \put(-0.2,0){\makebox(0,0)[r]{\bf -1.0}}
        \end{picture}}
   \put(0,1.5){\setlength{\unitlength}{1cm}\begin{picture}(0,0)(0,0)
        \put(0,0){\line(1,0){0.12}}
        \put(-0.2,0){\makebox(0,0)[r]{\bf 1.5}}
        \end{picture}}
   \put(0,-1.5){\setlength{\unitlength}{1cm}\begin{picture}(0,0)(0,0)
        \put(0,0){\line(1,0){0.12}}
        \put(-0.2,0){\makebox(0,0)[r]{\bf -1.5}}
        \end{picture}}
   \put(0,0.0){\setlength{\unitlength}{1cm}\begin{picture}(0,0)(0,0)
        \put(0,0){\line(1,0){0.12}}
        \put(-0.2,0){\makebox(0,0)[r]{\bf 0.0}}
        \end{picture}}
   \put(0,0.0){\setlength{\unitlength}{1cm}\begin{picture}(0,0)(0,0)
        \put(0,0){\line(1,0){0.12}}
        \put(-0.2,0){\makebox(0,0)[r]{\bf 0.0}}
        \end{picture}}
   \put(0,0.0){\setlength{\unitlength}{1cm}\begin{picture}(0,0)(0,0)
        \put(0,0){\line(1,0){0.12}}
        \end{picture}}
   \end{picture}}

   \put(7527.5,1176.470){\setlength{\unitlength}{1cm}\begin{picture}(0,0)(0,0)
        \put(0,0){\line(0,-1){0.2}}
   \end{picture}}
   \put(7892.5,1176.470){\setlength{\unitlength}{1cm}\begin{picture}(0,0)(0,0)
        \put(0,0){\line(0,-1){0.2}}
   \end{picture}}
   \put(8257.5,1176.470){\setlength{\unitlength}{1cm}\begin{picture}(0,0)(0,0)
        \put(0,0){\line(0,-1){0.2}}
  \end{picture}}
   \put(8622.5,1176.470){\setlength{\unitlength}{1cm}\begin{picture}(0,0)(0,0)
        \put(0,0){\line(0,-1){0.2}}
   \end{picture}}
   \put(8987.5,1176.470){\setlength{\unitlength}{1cm}\begin{picture}(0,0)(0,0)
        \put(0,0){\line(0,-1){0.2}}
   \end{picture}}

    \multiput(7250,0)(50,0){40}%
        {\setlength{\unitlength}{1cm}\begin{picture}(0,0)(0,0)
        \put(0,0){\line(0,1){0.12}}
    \end{picture}}
    \put(7500,0){\setlength{\unitlength}{1cm}\begin{picture}(0,0)(0,0)
        \put(0,0){\line(0,1){0.2}}
        \put(0,-0.2){\makebox(0,0)[t]{\bf 7500}}
    \end{picture}}
    \put(8000,0){\setlength{\unitlength}{1cm}\begin{picture}(0,0)(0,0)
        \put(0,0){\line(0,1){0.2}}
        \put(0,-0.2){\makebox(0,0)[t]{\bf 8000}}
    \end{picture}}
   \put(7750,0){\setlength{\unitlength}{1cm}\begin{picture}(0,0)(0,0)
        \put(0,0){\line(0,1){0.2}}
        \put(0,-0.2){\makebox(0,0)[t]{\bf 7750}}
    \end{picture}}
   \put(8250,0){\setlength{\unitlength}{1cm}\begin{picture}(0,0)(0,0)
        \put(0,0){\line(0,1){0.2}}
        \put(0,-0.2){\makebox(0,0)[t]{\bf 8250}}
    \end{picture}}
    \put(8500,0){\setlength{\unitlength}{1cm}\begin{picture}(0,0)(0,0)
        \put(0,0){\line(0,1){0.2}}
        \put(0,-0.2){\makebox(0,0)[t]{\bf 8500}}
    \end{picture}}
    \put(8750,0){\setlength{\unitlength}{1cm}\begin{picture}(0,0)(0,0)
       \put(0,0){\line(0,1){0.2}}
        \put(0,-0.2){\makebox(0,0)[t]{\bf 8750}}
    \end{picture}}
    \put(9000,0){\setlength{\unitlength}{1cm}\begin{picture}(0,0)(0,0)
       \put(0,0){\line(0,1){0.2}}
        \put(0,-0.2){\makebox(0,0)[t]{\bf 9000}}
    \end{picture}}

\punktc{09/06/89}{08:52}{7686.870}{  0.780}{0.006}{ 100}{1.68}{
576}{1.2CA}
\punktc{09/06/89}{09:02}{7686.876}{  0.775}{0.024}{ 100}{1.79}{
581}{1.2CA}
\punktc{09/06/89}{21:27}{7687.394}{  0.771}{0.018}{ 500}{1.80}{
1040}{1.2CA}
\punktc{09/02/90}{03:53}{7931.662}{ -0.956}{0.029}{ 500}{1.61}{
1322}{1.2CA}
\punktc{16/02/91}{00:46}{8303.532}{  0.978}{0.027}{ 100}{1.90}{
313}{1.2CA}
\punktc{16/02/91}{00:53}{8303.537}{  0.963}{0.014}{ 300}{1.81}{
425}{1.2CA}
\punktc{16/02/91}{02:10}{8303.590}{  0.943}{0.019}{ 100}{1.46}{
520}{1.2CA}
\punktc{16/02/91}{02:25}{8303.601}{  0.929}{0.009}{ 300}{1.42}{
435}{1.2CA}
\punktc{23/02/91}{02:44}{8310.614}{ -0.165}{0.017}{ 100}{1.38}{
367}{1.2CA}
\punktc{23/02/91}{02:50}{8310.619}{ -0.156}{0.013}{ 300}{1.40}{
501}{1.2CA}
\punktc{15/01/92}{02:04}{8636.587}{ -0.597}{0.025}{ 300}{1.58}{
857}{1.2CA}
\punktc{17/01/92}{00:47}{8638.533}{ -0.481}{0.025}{ 300}{2.09}{
1087}{1.2CA}
\punktc{17/01/92}{23:11}{8639.466}{ -0.415}{0.029}{ 300}{1.20}{
2738}{1.2CA}
\punktc{10/02/92}{04:36}{8662.692}{ -1.805}{0.014}{ 100}{1.05}{
312}{1.2CA}
\punktc{10/02/92}{05:09}{8662.715}{ -1.795}{0.019}{ 500}{1.32}{
563}{1.2CA}
\punktc{14/02/92}{04:01}{8666.667}{ -1.416}{0.018}{ 100}{1.38}{
311}{1.2CA}
\punktc{14/02/92}{04:09}{8666.673}{ -1.398}{0.016}{ 500}{1.30}{
540}{1.2CA}
\punktcc{15/02/92}{04:21}{8667.681}{ -1.405}{0.043}{ 100}{2.03}{
323}{1.2CA}
\punktc{15/02/92}{04:32}{8667.690}{ -1.397}{0.026}{ 500}{1.82}{
565}{1.2CA}
\punktcc{08/03/92}{06:39}{8689.778}{ -0.711}{0.051}{ 150}{1.78}{
227}{1.2CA}
\punktc{08/03/92}{06:46}{8689.782}{ -0.716}{0.030}{ 300}{1.38}{
270}{1.2CA}
\punktcc{22/03/92}{07:35}{8703.816}{ -0.533}{0.046}{ 200}{3.71}{
977}{1.2CA}
\punktc{22/03/92}{07:42}{8703.821}{ -0.489}{0.033}{ 500}{2.95}{
2199}{1.2CA}
\punktc{25/03/92}{06:13}{8706.760}{ -0.231}{0.037}{ 150}{1.28}{
263}{1.2CA}
\punktc{25/03/92}{06:17}{8706.762}{ -0.215}{0.043}{ 150}{1.42}{
265}{1.2CA}
\punktc{25/03/92}{06:20}{8706.764}{ -0.218}{0.028}{ 150}{1.32}{
260}{1.2CA}
\punktc{25/03/92}{06:20}{8706.764}{ -0.218}{0.028}{ 150}{1.32}{
260}{1.2CA}

\end{picture}}

\end{picture}

\addtocounter{figure}{-1}

\vspace*{1cm}
\caption{(continued)}
\end{figure*}

\noindent
{\bf 3C\,66A (0219+428)} is another BL\,Lac object at medium redshift.
It was included in the early Rosemary Hill sample; PPSL detected
variability with a total range of 1.2\,mag in their highly undersampled
lightcurve. Miller \& McGimsey (\cite{MG78}) searched for intraday
variability, with a negative result. Eight photometric data points obtained
in the early '80s by Corso et al.\ (\cite{CSD86}) yielded a 0.42\,mag range of
variability. In the Tuorla programme (Sillanp\"a\"a et al.\ \cite{SMV91}),
the authors found a variability range of 1.4\,mag caused by an outburst in
Oct.\ 1985. Our HQM lightcurve is well sampled in some parts, it shows one
sharp peak at the end of 1989; the curve is relatively smooth compared to
other BL\,Lac objects. On our CCD frames there
is a number excess of relatively bright galaxies inside $2\arcmin$.

\noindent
{\bf AO\,0235+164.} This BL\,Lac object has sometimes been proposed as a
very good microlensing candidate (Kayser \cite{Kay88}, Stickel et al.\
\cite{SFK88}) since it has two foreground galaxies at small angular
separations (see Table~1). WSLF present a lightcurve spanning from
1975 to 1986; they
measured a 5\,mag outburst in 1979 which was of similar shape than an
outburst in 1975; another strong event occurred in 1986/87. The HQM
lightcurve shows very rapid, large amplitude variations; two parts of it
are shown in Figs.~2 and 3 with higher resolution. The most rapid variation
is displayed in Fig.~2 where the object brightened by 1.60\,mag within 47.5
hours between Feb.~20 and 22, 1989. On Feb.~15, 1989 the object faded by
0.68\,mag within 18.5 hours. Another rapid brightening of 0.67\,mag within
28.5 hours occurred between Feb.~23 and 24, 1991 (Fig.~3). Making use of
the photometric sequence of McGimsey ({\cite{McG76}) and Smith et al.\
(\cite{SBHE85}), we were able to determine the reference magnitude of our
lightcurve, $R_0=16.23$.

\begin{figure*}

\vspace*{0.5cm}

\begin{picture}(17 ,5 )(-1,0)
\put(0,0){\setlength{\unitlength}{0.0085cm}%
\begin{picture}(2000, 588.235)(7250,0)
\put(7250,0){\framebox(2000, 588.235)[tl]{\begin{picture}(0,0)(0,0)
        \put(0,0){\makebox(0,0)[tr]{$\Delta R$\hspace*{0.2cm}}}
        \put(2000,0){\makebox(0,0)[tr]{\large{1308+326}\T{4}
                                 \hspace*{0.5cm}}}
        \put(2000,-
588.235){\setlength{\unitlength}{1cm}\begin{picture}(0,0)(0,0)
        \end{picture}}
    \end{picture}}}

\thicklines
\put(7250,0){\setlength{\unitlength}{2.5cm}\begin{picture}(0,0)(0,-1)
   \put(0,0){\setlength{\unitlength}{1cm}\begin{picture}(0,0)(0,0)
        \put(0,0){\line(1,0){0.3}}
        \end{picture}}
   \end{picture}}

\put(9250,0){\setlength{\unitlength}{2.5cm}\begin{picture}(0,0)(0,-1)
   \put(0,0){\setlength{\unitlength}{1cm}\begin{picture}(0,0)(0,0)
        \put(0,0){\line(-1,0){0.3}}
        \end{picture}}
   \end{picture}}

\thinlines
\put(7250,0){\setlength{\unitlength}{2.5cm}\begin{picture}(0,0)(0,-1)
   \multiput(0,0)(0,0.1){10}{\setlength{\unitlength}{1cm}%
\begin{picture}(0,0)(0,0)
        \put(0,0){\line(1,0){0.12}}
        \end{picture}}
   \end{picture}}

\put(7250,0){\setlength{\unitlength}{2.5cm}\begin{picture}(0,0)(0,-1)
   \multiput(0,0)(0,-0.1){10}{\setlength{\unitlength}{1cm}%
\begin{picture}(0,0)(0,0)
        \put(0,0){\line(1,0){0.12}}
        \end{picture}}
   \end{picture}}

\put(9250,0){\setlength{\unitlength}{2.5cm}\begin{picture}(0,0)(0,-1)
   \multiput(0,0)(0,0.1){10}{\setlength{\unitlength}{1cm}%
\begin{picture}(0,0)(0,0)
        \put(0,0){\line(-1,0){0.12}}
        \end{picture}}
   \end{picture}}

\put(9250,0){\setlength{\unitlength}{2.5cm}\begin{picture}(0,0)(0,-1)
   \multiput(0,0)(0,-0.1){10}{\setlength{\unitlength}{1cm}%
\begin{picture}(0,0)(0,0)
        \put(0,0){\line(-1,0){0.12}}
        \end{picture}}
   \end{picture}}

\put(7250,0){\setlength{\unitlength}{2.5cm}\begin{picture}(0,0)(0,-1)
   \put(0,0.2){\setlength{\unitlength}{1cm}\begin{picture}(0,0)(0,0)
        \put(0,0){\line(1,0){0.12}}
        \put(-0.2,0){\makebox(0,0)[r]{\bf 0.2}}
        \end{picture}}
   \put(0,0.4){\setlength{\unitlength}{1cm}\begin{picture}(0,0)(0,0)
        \put(0,0){\line(1,0){0.12}}
        \put(-0.2,0){\makebox(0,0)[r]{\bf 0.4}}
        \end{picture}}
   \put(0,-0.2){\setlength{\unitlength}{1cm}\begin{picture}(0,0)(0,0)
        \put(0,0){\line(1,0){0.12}}
        \put(-0.2,0){\makebox(0,0)[r]{\bf -0.2}}
        \end{picture}}
   \put(0,-0.4){\setlength{\unitlength}{1cm}\begin{picture}(0,0)(0,0)
        \put(0,0){\line(1,0){0.12}}
        \put(-0.2,0){\makebox(0,0)[r]{\bf -0.4}}
        \end{picture}}
   \put(0,0.6){\setlength{\unitlength}{1cm}\begin{picture}(0,0)(0,0)
        \put(0,0){\line(1,0){0.12}}
        \put(-0.2,0){\makebox(0,0)[r]{\bf 0.6}}
        \end{picture}}
   \put(0,-0.6){\setlength{\unitlength}{1cm}\begin{picture}(0,0)(0,0)
        \put(0,0){\line(1,0){0.12}}
        \put(-0.2,0){\makebox(0,0)[r]{\bf -0.6}}
        \end{picture}}
   \put(0,-0.8){\setlength{\unitlength}{1cm}\begin{picture}(0,0)(0,0)
        \put(0,0){\line(1,0){0.12}}
        \put(-0.2,0){\makebox(0,0)[r]{\bf -0.8}}
        \end{picture}}
   \put(0,0.0){\setlength{\unitlength}{1cm}\begin{picture}(0,0)(0,0)
        \put(0,0){\line(1,0){0.12}}
        \put(-0.2,0){\makebox(0,0)[r]{\bf 0.0}}
        \end{picture}}
   \put(0,0.8){\setlength{\unitlength}{1cm}\begin{picture}(0,0)(0,0)
        \put(0,0){\line(1,0){0.12}}
        \end{picture}}
   \end{picture}}

   \put(7527.5, 588.235){\setlength{\unitlength}{1cm}\begin{picture}(0,0)(0,0)
        \put(0,0){\line(0,-1){0.2}}
        \put(0,0.2){\makebox(0,0)[b]{\bf 1989}}
   \end{picture}}
   \put(7892.5, 588.235){\setlength{\unitlength}{1cm}\begin{picture}(0,0)(0,0)
        \put(0,0){\line(0,-1){0.2}}
        \put(0,0.2){\makebox(0,0)[b]{\bf 1990}}
   \end{picture}}
   \put(8257.5, 588.235){\setlength{\unitlength}{1cm}\begin{picture}(0,0)(0,0)
        \put(0,0){\line(0,-1){0.2}}
        \put(0,0.2){\makebox(0,0)[b]{\bf 1991}}
  \end{picture}}
   \put(8622.5, 588.235){\setlength{\unitlength}{1cm}\begin{picture}(0,0)(0,0)
        \put(0,0){\line(0,-1){0.2}}
        \put(0,0.2){\makebox(0,0)[b]{\bf 1992}}
   \end{picture}}
   \put(8987.5, 588.235){\setlength{\unitlength}{1cm}\begin{picture}(0,0)(0,0)
        \put(0,0){\line(0,-1){0.2}}
        \put(0,0.2){\makebox(0,0)[b]{\bf 1993}}
   \end{picture}}

    \multiput(7250,0)(50,0){40}%
        {\setlength{\unitlength}{1cm}\begin{picture}(0,0)(0,0)
        \put(0,0){\line(0,1){0.12}}
    \end{picture}}
    \put(7500,0){\setlength{\unitlength}{1cm}\begin{picture}(0,0)(0,0)
        \put(0,0){\line(0,1){0.2}}
    \end{picture}}
    \put(8000,0){\setlength{\unitlength}{1cm}\begin{picture}(0,0)(0,0)
        \put(0,0){\line(0,1){0.2}}
    \end{picture}}
   \put(7750,0){\setlength{\unitlength}{1cm}\begin{picture}(0,0)(0,0)
        \put(0,0){\line(0,1){0.2}}
    \end{picture}}
   \put(8250,0){\setlength{\unitlength}{1cm}\begin{picture}(0,0)(0,0)
        \put(0,0){\line(0,1){0.2}}
    \end{picture}}
    \put(8500,0){\setlength{\unitlength}{1cm}\begin{picture}(0,0)(0,0)
        \put(0,0){\line(0,1){0.2}}
    \end{picture}}
    \put(8750,0){\setlength{\unitlength}{1cm}\begin{picture}(0,0)(0,0)
       \put(0,0){\line(0,1){0.2}}
    \end{picture}}
    \put(9000,0){\setlength{\unitlength}{1cm}\begin{picture}(0,0)(0,0)
       \put(0,0){\line(0,1){0.2}}
    \end{picture}}

\punktg{23/06/89}{22:07}{7701.422}{ -0.356}{0.068}{1000}{2.54}{
1260}{2.2CA}
\punktg{24/06/89}{22:17}{7702.429}{ -0.412}{0.032}{1000}{1.40}{
1345}{2.2CA}
\punktg{09/02/90}{04:15}{7931.677}{  0.212}{0.032}{ 500}{1.19}{
1389}{2.2CA}
\punktgg{28/02/90}{03:22}{7950.641}{ -0.455}{0.125}{ 548}{1.33}{
664}{2.2CA}
\punktg{28/02/90}{03:35}{7950.649}{ -0.422}{0.070}{ 500}{1.44}{
717}{2.2CA}
\punktg{22/05/90}{23:10}{8034.466}{ -0.394}{0.039}{ 400}{2.41}{
1356}{2.2CA}
\punktgg{23/05/90}{22:45}{8035.448}{ -0.349}{0.099}{ 400}{2.69}{
1513}{2.2CA}
\punktg{25/05/90}{00:12}{8036.508}{ -0.403}{0.023}{ 400}{2.19}{
1548}{2.2CA}
\punktg{25/05/90}{00:38}{8036.526}{ -0.423}{0.006}{ 400}{1.94}{
1736}{2.2CA}
\punktg{03/02/91}{01:18}{8290.554}{  0.096}{0.087}{ 500}{2.72}{
1423}{2.2CA}
\punktg{03/02/91}{02:23}{8290.599}{  0.051}{0.069}{ 500}{2.57}{
1353}{2.2CA}
\punktg{03/02/91}{04:17}{8290.679}{  0.076}{0.083}{ 300}{2.65}{
837}{2.2CA}
\punktgg{05/02/91}{00:51}{8292.536}{  0.007}{0.288}{1801}{3.43}{
1943}{2.2CA}
\punktg{09/02/91}{01:32}{8296.564}{ -0.055}{0.070}{ 500}{2.53}{
554}{2.2CA}
\punktg{15/02/91}{01:21}{8302.557}{  0.009}{0.051}{ 500}{3.20}{
530}{2.2CA}
\punktg{15/02/91}{02:53}{8302.620}{ -0.010}{0.030}{ 500}{2.23}{
556}{2.2CA}
\punktg{15/02/91}{22:44}{8303.447}{  0.003}{0.071}{ 500}{2.37}{
587}{2.2CA}
\punktgg{25/02/91}{03:59}{8312.666}{ -0.082}{0.242}{1800}{3.27}{
1326}{2.2CA}
\punktgg{25/02/91}{03:59}{8312.666}{ -0.083}{0.243}{1800}{3.24}{
1361}{2.2CA}
\punktgg{26/02/91}{03:22}{8313.640}{ -0.047}{0.323}{1800}{3.14}{
2438}{2.2CA}
\punktgg{20/03/91}{01:06}{8335.546}{ -0.399}{0.159}{ 100}{1.55}{
306}{2.2CA}
\punktg{20/03/91}{01:27}{8335.560}{ -0.206}{0.034}{ 600}{1.48}{
543}{2.2CA}
\punktgg{22/05/91}{00:00}{8398.501}{  0.008}{0.159}{ 100}{1.75}{
397}{2.2CA}
\punktgg{22/05/91}{00:04}{8398.503}{ -0.018}{0.159}{ 500}{2.34}{
977}{2.2CA}
\punktgg{23/05/91}{22:48}{8400.450}{ -0.155}{0.058}{ 100}{1.01}{
473}{2.2CA}
\punktgg{23/05/91}{22:57}{8400.457}{ -0.098}{0.095}{ 150}{1.40}{
518}{2.2CA}
\punktgg{28/05/91}{22:42}{8405.446}{ -0.046}{0.085}{ 500}{1.59}{
2321}{2.2CA}
\punktg{07/02/92}{04:47}{8659.700}{ -0.279}{0.019}{ 500}{1.68}{
464}{2.2CA}
\punktg{08/02/92}{05:33}{8660.732}{ -0.239}{0.016}{ 500}{1.35}{
465}{2.2CA}
\punktg{09/02/92}{05:18}{8661.721}{ -0.199}{0.018}{ 500}{0.89}{
456}{2.2CA}
\punktg{10/02/92}{05:17}{8662.721}{ -0.142}{0.035}{ 100}{0.91}{
302}{2.2CA}
\punktg{10/02/92}{05:25}{8662.726}{ -0.146}{0.012}{ 500}{1.07}{
493}{2.2CA}
\punktg{11/02/92}{03:42}{8663.654}{ -0.118}{0.020}{ 100}{0.91}{
292}{2.2CA}
\punktg{11/02/92}{03:52}{8663.662}{ -0.087}{0.033}{ 500}{1.36}{
452}{2.2CA}
\punktg{11/02/92}{04:40}{8663.695}{ -0.149}{0.018}{ 500}{0.92}{
460}{2.2CA}
\punktg{14/02/92}{04:38}{8666.693}{ -0.044}{0.015}{ 100}{1.09}{
299}{2.2CA}
\punktg{14/02/92}{04:46}{8666.699}{  0.051}{0.013}{ 500}{1.17}{
485}{2.2CA}
\punktgg{15/02/92}{05:44}{8667.739}{ -0.028}{0.185}{ 100}{1.25}{
341}{2.2CA}
\punktg{15/02/92}{05:52}{8667.745}{  0.059}{0.025}{ 500}{1.41}{
1438}{2.2CA}
\punktgg{17/02/92}{03:32}{8669.648}{ -0.076}{0.177}{ 100}{1.22}{
786}{2.2CA}
\punktg{28/02/92}{03:21}{8680.640}{ -0.052}{0.060}{ 500}{2.66}{
550}{2.2CA}
\punktg{10/02/93}{04:41}{9028.695}{  0.892}{0.026}{ 200}{3.54}{
2595}{2.2CA}
\punktg{11/02/93}{03:13}{9029.634}{  0.844}{0.003}{ 200}{1.48}{
3937}{2.2CA}
\punktg{11/02/93}{03:21}{9029.640}{  0.824}{0.003}{ 300}{1.64}{
5299}{2.2CA}
\punktg{18/02/93}{05:03}{9036.711}{  0.380}{0.014}{ 200}{1.68}{
370}{2.2CA}
\punktg{18/02/93}{05:08}{9036.714}{  0.378}{0.023}{ 200}{1.51}{
452}{2.2CA}
\punktg{18/02/93}{05:08}{9036.714}{  0.378}{0.023}{ 200}{1.51}{
452}{2.2CA}

\end{picture}}

\end{picture}

\vspace*{-0.02cm}

\begin{picture}(17 ,7 )(-1,0)
\put(0,0){\setlength{\unitlength}{0.0085cm}%
\begin{picture}(2000, 823.529)(7250,0)
\put(7250,0){\framebox(2000, 823.529)[tl]{\begin{picture}(0,0)(0,0)
        \put(2000,0){\makebox(0,0)[tr]{\large{1638+398}\T{0.4}
                                 \hspace*{0.5cm}}}
        \put(2000,-
823.529){\setlength{\unitlength}{1cm}\begin{picture}(0,0)(0,0)
        \end{picture}}
    \end{picture}}}

\thicklines
\put(7250,0){\setlength{\unitlength}{2.5cm}\begin{picture}(0,0)(0,-1.4)
   \put(0,0){\setlength{\unitlength}{1cm}\begin{picture}(0,0)(0,0)
        \put(0,0){\line(1,0){0.3}}
        \end{picture}}
   \end{picture}}

\put(9250,0){\setlength{\unitlength}{2.5cm}\begin{picture}(0,0)(0,-1.4)
   \put(0,0){\setlength{\unitlength}{1cm}\begin{picture}(0,0)(0,0)
        \put(0,0){\line(-1,0){0.3}}
        \end{picture}}
   \end{picture}}

\thinlines
\put(7250,0){\setlength{\unitlength}{2.5cm}\begin{picture}(0,0)(0,-1.4)
   \multiput(0,0)(0,0.1){14}{\setlength{\unitlength}{1cm}%
\begin{picture}(0,0)(0,0)
        \put(0,0){\line(1,0){0.12}}
        \end{picture}}
   \end{picture}}

\put(7250,0){\setlength{\unitlength}{2.5cm}\begin{picture}(0,0)(0,-1.4)
   \multiput(0,0)(0,-0.1){14}{\setlength{\unitlength}{1cm}%
\begin{picture}(0,0)(0,0)
        \put(0,0){\line(1,0){0.12}}
        \end{picture}}
   \end{picture}}

\put(9250,0){\setlength{\unitlength}{2.5cm}\begin{picture}(0,0)(0,-1.4)
   \multiput(0,0)(0,0.1){14}{\setlength{\unitlength}{1cm}%
\begin{picture}(0,0)(0,0)
        \put(0,0){\line(-1,0){0.12}}
        \end{picture}}
   \end{picture}}

\put(9250,0){\setlength{\unitlength}{2.5cm}\begin{picture}(0,0)(0,-1.4)
   \multiput(0,0)(0,-0.1){14}{\setlength{\unitlength}{1cm}%
\begin{picture}(0,0)(0,0)
        \put(0,0){\line(-1,0){0.12}}
        \end{picture}}
   \end{picture}}

\put(7250,0){\setlength{\unitlength}{2.5cm}\begin{picture}(0,0)(0,-1.4)
   \put(0,0.5){\setlength{\unitlength}{1cm}\begin{picture}(0,0)(0,0)
        \put(0,0){\line(1,0){0.12}}
        \put(-0.2,0){\makebox(0,0)[r]{\bf 0.5}}
        \end{picture}}
   \put(0,0.0){\setlength{\unitlength}{1cm}\begin{picture}(0,0)(0,0)
        \put(0,0){\line(1,0){0.12}}
        \put(-0.2,0){\makebox(0,0)[r]{\bf 0.0}}
        \end{picture}}
   \put(0,-1.0){\setlength{\unitlength}{1cm}\begin{picture}(0,0)(0,0)
        \put(0,0){\line(1,0){0.12}}
        \put(-0.2,0){\makebox(0,0)[r]{\bf -1.0}}
        \end{picture}}
   \put(0,-0.5){\setlength{\unitlength}{1cm}\begin{picture}(0,0)(0,0)
        \put(0,0){\line(1,0){0.12}}
        \put(-0.2,0){\makebox(0,0)[r]{\bf -0.5}}
        \end{picture}}
   \put(0,1.0){\setlength{\unitlength}{1cm}\begin{picture}(0,0)(0,0)
        \put(0,0){\line(1,0){0.12}}
        \put(-0.2,0){\makebox(0,0)[r]{\bf 1.0}}
        \end{picture}}
   \put(0,0.0){\setlength{\unitlength}{1cm}\begin{picture}(0,0)(0,0)
        \put(0,0){\line(1,0){0.12}}
        \put(-0.2,0){\makebox(0,0)[r]{\bf 0.0}}
        \end{picture}}
   \put(0,0.0){\setlength{\unitlength}{1cm}\begin{picture}(0,0)(0,0)
        \put(0,0){\line(1,0){0.12}}
        \put(-0.2,0){\makebox(0,0)[r]{\bf 0.0}}
        \end{picture}}
   \put(0,0.0){\setlength{\unitlength}{1cm}\begin{picture}(0,0)(0,0)
        \put(0,0){\line(1,0){0.12}}
        \put(-0.2,0){\makebox(0,0)[r]{\bf 0.0}}
        \end{picture}}
   \put(0,0.0){\setlength{\unitlength}{1cm}\begin{picture}(0,0)(0,0)
        \put(0,0){\line(1,0){0.12}}
        \end{picture}}
   \end{picture}}

   \put(7527.5, 823.529){\setlength{\unitlength}{1cm}\begin{picture}(0,0)(0,0)
        \put(0,0){\line(0,-1){0.2}}
   \end{picture}}
   \put(7892.5, 823.529){\setlength{\unitlength}{1cm}\begin{picture}(0,0)(0,0)
        \put(0,0){\line(0,-1){0.2}}
   \end{picture}}
   \put(8257.5, 823.529){\setlength{\unitlength}{1cm}\begin{picture}(0,0)(0,0)
        \put(0,0){\line(0,-1){0.2}}
  \end{picture}}
   \put(8622.5, 823.529){\setlength{\unitlength}{1cm}\begin{picture}(0,0)(0,0)
        \put(0,0){\line(0,-1){0.2}}
   \end{picture}}
   \put(8987.5, 823.529){\setlength{\unitlength}{1cm}\begin{picture}(0,0)(0,0)
        \put(0,0){\line(0,-1){0.2}}
   \end{picture}}

    \multiput(7250,0)(50,0){40}%
        {\setlength{\unitlength}{1cm}\begin{picture}(0,0)(0,0)
        \put(0,0){\line(0,1){0.12}}
    \end{picture}}
    \put(7500,0){\setlength{\unitlength}{1cm}\begin{picture}(0,0)(0,0)
        \put(0,0){\line(0,1){0.2}}
    \end{picture}}
    \put(8000,0){\setlength{\unitlength}{1cm}\begin{picture}(0,0)(0,0)
        \put(0,0){\line(0,1){0.2}}
    \end{picture}}
   \put(7750,0){\setlength{\unitlength}{1cm}\begin{picture}(0,0)(0,0)
        \put(0,0){\line(0,1){0.2}}
    \end{picture}}
   \put(8250,0){\setlength{\unitlength}{1cm}\begin{picture}(0,0)(0,0)
        \put(0,0){\line(0,1){0.2}}
    \end{picture}}
    \put(8500,0){\setlength{\unitlength}{1cm}\begin{picture}(0,0)(0,0)
        \put(0,0){\line(0,1){0.2}}
    \end{picture}}
    \put(8750,0){\setlength{\unitlength}{1cm}\begin{picture}(0,0)(0,0)
       \put(0,0){\line(0,1){0.2}}
    \end{picture}}
    \put(9000,0){\setlength{\unitlength}{1cm}\begin{picture}(0,0)(0,0)
       \put(0,0){\line(0,1){0.2}}
    \end{picture}}

\punktd{08/10/88}{18:08}{7443.256}{  1.211}{0.028}{ 500}{1.42}{
782}{1.2CA}
\punktd{10/06/89}{00:19}{7687.513}{  0.123}{0.038}{1000}{1.80}{
1263}{1.2CA}
\punktdd{14/06/89}{23:41}{7692.487}{  0.178}{0.508}{ 500}{2.06}{
1639}{1.2CA}
\punktd{25/06/89}{02:35}{7702.608}{  0.027}{0.052}{ 500}{1.49}{
997}{1.2CA}
\punktd{20/07/89}{21:22}{7728.391}{  0.078}{0.025}{ 500}{1.16}{
1111}{1.2CA}
\punktd{21/07/89}{20:50}{7729.368}{ -0.010}{0.027}{ 500}{1.83}{
1095}{1.2CA}
\punktd{22/07/89}{21:12}{7730.384}{ -0.056}{0.031}{ 500}{1.27}{
775}{1.2CA}
\punktd{24/07/89}{00:30}{7731.521}{ -0.007}{0.026}{ 500}{2.11}{
1156}{1.2CA}
\punktd{24/07/89}{10:49}{7731.951}{ -0.032}{0.024}{ 500}{1.25}{
905}{1.2CA}
\punktd{24/07/89}{22:49}{7732.451}{ -0.011}{0.026}{ 500}{1.27}{
895}{1.2CA}
\punktd{10/08/89}{09:02}{7748.877}{  0.196}{0.073}{ 500}{1.93}{
1332}{1.2CA}
\punktd{11/08/89}{21:34}{7750.399}{  0.271}{0.027}{ 500}{2.07}{
1115}{1.2CA}
\punktd{12/08/89}{21:13}{7751.384}{  0.326}{0.020}{ 979}{1.85}{
1789}{1.2CA}
\punktd{13/08/89}{22:21}{7752.431}{  0.215}{0.025}{ 500}{1.82}{
1051}{1.2CA}
\punktd{14/08/89}{22:30}{7753.438}{  0.228}{0.024}{ 500}{1.76}{
1217}{1.2CA}
\punktd{15/08/89}{22:06}{7754.421}{  0.428}{0.016}{1000}{1.32}{
2461}{1.2CA}
\punktd{16/08/89}{21:44}{7755.406}{  0.417}{0.021}{1000}{1.59}{
3441}{1.2CA}
\punktd{03/09/89}{20:29}{7773.354}{ -0.065}{0.028}{ 500}{2.57}{
728}{1.2CA}
\punktd{22/05/90}{00:27}{8033.519}{  0.238}{0.011}{ 500}{1.37}{
1686}{1.2CA}
\punktd{22/05/90}{00:37}{8033.526}{  0.237}{0.009}{ 500}{1.32}{
1702}{1.2CA}
\punktd{24/05/90}{01:34}{8035.566}{  0.292}{0.020}{ 100}{3.23}{
605}{1.2CA}
\punktd{31/07/90}{23:31}{8104.480}{ -0.406}{0.008}{ 500}{1.61}{
842}{1.2CA}
\punktd{01/08/90}{22:43}{8105.447}{ -0.485}{0.010}{ 500}{1.50}{
991}{1.2CA}
\punktd{02/08/90}{22:11}{8106.424}{ -0.424}{0.071}{ 500}{1.65}{
1370}{1.2CA}
\punktdd{29/09/90}{19:26}{8164.310}{ -0.583}{0.272}{ 100}{2.46}{
698}{1.2CA}
\punktdd{29/09/90}{19:34}{8164.316}{ -0.552}{0.251}{ 500}{2.32}{
1918}{1.2CA}
\punktd{29/09/90}{19:34}{8164.316}{ -0.561}{0.010}{ 500}{2.33}{
1923}{1.2CA}
\punktd{16/02/91}{04:17}{8303.679}{ -0.282}{0.036}{ 500}{1.52}{
513}{1.2CA}
\punktd{23/02/91}{04:23}{8310.683}{ -0.167}{0.042}{ 100}{1.16}{
306}{1.2CA}
\punktd{23/02/91}{04:32}{8310.689}{ -0.094}{0.020}{ 500}{1.05}{
509}{1.2CA}
\punktd{25/02/91}{05:32}{8312.731}{  0.014}{0.016}{ 500}{3.38}{
1909}{1.2CA}
\punktd{23/05/91}{00:04}{8399.503}{ -0.081}{0.076}{ 500}{1.40}{
696}{1.2CA}
\punktdd{26/05/91}{00:04}{8402.503}{  0.037}{0.249}{ 500}{2.31}{
1573}{1.2CA}
\punktdd{29/07/91}{21:34}{8467.399}{ -0.389}{0.281}{ 100}{2.80}{
347}{1.2CA}
\punktdd{31/07/91}{21:51}{8469.411}{ -0.066}{0.039}{ 100}{1.30}{
327}{1.2CA}
\punktd{31/07/91}{22:00}{8469.417}{ -0.039}{0.018}{ 500}{1.37}{
559}{1.2CA}
\punktdd{01/08/91}{21:17}{8470.387}{  0.089}{0.030}{ 100}{1.01}{
306}{1.2CA}
\punktd{01/08/91}{21:26}{8470.393}{ -0.028}{0.022}{ 500}{1.30}{
459}{1.2CA}
\punktdd{02/08/91}{20:58}{8471.374}{  0.032}{0.042}{ 100}{1.37}{
299}{1.2CA}
\punktd{02/08/91}{21:09}{8471.382}{  0.001}{0.016}{ 500}{1.34}{
491}{1.2CA}
\punktdd{03/08/91}{22:09}{8472.423}{  0.100}{0.055}{ 100}{1.72}{
296}{1.2CA}
\punktd{03/08/91}{22:17}{8472.429}{  0.110}{0.014}{ 500}{1.30}{
473}{1.2CA}
\punktdd{04/08/91}{21:52}{8473.412}{  0.112}{0.041}{ 100}{1.58}{
293}{1.2CA}
\punktd{04/08/91}{22:01}{8473.418}{  0.122}{0.021}{ 500}{1.57}{
480}{1.2CA}
\punktdd{05/08/91}{21:38}{8474.402}{ -0.045}{0.074}{ 100}{1.39}{
562}{1.2CA}
\punktd{05/08/91}{21:46}{8474.408}{ -0.040}{0.016}{ 500}{1.37}{
452}{1.2CA}
\punktd{05/08/91}{23:44}{8474.489}{ -0.037}{0.020}{ 500}{1.39}{
460}{1.2CA}
\punktd{06/08/91}{20:21}{8475.348}{ -0.131}{0.026}{ 500}{1.55}{
577}{1.2CA}
\punktdd{07/08/91}{00:59}{8475.541}{ -0.147}{0.069}{ 100}{1.41}{
299}{1.2CA}
\punktd{07/08/91}{01:09}{8475.548}{ -0.202}{0.026}{ 500}{1.44}{
494}{1.2CA}
\punktdd{08/08/91}{20:29}{8477.354}{ -0.290}{0.103}{ 100}{2.05}{
311}{1.2CA}
\punktd{08/08/91}{20:40}{8477.361}{ -0.199}{0.036}{ 500}{1.96}{
470}{1.2CA}
\punktdd{09/08/91}{00:57}{8477.540}{ -0.178}{0.109}{ 500}{3.04}{
531}{1.2CA}
\punktd{10/08/91}{20:19}{8479.347}{ -0.370}{0.086}{ 500}{2.80}{
551}{1.2CA}
\punktdd{11/08/91}{21:02}{8480.377}{ -0.351}{0.104}{ 100}{1.86}{
305}{1.2CA}
\punktdd{20/08/91}{00:23}{8488.516}{ -0.267}{0.161}{ 500}{2.23}{
708}{1.2CA}
\punktdd{26/08/91}{20:22}{8495.349}{ -0.261}{0.116}{ 500}{1.44}{
1366}{1.2CA}
\punktdd{22/09/92}{19:41}{8888.321}{ -0.767}{0.203}{ 200}{2.74}{
527}{1.2CA}
\punktd{}{}{9127.601}{0.089}{0.015}{}{}{}{1.2CA}%

\end{picture}}

\end{picture}

\vspace*{-0.02cm}

\begin{picture}(17 ,8.75 )(-1,0)
\put(0,0){\setlength{\unitlength}{0.0085cm}%
\begin{picture}(2000,1029.411)(7250,0)
\put(7250,0){\framebox(2000,1029.411)[tl]{\begin{picture}(0,0)(0,0)
        \put(2000,0){\makebox(0,0)[tr]{\large{3C\,345}\T{0.4}
                                 \hspace*{0.5cm}}}

\put(2000,-1029.411){\setlength{\unitlength}{1cm}\begin{picture}(0,0)(0,0)
            \put(0,-1){\makebox(0,0)[br]{\bf J.D.\,2,440,000\,+}}
        \end{picture}}
    \end{picture}}}

\thicklines
\put(7250,0){\setlength{\unitlength}{2.5cm}\begin{picture}(0,0)(0,-2)
   \put(0,0){\setlength{\unitlength}{1cm}\begin{picture}(0,0)(0,0)
        \put(0,0){\line(1,0){0.3}}
        \end{picture}}
   \end{picture}}

\put(9250,0){\setlength{\unitlength}{2.5cm}\begin{picture}(0,0)(0,-2)
   \put(0,0){\setlength{\unitlength}{1cm}\begin{picture}(0,0)(0,0)
        \put(0,0){\line(-1,0){0.3}}
        \end{picture}}
   \end{picture}}

\thinlines
\put(7250,0){\setlength{\unitlength}{2.5cm}\begin{picture}(0,0)(0,-2)
   \multiput(0,0)(0,0.1){15}{\setlength{\unitlength}{1cm}%
\begin{picture}(0,0)(0,0)
        \put(0,0){\line(1,0){0.12}}
        \end{picture}}
   \end{picture}}

\put(7250,0){\setlength{\unitlength}{2.5cm}\begin{picture}(0,0)(0,-2)
   \multiput(0,0)(0,-0.1){20}{\setlength{\unitlength}{1cm}%
\begin{picture}(0,0)(0,0)
        \put(0,0){\line(1,0){0.12}}
        \end{picture}}
   \end{picture}}

\put(9250,0){\setlength{\unitlength}{2.5cm}\begin{picture}(0,0)(0,-2)
   \multiput(0,0)(0,0.1){15}{\setlength{\unitlength}{1cm}%
\begin{picture}(0,0)(0,0)
        \put(0,0){\line(-1,0){0.12}}
        \end{picture}}
   \end{picture}}

\put(9250,0){\setlength{\unitlength}{2.5cm}\begin{picture}(0,0)(0,-2)
   \multiput(0,0)(0,-0.1){20}{\setlength{\unitlength}{1cm}%
\begin{picture}(0,0)(0,0)
        \put(0,0){\line(-1,0){0.12}}
        \end{picture}}
   \end{picture}}

\put(7250,0){\setlength{\unitlength}{2.5cm}\begin{picture}(0,0)(0,-2)
   \put(0,0.5){\setlength{\unitlength}{1cm}\begin{picture}(0,0)(0,0)
        \put(0,0){\line(1,0){0.12}}
        \put(-0.2,0){\makebox(0,0)[r]{\bf 0.5}}
        \end{picture}}
   \put(0,1.0){\setlength{\unitlength}{1cm}\begin{picture}(0,0)(0,0)
        \put(0,0){\line(1,0){0.12}}
        \put(-0.2,0){\makebox(0,0)[r]{\bf 1.0}}
        \end{picture}}
   \put(0,-0.5){\setlength{\unitlength}{1cm}\begin{picture}(0,0)(0,0)
        \put(0,0){\line(1,0){0.12}}
        \put(-0.2,0){\makebox(0,0)[r]{\bf -0.5}}
        \end{picture}}
   \put(0,-1.0){\setlength{\unitlength}{1cm}\begin{picture}(0,0)(0,0)
        \put(0,0){\line(1,0){0.12}}
        \put(-0.2,0){\makebox(0,0)[r]{\bf -1.0}}
        \end{picture}}
   \put(0,-1.5){\setlength{\unitlength}{1cm}\begin{picture}(0,0)(0,0)
        \put(0,0){\line(1,0){0.12}}
        \put(-0.2,0){\makebox(0,0)[r]{\bf -1.5}}
        \end{picture}}
   \put(0,-1.5){\setlength{\unitlength}{1cm}\begin{picture}(0,0)(0,0)
        \put(0,0){\line(1,0){0.12}}
        \put(-0.2,0){\makebox(0,0)[r]{\bf -1.5}}
        \end{picture}}
   \put(0,0.0){\setlength{\unitlength}{1cm}\begin{picture}(0,0)(0,0)
        \put(0,0){\line(1,0){0.12}}
        \put(-0.2,0){\makebox(0,0)[r]{\bf 0.0}}
        \end{picture}}
   \put(0,0.0){\setlength{\unitlength}{1cm}\begin{picture}(0,0)(0,0)
        \put(0,0){\line(1,0){0.12}}
        \put(-0.2,0){\makebox(0,0)[r]{\bf 0.0}}
        \end{picture}}
   \put(0,0.0){\setlength{\unitlength}{1cm}\begin{picture}(0,0)(0,0)
        \put(0,0){\line(1,0){0.12}}
        \end{picture}}
   \end{picture}}

   \put(7527.5,1029.411){\setlength{\unitlength}{1cm}\begin{picture}(0,0)(0,0)
        \put(0,0){\line(0,-1){0.2}}
   \end{picture}}
   \put(7892.5,1029.411){\setlength{\unitlength}{1cm}\begin{picture}(0,0)(0,0)
        \put(0,0){\line(0,-1){0.2}}
   \end{picture}}
   \put(8257.5,1029.411){\setlength{\unitlength}{1cm}\begin{picture}(0,0)(0,0)
        \put(0,0){\line(0,-1){0.2}}
  \end{picture}}
   \put(8622.5,1029.411){\setlength{\unitlength}{1cm}\begin{picture}(0,0)(0,0)
        \put(0,0){\line(0,-1){0.2}}
   \end{picture}}
   \put(8987.5,1029.411){\setlength{\unitlength}{1cm}\begin{picture}(0,0)(0,0)
        \put(0,0){\line(0,-1){0.2}}
   \end{picture}}

    \multiput(7250,0)(50,0){40}%
        {\setlength{\unitlength}{1cm}\begin{picture}(0,0)(0,0)
        \put(0,0){\line(0,1){0.12}}
    \end{picture}}
    \put(7500,0){\setlength{\unitlength}{1cm}\begin{picture}(0,0)(0,0)
        \put(0,0){\line(0,1){0.2}}
        \put(0,-0.2){\makebox(0,0)[t]{\bf 7500}}
    \end{picture}}
    \put(8000,0){\setlength{\unitlength}{1cm}\begin{picture}(0,0)(0,0)
        \put(0,0){\line(0,1){0.2}}
        \put(0,-0.2){\makebox(0,0)[t]{\bf 8000}}
    \end{picture}}
   \put(7750,0){\setlength{\unitlength}{1cm}\begin{picture}(0,0)(0,0)
        \put(0,0){\line(0,1){0.2}}
        \put(0,-0.2){\makebox(0,0)[t]{\bf 7750}}
    \end{picture}}
   \put(8250,0){\setlength{\unitlength}{1cm}\begin{picture}(0,0)(0,0)
        \put(0,0){\line(0,1){0.2}}
        \put(0,-0.2){\makebox(0,0)[t]{\bf 8250}}
    \end{picture}}
    \put(8500,0){\setlength{\unitlength}{1cm}\begin{picture}(0,0)(0,0)
        \put(0,0){\line(0,1){0.2}}
        \put(0,-0.2){\makebox(0,0)[t]{\bf 8500}}
    \end{picture}}
    \put(8750,0){\setlength{\unitlength}{1cm}\begin{picture}(0,0)(0,0)
       \put(0,0){\line(0,1){0.2}}
        \put(0,-0.2){\makebox(0,0)[t]{\bf 8750}}
    \end{picture}}
    \put(9000,0){\setlength{\unitlength}{1cm}\begin{picture}(0,0)(0,0)
       \put(0,0){\line(0,1){0.2}}
        \put(0,-0.2){\makebox(0,0)[t]{\bf 9000}}
    \end{picture}}

\punktc{05/09/88}{21:33}{7410.398}{ -1.202}{0.026}{ 500}{1.57}{
670}{1.2CA}
\punktc{07/09/88}{19:58}{7412.332}{ -1.306}{0.024}{ 500}{1.29}{
1116}{1.2CA}
\punktcc{03/10/88}{18:28}{7438.270}{ -1.343}{0.178}{ 100}{1.63}{
604}{1.2CA}
\punktc{03/10/88}{18:42}{7438.280}{ -1.193}{0.022}{ 500}{1.56}{
849}{1.2CA}
\punktc{07/10/88}{17:50}{7442.244}{ -1.179}{0.025}{ 500}{1.47}{
1453}{1.2CA}
\punktc{10/10/88}{18:27}{7445.269}{ -1.253}{0.031}{ 500}{2.04}{
759}{1.2CA}
\punktcc{13/10/88}{17:23}{7448.225}{ -1.282}{0.124}{ 500}{3.24}{
1165}{1.2CA}
\punktc{09/06/89}{23:02}{7687.460}{ -1.587}{0.014}{ 500}{1.61}{
762}{1.2CA}
\punktc{10/06/89}{23:48}{7688.492}{ -1.590}{0.015}{ 500}{1.48}{
879}{1.2CA}
\punktc{12/06/89}{02:22}{7689.599}{ -1.627}{0.022}{ 100}{1.46}{
531}{1.2CA}
\punktc{12/06/89}{02:35}{7689.608}{ -1.601}{0.016}{1000}{1.57}{
1515}{1.2CA}
\punktc{13/06/89}{00:10}{7690.507}{ -1.600}{0.011}{1000}{1.34}{
1510}{1.2CA}
\punktc{14/06/89}{23:27}{7692.478}{ -1.672}{0.067}{ 500}{1.83}{
1423}{1.2CA}
\punktc{24/06/89}{00:09}{7701.507}{ -1.633}{0.022}{ 500}{1.69}{
1062}{1.2CA}
\punktc{25/06/89}{00:40}{7702.528}{ -1.687}{0.035}{ 500}{1.18}{
1016}{1.2CA}
\punktc{26/06/89}{00:57}{7703.540}{ -1.689}{0.022}{ 500}{1.23}{
902}{1.2CA}
\punktc{27/06/89}{00:59}{7704.542}{ -1.654}{0.020}{ 500}{1.63}{
885}{1.2CA}
\punktc{28/06/89}{01:38}{7705.569}{ -1.683}{0.028}{ 500}{2.08}{
806}{1.2CA}
\punktc{20/07/89}{21:44}{7728.406}{ -1.701}{0.018}{ 500}{1.25}{
1292}{1.2CA}
\punktc{20/07/89}{23:41}{7728.487}{ -1.665}{0.018}{ 500}{1.54}{
2190}{1.2CA}
\punktc{21/07/89}{21:05}{7729.379}{ -1.695}{0.024}{ 500}{1.55}{
1144}{1.2CA}
\punktc{22/07/89}{21:26}{7730.393}{ -1.702}{0.020}{ 500}{1.28}{
784}{1.2CA}
\punktc{24/07/89}{01:02}{7731.543}{ -1.617}{0.021}{ 500}{2.08}{
1254}{1.2CA}
\punktc{24/07/89}{23:07}{7732.464}{ -1.714}{0.013}{ 500}{1.33}{
1244}{1.2CA}
\punktc{10/08/89}{21:21}{7749.390}{ -1.735}{0.030}{ 500}{1.41}{
1866}{1.2CA}
\punktc{11/08/89}{21:58}{7750.416}{ -1.679}{0.015}{1000}{1.98}{
1426}{1.2CA}
\punktc{13/08/89}{22:35}{7752.442}{ -1.707}{0.017}{ 500}{1.53}{
1137}{1.2CA}
\punktc{14/08/89}{23:01}{7753.459}{ -1.661}{0.062}{ 500}{1.63}{
1232}{1.2CA}
\punktc{15/08/89}{22:23}{7754.433}{ -1.676}{0.062}{ 500}{1.35}{
1452}{1.2CA}
\punktc{16/08/89}{22:05}{7755.420}{ -1.733}{0.009}{1000}{1.66}{
3532}{1.2CA}
\punktc{02/09/89}{20:12}{7772.342}{ -1.728}{0.021}{ 500}{1.75}{
744}{1.2CA}
\punktc{03/09/89}{20:44}{7773.364}{ -1.689}{0.017}{ 500}{2.63}{
754}{1.2CA}
\punktcc{01/10/89}{07:34}{7800.816}{ -1.546}{0.122}{ 500}{2.89}{
570}{1.2CA}
\punktc{16/10/89}{20:10}{7816.341}{ -1.558}{0.012}{1000}{3.12}{
1603}{1.2CA}
\punktc{31/10/89}{18:47}{7831.283}{ -1.558}{0.016}{ 500}{2.77}{
617}{1.2CA}
\punktc{01/11/89}{18:33}{7832.273}{ -1.552}{0.014}{ 500}{1.73}{
614}{1.2CA}
\punktc{20/12/89}{05:27}{7880.728}{ -1.390}{0.037}{ 500}{3.05}{
719}{1.2CA}
\punktc{23/05/90}{02:32}{8034.606}{ -1.523}{0.040}{ 400}{2.06}{
1507}{1.2CA}
\punktc{31/07/90}{23:14}{8104.469}{ -1.402}{0.007}{ 500}{1.61}{
865}{1.2CA}
\punktc{01/08/90}{22:58}{8105.458}{ -1.344}{0.082}{ 500}{1.53}{
1041}{1.2CA}
\punktc{01/08/90}{22:58}{8105.458}{ -1.380}{0.008}{ 500}{1.51}{
1029}{1.2CA}
\punktc{04/08/90}{21:30}{8108.396}{ -1.384}{0.027}{ 500}{2.54}{
1141}{1.2CA}
\punktc{13/08/90}{01:37}{8116.567}{ -1.199}{0.038}{ 120}{2.26}{
1899}{1.2CA}
\punktcc{13/08/90}{20:36}{8117.359}{ -1.294}{0.084}{ 300}{1.21}{
1120}{1.2CA}
\punktcc{13/08/90}{23:43}{8117.488}{ -1.291}{0.081}{ 199}{1.40}{
1279}{1.2CA}
\punktcc{15/08/90}{20:03}{8119.336}{ -1.293}{0.078}{  50}{0.96}{
621}{1.2CA}
\punktc{24/09/90}{19:44}{8159.322}{ -1.199}{0.020}{ 500}{2.84}{
1259}{1.2CA}
\punktc{25/09/90}{19:16}{8160.303}{ -1.213}{0.016}{ 500}{2.06}{
992}{1.2CA}
\punktc{25/09/90}{19:31}{8160.314}{ -1.149}{0.064}{ 100}{2.55}{
929}{1.2CA}
\punktc{26/09/90}{19:58}{8161.332}{ -1.103}{0.069}{ 100}{2.00}{
751}{1.2CA}
\punktc{26/09/90}{20:03}{8161.336}{ -1.209}{0.033}{ 100}{2.03}{
666}{1.2CA}
\punktc{26/09/90}{20:06}{8161.338}{ -1.259}{0.025}{ 100}{1.98}{
639}{1.2CA}
\punktc{26/09/90}{20:13}{8161.342}{ -1.254}{0.026}{ 100}{2.01}{
641}{1.2CA}
\punktc{26/09/90}{20:28}{8161.353}{ -1.212}{0.036}{ 500}{2.46}{
1401}{1.2CA}
\punktc{29/09/90}{19:10}{8164.299}{ -1.162}{0.047}{ 100}{2.46}{
749}{1.2CA}
\punktc{30/09/90}{19:06}{8165.296}{ -1.236}{0.044}{ 100}{2.62}{
680}{1.2CA}
\punktc{30/09/90}{19:16}{8165.303}{ -1.175}{0.024}{ 500}{2.38}{
1480}{1.2CA}
\punktc{30/09/90}{19:18}{8165.304}{ -1.183}{0.025}{ 500}{2.55}{
1418}{1.2CA}
\punktc{01/10/90}{19:06}{8166.296}{ -1.240}{0.040}{ 100}{2.61}{
681}{1.2CA}
\punktc{18/10/90}{19:58}{8183.333}{ -1.105}{0.016}{ 500}{2.36}{
844}{1.2CA}
\punktc{19/10/90}{19:50}{8184.327}{ -1.093}{0.024}{ 100}{2.42}{
552}{1.2CA}
\punktc{19/10/90}{19:58}{8184.333}{ -1.107}{0.017}{ 500}{2.36}{
843}{1.2CA}
\punktc{03/02/91}{04:28}{8290.687}{ -0.297}{0.036}{ 100}{2.89}{
582}{1.2CA}
\punktc{03/02/91}{04:37}{8290.692}{ -0.301}{0.012}{ 500}{2.48}{
1033}{1.2CA}
\punktc{10/02/91}{05:26}{8297.727}{ -0.154}{0.015}{ 100}{1.91}{
319}{1.2CA}
\punktc{10/02/91}{05:29}{8297.729}{ -0.170}{0.012}{ 100}{1.76}{
317}{1.2CA}
\punktc{10/02/91}{05:38}{8297.735}{ -0.150}{0.025}{ 500}{2.31}{
547}{1.2CA}
\punktc{15/02/91}{04:27}{8302.686}{ -0.180}{0.017}{ 100}{2.38}{
310}{1.2CA}
\punktc{15/02/91}{04:35}{8302.691}{ -0.185}{0.016}{ 500}{2.51}{
509}{1.2CA}
\punktc{16/02/91}{03:48}{8303.659}{  0.113}{0.014}{ 100}{1.35}{
327}{1.2CA}
\punktc{16/02/91}{03:54}{8303.663}{  0.118}{0.015}{ 300}{1.38}{
438}{1.2CA}
\punktc{20/02/91}{00:00}{8307.500}{  0.363}{0.054}{ 121}{2.19}{
379}{1.2CA}
\punktc{20/02/91}{02:55}{8307.622}{  0.370}{0.053}{ 600}{2.75}{
520}{1.2CA}
\punktc{20/02/91}{04:33}{8307.690}{  0.425}{0.065}{ 600}{2.23}{
532}{1.2CA}
\punktc{21/02/91}{02:37}{8308.610}{  0.508}{0.064}{ 900}{2.30}{
725}{1.2CA}
\punktc{21/02/91}{04:19}{8308.680}{  0.515}{0.054}{ 600}{2.16}{
578}{1.2CA}
\punktc{21/02/91}{05:26}{8308.727}{  0.474}{0.102}{ 600}{2.14}{
3528}{1.2CA}
\punktc{22/02/91}{04:56}{8309.706}{  0.500}{0.042}{ 100}{4.72}{
341}{1.2CA}
\punktc{22/02/91}{05:05}{8309.712}{  0.501}{0.025}{ 500}{4.76}{
670}{1.2CA}
\punktc{22/02/91}{05:56}{8309.748}{  0.515}{0.037}{ 150}{4.03}{
624}{1.2CA}
\punktc{23/02/91}{01:43}{8310.572}{  0.734}{0.017}{ 100}{1.99}{
410}{1.2CA}
\punktc{23/02/91}{01:49}{8310.576}{  0.745}{0.014}{ 300}{1.76}{
550}{1.2CA}
\punktc{23/02/91}{04:02}{8310.669}{  0.794}{0.012}{ 100}{1.38}{
310}{1.2CA}
\punktc{23/02/91}{04:12}{8310.675}{  0.802}{0.009}{ 300}{1.39}{
430}{1.2CA}
\punktc{23/02/91}{05:41}{8310.737}{  0.806}{0.008}{ 100}{1.09}{
317}{1.2CA}
\punktc{23/02/91}{05:47}{8310.742}{  0.808}{0.011}{ 300}{1.12}{
522}{1.2CA}
\punktc{24/02/91}{00:20}{8311.514}{  0.852}{0.013}{ 300}{1.94}{
1012}{1.2CA}
\punktc{25/02/91}{01:18}{8312.555}{  0.703}{0.054}{ 600}{3.14}{
954}{1.2CA}
\punktcc{25/02/91}{04:23}{8312.683}{  0.668}{0.071}{ 600}{3.44}{
553}{1.2CA}
\punktc{25/02/91}{04:45}{8312.699}{  0.704}{0.042}{ 100}{3.18}{
629}{1.2CA}
\punktc{25/02/91}{04:52}{8312.703}{  0.678}{0.061}{ 300}{4.12}{
1549}{1.2CA}
\punktcc{25/02/91}{04:57}{8312.707}{  0.679}{0.098}{ 200}{4.21}{
1263}{1.2CA}
\punktcc{25/02/91}{05:20}{8312.722}{  0.682}{0.091}{ 600}{3.40}{
2033}{1.2CA}
\punktc{25/02/91}{05:49}{8312.743}{  0.763}{0.053}{ 200}{3.06}{
1257}{1.2CA}
\punktc{26/02/91}{01:45}{8313.573}{  0.636}{0.055}{ 600}{2.64}{
1104}{1.2CA}
\punktc{26/02/91}{03:45}{8313.657}{  0.609}{0.049}{ 600}{2.89}{
830}{1.2CA}
\punktc{26/02/91}{05:05}{8313.712}{  0.610}{0.061}{ 600}{2.86}{
718}{1.2CA}
\punktcc{27/02/91}{01:53}{8314.578}{  0.650}{0.125}{ 603}{3.40}{
1202}{1.2CA}
\punktcc{27/02/91}{03:10}{8314.632}{  0.657}{0.121}{ 600}{3.18}{
1071}{1.2CA}
\punktcc{27/02/91}{04:28}{8314.686}{  0.631}{0.150}{ 600}{3.85}{
878}{1.2CA}
\punktcc{28/02/91}{03:53}{8315.662}{  0.392}{0.352}{ 100}{3.35}{
1326}{1.2CA}
\punktcc{17/03/91}{02:07}{8332.589}{ -0.232}{0.109}{ 100}{5.07}{
309}{1.2CA}
\punktc{19/03/91}{02:58}{8334.624}{ -0.118}{0.019}{ 100}{1.77}{
306}{1.2CA}
\punktc{19/03/91}{03:08}{8334.631}{ -0.152}{0.014}{ 208}{1.71}{
365}{1.2CA}
\punktc{20/03/91}{02:22}{8335.599}{ -0.097}{0.020}{ 100}{1.11}{
306}{1.2CA}
\punktc{20/03/91}{02:41}{8335.612}{ -0.095}{0.024}{ 500}{1.42}{
496}{1.2CA}
\punktc{23/05/91}{00:00}{8399.501}{ -0.697}{0.003}{ 100}{1.92}{
377}{1.2CA}
\punktc{23/05/91}{00:01}{8399.501}{ -0.686}{0.009}{ 150}{1.60}{
435}{1.2CA}
\punktc{23/05/91}{00:04}{8399.503}{ -0.686}{0.006}{ 500}{1.81}{
837}{1.2CA}
\punktc{23/05/91}{01:53}{8399.579}{ -0.686}{0.027}{ 100}{1.97}{
313}{1.2CA}
\punktc{23/05/91}{02:07}{8399.588}{ -0.688}{0.031}{ 500}{1.92}{
502}{1.2CA}
\punktc{24/05/91}{00:12}{8400.509}{ -0.734}{0.022}{ 100}{1.21}{
407}{1.2CA}
\punktc{24/05/91}{00:22}{8400.515}{ -0.720}{0.013}{ 500}{1.13}{
972}{1.2CA}
\punktc{24/05/91}{01:22}{8400.557}{ -0.729}{0.017}{ 100}{1.16}{
345}{1.2CA}
\punktc{25/05/91}{00:03}{8401.502}{ -0.752}{0.047}{ 100}{2.26}{
430}{1.2CA}
\punktc{25/05/91}{00:07}{8401.505}{ -0.741}{0.040}{ 300}{2.02}{
760}{1.2CA}
\punktc{25/05/91}{23:58}{8402.499}{ -0.728}{0.042}{ 100}{2.09}{
431}{1.2CA}
\punktc{26/05/91}{00:18}{8402.513}{ -0.853}{0.057}{ 100}{2.08}{
513}{1.2CA}
\punktc{26/05/91}{00:26}{8402.519}{ -0.803}{0.036}{ 500}{2.42}{
1455}{1.2CA}
\punktc{27/05/91}{00:36}{8403.525}{ -0.778}{0.069}{ 100}{2.12}{
619}{1.2CA}
\punktc{27/05/91}{00:44}{8403.531}{ -0.820}{0.036}{ 500}{2.16}{
2029}{1.2CA}
\punktc{27/05/91}{02:41}{8403.612}{ -0.823}{0.043}{ 100}{1.59}{
554}{1.2CA}
\punktc{27/05/91}{02:47}{8403.616}{ -0.818}{0.025}{ 300}{1.37}{
1114}{1.2CA}
\punktcc{28/05/91}{00:21}{8404.515}{ -0.747}{0.165}{ 100}{2.45}{
715}{1.2CA}
\punktc{28/05/91}{00:29}{8404.520}{ -0.829}{0.083}{ 500}{2.31}{
2721}{1.2CA}
\punktcc{28/05/91}{01:25}{8404.559}{ -0.811}{0.183}{ 100}{2.27}{
877}{1.2CA}
\punktcc{28/05/91}{01:31}{8404.563}{ -0.872}{0.217}{ 300}{2.02}{
2824}{1.2CA}
\punktc{29/05/91}{00:39}{8405.527}{ -0.831}{0.033}{ 100}{1.17}{
730}{1.2CA}
\punktc{29/05/91}{00:47}{8405.533}{ -0.821}{0.033}{ 500}{1.29}{
2844}{1.2CA}
\punktc{23/07/91}{20:33}{8461.357}{ -0.258}{0.013}{ 100}{1.40}{
571}{1.2CA}
\punktc{23/07/91}{20:38}{8461.360}{ -0.261}{0.018}{ 250}{1.95}{
952}{1.2CA}
\punktc{23/07/91}{20:43}{8461.364}{ -0.250}{0.012}{ 300}{1.43}{
1006}{1.2CA}
\punktc{23/07/91}{22:10}{8461.424}{ -0.251}{0.018}{ 100}{1.33}{
522}{1.2CA}
\punktc{23/07/91}{22:18}{8461.430}{ -0.267}{0.013}{ 500}{1.34}{
1523}{1.2CA}
\punktc{24/07/91}{20:22}{8462.349}{ -0.243}{0.024}{ 100}{1.36}{
938}{1.2CA}
\punktc{24/07/91}{20:26}{8462.352}{ -0.225}{0.014}{  98}{1.43}{
732}{1.2CA}
\punktc{24/07/91}{20:32}{8462.356}{ -0.239}{0.014}{ 500}{1.56}{
2028}{1.2CA}
\punktc{24/07/91}{22:15}{8462.428}{ -0.248}{0.015}{ 100}{1.57}{
565}{1.2CA}
\punktc{24/07/91}{22:23}{8462.433}{ -0.243}{0.014}{ 500}{1.58}{
1787}{1.2CA}
\punktc{25/07/91}{20:32}{8463.356}{ -0.237}{0.012}{ 100}{1.50}{
548}{1.2CA}
\punktc{25/07/91}{20:41}{8463.362}{ -0.238}{0.010}{ 500}{1.68}{
1440}{1.2CA}
\punktc{25/07/91}{23:37}{8463.485}{ -0.253}{0.012}{ 300}{1.19}{
1161}{1.2CA}
\punktc{26/07/91}{20:19}{8464.347}{ -0.242}{0.016}{ 150}{1.34}{
3124}{1.2CA}
\punktc{26/07/91}{20:39}{8464.361}{ -0.232}{0.029}{ 300}{2.03}{
3029}{1.2CA}
\punktc{26/07/91}{20:51}{8464.369}{ -0.237}{0.027}{ 300}{1.28}{
3094}{1.2CA}
\punktc{26/07/91}{22:10}{8464.424}{ -0.241}{0.033}{ 300}{1.17}{
4032}{1.2CA}
\punktc{26/07/91}{22:16}{8464.428}{ -0.239}{0.021}{ 200}{1.09}{
2779}{1.2CA}
\punktc{28/07/91}{20:56}{8466.372}{ -0.183}{0.021}{ 100}{1.95}{
434}{1.2CA}
\punktc{28/07/91}{21:00}{8466.375}{ -0.175}{0.022}{ 150}{2.17}{
542}{1.2CA}
\punktc{28/07/91}{21:05}{8466.379}{ -0.163}{0.017}{ 300}{2.02}{
835}{1.2CA}
\punktc{29/07/91}{20:58}{8467.374}{ -0.147}{0.020}{ 100}{3.11}{
334}{1.2CA}
\punktc{29/07/91}{21:08}{8467.381}{ -0.159}{0.017}{ 500}{3.32}{
704}{1.2CA}
\punktc{29/07/91}{22:51}{8467.453}{ -0.162}{0.019}{ 300}{3.26}{
662}{1.2CA}
\punktc{30/07/91}{21:21}{8468.390}{ -0.145}{0.021}{ 150}{2.93}{
421}{1.2CA}
\punktc{30/07/91}{21:29}{8468.396}{ -0.140}{0.026}{ 500}{2.62}{
851}{1.2CA}
\punktc{30/07/91}{23:42}{8468.488}{ -0.143}{0.019}{ 300}{3.04}{
817}{1.2CA}
\punktc{31/07/91}{00:05}{8468.504}{ -0.148}{0.023}{ 300}{3.03}{
835}{1.2CA}
\punktc{31/07/91}{20:20}{8469.347}{ -0.150}{0.013}{ 100}{1.31}{
544}{1.2CA}
\punktc{31/07/91}{20:27}{8469.353}{ -0.166}{0.011}{ 500}{1.49}{
1041}{1.2CA}
\punktc{31/07/91}{23:07}{8469.463}{ -0.163}{0.011}{ 300}{1.14}{
504}{1.2CA}
\punktc{01/08/91}{20:27}{8470.352}{ -0.107}{0.017}{ 100}{1.05}{
356}{1.2CA}
\punktc{01/08/91}{20:34}{8470.357}{ -0.118}{0.031}{ 300}{1.12}{
436}{1.2CA}
\punktc{01/08/91}{22:18}{8470.430}{ -0.115}{0.014}{ 100}{1.12}{
315}{1.2CA}
\punktc{01/08/91}{22:24}{8470.434}{ -0.112}{0.031}{ 300}{1.05}{
447}{1.2CA}
\punktc{02/08/91}{20:24}{8471.351}{ -0.094}{0.015}{ 100}{1.32}{
359}{1.2CA}
\punktc{02/08/91}{20:32}{8471.356}{ -0.079}{0.010}{ 500}{1.32}{
593}{1.2CA}
\punktc{02/08/91}{22:13}{8471.426}{ -0.075}{0.011}{ 100}{1.09}{
306}{1.2CA}
\punktc{02/08/91}{22:20}{8471.431}{ -0.067}{0.011}{ 500}{1.07}{
524}{1.2CA}
\punktc{03/08/91}{20:13}{8472.343}{ -0.038}{0.039}{ 100}{1.71}{
539}{1.2CA}
\punktc{03/08/91}{20:21}{8472.349}{ -0.024}{0.023}{ 500}{1.33}{
840}{1.2CA}
\punktc{03/08/91}{23:18}{8472.471}{ -0.019}{0.013}{ 100}{1.44}{
307}{1.2CA}
\punktc{03/08/91}{23:26}{8472.477}{ -0.021}{0.011}{ 500}{1.45}{
532}{1.2CA}
\punktc{04/08/91}{20:05}{8473.337}{ -0.001}{0.017}{ 100}{1.37}{
1329}{1.2CA}
\punktc{04/08/91}{20:14}{8473.344}{ -0.008}{0.024}{ 500}{1.42}{
1545}{1.2CA}
\punktc{04/08/91}{22:17}{8473.429}{  0.013}{0.009}{ 100}{1.29}{
296}{1.2CA}
\punktc{04/08/91}{22:26}{8473.435}{  0.003}{0.024}{ 500}{1.34}{
476}{1.2CA}
\punktc{05/08/91}{00:27}{8473.519}{  0.013}{0.013}{ 100}{1.28}{
324}{1.2CA}
\punktc{05/08/91}{00:39}{8473.527}{  0.009}{0.025}{ 500}{1.73}{
732}{1.2CA}
\punktc{05/08/91}{20:10}{8474.340}{  0.018}{0.026}{ 100}{1.45}{
1467}{1.2CA}
\punktc{05/08/91}{20:20}{8474.348}{  0.004}{0.022}{ 500}{1.47}{
841}{1.2CA}
\punktc{05/08/91}{21:54}{8474.413}{  0.000}{0.014}{ 100}{1.59}{
480}{1.2CA}
\punktc{05/08/91}{22:05}{8474.420}{  0.003}{0.011}{ 500}{1.21}{
507}{1.2CA}
\punktc{05/08/91}{23:45}{8474.490}{  0.010}{0.011}{ 500}{1.35}{
473}{1.2CA}
\punktc{05/08/91}{23:52}{8474.495}{  0.014}{0.014}{ 100}{1.28}{
291}{1.2CA}
\punktc{06/08/91}{19:52}{8475.328}{  0.030}{0.035}{ 100}{1.59}{
3528}{1.2CA}
\punktc{06/08/91}{20:03}{8475.336}{  0.060}{0.018}{ 500}{1.94}{
2816}{1.2CA}
\punktc{06/08/91}{21:31}{8475.397}{  0.055}{0.051}{ 100}{1.23}{
289}{1.2CA}
\punktcc{06/08/91}{21:41}{8475.404}{  0.160}{0.092}{ 500}{1.55}{
418}{1.2CA}
\punktc{06/08/91}{22:27}{8475.436}{  0.138}{0.057}{ 500}{1.33}{
429}{1.2CA}
\punktc{06/08/91}{22:49}{8475.451}{  0.069}{0.024}{ 500}{1.39}{
459}{1.2CA}
\punktc{07/08/91}{00:43}{8475.530}{  0.103}{0.023}{ 100}{1.21}{
304}{1.2CA}
\punktc{07/08/91}{00:52}{8475.536}{  0.176}{0.055}{ 500}{1.45}{
479}{1.2CA}
\punktc{07/08/91}{01:16}{8475.553}{  0.135}{0.028}{  50}{1.39}{
276}{1.2CA}
\punktc{07/08/91}{01:28}{8475.562}{  0.097}{0.015}{ 300}{1.57}{
416}{1.2CA}
\punktc{07/08/91}{20:06}{8476.338}{  0.183}{0.012}{ 100}{1.60}{
750}{1.2CA}
\punktc{07/08/91}{20:17}{8476.345}{  0.196}{0.021}{ 500}{1.76}{
948}{1.2CA}
\punktc{07/08/91}{23:09}{8476.465}{  0.242}{0.013}{ 100}{2.13}{
351}{1.2CA}
\punktc{07/08/91}{23:17}{8476.470}{  0.225}{0.021}{ 500}{1.68}{
646}{1.2CA}
\punktc{08/08/91}{20:10}{8477.340}{  0.197}{0.015}{ 100}{1.89}{
766}{1.2CA}
\punktc{08/08/91}{20:22}{8477.349}{  0.195}{0.021}{ 500}{1.98}{
676}{1.2CA}
\punktc{08/08/91}{22:21}{8477.431}{  0.208}{0.016}{ 100}{2.08}{
297}{1.2CA}
\punktc{08/08/91}{22:29}{8477.437}{  0.200}{0.022}{ 500}{2.23}{
486}{1.2CA}
\punktc{09/08/91}{00:22}{8477.516}{  0.217}{0.022}{ 100}{2.62}{
306}{1.2CA}
\punktc{09/08/91}{00:36}{8477.525}{  0.205}{0.014}{ 500}{2.75}{
526}{1.2CA}
\punktc{09/08/91}{21:00}{8478.375}{  0.292}{0.059}{ 100}{4.15}{
588}{1.2CA}
\punktc{09/08/91}{21:10}{8478.382}{  0.291}{0.020}{ 500}{4.22}{
711}{1.2CA}
\punktcc{10/08/91}{19:49}{8479.326}{  0.288}{0.080}{ 100}{2.79}{
4771}{1.2CA}
\punktc{10/08/91}{20:00}{8479.334}{  0.253}{0.038}{ 500}{3.11}{
2875}{1.2CA}
\punktc{11/08/91}{00:21}{8479.515}{  0.294}{0.027}{ 100}{2.72}{
325}{1.2CA}
\punktc{11/08/91}{00:29}{8479.520}{  0.270}{0.029}{ 500}{3.19}{
571}{1.2CA}
\punktc{11/08/91}{20:47}{8480.366}{  0.255}{0.013}{ 100}{1.74}{
309}{1.2CA}
\punktc{11/08/91}{20:56}{8480.372}{  0.254}{0.021}{ 500}{1.99}{
505}{1.2CA}
\punktc{12/08/91}{00:06}{8480.505}{  0.268}{0.013}{ 100}{1.49}{
328}{1.2CA}
\punktc{12/08/91}{00:13}{8480.509}{  0.265}{0.029}{ 300}{2.40}{
476}{1.2CA}
\punktc{12/08/91}{20:05}{8481.337}{  0.299}{0.011}{ 100}{1.35}{
443}{1.2CA}
\punktc{12/08/91}{20:14}{8481.344}{  0.293}{0.022}{ 500}{1.38}{
601}{1.2CA}
\punktc{12/08/91}{22:22}{8481.433}{  0.281}{0.015}{ 100}{1.07}{
341}{1.2CA}
\punktc{12/08/91}{22:31}{8481.439}{  0.300}{0.014}{ 500}{1.30}{
557}{1.2CA}
\punktc{16/08/91}{00:26}{8484.519}{  0.293}{0.050}{ 300}{1.71}{
420}{1.2CA}
\punktc{18/08/91}{22:27}{8487.436}{  0.230}{0.027}{ 300}{1.54}{
532}{1.2CA}
\punktc{19/08/91}{19:51}{8488.328}{  0.261}{0.019}{ 500}{1.13}{
4967}{1.2CA}
\punktc{19/08/91}{20:51}{8488.369}{  0.275}{0.018}{ 100}{1.15}{
390}{1.2CA}
\punktc{19/08/91}{20:59}{8488.375}{  0.257}{0.017}{ 500}{1.26}{
939}{1.2CA}
\punktc{19/08/91}{23:57}{8488.498}{  0.240}{0.036}{ 100}{1.81}{
360}{1.2CA}
\punktc{20/08/91}{00:05}{8488.504}{  0.234}{0.024}{ 500}{2.06}{
804}{1.2CA}
\punktc{21/08/91}{20:56}{8490.372}{  0.190}{0.030}{ 100}{1.41}{
584}{1.2CA}
\punktc{21/08/91}{21:07}{8490.380}{  0.172}{0.030}{ 500}{1.74}{
2669}{1.2CA}
\punktc{22/08/91}{19:38}{8491.318}{  0.141}{0.098}{ 100}{2.74}{
2017}{1.2CA}
\punktc{22/08/91}{19:46}{8491.324}{  0.159}{0.031}{ 500}{2.76}{
2929}{1.2CA}
\punktc{23/08/91}{19:41}{8492.321}{  0.165}{0.055}{ 100}{1.78}{
1535}{1.2CA}
\punktc{23/08/91}{19:53}{8492.329}{  0.184}{0.021}{ 500}{2.04}{
1892}{1.2CA}
\punktc{23/08/91}{22:14}{8492.427}{  0.172}{0.032}{ 100}{1.96}{
531}{1.2CA}
\punktc{23/08/91}{22:24}{8492.434}{  0.188}{0.026}{ 500}{2.02}{
1751}{1.2CA}
\punktc{24/08/91}{19:52}{8493.328}{  0.210}{0.027}{ 100}{1.37}{
749}{1.2CA}
\punktc{24/08/91}{20:00}{8493.333}{  0.221}{0.016}{ 500}{1.26}{
1875}{1.2CA}
\punktc{25/08/91}{19:42}{8494.321}{  0.362}{0.053}{ 100}{1.23}{
704}{1.2CA}
\punktc{25/08/91}{19:45}{8494.323}{  0.252}{0.022}{ 100}{1.17}{
847}{1.2CA}
\punktc{25/08/91}{19:56}{8494.331}{  0.252}{0.018}{ 500}{1.37}{
1801}{1.2CA}
\punktc{25/08/91}{22:14}{8494.427}{  0.262}{0.037}{ 100}{1.60}{
992}{1.2CA}
\punktc{25/08/91}{22:27}{8494.436}{  0.254}{0.016}{ 500}{1.57}{
3091}{1.2CA}
\punktc{26/08/91}{19:37}{8495.318}{  0.270}{0.029}{ 100}{1.53}{
1000}{1.2CA}
\punktc{26/08/91}{19:45}{8495.323}{  0.282}{0.021}{ 500}{1.62}{
1967}{1.2CA}
\punktc{26/08/91}{22:56}{8495.456}{  0.285}{0.039}{ 100}{1.85}{
731}{1.2CA}
\punktc{26/08/91}{23:04}{8495.461}{  0.278}{0.034}{ 500}{2.05}{
2552}{1.2CA}
\punktc{27/08/91}{19:41}{8496.321}{  0.244}{0.028}{ 200}{2.22}{
885}{1.2CA}
\punktc{27/08/91}{19:51}{8496.327}{  0.238}{0.021}{ 500}{2.13}{
964}{1.2CA}
\punktc{28/08/91}{19:39}{8497.319}{  0.262}{0.049}{ 100}{2.56}{
516}{1.2CA}
\punktc{28/08/91}{19:46}{8497.324}{  0.290}{0.042}{ 472}{2.52}{
897}{1.2CA}
\punktc{18/09/91}{19:34}{8518.316}{  0.417}{0.029}{ 100}{1.79}{
420}{1.2CA}
\punktc{18/09/91}{19:45}{8518.323}{  0.440}{0.017}{ 500}{2.10}{
1073}{1.2CA}
\punktc{18/09/91}{20:00}{8518.334}{  0.443}{0.012}{ 500}{1.74}{
1076}{1.2CA}
\punktc{19/09/91}{19:55}{8519.330}{  0.655}{0.017}{ 500}{2.05}{
1220}{1.2CA}
\punktc{20/09/91}{19:31}{8520.314}{  0.926}{0.006}{ 500}{1.32}{
1649}{1.2CA}
\punktc{20/09/91}{20:00}{8520.334}{  0.929}{0.007}{ 500}{1.48}{
1589}{1.2CA}
\punktc{20/09/91}{21:10}{8520.382}{  0.903}{0.052}{ 546}{2.04}{
3534}{1.2CA}
\punktc{21/09/91}{19:23}{8521.308}{  0.959}{0.010}{ 500}{1.54}{
1524}{1.2CA}
\punktc{22/09/91}{19:15}{8522.302}{  1.220}{0.012}{ 500}{1.71}{
1490}{1.2CA}
\punktc{22/09/91}{20:57}{8522.373}{  1.244}{0.015}{ 500}{1.76}{
1985}{1.2CA}
\punktcc{03/10/91}{18:30}{8533.271}{  1.256}{0.278}{1740}{4.91}{
1656}{1.2CA}
\punktcc{04/10/91}{19:12}{8534.300}{  1.175}{0.171}{ 900}{4.80}{
1106}{1.2CA}
\punktc{09/10/91}{18:40}{8539.278}{  1.022}{0.077}{ 600}{2.50}{
1165}{1.2CA}
\punktc{10/10/91}{18:34}{8540.274}{  1.044}{0.065}{ 600}{3.06}{
976}{1.2CA}
\punktcc{11/10/91}{19:01}{8541.293}{  0.962}{0.239}{ 900}{4.46}{
1592}{1.2CA}
\punktcc{12/10/91}{18:01}{8542.251}{  0.769}{0.148}{ 600}{3.11}{
1258}{1.2CA}
\punktc{16/10/91}{20:07}{8546.339}{  0.611}{0.034}{ 150}{4.25}{
515}{1.2CA}
\punktc{16/10/91}{20:11}{8546.342}{  0.640}{0.042}{ 100}{3.76}{
428}{1.2CA}
\punktc{17/10/91}{18:43}{8547.280}{  0.561}{0.041}{ 100}{2.58}{
621}{1.2CA}
\punktc{17/10/91}{18:52}{8547.286}{  0.558}{0.016}{ 500}{2.50}{
1290}{1.2CA}
\punktc{18/10/91}{18:57}{8548.290}{  0.524}{0.043}{ 100}{2.29}{
546}{1.2CA}
\punktcc{18/10/91}{19:05}{8548.296}{  0.560}{0.117}{ 160}{2.68}{
913}{1.2CA}
\punktc{19/10/91}{18:53}{8549.287}{  0.562}{0.018}{ 100}{1.82}{
485}{1.2CA}
\punktc{19/10/91}{19:03}{8549.294}{  0.552}{0.009}{ 500}{1.93}{
1463}{1.2CA}
\punktc{20/10/91}{18:32}{8550.273}{  0.628}{0.009}{ 100}{1.63}{
468}{1.2CA}
\punktc{20/10/91}{18:48}{8550.283}{  0.606}{0.020}{ 500}{1.54}{
1233}{1.2CA}
\punktc{20/10/91}{20:01}{8550.335}{  0.611}{0.010}{ 100}{1.59}{
545}{1.2CA}
\punktc{20/10/91}{20:10}{8550.340}{  0.606}{0.014}{ 500}{1.65}{
1770}{1.2CA}
\punktc{22/10/91}{19:37}{8552.318}{  0.514}{0.070}{ 100}{4.09}{
757}{1.2CA}
\punktc{22/10/91}{19:45}{8552.323}{  0.582}{0.056}{ 500}{4.45}{
2852}{1.2CA}
\punktc{26/10/91}{19:01}{8556.293}{  0.550}{0.023}{ 100}{3.36}{
341}{1.2CA}
\punktc{26/10/91}{19:10}{8556.299}{  0.530}{0.020}{ 500}{4.05}{
698}{1.2CA}
\punktc{28/10/91}{19:12}{8558.300}{  0.812}{0.017}{ 100}{1.15}{
503}{1.2CA}
\punktc{28/10/91}{19:21}{8558.307}{  0.791}{0.008}{ 500}{1.19}{
1446}{1.2CA}
\punktc{28/10/91}{19:50}{8558.327}{  0.793}{0.012}{ 500}{1.31}{
1040}{1.2CA}
\punktc{08/02/92}{06:06}{8660.755}{ -0.317}{0.016}{ 500}{1.39}{
1153}{1.2CA}
\punktc{09/02/92}{05:40}{8661.736}{ -0.322}{0.031}{ 500}{1.11}{
499}{1.2CA}
\punktc{10/02/92}{05:35}{8662.733}{ -0.324}{0.010}{ 100}{1.04}{
307}{1.2CA}
\punktc{10/02/92}{05:44}{8662.739}{ -0.154}{0.070}{ 500}{0.97}{
531}{1.2CA}
\punktc{11/02/92}{05:10}{8663.716}{ -0.276}{0.064}{ 100}{1.17}{
302}{1.2CA}
\punktc{11/02/92}{05:19}{8663.722}{ -0.256}{0.031}{ 500}{0.96}{
484}{1.2CA}
\punktc{12/02/92}{05:46}{8664.740}{ -0.275}{0.012}{ 100}{1.21}{
313}{1.2CA}
\punktc{12/02/92}{05:56}{8664.747}{ -0.276}{0.021}{ 442}{1.48}{
671}{1.2CA}
\punktc{12/02/92}{06:04}{8664.753}{ -0.281}{0.015}{ 389}{1.26}{
1257}{1.2CA}
\punktc{14/02/92}{05:29}{8666.729}{ -0.338}{0.012}{ 100}{1.01}{
302}{1.2CA}
\punktc{14/02/92}{05:37}{8666.734}{ -0.340}{0.029}{ 500}{1.21}{
499}{1.2CA}
\punktc{15/02/92}{05:28}{8667.728}{ -0.327}{0.011}{ 100}{1.31}{
307}{1.2CA}
\punktc{15/02/92}{05:36}{8667.734}{ -0.313}{0.029}{ 500}{1.26}{
569}{1.2CA}
\punktc{16/02/92}{04:12}{8668.675}{ -0.317}{0.025}{ 100}{1.56}{
536}{1.2CA}
\punktc{16/02/92}{04:19}{8668.680}{ -0.306}{0.012}{ 500}{1.44}{
1428}{1.2CA}
\punktc{17/02/92}{04:32}{8669.689}{ -0.225}{0.020}{ 100}{1.27}{
468}{1.2CA}
\punktc{17/02/92}{04:47}{8669.700}{ -0.264}{0.017}{ 500}{1.31}{
1645}{1.2CA}
\punktc{28/02/92}{05:41}{8680.737}{ -0.071}{0.021}{ 500}{2.40}{
612}{1.2CA}
\punktc{05/06/92}{00:00}{8778.501}{  0.213}{0.054}{ 110}{3.11}{
530}{1.2CA}
\punktc{06/06/92}{08:36}{8779.859}{  0.215}{0.083}{ 110}{2.72}{
1012}{1.2CA}
\punktc{06/06/92}{22:21}{8780.432}{  0.165}{0.054}{ 200}{2.50}{
504}{1.2CA}
\punktc{06/06/92}{22:32}{8780.439}{  0.167}{0.052}{ 400}{3.14}{
714}{1.2CA}
\punktc{07/06/92}{00:17}{8780.512}{  0.125}{0.031}{ 300}{2.24}{
533}{1.2CA}
\punktc{08/06/92}{23:50}{8782.493}{  0.098}{0.032}{ 300}{1.52}{
761}{1.2CA}
\punktc{15/06/92}{14:55}{8789.122}{  0.342}{0.045}{ 120}{3.05}{
10546}{1.2CA}
\punktc{15/06/92}{23:26}{8789.477}{  0.327}{0.045}{  60}{3.04}{
4561}{1.2CA}
\punktc{20/09/92}{19:26}{8886.310}{ -0.637}{0.025}{ 100}{0.85}{
420}{1.2CA}
\punktc{20/09/92}{19:36}{8886.317}{ -0.645}{0.028}{ 300}{1.02}{
648}{1.2CA}
\punktc{21/09/92}{21:20}{8887.390}{ -0.630}{0.037}{ 150}{1.82}{
493}{1.2CA}
\punktc{21/09/92}{21:24}{8887.392}{ -0.632}{0.030}{ 100}{1.74}{
420}{1.2CA}
\punktc{22/09/92}{19:21}{8888.307}{ -0.650}{0.055}{ 100}{2.35}{
417}{1.2CA}
\punktc{22/09/92}{19:27}{8888.311}{ -0.637}{0.029}{ 300}{2.36}{
739}{1.2CA}
\punktc{23/09/92}{19:41}{8889.320}{ -0.643}{0.034}{ 100}{1.65}{
467}{1.2CA}
\punktc{23/09/92}{19:43}{8889.322}{ -0.640}{0.033}{ 100}{1.51}{
459}{1.2CA}
\punktc{24/09/92}{20:49}{8890.368}{ -0.632}{0.025}{ 100}{0.84}{
432}{1.2CA}
\punktc{24/09/92}{20:52}{8890.370}{ -0.628}{0.021}{  50}{0.81}{
350}{1.2CA}
\punktc{24/09/92}{20:52}{8890.370}{ -0.628}{0.021}{  50}{0.81}{
350}{1.2CA}
\punktc{28/12/92}{05:15}{8984.719}{ -1.263}{0.019}{1800}{3.48}{
1396}{1.2CA}
\punktcc{02/01/93}{05:27}{8989.727}{ -1.243}{0.018}{1800}{5.28}{
1336}{1.2CA}
\punktc{04/01/93}{03:17}{8991.637}{ -1.252}{0.024}{1800}{3.81}{
1523}{1.2CA}
\punktc{08/02/93}{04:42}{9026.696}{ -1.298}{0.040}{ 200}{1.87}{
2138}{1.2CA}
\punktc{08/02/93}{04:46}{9026.699}{ -1.260}{0.035}{ 200}{1.73}{
2007}{1.2CA}
\punktc{08/02/93}{04:50}{9026.702}{ -1.256}{0.010}{ 100}{1.71}{
920}{1.2CA}
\punktc{09/02/93}{05:32}{9027.731}{ -1.277}{0.035}{ 200}{1.69}{
1760}{1.2CA}
\punktc{17/02/93}{04:20}{9035.681}{ -1.257}{0.040}{ 200}{1.65}{
532}{1.2CA}
\punktc{17/02/93}{04:25}{9035.684}{ -1.263}{0.038}{ 200}{1.50}{
539}{1.2CA}
\punktc{18/02/93}{05:41}{9036.737}{ -1.270}{0.041}{ 200}{1.64}{
558}{1.2CA}
\punktc{18/02/93}{05:45}{9036.740}{ -1.275}{0.038}{ 200}{1.48}{
603}{1.2CA}
\punktc{19/05/93}{22:32}{9127.439}{ -1.250}{0.015}{ 144}{2.78}{
363}{1.2CA}
\punktc{19/05/93}{22:42}{9127.446}{ -1.287}{0.012}{ 500}{2.54}{
622}{1.2CA}
\punktc{22/05/93}{01:21}{9129.557}{ -1.275}{0.009}{ 277}{1.58}{
411}{1.2CA}
\punktc{23/05/93}{03:15}{9130.636}{ -1.297}{0.009}{ 100}{1.65}{
335}{1.2CA}
\punktc{23/05/93}{03:19}{9130.638}{ -1.289}{0.008}{ 143}{1.48}{
428}{1.2CA}
\punktc{23/05/93}{03:29}{9130.645}{ -1.277}{0.009}{ 300}{1.52}{
422}{1.2CA}
\punktc{}{}{9195.462}{-1.248}{0.012}{}{}{}{1.2CA}%

\end{picture}}

\end{picture}

\addtocounter{figure}{-1}

\vspace*{1cm}
\caption{(continued)}
\end{figure*}

\begin{figure*}

\vspace*{0.5cm}

\begin{picture}(17 ,5 )(-1,0)
\put(0,0){\setlength{\unitlength}{0.0085cm}%
\begin{picture}(2000, 588.235)(7250,0)
\put(7250,0){\framebox(2000, 588.235)[tl]{\begin{picture}(0,0)(0,0)
        \put(0,0){\makebox(0,0)[tr]{$\Delta R$\hspace*{0.2cm}}}
        \put(2000,0){\makebox(0,0)[tr]{\large{1749+701}\T{0.4}
                                 \hspace*{0.5cm}}}
        \put(2000,-
588.235){\setlength{\unitlength}{1cm}\begin{picture}(0,0)(0,0)
        \end{picture}}
    \end{picture}}}

\thicklines
\put(7250,0){\setlength{\unitlength}{2.5cm}\begin{picture}(0,0)(0,-1)
   \put(0,0){\setlength{\unitlength}{1cm}\begin{picture}(0,0)(0,0)
        \put(0,0){\line(1,0){0.3}}
        \end{picture}}
   \end{picture}}

\put(9250,0){\setlength{\unitlength}{2.5cm}\begin{picture}(0,0)(0,-1)
   \put(0,0){\setlength{\unitlength}{1cm}\begin{picture}(0,0)(0,0)
        \put(0,0){\line(-1,0){0.3}}
        \end{picture}}
   \end{picture}}

\thinlines
\put(7250,0){\setlength{\unitlength}{2.5cm}\begin{picture}(0,0)(0,-1)
   \multiput(0,0)(0,0.1){10}{\setlength{\unitlength}{1cm}%
\begin{picture}(0,0)(0,0)
        \put(0,0){\line(1,0){0.12}}
        \end{picture}}
   \end{picture}}

\put(7250,0){\setlength{\unitlength}{2.5cm}\begin{picture}(0,0)(0,-1)
   \multiput(0,0)(0,-0.1){10}{\setlength{\unitlength}{1cm}%
\begin{picture}(0,0)(0,0)
        \put(0,0){\line(1,0){0.12}}
        \end{picture}}
   \end{picture}}

\put(9250,0){\setlength{\unitlength}{2.5cm}\begin{picture}(0,0)(0,-1)
   \multiput(0,0)(0,0.1){10}{\setlength{\unitlength}{1cm}%
\begin{picture}(0,0)(0,0)
        \put(0,0){\line(-1,0){0.12}}
        \end{picture}}
   \end{picture}}

\put(9250,0){\setlength{\unitlength}{2.5cm}\begin{picture}(0,0)(0,-1)
   \multiput(0,0)(0,-0.1){10}{\setlength{\unitlength}{1cm}%
\begin{picture}(0,0)(0,0)
        \put(0,0){\line(-1,0){0.12}}
        \end{picture}}
   \end{picture}}

\put(7250,0){\setlength{\unitlength}{2.5cm}\begin{picture}(0,0)(0,-1)
   \put(0,0.2){\setlength{\unitlength}{1cm}\begin{picture}(0,0)(0,0)
        \put(0,0){\line(1,0){0.12}}
        \put(-0.2,0){\makebox(0,0)[r]{\bf 0.2}}
        \end{picture}}
   \put(0,0.4){\setlength{\unitlength}{1cm}\begin{picture}(0,0)(0,0)
        \put(0,0){\line(1,0){0.12}}
        \put(-0.2,0){\makebox(0,0)[r]{\bf 0.4}}
        \end{picture}}
   \put(0,-0.2){\setlength{\unitlength}{1cm}\begin{picture}(0,0)(0,0)
        \put(0,0){\line(1,0){0.12}}
        \put(-0.2,0){\makebox(0,0)[r]{\bf -0.2}}
        \end{picture}}
   \put(0,-0.4){\setlength{\unitlength}{1cm}\begin{picture}(0,0)(0,0)
        \put(0,0){\line(1,0){0.12}}
        \put(-0.2,0){\makebox(0,0)[r]{\bf -0.4}}
        \end{picture}}
   \put(0,0.6){\setlength{\unitlength}{1cm}\begin{picture}(0,0)(0,0)
        \put(0,0){\line(1,0){0.12}}
        \put(-0.2,0){\makebox(0,0)[r]{\bf 0.6}}
        \end{picture}}
   \put(0,-0.6){\setlength{\unitlength}{1cm}\begin{picture}(0,0)(0,0)
        \put(0,0){\line(1,0){0.12}}
        \put(-0.2,0){\makebox(0,0)[r]{\bf -0.6}}
        \end{picture}}
   \put(0,-0.8){\setlength{\unitlength}{1cm}\begin{picture}(0,0)(0,0)
        \put(0,0){\line(1,0){0.12}}
        \put(-0.2,0){\makebox(0,0)[r]{\bf -0.8}}
        \end{picture}}
   \put(0,0.0){\setlength{\unitlength}{1cm}\begin{picture}(0,0)(0,0)
        \put(0,0){\line(1,0){0.12}}
        \put(-0.2,0){\makebox(0,0)[r]{\bf 0.0}}
        \end{picture}}
   \put(0,0.8){\setlength{\unitlength}{1cm}\begin{picture}(0,0)(0,0)
        \put(0,0){\line(1,0){0.12}}
        \end{picture}}
   \end{picture}}

   \put(7527.5, 588.235){\setlength{\unitlength}{1cm}\begin{picture}(0,0)(0,0)
        \put(0,0){\line(0,-1){0.2}}
        \put(0,0.2){\makebox(0,0)[b]{\bf 1989}}
   \end{picture}}
   \put(7892.5, 588.235){\setlength{\unitlength}{1cm}\begin{picture}(0,0)(0,0)
        \put(0,0){\line(0,-1){0.2}}
        \put(0,0.2){\makebox(0,0)[b]{\bf 1990}}
   \end{picture}}
   \put(8257.5, 588.235){\setlength{\unitlength}{1cm}\begin{picture}(0,0)(0,0)
        \put(0,0){\line(0,-1){0.2}}
        \put(0,0.2){\makebox(0,0)[b]{\bf 1991}}
  \end{picture}}
   \put(8622.5, 588.235){\setlength{\unitlength}{1cm}\begin{picture}(0,0)(0,0)
        \put(0,0){\line(0,-1){0.2}}
        \put(0,0.2){\makebox(0,0)[b]{\bf 1992}}
   \end{picture}}
   \put(8987.5, 588.235){\setlength{\unitlength}{1cm}\begin{picture}(0,0)(0,0)
        \put(0,0){\line(0,-1){0.2}}
        \put(0,0.2){\makebox(0,0)[b]{\bf 1993}}
   \end{picture}}

    \multiput(7250,0)(50,0){40}%
        {\setlength{\unitlength}{1cm}\begin{picture}(0,0)(0,0)
        \put(0,0){\line(0,1){0.12}}
    \end{picture}}
    \put(7500,0){\setlength{\unitlength}{1cm}\begin{picture}(0,0)(0,0)
        \put(0,0){\line(0,1){0.2}}
    \end{picture}}
    \put(8000,0){\setlength{\unitlength}{1cm}\begin{picture}(0,0)(0,0)
        \put(0,0){\line(0,1){0.2}}
    \end{picture}}
   \put(7750,0){\setlength{\unitlength}{1cm}\begin{picture}(0,0)(0,0)
        \put(0,0){\line(0,1){0.2}}
    \end{picture}}
   \put(8250,0){\setlength{\unitlength}{1cm}\begin{picture}(0,0)(0,0)
        \put(0,0){\line(0,1){0.2}}
    \end{picture}}
    \put(8500,0){\setlength{\unitlength}{1cm}\begin{picture}(0,0)(0,0)
        \put(0,0){\line(0,1){0.2}}
    \end{picture}}
    \put(8750,0){\setlength{\unitlength}{1cm}\begin{picture}(0,0)(0,0)
       \put(0,0){\line(0,1){0.2}}
    \end{picture}}
    \put(9000,0){\setlength{\unitlength}{1cm}\begin{picture}(0,0)(0,0)
       \put(0,0){\line(0,1){0.2}}
    \end{picture}}

\punktg{11/06/89}{02:17}{7688.596}{  0.158}{0.009}{ 500}{1.72}{
1015}{1.2CA}
\punktg{12/06/89}{23:52}{7690.495}{  0.275}{0.009}{ 500}{1.25}{
1049}{1.2CA}
\punktg{24/06/89}{02:16}{7701.594}{  0.410}{0.038}{ 300}{1.46}{
822}{1.2CA}
\punktg{24/06/89}{02:22}{7701.599}{  0.412}{0.013}{ 300}{1.36}{
818}{1.2CA}
\punktg{21/07/89}{21:28}{7729.395}{  0.143}{0.011}{1000}{1.77}{
1504}{1.2CA}
\punktg{24/07/89}{01:48}{7731.575}{ -0.054}{0.014}{ 500}{1.74}{
1050}{1.2CA}
\punktg{13/08/89}{22:49}{7752.451}{ -0.093}{0.019}{ 500}{1.37}{
1159}{1.2CA}
\punktg{15/08/89}{22:37}{7754.443}{ -0.077}{0.012}{ 500}{1.30}{
1706}{1.2CA}
\punktg{02/09/89}{21:58}{7772.416}{  0.199}{0.015}{ 500}{1.56}{
748}{1.2CA}
\punktg{16/10/89}{21:12}{7816.384}{  0.226}{0.014}{1000}{3.17}{
1469}{1.2CA}
\punktg{01/11/89}{19:23}{7832.308}{ -0.466}{0.062}{ 400}{1.89}{
570}{1.2CA}
\punktg{22/05/90}{01:38}{8033.569}{  0.227}{0.009}{ 250}{1.71}{
1148}{1.2CA}
\punktg{01/08/90}{00:15}{8104.511}{ -0.107}{0.010}{ 500}{1.66}{
932}{1.2CA}
\punktg{03/08/90}{21:50}{8107.410}{ -0.268}{0.032}{ 500}{2.29}{
1075}{1.2CA}
\punktg{13/08/90}{01:47}{8116.575}{  0.093}{0.037}{ 239}{2.01}{
2223}{1.2CA}
\punktg{13/08/90}{20:45}{8117.365}{  0.103}{0.027}{ 100}{1.31}{
407}{1.2CA}
\punktg{14/08/90}{08:45}{8117.865}{  0.108}{0.023}{ 100}{1.33}{
415}{1.2CA}
\punktg{14/08/90}{20:04}{8118.337}{  0.056}{0.018}{ 100}{1.70}{
1918}{1.2CA}
\punktg{15/08/90}{08:04}{8118.837}{  0.067}{0.021}{ 100}{1.70}{
1948}{1.2CA}
\punktg{15/08/90}{20:09}{8119.340}{  0.213}{0.012}{ 100}{0.86}{
793}{1.2CA}
\punktg{16/08/90}{08:09}{8119.840}{  0.188}{0.021}{ 100}{0.87}{
791}{1.2CA}
\punktg{24/09/90}{20:27}{8159.353}{ -0.016}{0.030}{ 500}{4.42}{
988}{1.2CA}
\punktg{25/09/90}{19:57}{8160.331}{ -0.089}{0.009}{ 500}{1.82}{
1023}{1.2CA}
\punktg{26/09/90}{19:57}{8161.331}{ -0.089}{0.008}{ 500}{1.82}{
1024}{1.2CA}
\punktg{29/09/90}{20:10}{8164.340}{ -0.008}{0.078}{ 100}{2.51}{
887}{1.2CA}
\punktg{29/09/90}{20:19}{8164.347}{  0.025}{0.028}{ 500}{2.52}{
1967}{1.2CA}
\punktgg{30/09/90}{19:49}{8165.326}{  0.039}{0.059}{ 100}{1.94}{
689}{1.2CA}
\punktg{30/09/90}{20:04}{8165.336}{  0.007}{0.016}{ 500}{1.87}{
1806}{1.2CA}
\punktg{16/02/91}{03:40}{8303.653}{  0.211}{0.043}{ 300}{2.47}{
514}{1.2CA}
\punktg{23/02/91}{04:47}{8310.700}{ -0.084}{0.012}{ 300}{1.24}{
446}{1.2CA}
\punktgg{23/05/91}{00:00}{8399.501}{ -0.017}{0.040}{ 100}{2.04}{
338}{1.2CA}
\punktg{23/05/91}{00:04}{8399.503}{ -0.036}{0.020}{ 500}{2.03}{
628}{1.2CA}
\punktgg{25/05/91}{00:57}{8401.540}{ -0.116}{0.040}{ 100}{1.96}{
412}{1.2CA}
\punktg{25/05/91}{01:06}{8401.546}{ -0.096}{0.024}{ 500}{2.30}{
1070}{1.2CA}
\punktgg{26/05/91}{02:11}{8402.592}{ -0.100}{0.049}{ 100}{2.35}{
408}{1.2CA}
\punktgg{26/05/91}{02:15}{8402.594}{ -0.136}{0.063}{  50}{2.27}{
334}{1.2CA}
\punktg{26/05/91}{02:20}{8402.597}{ -0.109}{0.031}{ 300}{2.35}{
662}{1.2CA}
\punktg{27/05/91}{01:14}{8403.552}{  0.057}{0.049}{ 100}{1.93}{
592}{1.2CA}
\punktgg{29/05/91}{02:41}{8405.612}{  0.117}{0.050}{ 100}{1.54}{
753}{1.2CA}
\punktg{29/05/91}{02:48}{8405.617}{  0.154}{0.042}{ 300}{1.58}{
1686}{1.2CA}
\punktg{23/07/91}{21:26}{8461.393}{ -0.301}{0.028}{ 100}{1.11}{
767}{1.2CA}
\punktg{23/07/91}{21:39}{8461.402}{ -0.304}{0.013}{ 500}{1.54}{
1448}{1.2CA}
\punktg{26/07/91}{00:03}{8463.503}{ -0.160}{0.012}{ 100}{1.21}{
598}{1.2CA}
\punktg{26/07/91}{00:10}{8463.507}{ -0.161}{0.012}{ 300}{1.28}{
1234}{1.2CA}
\punktg{26/07/91}{23:14}{8464.468}{ -0.240}{0.019}{ 300}{1.51}{
4247}{1.2CA}
\punktg{28/07/91}{21:45}{8466.407}{ -0.224}{0.023}{ 500}{2.01}{
1833}{1.2CA}
\punktg{29/07/91}{22:14}{8467.427}{ -0.227}{0.035}{ 100}{3.01}{
380}{1.2CA}
\punktg{29/07/91}{22:22}{8467.432}{ -0.234}{0.017}{ 500}{2.60}{
982}{1.2CA}
\punktg{30/07/91}{22:41}{8468.445}{ -0.278}{0.026}{ 100}{2.40}{
446}{1.2CA}
\punktg{30/07/91}{22:49}{8468.451}{ -0.244}{0.025}{ 300}{2.80}{
829}{1.2CA}
\punktgg{31/07/91}{00:32}{8468.523}{ -0.248}{0.073}{ 100}{2.54}{
474}{1.2CA}
\punktg{31/07/91}{00:41}{8468.529}{ -0.273}{0.049}{ 300}{3.43}{
985}{1.2CA}
\punktg{31/07/91}{22:26}{8469.435}{ -0.190}{0.008}{ 100}{0.90}{
369}{1.2CA}
\punktg{31/07/91}{22:31}{8469.438}{ -0.204}{0.006}{ 100}{1.01}{
352}{1.2CA}
\punktg{31/07/91}{22:36}{8469.442}{ -0.202}{0.011}{ 300}{1.00}{
502}{1.2CA}
\punktg{31/07/91}{23:36}{8469.484}{ -0.220}{0.010}{ 300}{1.08}{
558}{1.2CA}
\punktg{01/08/91}{21:48}{8470.409}{ -0.302}{0.012}{ 100}{0.94}{
312}{1.2CA}
\punktg{01/08/91}{21:54}{8470.413}{ -0.271}{0.011}{ 300}{1.03}{
414}{1.2CA}
\punktg{02/08/91}{00:58}{8470.541}{ -0.257}{0.014}{ 100}{0.99}{
378}{1.2CA}
\punktg{02/08/91}{01:04}{8470.545}{ -0.286}{0.009}{ 300}{1.06}{
582}{1.2CA}
\punktg{02/08/91}{21:36}{8471.400}{ -0.318}{0.011}{ 100}{1.04}{
316}{1.2CA}
\punktg{02/08/91}{21:43}{8471.405}{ -0.321}{0.007}{ 500}{1.36}{
523}{1.2CA}
\punktg{02/08/91}{23:39}{8471.486}{ -0.314}{0.009}{ 100}{1.01}{
342}{1.2CA}
\punktg{02/08/91}{23:48}{8471.492}{ -0.313}{0.008}{ 500}{1.24}{
636}{1.2CA}
\punktg{03/08/91}{22:44}{8472.447}{ -0.177}{0.010}{ 100}{1.26}{
314}{1.2CA}
\punktg{03/08/91}{22:51}{8472.453}{ -0.171}{0.009}{ 500}{1.44}{
548}{1.2CA}
\punktg{04/08/91}{23:44}{8473.489}{ -0.169}{0.011}{ 100}{1.06}{
347}{1.2CA}
\punktg{04/08/91}{23:53}{8473.496}{ -0.157}{0.010}{ 500}{1.20}{
542}{1.2CA}
\punktg{05/08/91}{23:02}{8474.460}{ -0.168}{0.013}{ 100}{1.22}{
306}{1.2CA}
\punktg{05/08/91}{23:10}{8474.466}{ -0.164}{0.011}{ 500}{1.20}{
467}{1.2CA}
\punktg{06/08/91}{23:32}{8475.481}{ -0.149}{0.010}{ 100}{1.40}{
304}{1.2CA}
\punktg{06/08/91}{23:42}{8475.488}{ -0.157}{0.011}{ 500}{1.44}{
525}{1.2CA}
\punktg{07/08/91}{22:50}{8476.452}{ -0.059}{0.009}{ 100}{1.41}{
386}{1.2CA}
\punktg{07/08/91}{23:02}{8476.460}{ -0.062}{0.015}{ 500}{1.99}{
720}{1.2CA}
\punktgg{08/08/91}{22:54}{8477.455}{ -0.128}{0.071}{ 100}{2.05}{
312}{1.2CA}
\punktg{08/08/91}{23:02}{8477.460}{ -0.125}{0.012}{ 500}{2.09}{
559}{1.2CA}
\punktg{11/08/91}{23:43}{8480.489}{  0.030}{0.008}{ 300}{1.37}{
515}{1.2CA}
\punktg{11/08/91}{23:54}{8480.496}{  0.040}{0.009}{ 100}{1.36}{
338}{1.2CA}
\punktg{12/08/91}{22:52}{8481.453}{  0.037}{0.012}{ 100}{1.00}{
361}{1.2CA}
\punktg{12/08/91}{22:58}{8481.457}{  0.058}{0.009}{ 300}{1.30}{
598}{1.2CA}
\punktg{19/08/91}{22:02}{8488.418}{ -0.090}{0.018}{ 100}{1.12}{
358}{1.2CA}
\punktg{19/08/91}{22:14}{8488.427}{ -0.066}{0.019}{ 500}{1.14}{
773}{1.2CA}
\punktgg{22/08/91}{01:16}{8490.553}{ -0.013}{0.083}{ 100}{2.39}{
476}{1.2CA}
\punktg{22/08/91}{01:24}{8490.559}{ -0.050}{0.039}{ 500}{2.36}{
1337}{1.2CA}
\punktgg{22/08/91}{21:05}{8491.379}{ -0.048}{0.087}{ 100}{2.28}{
1194}{1.2CA}
\punktg{22/08/91}{21:13}{8491.385}{ -0.034}{0.027}{ 500}{2.24}{
1397}{1.2CA}
\punktg{24/08/91}{22:59}{8493.458}{ -0.134}{0.036}{ 100}{1.54}{
728}{1.2CA}
\punktg{24/08/91}{23:07}{8493.463}{ -0.150}{0.025}{ 500}{1.81}{
2561}{1.2CA}
\punktg{25/08/91}{21:20}{8494.390}{ -0.132}{0.026}{ 100}{1.32}{
604}{1.2CA}
\punktg{25/08/91}{21:28}{8494.395}{ -0.133}{0.020}{ 500}{1.42}{
2837}{1.2CA}
\punktgg{26/08/91}{21:41}{8495.404}{  0.034}{0.056}{ 100}{2.07}{
631}{1.2CA}
\punktg{26/08/91}{21:49}{8495.409}{  0.027}{0.039}{ 500}{2.51}{
2091}{1.2CA}
\punktg{18/09/91}{21:17}{8518.387}{  0.442}{0.009}{ 100}{1.41}{
413}{1.2CA}
\punktg{18/09/91}{21:42}{8518.404}{  0.441}{0.011}{ 500}{1.43}{
1031}{1.2CA}
\punktg{17/10/91}{19:51}{8547.327}{ -0.049}{0.038}{ 100}{2.57}{
418}{1.2CA}
\punktg{17/10/91}{20:01}{8547.334}{ -0.056}{0.021}{ 500}{2.31}{
1122}{1.2CA}
\punktg{20/10/91}{20:43}{8550.364}{  0.188}{0.019}{ 100}{1.10}{
497}{1.2CA}
\punktg{20/10/91}{20:53}{8550.371}{  0.171}{0.007}{ 500}{1.21}{
1368}{1.2CA}
\punktg{26/10/91}{19:56}{8556.331}{  0.507}{0.028}{ 100}{3.21}{
373}{1.2CA}
\punktg{26/10/91}{20:04}{8556.336}{  0.455}{0.017}{ 500}{3.26}{
892}{1.2CA}
\punktg{28/10/91}{20:32}{8558.356}{  0.405}{0.014}{ 100}{1.12}{
366}{1.2CA}
\punktg{28/10/91}{20:37}{8558.359}{  0.393}{0.012}{ 103}{1.17}{
353}{1.2CA}
\punktg{28/10/91}{20:43}{8558.363}{  0.413}{0.009}{ 500}{1.46}{
675}{1.2CA}
\punktg{22/09/92}{20:01}{8888.334}{  0.603}{0.055}{ 150}{2.27}{
487}{1.2CA}
\punktg{22/09/92}{20:05}{8888.337}{  0.531}{0.051}{ 150}{2.64}{
480}{1.2CA}
\punktg{11/02/93}{04:31}{9029.689}{  0.360}{0.029}{ 250}{2.04}{
5651}{1.2CA}
\punktg{11/02/93}{04:37}{9029.693}{  0.421}{0.038}{ 250}{2.75}{
5837}{1.2CA}
\punktg{11/02/93}{04:37}{9029.693}{  0.421}{0.038}{ 250}{2.75}{
5837}{1.2CA}

\end{picture}}

\end{picture}

\vspace*{-0.02cm}

\begin{picture}(17 ,5 )(-1,0)
\put(0,0){\setlength{\unitlength}{0.0085cm}%
\begin{picture}(2000, 588.235)(7250,0)
\put(7250,0){\framebox(2000, 588.235)[tl]{\begin{picture}(0,0)(0,0)
        \put(2000,0){\makebox(0,0)[tr]{\large{1823+568}\T{4}
                                 \hspace*{0.5cm}}}
        \put(2000,-
588.235){\setlength{\unitlength}{1cm}\begin{picture}(0,0)(0,0)
        \end{picture}}
    \end{picture}}}

\thicklines
\put(7250,0){\setlength{\unitlength}{2.5cm}\begin{picture}(0,0)(0,-1)
   \put(0,0){\setlength{\unitlength}{1cm}\begin{picture}(0,0)(0,0)
        \put(0,0){\line(1,0){0.3}}
        \end{picture}}
   \end{picture}}

\put(9250,0){\setlength{\unitlength}{2.5cm}\begin{picture}(0,0)(0,-1)
   \put(0,0){\setlength{\unitlength}{1cm}\begin{picture}(0,0)(0,0)
        \put(0,0){\line(-1,0){0.3}}
        \end{picture}}
   \end{picture}}

\thinlines
\put(7250,0){\setlength{\unitlength}{2.5cm}\begin{picture}(0,0)(0,-1)
   \multiput(0,0)(0,0.1){10}{\setlength{\unitlength}{1cm}%
\begin{picture}(0,0)(0,0)
        \put(0,0){\line(1,0){0.12}}
        \end{picture}}
   \end{picture}}

\put(7250,0){\setlength{\unitlength}{2.5cm}\begin{picture}(0,0)(0,-1)
   \multiput(0,0)(0,-0.1){10}{\setlength{\unitlength}{1cm}%
\begin{picture}(0,0)(0,0)
        \put(0,0){\line(1,0){0.12}}
        \end{picture}}
   \end{picture}}

\put(9250,0){\setlength{\unitlength}{2.5cm}\begin{picture}(0,0)(0,-1)
   \multiput(0,0)(0,0.1){10}{\setlength{\unitlength}{1cm}%
\begin{picture}(0,0)(0,0)
        \put(0,0){\line(-1,0){0.12}}
        \end{picture}}
   \end{picture}}

\put(9250,0){\setlength{\unitlength}{2.5cm}\begin{picture}(0,0)(0,-1)
   \multiput(0,0)(0,-0.1){10}{\setlength{\unitlength}{1cm}%
\begin{picture}(0,0)(0,0)
        \put(0,0){\line(-1,0){0.12}}
        \end{picture}}
   \end{picture}}

\put(7250,0){\setlength{\unitlength}{2.5cm}\begin{picture}(0,0)(0,-1)
   \put(0,0.2){\setlength{\unitlength}{1cm}\begin{picture}(0,0)(0,0)
        \put(0,0){\line(1,0){0.12}}
        \put(-0.2,0){\makebox(0,0)[r]{\bf 0.2}}
        \end{picture}}
   \put(0,0.4){\setlength{\unitlength}{1cm}\begin{picture}(0,0)(0,0)
        \put(0,0){\line(1,0){0.12}}
        \put(-0.2,0){\makebox(0,0)[r]{\bf 0.4}}
        \end{picture}}
   \put(0,-0.2){\setlength{\unitlength}{1cm}\begin{picture}(0,0)(0,0)
        \put(0,0){\line(1,0){0.12}}
        \put(-0.2,0){\makebox(0,0)[r]{\bf -0.2}}
        \end{picture}}
   \put(0,-0.4){\setlength{\unitlength}{1cm}\begin{picture}(0,0)(0,0)
        \put(0,0){\line(1,0){0.12}}
        \put(-0.2,0){\makebox(0,0)[r]{\bf -0.4}}
        \end{picture}}
   \put(0,0.6){\setlength{\unitlength}{1cm}\begin{picture}(0,0)(0,0)
        \put(0,0){\line(1,0){0.12}}
        \put(-0.2,0){\makebox(0,0)[r]{\bf 0.6}}
        \end{picture}}
   \put(0,-0.6){\setlength{\unitlength}{1cm}\begin{picture}(0,0)(0,0)
        \put(0,0){\line(1,0){0.12}}
        \put(-0.2,0){\makebox(0,0)[r]{\bf -0.6}}
        \end{picture}}
   \put(0,-0.8){\setlength{\unitlength}{1cm}\begin{picture}(0,0)(0,0)
        \put(0,0){\line(1,0){0.12}}
        \put(-0.2,0){\makebox(0,0)[r]{\bf -0.8}}
        \end{picture}}
   \put(0,0.0){\setlength{\unitlength}{1cm}\begin{picture}(0,0)(0,0)
        \put(0,0){\line(1,0){0.12}}
        \put(-0.2,0){\makebox(0,0)[r]{\bf 0.0}}
        \end{picture}}
   \put(0,0.8){\setlength{\unitlength}{1cm}\begin{picture}(0,0)(0,0)
        \put(0,0){\line(1,0){0.12}}
        \put(-0.2,0){\makebox(0,0)[r]{\bf 0.8}}
        \end{picture}}
   \end{picture}}

   \put(7527.5, 588.235){\setlength{\unitlength}{1cm}\begin{picture}(0,0)(0,0)
        \put(0,0){\line(0,-1){0.2}}
   \end{picture}}
   \put(7892.5, 588.235){\setlength{\unitlength}{1cm}\begin{picture}(0,0)(0,0)
        \put(0,0){\line(0,-1){0.2}}
   \end{picture}}
   \put(8257.5, 588.235){\setlength{\unitlength}{1cm}\begin{picture}(0,0)(0,0)
        \put(0,0){\line(0,-1){0.2}}
  \end{picture}}
   \put(8622.5, 588.235){\setlength{\unitlength}{1cm}\begin{picture}(0,0)(0,0)
        \put(0,0){\line(0,-1){0.2}}
   \end{picture}}
   \put(8987.5, 588.235){\setlength{\unitlength}{1cm}\begin{picture}(0,0)(0,0)
        \put(0,0){\line(0,-1){0.2}}
   \end{picture}}

    \multiput(7250,0)(50,0){40}%
        {\setlength{\unitlength}{1cm}\begin{picture}(0,0)(0,0)
        \put(0,0){\line(0,1){0.12}}
    \end{picture}}
    \put(7500,0){\setlength{\unitlength}{1cm}\begin{picture}(0,0)(0,0)
        \put(0,0){\line(0,1){0.2}}
    \end{picture}}
    \put(8000,0){\setlength{\unitlength}{1cm}\begin{picture}(0,0)(0,0)
        \put(0,0){\line(0,1){0.2}}
    \end{picture}}
   \put(7750,0){\setlength{\unitlength}{1cm}\begin{picture}(0,0)(0,0)
        \put(0,0){\line(0,1){0.2}}
    \end{picture}}
   \put(8250,0){\setlength{\unitlength}{1cm}\begin{picture}(0,0)(0,0)
        \put(0,0){\line(0,1){0.2}}
    \end{picture}}
    \put(8500,0){\setlength{\unitlength}{1cm}\begin{picture}(0,0)(0,0)
        \put(0,0){\line(0,1){0.2}}
    \end{picture}}
    \put(8750,0){\setlength{\unitlength}{1cm}\begin{picture}(0,0)(0,0)
       \put(0,0){\line(0,1){0.2}}
    \end{picture}}
    \put(9000,0){\setlength{\unitlength}{1cm}\begin{picture}(0,0)(0,0)
       \put(0,0){\line(0,1){0.2}}
    \end{picture}}

\punktg{25/09/90}{20:20}{8160.348}{ -0.631}{0.035}{ 100}{1.79}{
606}{1.2CA}
\punktg{25/09/90}{20:27}{8160.353}{ -0.613}{0.018}{ 500}{1.91}{
897}{1.2CA}
\punktg{26/09/90}{20:20}{8161.348}{ -0.624}{0.037}{ 100}{1.78}{
605}{1.2CA}
\punktg{26/09/90}{20:27}{8161.353}{ -0.591}{0.018}{ 500}{1.90}{
895}{1.2CA}
\punktgg{26/05/91}{00:02}{8402.501}{  0.249}{0.303}{ 250}{2.22}{
1915}{1.2CA}
\punktgg{27/05/91}{01:48}{8403.575}{  0.281}{0.118}{ 100}{1.88}{
570}{1.2CA}
\punktgg{27/05/91}{01:52}{8403.578}{  0.304}{0.106}{ 100}{1.87}{
561}{1.2CA}
\punktg{27/05/91}{01:56}{8403.581}{  0.325}{0.097}{ 300}{2.18}{
1083}{1.2CA}
\punktg{24/07/91}{22:50}{8462.452}{ -0.436}{0.034}{ 500}{1.70}{
1500}{1.2CA}
\punktgg{26/07/91}{01:23}{8463.558}{ -0.255}{0.133}{ 100}{1.22}{
582}{1.2CA}
\punktg{26/07/91}{01:32}{8463.564}{ -0.299}{0.025}{ 500}{1.38}{
1819}{1.2CA}
\punktg{26/07/91}{23:32}{8464.481}{  0.121}{0.027}{ 300}{1.33}{
4420}{1.2CA}
\punktg{26/07/91}{23:39}{8464.486}{  0.142}{0.027}{ 300}{1.35}{
4268}{1.2CA}
\punktg{28/07/91}{21:19}{8466.388}{  0.012}{0.041}{ 300}{1.86}{
1036}{1.2CA}
\punktgg{29/07/91}{00:48}{8466.533}{ -0.135}{0.500}{ 150}{2.32}{
661}{1.2CA}
\punktg{29/07/91}{00:56}{8466.539}{ -0.194}{0.031}{ 500}{2.30}{
1489}{1.2CA}
\punktgg{29/07/91}{22:29}{8467.437}{ -0.197}{0.130}{ 100}{2.60}{
409}{1.2CA}
\punktg{29/07/91}{22:37}{8467.443}{ -0.243}{0.055}{ 500}{2.94}{
886}{1.2CA}
\punktg{30/07/91}{23:17}{8468.471}{ -0.081}{0.077}{ 300}{3.34}{
875}{1.2CA}
\punktg{30/07/91}{23:25}{8468.476}{ -0.195}{0.078}{ 300}{3.19}{
893}{1.2CA}
\punktg{31/07/91}{22:48}{8469.450}{  0.023}{0.010}{ 300}{1.13}{
478}{1.2CA}
\punktg{31/07/91}{22:54}{8469.454}{  0.003}{0.065}{ 300}{1.13}{
483}{1.2CA}
\punktg{31/07/91}{23:45}{8469.490}{  0.085}{0.013}{ 100}{1.05}{
364}{1.2CA}
\punktg{31/07/91}{23:52}{8469.495}{  0.075}{0.063}{ 300}{1.08}{
543}{1.2CA}
\punktg{01/08/91}{22:38}{8470.444}{ -0.149}{0.053}{ 500}{1.13}{
591}{1.2CA}
\punktg{01/08/91}{22:44}{8470.448}{ -0.143}{0.013}{ 150}{0.95}{
359}{1.2CA}
\punktg{02/08/91}{22:28}{8471.437}{ -0.257}{0.024}{ 100}{1.09}{
304}{1.2CA}
\punktg{02/08/91}{22:37}{8471.443}{ -0.257}{0.008}{ 500}{1.05}{
527}{1.2CA}
\punktg{03/08/91}{23:02}{8472.460}{ -0.360}{0.018}{ 100}{1.22}{
299}{1.2CA}
\punktg{03/08/91}{23:10}{8472.466}{ -0.395}{0.062}{ 500}{1.38}{
508}{1.2CA}
\punktg{05/08/91}{23:19}{8474.472}{ -0.434}{0.014}{ 100}{1.04}{
291}{1.2CA}
\punktg{05/08/91}{23:28}{8474.478}{ -0.437}{0.064}{ 500}{1.11}{
444}{1.2CA}
\punktgg{07/08/91}{00:08}{8475.506}{ -0.425}{0.048}{ 100}{1.04}{
288}{1.2CA}
\punktg{07/08/91}{00:16}{8475.512}{ -0.477}{0.008}{ 500}{1.17}{
473}{1.2CA}
\punktg{08/08/91}{00:18}{8476.513}{ -0.380}{0.047}{ 100}{1.63}{
376}{1.2CA}
\punktg{08/08/91}{00:28}{8476.520}{ -0.438}{0.062}{ 500}{1.46}{
599}{1.2CA}
\punktgg{08/08/91}{23:26}{8477.476}{ -0.270}{0.137}{ 100}{1.86}{
317}{1.2CA}
\punktg{08/08/91}{23:34}{8477.482}{ -0.221}{0.059}{ 500}{2.14}{
471}{1.2CA}
\punktgg{09/08/91}{22:09}{8478.423}{  0.207}{0.074}{ 100}{3.11}{
379}{1.2CA}
\punktg{09/08/91}{22:22}{8478.432}{  0.150}{0.025}{1000}{3.20}{
825}{1.2CA}
\punktg{09/08/91}{23:23}{8478.475}{  0.230}{0.044}{1000}{3.85}{
1177}{1.2CA}
\punktg{10/08/91}{00:09}{8478.507}{  0.245}{0.027}{1000}{3.59}{
895}{1.2CA}
\punktg{10/08/91}{00:27}{8478.519}{  0.263}{0.027}{1000}{3.37}{
914}{1.2CA}
\punktg{10/08/91}{01:33}{8478.565}{  0.255}{0.021}{1000}{3.08}{
933}{1.2CA}
\punktg{10/08/91}{01:52}{8478.578}{  0.272}{0.027}{1000}{3.50}{
974}{1.2CA}
\punktg{10/08/91}{20:31}{8479.355}{  0.286}{0.041}{ 100}{2.62}{
321}{1.2CA}
\punktg{10/08/91}{23:11}{8479.466}{  0.237}{0.044}{ 100}{2.91}{
324}{1.2CA}
\punktg{10/08/91}{23:33}{8479.482}{  0.294}{0.022}{ 500}{2.63}{
521}{1.2CA}
\punktg{11/08/91}{00:10}{8479.508}{  0.327}{0.048}{ 500}{2.67}{
519}{1.2CA}
\punktgg{11/08/91}{19:53}{8480.329}{  0.106}{0.093}{ 100}{1.65}{
2167}{1.2CA}
\punktg{11/08/91}{20:13}{8480.343}{  0.137}{0.011}{ 500}{1.59}{
870}{1.2CA}
\punktg{11/08/91}{22:52}{8480.453}{  0.182}{0.022}{ 100}{1.73}{
330}{1.2CA}
\punktg{11/08/91}{22:59}{8480.458}{  0.194}{0.010}{ 500}{1.63}{
556}{1.2CA}
\punktg{12/08/91}{00:34}{8480.524}{  0.264}{0.018}{ 100}{1.50}{
323}{1.2CA}
\punktg{12/08/91}{00:42}{8480.529}{  0.265}{0.007}{ 500}{1.41}{
597}{1.2CA}
\punktg{12/08/91}{21:01}{8481.376}{  0.256}{0.019}{ 100}{1.55}{
328}{1.2CA}
\punktg{12/08/91}{21:11}{8481.383}{  0.257}{0.045}{ 500}{1.75}{
491}{1.2CA}
\punktg{12/08/91}{23:37}{8481.484}{  0.252}{0.021}{ 100}{1.60}{
352}{1.2CA}
\punktg{12/08/91}{23:47}{8481.491}{  0.283}{0.009}{ 500}{1.65}{
534}{1.2CA}
\punktg{19/08/91}{22:40}{8488.445}{ -0.288}{0.034}{ 100}{1.11}{
342}{1.2CA}
\punktg{19/08/91}{22:48}{8488.450}{ -0.259}{0.019}{ 500}{1.18}{
705}{1.2CA}
\punktg{19/08/91}{23:13}{8488.468}{ -0.282}{0.030}{ 100}{1.17}{
333}{1.2CA}
\punktg{21/08/91}{20:24}{8490.351}{ -0.077}{0.039}{ 500}{1.37}{
1730}{1.2CA}
\punktgg{22/08/91}{20:03}{8491.336}{ -0.253}{0.106}{ 500}{2.75}{
1290}{1.2CA}
\punktgg{23/08/91}{21:35}{8492.400}{ -0.247}{0.099}{ 100}{1.78}{
495}{1.2CA}
\punktg{23/08/91}{21:43}{8492.405}{ -0.226}{0.059}{ 500}{1.81}{
1500}{1.2CA}
\punktgg{24/08/91}{00:49}{8492.534}{ -0.244}{0.108}{ 100}{1.55}{
548}{1.2CA}
\punktg{24/08/91}{00:57}{8492.540}{ -0.191}{0.060}{ 500}{1.72}{
1735}{1.2CA}
\punktgg{24/08/91}{21:06}{8493.379}{ -0.351}{0.070}{ 100}{1.38}{
506}{1.2CA}
\punktg{24/08/91}{21:13}{8493.385}{ -0.293}{0.035}{ 500}{1.41}{
1603}{1.2CA}
\punktgg{25/08/91}{21:37}{8494.401}{ -0.272}{0.342}{ 100}{1.60}{
1212}{1.2CA}
\punktg{25/08/91}{21:45}{8494.407}{ -0.267}{0.077}{ 500}{1.42}{
3502}{1.2CA}
\punktgg{26/08/91}{22:26}{8495.435}{ -0.312}{0.202}{ 500}{2.93}{
2015}{1.2CA}
\punktgg{27/08/91}{20:36}{8496.358}{ -0.235}{0.260}{ 100}{2.84}{
421}{1.2CA}
\punktg{27/08/91}{20:46}{8496.366}{ -0.247}{0.062}{ 500}{2.78}{
1047}{1.2CA}
\punktg{17/10/91}{20:28}{8547.353}{  0.260}{0.086}{ 100}{2.70}{
417}{1.2CA}
\punktg{17/10/91}{20:35}{8547.358}{  0.229}{0.056}{ 500}{2.40}{
1068}{1.2CA}
\punktg{20/10/91}{21:09}{8550.381}{  0.053}{0.021}{ 100}{1.27}{
503}{1.2CA}
\punktg{20/10/91}{21:16}{8550.386}{  0.003}{0.020}{ 500}{1.32}{
1547}{1.2CA}
\punktg{26/10/91}{19:31}{8556.314}{  0.215}{0.053}{ 500}{3.63}{
641}{1.2CA}
\punktgg{27/10/91}{21:00}{8557.375}{  0.252}{0.118}{ 100}{1.95}{
597}{1.2CA}
\punktg{27/10/91}{21:08}{8557.381}{  0.267}{0.050}{ 500}{1.94}{
778}{1.2CA}
\punktgg{28/10/91}{20:52}{8558.369}{  0.360}{0.109}{ 100}{1.17}{
338}{1.2CA}
\punktg{28/10/91}{20:59}{8558.375}{  0.297}{0.050}{ 500}{1.58}{
684}{1.2CA}
\punktg{21/09/92}{21:35}{8887.400}{  0.565}{0.026}{ 100}{1.69}{
403}{1.2CA}
\punktg{21/09/92}{21:38}{8887.402}{  0.576}{0.032}{ 100}{1.64}{
399}{1.2CA}
\punktg{21/09/92}{21:38}{8887.402}{  0.576}{0.032}{ 100}{1.64}{
399}{1.2CA}
\punktg{}{}{9127.641}{0.930}{0.008}{}{}{}{1.2CA}%
\punktg{}{}{9129.632}{0.806}{0.009}{}{}{}{1.2CA}%

\end{picture}}

\end{picture}

\vspace*{-0.02cm}

\begin{picture}(17 ,10 )(-1,0)
\put(0,0){\setlength{\unitlength}{0.0085cm}%
\begin{picture}(2000,1176.470)(7250,0)
\put(7250,0){\framebox(2000,1176.470)[tl]{\begin{picture}(0,0)(0,0)
        \put(2000,0){\makebox(0,0)[tr]{\large{2223$-$052}\T{0.4}
                                 \hspace*{0.5cm}}}

\put(2000,-1176.470){\setlength{\unitlength}{1cm}\begin{picture}(0,0)(0,0)
            \put(0,-1){\makebox(0,0)[br]{\bf J.D.\,2,440,000\,+}}
        \end{picture}}
    \end{picture}}}

\thicklines
\put(7250,0){\setlength{\unitlength}{2.5cm}\begin{picture}(0,0)(0,-2)
   \put(0,0){\setlength{\unitlength}{1cm}\begin{picture}(0,0)(0,0)
        \put(0,0){\line(1,0){0.3}}
        \end{picture}}
   \end{picture}}

\put(9250,0){\setlength{\unitlength}{2.5cm}\begin{picture}(0,0)(0,-2)
   \put(0,0){\setlength{\unitlength}{1cm}\begin{picture}(0,0)(0,0)
        \put(0,0){\line(-1,0){0.3}}
        \end{picture}}
   \end{picture}}

\thinlines
\put(7250,0){\setlength{\unitlength}{2.5cm}\begin{picture}(0,0)(0,-2)
   \multiput(0,0)(0,0.1){20}{\setlength{\unitlength}{1cm}%
\begin{picture}(0,0)(0,0)
        \put(0,0){\line(1,0){0.12}}
        \end{picture}}
   \end{picture}}

\put(7250,0){\setlength{\unitlength}{2.5cm}\begin{picture}(0,0)(0,-2)
   \multiput(0,0)(0,-0.1){20}{\setlength{\unitlength}{1cm}%
\begin{picture}(0,0)(0,0)
        \put(0,0){\line(1,0){0.12}}
        \end{picture}}
   \end{picture}}

\put(9250,0){\setlength{\unitlength}{2.5cm}\begin{picture}(0,0)(0,-2)
   \multiput(0,0)(0,0.1){20}{\setlength{\unitlength}{1cm}%
\begin{picture}(0,0)(0,0)
        \put(0,0){\line(-1,0){0.12}}
        \end{picture}}
   \end{picture}}

\put(9250,0){\setlength{\unitlength}{2.5cm}\begin{picture}(0,0)(0,-2)
   \multiput(0,0)(0,-0.1){20}{\setlength{\unitlength}{1cm}%
\begin{picture}(0,0)(0,0)
        \put(0,0){\line(-1,0){0.12}}
        \end{picture}}
   \end{picture}}

\put(7250,0){\setlength{\unitlength}{2.5cm}\begin{picture}(0,0)(0,-2)
   \put(0,0.5){\setlength{\unitlength}{1cm}\begin{picture}(0,0)(0,0)
        \put(0,0){\line(1,0){0.12}}
        \put(-0.2,0){\makebox(0,0)[r]{\bf 0.5}}
        \end{picture}}
   \put(0,1.0){\setlength{\unitlength}{1cm}\begin{picture}(0,0)(0,0)
        \put(0,0){\line(1,0){0.12}}
        \put(-0.2,0){\makebox(0,0)[r]{\bf 1.0}}
        \end{picture}}
   \put(0,-0.5){\setlength{\unitlength}{1cm}\begin{picture}(0,0)(0,0)
        \put(0,0){\line(1,0){0.12}}
        \put(-0.2,0){\makebox(0,0)[r]{\bf -0.5}}
        \end{picture}}
   \put(0,-1.0){\setlength{\unitlength}{1cm}\begin{picture}(0,0)(0,0)
        \put(0,0){\line(1,0){0.12}}
        \put(-0.2,0){\makebox(0,0)[r]{\bf -1.0}}
        \end{picture}}
   \put(0,1.5){\setlength{\unitlength}{1cm}\begin{picture}(0,0)(0,0)
        \put(0,0){\line(1,0){0.12}}
        \put(-0.2,0){\makebox(0,0)[r]{\bf 1.5}}
        \end{picture}}
   \put(0,-1.5){\setlength{\unitlength}{1cm}\begin{picture}(0,0)(0,0)
        \put(0,0){\line(1,0){0.12}}
        \put(-0.2,0){\makebox(0,0)[r]{\bf -1.5}}
        \end{picture}}
   \put(0,0.0){\setlength{\unitlength}{1cm}\begin{picture}(0,0)(0,0)
        \put(0,0){\line(1,0){0.12}}
        \put(-0.2,0){\makebox(0,0)[r]{\bf 0.0}}
        \end{picture}}
   \put(0,0.0){\setlength{\unitlength}{1cm}\begin{picture}(0,0)(0,0)
        \put(0,0){\line(1,0){0.12}}
        \put(-0.2,0){\makebox(0,0)[r]{\bf 0.0}}
        \end{picture}}
   \put(0,0.0){\setlength{\unitlength}{1cm}\begin{picture}(0,0)(0,0)
        \put(0,0){\line(1,0){0.12}}
        \end{picture}}
   \end{picture}}

   \put(7527.5,1176.470){\setlength{\unitlength}{1cm}\begin{picture}(0,0)(0,0)
        \put(0,0){\line(0,-1){0.2}}
   \end{picture}}
   \put(7892.5,1176.470){\setlength{\unitlength}{1cm}\begin{picture}(0,0)(0,0)
        \put(0,0){\line(0,-1){0.2}}
   \end{picture}}
   \put(8257.5,1176.470){\setlength{\unitlength}{1cm}\begin{picture}(0,0)(0,0)
        \put(0,0){\line(0,-1){0.2}}
  \end{picture}}
   \put(8622.5,1176.470){\setlength{\unitlength}{1cm}\begin{picture}(0,0)(0,0)
        \put(0,0){\line(0,-1){0.2}}
   \end{picture}}
   \put(8987.5,1176.470){\setlength{\unitlength}{1cm}\begin{picture}(0,0)(0,0)
        \put(0,0){\line(0,-1){0.2}}
   \end{picture}}

    \multiput(7250,0)(50,0){40}%
        {\setlength{\unitlength}{1cm}\begin{picture}(0,0)(0,0)
        \put(0,0){\line(0,1){0.12}}
    \end{picture}}
    \put(7500,0){\setlength{\unitlength}{1cm}\begin{picture}(0,0)(0,0)
        \put(0,0){\line(0,1){0.2}}
        \put(0,-0.2){\makebox(0,0)[t]{\bf 7500}}
    \end{picture}}
    \put(8000,0){\setlength{\unitlength}{1cm}\begin{picture}(0,0)(0,0)
        \put(0,0){\line(0,1){0.2}}
        \put(0,-0.2){\makebox(0,0)[t]{\bf 8000}}
    \end{picture}}
   \put(7750,0){\setlength{\unitlength}{1cm}\begin{picture}(0,0)(0,0)
        \put(0,0){\line(0,1){0.2}}
        \put(0,-0.2){\makebox(0,0)[t]{\bf 7750}}
    \end{picture}}
   \put(8250,0){\setlength{\unitlength}{1cm}\begin{picture}(0,0)(0,0)
        \put(0,0){\line(0,1){0.2}}
        \put(0,-0.2){\makebox(0,0)[t]{\bf 8250}}
    \end{picture}}
    \put(8500,0){\setlength{\unitlength}{1cm}\begin{picture}(0,0)(0,0)
        \put(0,0){\line(0,1){0.2}}
        \put(0,-0.2){\makebox(0,0)[t]{\bf 8500}}
    \end{picture}}
    \put(8750,0){\setlength{\unitlength}{1cm}\begin{picture}(0,0)(0,0)
       \put(0,0){\line(0,1){0.2}}
        \put(0,-0.2){\makebox(0,0)[t]{\bf 8750}}
    \end{picture}}
    \put(9000,0){\setlength{\unitlength}{1cm}\begin{picture}(0,0)(0,0)
       \put(0,0){\line(0,1){0.2}}
        \put(0,-0.2){\makebox(0,0)[t]{\bf 9000}}
    \end{picture}}

\punktc{07/09/88}{22:11}{7412.425}{  0.921}{0.014}{ 500}{1.37}{
657}{1.2CA}
\punktc{04/10/88}{21:33}{7439.398}{  1.141}{0.026}{ 500}{1.66}{
585}{1.2CA}
\punktc{05/10/88}{21:11}{7440.383}{  0.789}{0.035}{ 500}{2.47}{
655}{1.2CA}
\punktc{05/10/88}{21:24}{7440.392}{  0.764}{0.038}{ 500}{2.37}{
681}{1.2CA}
\punktc{05/10/88}{21:37}{7440.401}{  0.763}{0.029}{ 500}{2.17}{
714}{1.2CA}
\punktc{07/10/88}{21:03}{7442.378}{  0.931}{0.024}{ 500}{1.59}{
621}{1.2CA}
\punktc{08/10/88}{20:38}{7443.360}{  0.981}{0.025}{ 500}{1.28}{
647}{1.2CA}
\punktc{10/10/88}{20:10}{7445.341}{  0.539}{0.030}{ 500}{2.16}{
636}{1.2CA}
\punktc{16/10/88}{22:40}{7451.445}{  0.465}{0.030}{ 500}{2.30}{
640}{1.2CA}
\punktc{11/06/89}{02:45}{7688.615}{  1.746}{0.020}{ 100}{2.39}{
486}{1.2CA}
\punktc{11/06/89}{02:51}{7688.619}{  1.768}{0.036}{ 500}{1.91}{
758}{1.2CA}
\punktcc{13/06/89}{03:04}{7690.628}{  1.603}{0.046}{ 100}{1.81}{
485}{1.2CA}
\punktc{13/06/89}{03:14}{7690.635}{  1.596}{0.022}{ 500}{1.87}{
748}{1.2CA}
\punktc{25/06/89}{14:16}{7703.095}{ -0.026}{0.011}{ 500}{1.44}{
1476}{1.2CA}
\punktc{27/06/89}{02:23}{7704.600}{  0.035}{0.039}{ 500}{1.92}{
840}{1.2CA}
\punktc{11/08/89}{00:24}{7749.517}{ -0.287}{0.008}{1000}{1.55}{
1043}{1.2CA}
\punktc{11/08/89}{01:19}{7749.555}{ -0.298}{0.013}{1000}{1.62}{
1125}{1.2CA}
\punktc{11/08/89}{01:59}{7749.583}{ -0.317}{0.017}{1000}{1.74}{
960}{1.2CA}
\punktc{13/08/89}{00:54}{7751.538}{ -0.272}{0.010}{ 500}{1.48}{
786}{1.2CA}
\punktc{01/09/89}{00:57}{7770.540}{  0.463}{0.011}{ 500}{2.75}{
1028}{1.2CA}
\punktc{02/09/89}{00:13}{7771.509}{  0.567}{0.009}{ 500}{2.44}{
884}{1.2CA}
\punktc{03/09/89}{00:11}{7772.508}{  0.704}{0.028}{ 500}{1.93}{
778}{1.2CA}
\punktc{03/09/89}{23:40}{7773.486}{  0.929}{0.030}{ 500}{1.90}{
755}{1.2CA}
\punktc{01/10/89}{09:08}{7800.881}{  0.373}{0.034}{ 500}{2.25}{
531}{1.2CA}
\punktc{16/10/89}{22:18}{7816.430}{  0.088}{0.014}{1000}{2.36}{
1789}{1.2CA}
\punktc{31/10/89}{21:11}{7831.383}{ -0.221}{0.056}{ 500}{1.95}{
571}{1.2CA}
\punktc{01/11/89}{20:15}{7832.344}{ -0.295}{0.059}{ 500}{1.36}{
618}{1.2CA}
\punktc{02/11/89}{19:55}{7833.330}{ -0.326}{0.084}{ 500}{1.79}{
567}{1.2CA}
\punktc{15/12/89}{18:45}{7876.281}{  1.325}{0.023}{ 100}{2.42}{
538}{1.2CA}
\punktc{15/12/89}{18:53}{7876.287}{  1.311}{0.007}{ 500}{2.45}{
702}{1.2CA}
\punktc{20/12/89}{18:38}{7881.277}{  1.136}{0.035}{ 542}{2.00}{
705}{1.2CA}
\punktc{24/05/90}{03:31}{8035.647}{ -0.438}{0.016}{ 400}{3.12}{
2469}{1.2CA}
\punktc{31/07/90}{01:05}{8103.546}{  0.400}{0.012}{ 500}{1.71}{
629}{1.2CA}
\punktcc{24/09/90}{21:59}{8159.417}{ -1.285}{0.073}{ 100}{3.10}{
556}{1.2CA}
\punktc{24/09/90}{22:07}{8159.422}{ -1.300}{0.021}{ 500}{2.38}{
847}{1.2CA}
\punktcc{25/09/90}{21:59}{8160.417}{ -1.308}{0.073}{ 100}{3.08}{
555}{1.2CA}
\punktc{25/09/90}{22:07}{8160.422}{ -1.309}{0.020}{ 500}{2.36}{
848}{1.2CA}
\punktc{16/10/90}{22:26}{8181.435}{ -1.613}{0.063}{ 500}{3.01}{
840}{1.2CA}
\punktcc{17/10/90}{22:15}{8182.427}{ -1.604}{0.094}{ 100}{3.18}{
550}{1.2CA}
\punktc{17/10/90}{22:26}{8182.435}{ -1.619}{0.063}{ 500}{3.02}{
840}{1.2CA}
\punktc{18/10/90}{19:38}{8183.319}{ -1.574}{0.022}{ 500}{2.48}{
798}{1.2CA}
\punktc{18/10/90}{19:38}{8183.319}{ -1.584}{0.005}{ 500}{2.48}{
798}{1.2CA}
\punktcc{19/10/90}{19:21}{8184.307}{ -1.645}{0.054}{ 100}{2.43}{
550}{1.2CA}
\punktc{19/10/90}{19:38}{8184.319}{ -1.577}{0.061}{ 500}{2.45}{
797}{1.2CA}
\punktc{19/10/90}{21:33}{8184.398}{ -1.584}{0.014}{ 500}{1.54}{
845}{1.2CA}
\punktc{20/10/90}{21:16}{8185.387}{ -1.585}{0.028}{ 100}{1.58}{
548}{1.2CA}
\punktc{20/10/90}{21:33}{8185.398}{ -1.589}{0.013}{ 500}{1.53}{
845}{1.2CA}
\punktcc{24/05/91}{03:55}{8400.664}{ -0.911}{0.106}{ 100}{1.44}{
617}{1.2CA}
\punktcc{24/05/91}{03:58}{8400.665}{ -0.841}{0.104}{ 100}{1.83}{
734}{1.2CA}
\punktcc{25/05/91}{03:10}{8401.633}{ -1.069}{0.071}{ 100}{1.95}{
328}{1.2CA}
\punktc{25/05/91}{03:20}{8401.640}{ -1.064}{0.105}{ 300}{2.60}{
463}{1.2CA}
\punktcc{27/05/91}{03:57}{8403.665}{ -1.051}{0.138}{ 100}{1.93}{
881}{1.2CA}
\punktcc{25/07/91}{01:02}{8462.543}{ -1.360}{0.137}{ 100}{1.48}{
716}{1.2CA}
\punktc{25/07/91}{01:12}{8462.551}{ -1.359}{0.043}{ 500}{1.56}{
2348}{1.2CA}
\punktcc{26/07/91}{01:53}{8463.579}{ -1.330}{0.139}{ 100}{1.39}{
1048}{1.2CA}
\punktc{01/08/91}{01:16}{8469.553}{ -1.233}{0.029}{ 100}{1.20}{
521}{1.2CA}
\punktc{01/08/91}{01:25}{8469.559}{ -1.273}{0.059}{ 500}{1.29}{
1579}{1.2CA}
\punktc{02/08/91}{00:23}{8470.517}{ -1.245}{0.059}{ 500}{1.60}{
972}{1.2CA}
\punktc{02/08/91}{23:23}{8471.475}{ -1.282}{0.030}{ 100}{1.54}{
361}{1.2CA}
\punktc{02/08/91}{23:31}{8471.480}{ -1.252}{0.021}{ 500}{1.75}{
776}{1.2CA}
\punktc{04/08/91}{01:34}{8472.566}{ -1.138}{0.026}{ 100}{1.47}{
324}{1.2CA}
\punktc{04/08/91}{01:43}{8472.572}{ -1.177}{0.012}{ 500}{1.28}{
627}{1.2CA}
\punktc{05/08/91}{01:58}{8473.582}{ -1.094}{0.023}{ 100}{1.35}{
334}{1.2CA}
\punktc{05/08/91}{02:08}{8473.589}{ -1.127}{0.010}{ 500}{1.32}{
661}{1.2CA}
\punktcc{06/08/91}{00:28}{8474.520}{ -1.044}{0.121}{ 100}{1.05}{
295}{1.2CA}
\punktc{06/08/91}{00:47}{8474.533}{ -1.039}{0.007}{ 500}{1.07}{
471}{1.2CA}
\punktc{07/08/91}{02:28}{8475.603}{ -1.039}{0.013}{ 100}{1.06}{
300}{1.2CA}
\punktc{07/08/91}{02:36}{8475.609}{ -1.029}{0.008}{ 500}{1.23}{
506}{1.2CA}
\punktc{08/08/91}{01:09}{8476.548}{ -0.941}{0.037}{ 100}{2.27}{
336}{1.2CA}
\punktc{08/08/91}{01:17}{8476.554}{ -0.970}{0.027}{ 500}{3.12}{
532}{1.2CA}
\punktcc{11/08/91}{00:55}{8479.539}{ -0.617}{0.134}{ 100}{3.76}{
313}{1.2CA}
\punktc{12/08/91}{01:43}{8480.572}{ -0.713}{0.020}{ 100}{1.46}{
324}{1.2CA}
\punktc{12/08/91}{01:50}{8480.577}{ -0.717}{0.009}{ 500}{1.55}{
590}{1.2CA}
\punktc{13/08/91}{01:23}{8481.558}{ -0.672}{0.019}{ 500}{1.75}{
526}{1.2CA}
\punktc{14/08/91}{03:53}{8482.662}{ -0.731}{0.050}{ 300}{1.85}{
448}{1.2CA}
\punktc{16/08/91}{03:33}{8484.648}{ -0.774}{0.104}{ 300}{2.05}{
412}{1.2CA}
\punktc{17/08/91}{01:01}{8485.543}{ -0.865}{0.026}{ 300}{1.63}{
403}{1.2CA}
\punktcc{18/08/91}{04:33}{8486.690}{ -0.794}{0.145}{ 300}{2.18}{
1666}{1.2CA}
\punktc{18/08/91}{23:16}{8487.470}{ -0.666}{0.057}{ 300}{2.28}{
493}{1.2CA}
\punktc{20/08/91}{01:14}{8488.552}{ -0.451}{0.031}{ 100}{1.33}{
301}{1.2CA}
\punktc{20/08/91}{01:22}{8488.557}{ -0.425}{0.017}{ 500}{1.20}{
493}{1.2CA}
\punktcc{26/10/91}{21:58}{8556.416}{ -1.248}{0.121}{ 100}{2.61}{
449}{1.2CA}
\punktc{26/10/91}{22:08}{8556.423}{ -1.260}{0.073}{ 500}{2.93}{
1345}{1.2CA}
\punktc{27/10/91}{22:37}{8557.443}{ -1.233}{0.029}{ 100}{1.38}{
400}{1.2CA}
\punktc{27/10/91}{22:47}{8557.450}{ -1.233}{0.024}{ 500}{1.65}{
985}{1.2CA}
\punktc{28/10/91}{22:39}{8558.444}{ -1.338}{0.033}{ 100}{1.44}{
377}{1.2CA}
\punktc{28/10/91}{22:46}{8558.449}{ -1.285}{0.064}{ 500}{1.91}{
871}{1.2CA}
\punktc{28/10/91}{22:46}{8558.449}{ -1.285}{0.064}{ 500}{1.91}{
871}{1.2CA}

\end{picture}}

\end{picture}

\addtocounter{figure}{-1}

\vspace*{1cm}
\caption{(continued)}
\end{figure*}

\noindent
{\bf PKS\,0735+178} is a BL\,Lac object with an unknown redshift. The
Rosemary Hill data reported by WSLF suggest long-term trends with
amplitudes of about 2\,mag with more rapid flares of 1\,--\,1.5\,mag
superimposed. The historical lightcurve of Pollock (\cite{Pol75}) shows a
2.2\,mag total range of variability. The HQM lightcurve shows one flare in
early 1991; the preceeding state has been monitored with higher resolution
and is shown in Fig.~5. Making use of
the photometric sequence of McGimsey ({\cite{McG76}) and Smith et al.\
(\cite{SBHE85}), we were able to determine the reference magnitude of our
lightcurve, $R_0=15.24$.

\begin{figure}

\vspace*{0.5cm}

\begin{picture}(7.8 ,4.5 )(-1,0)
\put(0,0){\setlength{\unitlength}{0.4cm}%
\begin{picture}(18,  11.250)(870,0)
\put(870,0){\framebox(18,  11.250)[tl]{\begin{picture}(0,0)(0,0)
        \put(0,0){\makebox(0,0)[tr]{$\Delta R$\hspace*{0.2cm}}}
        \put(0,-10.25){\makebox(0,0)[bl]{\hspace*{0.5cm}
                           \large{0235+164}}}
        \put(18,-
11.250){\setlength{\unitlength}{1cm}\begin{picture}(0,0)(0,0)
            \put(0,-1){\makebox(0,0)[br]{\bf J.D.\,2,440,000\,+}}
        \end{picture}}
    \end{picture}}}

\thicklines
\put(870,0){\setlength{\unitlength}{2.5cm}\begin{picture}(0,0)(0,-1.2)
   \put(0,0){\setlength{\unitlength}{1cm}\begin{picture}(0,0)(0,0)
        \put(0,0){\line(1,0){0.3}}
        \end{picture}}
   \end{picture}}

\put(888,0){\setlength{\unitlength}{2.5cm}\begin{picture}(0,0)(0,-1.2)
   \put(0,0){\setlength{\unitlength}{1cm}\begin{picture}(0,0)(0,0)
        \put(0,0){\line(-1,0){0.3}}
        \end{picture}}
   \end{picture}}

\thinlines
\put(870,0){\setlength{\unitlength}{2.5cm}\begin{picture}(0,0)(0,-1.2)
   \multiput(0,0)(0,0.1){6}{\setlength{\unitlength}{1cm}%
\begin{picture}(0,0)(0,0)
        \put(0,0){\line(1,0){0.12}}
        \end{picture}}
   \end{picture}}

\put(870,0){\setlength{\unitlength}{2.5cm}\begin{picture}(0,0)(0,-1.2)
   \multiput(0,0)(0,-0.1){12}{\setlength{\unitlength}{1cm}%
\begin{picture}(0,0)(0,0)
        \put(0,0){\line(1,0){0.12}}
        \end{picture}}
   \end{picture}}

\put(888,0){\setlength{\unitlength}{2.5cm}\begin{picture}(0,0)(0,-1.2)
   \multiput(0,0)(0,0.1){6}{\setlength{\unitlength}{1cm}%
\begin{picture}(0,0)(0,0)
        \put(0,0){\line(-1,0){0.12}}
        \end{picture}}
   \end{picture}}

\put(888,0){\setlength{\unitlength}{2.5cm}\begin{picture}(0,0)(0,-1.2)
   \multiput(0,0)(0,-0.1){12}{\setlength{\unitlength}{1cm}%
\begin{picture}(0,0)(0,0)
        \put(0,0){\line(-1,0){0.12}}
        \end{picture}}
   \end{picture}}

\put(870,0){\setlength{\unitlength}{2.5cm}\begin{picture}(0,0)(0,-1.2)
   \put(0,0.2){\setlength{\unitlength}{1cm}\begin{picture}(0,0)(0,0)
        \put(0,0){\line(1,0){0.12}}
        \put(-0.2,0){\makebox(0,0)[r]{\bf 0.2}}
        \end{picture}}
   \put(0,0.0){\setlength{\unitlength}{1cm}\begin{picture}(0,0)(0,0)
        \put(0,0){\line(1,0){0.12}}
        \put(-0.2,0){\makebox(0,0)[r]{\bf 0.0}}
        \end{picture}}
   \put(0,-0.2){\setlength{\unitlength}{1cm}\begin{picture}(0,0)(0,0)
        \put(0,0){\line(1,0){0.12}}
        \put(-0.2,0){\makebox(0,0)[r]{\bf -0.2}}
        \end{picture}}
   \put(0,-0.4){\setlength{\unitlength}{1cm}\begin{picture}(0,0)(0,0)
        \put(0,0){\line(1,0){0.12}}
        \put(-0.2,0){\makebox(0,0)[r]{\bf -0.4}}
        \end{picture}}
   \put(0,-0.6){\setlength{\unitlength}{1cm}\begin{picture}(0,0)(0,0)
        \put(0,0){\line(1,0){0.12}}
        \put(-0.2,0){\makebox(0,0)[r]{\bf -0.6}}
        \end{picture}}
   \put(0,-0.8){\setlength{\unitlength}{1cm}\begin{picture}(0,0)(0,0)
        \put(0,0){\line(1,0){0.12}}
        \put(-0.2,0){\makebox(0,0)[r]{\bf -0.8}}
        \end{picture}}
   \put(0,-1.0){\setlength{\unitlength}{1cm}\begin{picture}(0,0)(0,0)
        \put(0,0){\line(1,0){0.12}}
        \put(-0.2,0){\makebox(0,0)[r]{\bf -1.0}}
        \end{picture}}
   \put(0,0.0){\setlength{\unitlength}{1cm}\begin{picture}(0,0)(0,0)
        \put(0,0){\line(1,0){0.12}}
        \put(-0.2,0){\makebox(0,0)[r]{\bf 0.0}}
        \end{picture}}
   \put(0,0.4){\setlength{\unitlength}{1cm}\begin{picture}(0,0)(0,0)
        \put(0,0){\line(1,0){0.12}}
        \end{picture}}
   \end{picture}}

   \put(875.5,  11.250){\setlength{\unitlength}{1cm}\begin{picture}(0,0)(0,0)
        \put(0,0){\line(0,-1){0.2}}
        \put(0,0.2){\makebox(0,0)[b]{\bf 1989 Feb.~15.0}}
   \end{picture}}

    \multiput(870,0)(1,0){18}%
        {\setlength{\unitlength}{1cm}\begin{picture}(0,0)(0,0)
        \put(0,0){\line(0,1){0.12}}
    \end{picture}}
    \put(875,0){\setlength{\unitlength}{1cm}\begin{picture}(0,0)(0,0)
        \put(0,0){\line(0,1){0.2}}
        \put(0,-0.2){\makebox(0,0)[t]{\bf 7875}}
    \end{picture}}
    \put(875,0){\setlength{\unitlength}{1cm}\begin{picture}(0,0)(0,0)
        \put(0,0){\line(0,1){0.2}}
        \put(0,-0.2){\makebox(0,0)[t]{\bf 7875}}
    \end{picture}}
   \put(880,0){\setlength{\unitlength}{1cm}\begin{picture}(0,0)(0,0)
        \put(0,0){\line(0,1){0.2}}
        \put(0,-0.2){\makebox(0,0)[t]{\bf 7880}}
    \end{picture}}
   \put(880,0){\setlength{\unitlength}{1cm}\begin{picture}(0,0)(0,0)
        \put(0,0){\line(0,1){0.2}}
        \put(0,-0.2){\makebox(0,0)[t]{\bf 7880}}
    \end{picture}}
    \put(885,0){\setlength{\unitlength}{1cm}\begin{picture}(0,0)(0,0)
        \put(0,0){\line(0,1){0.2}}
        \put(0,-0.2){\makebox(0,0)[t]{\bf 7885}}
    \end{picture}}
    \put(885,0){\setlength{\unitlength}{1cm}\begin{picture}(0,0)(0,0)
       \put(0,0){\line(0,1){0.2}}
        \put(0,-0.2){\makebox(0,0)[t]{\bf 7885}}
    \end{picture}}
    \put(885,0){\setlength{\unitlength}{1cm}\begin{picture}(0,0)(0,0)
       \put(0,0){\line(0,1){0.2}}
        \put(0,-0.2){\makebox(0,0)[t]{\bf 7885}}
    \end{picture}}

\punkth{13/12/89}{19:54}{874.330}{ -0.195}{0.035}{ 500}{2.20}{
2396}{1.2CA}
\punkth{13/12/89}{23:33}{874.481}{  0.041}{0.047}{ 500}{2.02}{
4169}{1.2CA}
\punkth{14/12/89}{01:02}{874.543}{ -0.095}{0.036}{ 100}{1.89}{
1069}{1.2CA}
\punkth{14/12/89}{18:35}{875.274}{  0.413}{0.012}{ 500}{3.09}{
660}{1.2CA}
\punkth{15/12/89}{00:52}{875.536}{  0.356}{0.026}{ 500}{3.37}{
1648}{1.2CA}
\punkth{15/12/89}{19:37}{876.318}{ -0.324}{0.020}{ 100}{1.94}{
502}{1.2CA}
\punkth{15/12/89}{19:44}{876.323}{ -0.337}{0.047}{ 500}{2.08}{
606}{1.2CA}
\punkth{15/12/89}{20:23}{876.350}{ -0.392}{0.017}{ 500}{2.55}{
674}{1.2CA}
\punkth{15/12/89}{20:51}{876.369}{ -0.469}{0.013}{ 500}{2.35}{
964}{1.2CA}
\punkth{15/12/89}{20:51}{876.369}{ -0.475}{0.005}{ 500}{2.52}{
807}{1.2CA}
\punkth{19/12/89}{19:03}{880.294}{ -0.611}{0.019}{ 500}{1.55}{
629}{1.2CA}
\punkth{19/12/89}{21:04}{880.378}{ -0.606}{0.022}{ 250}{1.74}{
534}{1.2CA}
\punkth{20/12/89}{18:55}{881.289}{ -0.854}{0.026}{ 500}{2.05}{
630}{1.2CA}
\punkth{20/12/89}{20:24}{881.350}{ -1.097}{0.014}{ 500}{1.87}{
629}{1.2CA}
\punkth{21/12/89}{17:58}{882.249}{ -0.299}{0.030}{ 500}{3.97}{
646}{1.2CA}
\punkth{22/12/89}{19:45}{883.323}{  0.500}{0.034}{ 500}{3.11}{
611}{1.2CA}
\punkth{22/12/89}{21:15}{883.386}{  0.376}{0.012}{ 500}{2.75}{
616}{1.2CA}

\end{picture}}

\end{picture}

\vspace*{1cm}
\caption{Lightcurve of AO\,0235+164 in February 1989. Between Feb.~20 and
22, a 1.60\,mag brightening occurred within 47.5 hours}
\end{figure}
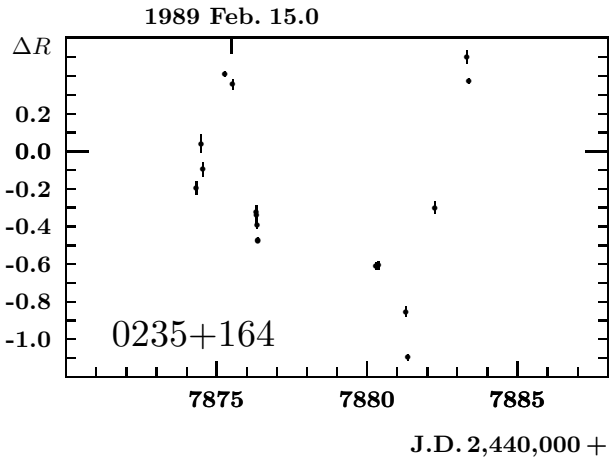

\begin{figure}

\vspace*{0.5cm}

\begin{picture}(7.8 ,4.5 )(-1,0)
\put(0,0){\setlength{\unitlength}{0.2cm}%
\begin{picture}(38,  22.500)(285,0)
\put(285,0){\framebox(38,  22.500)[tl]{\begin{picture}(0,0)(0,0)
        \put(0,0){\makebox(0,0)[tr]{$\Delta R$\hspace*{0.2cm}}}
        \put(38,0){\makebox(0,0)[tr]{\large{0235+164}\T{0.4}
                                 \hspace*{0.5cm}}}
        \put(38,-
22.500){\setlength{\unitlength}{1cm}\begin{picture}(0,0)(0,0)
            \put(0,-1){\makebox(0,0)[br]{\bf J.D.\,2,440,000\,+}}
        \end{picture}}
    \end{picture}}}

\thicklines
\put(285,0){\setlength{\unitlength}{2.5cm}\begin{picture}(0,0)(0,-1.2)
   \put(0,0){\setlength{\unitlength}{1cm}\begin{picture}(0,0)(0,0)
        \put(0,0){\line(1,0){0.3}}
        \end{picture}}
   \end{picture}}

\put(323,0){\setlength{\unitlength}{2.5cm}\begin{picture}(0,0)(0,-1.2)
   \put(0,0){\setlength{\unitlength}{1cm}\begin{picture}(0,0)(0,0)
        \put(0,0){\line(-1,0){0.3}}
        \end{picture}}
   \end{picture}}

\thinlines
\put(285,0){\setlength{\unitlength}{2.5cm}\begin{picture}(0,0)(0,-1.2)
   \multiput(0,0)(0,0.1){6}{\setlength{\unitlength}{1cm}%
\begin{picture}(0,0)(0,0)
        \put(0,0){\line(1,0){0.12}}
        \end{picture}}
   \end{picture}}

\put(285,0){\setlength{\unitlength}{2.5cm}\begin{picture}(0,0)(0,-1.2)
   \multiput(0,0)(0,-0.1){12}{\setlength{\unitlength}{1cm}%
\begin{picture}(0,0)(0,0)
        \put(0,0){\line(1,0){0.12}}
        \end{picture}}
   \end{picture}}

\put(323,0){\setlength{\unitlength}{2.5cm}\begin{picture}(0,0)(0,-1.2)
   \multiput(0,0)(0,0.1){6}{\setlength{\unitlength}{1cm}%
\begin{picture}(0,0)(0,0)
        \put(0,0){\line(-1,0){0.12}}
        \end{picture}}
   \end{picture}}

\put(323,0){\setlength{\unitlength}{2.5cm}\begin{picture}(0,0)(0,-1.2)
   \multiput(0,0)(0,-0.1){12}{\setlength{\unitlength}{1cm}%
\begin{picture}(0,0)(0,0)
        \put(0,0){\line(-1,0){0.12}}
        \end{picture}}
   \end{picture}}

\put(285,0){\setlength{\unitlength}{2.5cm}\begin{picture}(0,0)(0,-1.2)
   \put(0,0.2){\setlength{\unitlength}{1cm}\begin{picture}(0,0)(0,0)
        \put(0,0){\line(1,0){0.12}}
        \put(-0.2,0){\makebox(0,0)[r]{\bf 0.2}}
        \end{picture}}
   \put(0,0.0){\setlength{\unitlength}{1cm}\begin{picture}(0,0)(0,0)
        \put(0,0){\line(1,0){0.12}}
        \put(-0.2,0){\makebox(0,0)[r]{\bf 0.0}}
        \end{picture}}
   \put(0,-0.2){\setlength{\unitlength}{1cm}\begin{picture}(0,0)(0,0)
        \put(0,0){\line(1,0){0.12}}
        \put(-0.2,0){\makebox(0,0)[r]{\bf -0.2}}
        \end{picture}}
   \put(0,-0.4){\setlength{\unitlength}{1cm}\begin{picture}(0,0)(0,0)
        \put(0,0){\line(1,0){0.12}}
        \put(-0.2,0){\makebox(0,0)[r]{\bf -0.4}}
        \end{picture}}
   \put(0,-0.6){\setlength{\unitlength}{1cm}\begin{picture}(0,0)(0,0)
        \put(0,0){\line(1,0){0.12}}
        \put(-0.2,0){\makebox(0,0)[r]{\bf -0.6}}
        \end{picture}}
   \put(0,-0.8){\setlength{\unitlength}{1cm}\begin{picture}(0,0)(0,0)
        \put(0,0){\line(1,0){0.12}}
        \put(-0.2,0){\makebox(0,0)[r]{\bf -0.8}}
        \end{picture}}
   \put(0,-1.0){\setlength{\unitlength}{1cm}\begin{picture}(0,0)(0,0)
        \put(0,0){\line(1,0){0.12}}
        \put(-0.2,0){\makebox(0,0)[r]{\bf -1.0}}
        \end{picture}}
   \put(0,0.0){\setlength{\unitlength}{1cm}\begin{picture}(0,0)(0,0)
        \put(0,0){\line(1,0){0.12}}
        \put(-0.2,0){\makebox(0,0)[r]{\bf 0.0}}
        \end{picture}}
   \put(0,0.4){\setlength{\unitlength}{1cm}\begin{picture}(0,0)(0,0)
        \put(0,0){\line(1,0){0.12}}
        \end{picture}}
   \end{picture}}

   \put(302.5,  22.500){\setlength{\unitlength}{1cm}\begin{picture}(0,0)(0,0)
        \put(0,0){\line(0,-1){0.2}}
        \put(0,0.2){\makebox(0,0)[b]{\bf 1991 Feb.~15.0}}
   \end{picture}}

    \multiput(285,0)(1,0){38}%
        {\setlength{\unitlength}{1cm}\begin{picture}(0,0)(0,0)
        \put(0,0){\line(0,1){0.12}}
    \end{picture}}
    \put(290,0){\setlength{\unitlength}{1cm}\begin{picture}(0,0)(0,0)
        \put(0,0){\line(0,1){0.2}}
        \put(0,-0.2){\makebox(0,0)[t]{\bf 8290}}
    \end{picture}}
    \put(290,0){\setlength{\unitlength}{1cm}\begin{picture}(0,0)(0,0)
        \put(0,0){\line(0,1){0.2}}
        \put(0,-0.2){\makebox(0,0)[t]{\bf 8290}}
    \end{picture}}
   \put(300,0){\setlength{\unitlength}{1cm}\begin{picture}(0,0)(0,0)
        \put(0,0){\line(0,1){0.2}}
        \put(0,-0.2){\makebox(0,0)[t]{\bf 8300}}
    \end{picture}}
   \put(300,0){\setlength{\unitlength}{1cm}\begin{picture}(0,0)(0,0)
        \put(0,0){\line(0,1){0.2}}
        \put(0,-0.2){\makebox(0,0)[t]{\bf 8300}}
    \end{picture}}
    \put(310,0){\setlength{\unitlength}{1cm}\begin{picture}(0,0)(0,0)
        \put(0,0){\line(0,1){0.2}}
        \put(0,-0.2){\makebox(0,0)[t]{\bf 8310}}
    \end{picture}}
    \put(310,0){\setlength{\unitlength}{1cm}\begin{picture}(0,0)(0,0)
       \put(0,0){\line(0,1){0.2}}
        \put(0,-0.2){\makebox(0,0)[t]{\bf 8310}}
    \end{picture}}
    \put(310,0){\setlength{\unitlength}{1cm}\begin{picture}(0,0)(0,0)
       \put(0,0){\line(0,1){0.2}}
        \put(0,-0.2){\makebox(0,0)[t]{\bf 8310}}
    \end{picture}}

\punkth{31/01/91}{20:07}{288.339}{ -0.355}{0.035}{ 500}{3.70}{
1563}{1.2CA}
\punkthh{01/02/91}{20:18}{289.346}{ -0.238}{0.063}{1801}{3.19}{
2798}{1.2CA}
\punkthh{02/02/91}{19:04}{290.295}{ -0.239}{0.074}{ 100}{4.81}{
345}{1.2CA}
\punkth{02/02/91}{19:12}{290.300}{ -0.125}{0.026}{ 500}{4.58}{
791}{1.2CA}
\punkth{02/02/91}{19:41}{290.321}{ -0.164}{0.040}{ 100}{3.90}{
378}{1.2CA}
\punkth{02/02/91}{19:46}{290.324}{ -0.126}{0.027}{ 270}{4.07}{
545}{1.2CA}
\punkth{02/02/91}{19:52}{290.328}{ -0.134}{0.024}{ 300}{3.79}{
584}{1.2CA}
\punkth{02/02/91}{21:37}{290.401}{ -0.112}{0.038}{ 300}{3.05}{
679}{1.2CA}
\punkth{04/02/91}{18:05}{292.254}{  0.115}{0.049}{1800}{3.69}{
1797}{1.2CA}
\punkthh{07/02/91}{18:34}{295.274}{  0.271}{0.056}{ 100}{4.02}{
647}{1.2CA}
\punkth{07/02/91}{18:47}{295.283}{  0.267}{0.023}{ 300}{4.02}{
730}{1.2CA}
\punkth{07/02/91}{20:03}{295.336}{  0.264}{0.034}{ 150}{3.90}{
593}{1.2CA}
\punkth{07/02/91}{20:11}{295.341}{  0.236}{0.017}{ 500}{3.92}{
806}{1.2CA}
\punkth{09/02/91}{19:18}{297.304}{ -0.002}{0.027}{ 250}{1.65}{
475}{1.2CA}
\punkth{09/02/91}{20:10}{297.341}{ -0.008}{0.017}{ 150}{1.95}{
408}{1.2CA}
\punkth{09/02/91}{20:15}{297.344}{ -0.040}{0.026}{ 100}{1.97}{
379}{1.2CA}
\punkth{09/02/91}{20:20}{297.347}{ -0.055}{0.022}{ 300}{2.01}{
520}{1.2CA}
\punkth{10/02/91}{18:49}{298.284}{  0.053}{0.059}{ 150}{2.33}{
547}{1.2CA}
\punkth{11/02/91}{19:48}{299.325}{ -0.764}{0.039}{ 500}{4.14}{
816}{1.2CA}
\punkth{14/02/91}{18:49}{302.284}{ -0.577}{0.059}{ 500}{4.00}{
1676}{1.2CA}
\punkth{14/02/91}{19:31}{302.313}{ -0.554}{0.045}{ 100}{3.07}{
395}{1.2CA}
\punkth{14/02/91}{19:38}{302.318}{ -0.590}{0.024}{ 500}{3.25}{
902}{1.2CA}
\punkth{14/02/91}{21:07}{302.380}{ -0.708}{0.045}{ 300}{3.31}{
735}{1.2CA}
\punkthh{15/02/91}{18:50}{303.285}{ -0.952}{0.091}{ 100}{3.88}{
403}{1.2CA}
\punkth{15/02/91}{19:04}{303.295}{ -1.037}{0.020}{ 100}{1.81}{
391}{1.2CA}
\punkth{15/02/91}{19:11}{303.300}{ -1.072}{0.011}{ 344}{1.74}{
656}{1.2CA}
\punkthh{15/02/91}{20:57}{303.374}{ -1.031}{0.042}{ 100}{2.10}{
621}{1.2CA}
\punkth{15/02/91}{21:03}{303.378}{ -1.033}{0.017}{ 300}{2.14}{
645}{1.2CA}
\punkth{22/02/91}{18:11}{310.258}{ -1.074}{0.072}{1800}{3.05}{
5205}{1.2CA}
\punkthh{22/02/91}{18:54}{310.288}{ -1.047}{0.089}{ 100}{1.84}{
1507}{1.2CA}
\punkth{23/02/91}{19:11}{311.299}{ -0.720}{0.025}{ 100}{1.37}{
759}{1.2CA}
\punkth{23/02/91}{19:17}{311.304}{ -0.746}{0.016}{ 300}{1.24}{
1716}{1.2CA}
\punkth{24/02/91}{18:32}{312.272}{ -0.123}{0.082}{1800}{2.89}{
3299}{1.2CA}
\punkth{24/02/91}{23:38}{312.485}{ -0.098}{0.023}{ 100}{1.53}{
732}{1.2CA}
\punkth{24/02/91}{23:44}{312.489}{ -0.052}{0.019}{ 300}{1.48}{
1531}{1.2CA}
\punkth{26/02/91}{18:44}{314.281}{ -0.269}{0.054}{1801}{4.03}{
4921}{1.2CA}

\end{picture}}

\end{picture}

\vspace*{1cm}
\caption{Lightcurve of AO\,0235+164 in February 1991}
\end{figure}
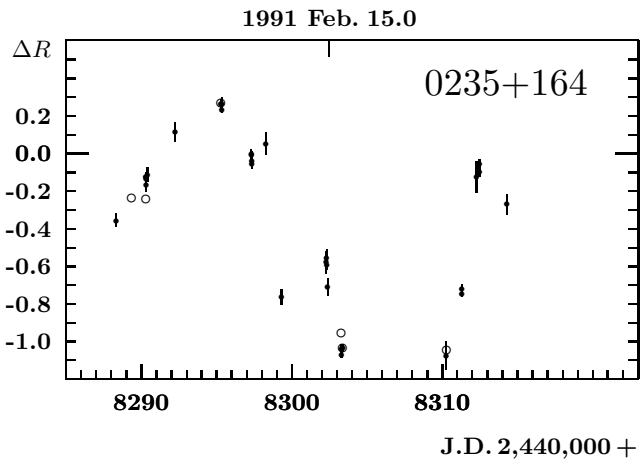

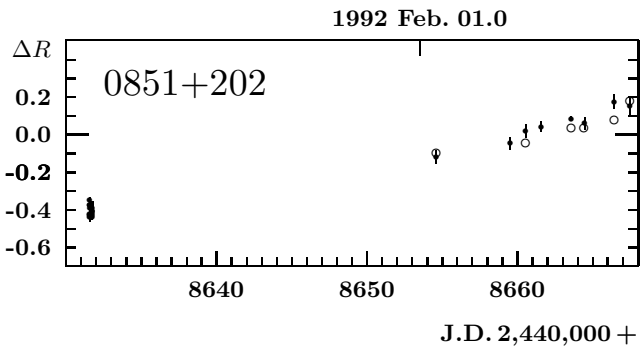
\begin{figure}

\vspace*{0.5cm}

\begin{picture}(7.8 ,3 )(-1,0)
\put(0,0){\setlength{\unitlength}{0.2cm}%
\begin{picture}(38,  15.000)(630,0)
\put(630,0){\framebox(38,  15.000)[tl]{\begin{picture}(0,0)(0,0)
        \put(0,0){\makebox(0,0)[tr]{$\Delta R$\hspace*{0.2cm}}}
        \put(0,0){\makebox(0,0)[tl]{\hspace*{0.5cm}\large{0851+202}\T{0.4}
                                 }}
        \put(38,-
15.000){\setlength{\unitlength}{1cm}\begin{picture}(0,0)(0,0)
            \put(0,-1){\makebox(0,0)[br]{\bf J.D.\,2,440,000\,+}}
        \end{picture}}
    \end{picture}}}

\thicklines
\put(630,0){\setlength{\unitlength}{2.5cm}\begin{picture}(0,0)(0,-0.7)
   \put(0,0){\setlength{\unitlength}{1cm}\begin{picture}(0,0)(0,0)
        \put(0,0){\line(1,0){0.3}}
        \end{picture}}
   \end{picture}}

\put(668,0){\setlength{\unitlength}{2.5cm}\begin{picture}(0,0)(0,-0.7)
   \put(0,0){\setlength{\unitlength}{1cm}\begin{picture}(0,0)(0,0)
        \put(0,0){\line(-1,0){0.3}}
        \end{picture}}
   \end{picture}}

\thinlines
\put(630,0){\setlength{\unitlength}{2.5cm}\begin{picture}(0,0)(0,-0.7)
   \multiput(0,0)(0,0.1){5}{\setlength{\unitlength}{1cm}%
\begin{picture}(0,0)(0,0)
        \put(0,0){\line(1,0){0.12}}
        \end{picture}}
   \end{picture}}

\put(630,0){\setlength{\unitlength}{2.5cm}\begin{picture}(0,0)(0,-0.7)
   \multiput(0,0)(0,-0.1){7}{\setlength{\unitlength}{1cm}%
\begin{picture}(0,0)(0,0)
        \put(0,0){\line(1,0){0.12}}
        \end{picture}}
   \end{picture}}

\put(668,0){\setlength{\unitlength}{2.5cm}\begin{picture}(0,0)(0,-0.7)
   \multiput(0,0)(0,0.1){5}{\setlength{\unitlength}{1cm}%
\begin{picture}(0,0)(0,0)
        \put(0,0){\line(-1,0){0.12}}
        \end{picture}}
   \end{picture}}

\put(668,0){\setlength{\unitlength}{2.5cm}\begin{picture}(0,0)(0,-0.7)
   \multiput(0,0)(0,-0.1){7}{\setlength{\unitlength}{1cm}%
\begin{picture}(0,0)(0,0)
        \put(0,0){\line(-1,0){0.12}}
        \end{picture}}
   \end{picture}}

\put(630,0){\setlength{\unitlength}{2.5cm}\begin{picture}(0,0)(0,-0.7)
   \put(0,-0.4){\setlength{\unitlength}{1cm}\begin{picture}(0,0)(0,0)
        \put(0,0){\line(1,0){0.12}}
        \put(-0.2,0){\makebox(0,0)[r]{\bf -0.4}}
        \end{picture}}
   \put(0,-0.6){\setlength{\unitlength}{1cm}\begin{picture}(0,0)(0,0)
        \put(0,0){\line(1,0){0.12}}
        \put(-0.2,0){\makebox(0,0)[r]{\bf -0.6}}
        \end{picture}}
   \put(0,-0.2){\setlength{\unitlength}{1cm}\begin{picture}(0,0)(0,0)
        \put(0,0){\line(1,0){0.12}}
        \put(-0.2,0){\makebox(0,0)[r]{\bf -0.2}}
        \end{picture}}
   \put(0,-0.2){\setlength{\unitlength}{1cm}\begin{picture}(0,0)(0,0)
        \put(0,0){\line(1,0){0.12}}
        \put(-0.2,0){\makebox(0,0)[r]{\bf -0.2}}
        \end{picture}}
   \put(0,-0.2){\setlength{\unitlength}{1cm}\begin{picture}(0,0)(0,0)
        \put(0,0){\line(1,0){0.12}}
        \put(-0.2,0){\makebox(0,0)[r]{\bf -0.2}}
        \end{picture}}
   \put(0,0.4){\setlength{\unitlength}{1cm}\begin{picture}(0,0)(0,0)
        \put(0,0){\line(1,0){0.12}}
        \end{picture}}
   \put(0,0.0){\setlength{\unitlength}{1cm}\begin{picture}(0,0)(0,0)
        \put(0,0){\line(1,0){0.12}}
        \put(-0.2,0){\makebox(0,0)[r]{\bf 0.0}}
        \end{picture}}
   \put(0,0.2){\setlength{\unitlength}{1cm}\begin{picture}(0,0)(0,0)
        \put(0,0){\line(1,0){0.12}}
        \put(-0.2,0){\makebox(0,0)[r]{\bf 0.2}}
        \end{picture}}
   \put(0,0.2){\setlength{\unitlength}{1cm}\begin{picture}(0,0)(0,0)
        \put(0,0){\line(1,0){0.12}}
        \end{picture}}
   \end{picture}}

   \put(653.5,  15.000){\setlength{\unitlength}{1cm}\begin{picture}(0,0)(0,0)
        \put(0,0){\line(0,-1){0.2}}
        \put(0,0.2){\makebox(0,0)[b]{\bf 1992 Feb.~01.0}}
   \end{picture}}

    \multiput(630,0)(1,0){38}%
        {\setlength{\unitlength}{1cm}\begin{picture}(0,0)(0,0)
        \put(0,0){\line(0,1){0.12}}
    \end{picture}}
    \put(640,0){\setlength{\unitlength}{1cm}\begin{picture}(0,0)(0,0)
        \put(0,0){\line(0,1){0.2}}
        \put(0,-0.2){\makebox(0,0)[t]{\bf 8640}}
    \end{picture}}
    \put(650,0){\setlength{\unitlength}{1cm}\begin{picture}(0,0)(0,0)
        \put(0,0){\line(0,1){0.2}}
        \put(0,-0.2){\makebox(0,0)[t]{\bf 8650}}
    \end{picture}}
    \put(660,0){\setlength{\unitlength}{1cm}\begin{picture}(0,0)(0,0)
       \put(0,0){\line(0,1){0.2}}
        \put(0,-0.2){\makebox(0,0)[t]{\bf 8660}}
    \end{picture}}

\punktn{10/01/92}{01:27}{631.561}{ -0.422}{0.009}{ 240}{2.48}{
1433}{1.2CA}
\punktn{10/01/92}{01:32}{631.564}{ -0.345}{0.010}{ 240}{2.40}{
1442}{1.2CA}
\punktn{10/01/92}{01:32}{631.564}{ -0.437}{0.021}{ 240}{2.35}{
1437}{1.2CA}
\punktn{10/01/92}{01:36}{631.567}{ -0.425}{0.017}{ 240}{2.42}{
1472}{1.2CA}
\punktn{10/01/92}{01:37}{631.567}{ -0.374}{0.019}{ 240}{2.43}{
1470}{1.2CA}
\punktn{10/01/92}{01:41}{631.571}{ -0.391}{0.017}{ 240}{2.46}{
1478}{1.2CA}
\punktn{10/01/92}{01:41}{631.571}{ -0.375}{0.017}{ 240}{2.48}{
1479}{1.2CA}
\punktn{10/01/92}{05:26}{631.727}{ -0.402}{0.017}{ 240}{3.22}{
2001}{1.2CA}
\punktn{10/01/92}{05:26}{631.727}{ -0.414}{0.016}{ 240}{3.21}{
2000}{1.2CA}
\punktn{10/01/92}{05:32}{631.731}{ -0.426}{0.018}{ 300}{3.45}{
2442}{1.2CA}
\punktn{10/01/92}{05:32}{631.731}{ -0.439}{0.011}{ 300}{3.62}{
2425}{1.2CA}
\punktn{10/01/92}{05:38}{631.735}{ -0.388}{0.030}{ 300}{3.70}{
2444}{1.2CA}
\punktn{10/01/92}{05:38}{631.735}{ -0.402}{0.013}{ 300}{3.84}{
2415}{1.2CA}
\punktnm{02/02/92}{01:53}{654.579}{ -0.095}{0.041}{ 100}{1.65}{
303}{1.2CA}
\punktn{02/02/92}{02:01}{654.584}{ -0.116}{0.035}{ 500}{1.95}{
518}{1.2CA}
\punktn{07/02/92}{00:23}{659.516}{ -0.046}{0.031}{ 100}{1.35}{
302}{1.2CA}
\punktnm{08/02/92}{00:53}{660.537}{ -0.046}{0.029}{ 100}{1.28}{
307}{1.2CA}
\punktn{08/02/92}{01:04}{660.545}{  0.020}{0.034}{ 500}{1.47}{
504}{1.2CA}
\punktn{09/02/92}{01:42}{661.571}{  0.042}{0.026}{ 500}{1.49}{
504}{1.2CA}
\punktnm{11/02/92}{01:21}{663.557}{  0.034}{0.037}{ 100}{0.98}{
307}{1.2CA}
\punktn{11/02/92}{01:29}{663.562}{  0.086}{0.010}{ 500}{0.98}{
512}{1.2CA}
\punktnm{11/02/92}{22:04}{664.420}{  0.036}{0.039}{ 100}{1.06}{
354}{1.2CA}
\punktn{11/02/92}{22:58}{664.457}{  0.062}{0.032}{ 500}{1.17}{
679}{1.2CA}
\punktnm{13/02/92}{22:13}{666.426}{  0.081}{0.037}{ 100}{1.61}{
544}{1.2CA}
\punktn{13/02/92}{22:20}{666.431}{  0.176}{0.036}{ 500}{2.05}{
1629}{1.2CA}
\punktnm{14/02/92}{23:24}{667.475}{  0.180}{0.039}{ 100}{0.96}{
852}{1.2CA}
\punktn{14/02/92}{23:32}{667.481}{  0.154}{0.047}{ 500}{1.22}{
3042}{1.2CA}

\end{picture}}

\end{picture}

\vspace*{1cm}
\caption{HQM-lightcurve of OJ\,287 (0851+202) in winter 1992}
\end{figure}

\noindent
{\bf 4C\,71.07 (0836+710).} The HQM lightcurve has been discussed in detail
in von~Linde et al.\ (\cite{LBSG93}). The lightcurve shown in Fig.~1
includes some more recent data points indicating that the object has
reached roughly the same brightness stage than before the flare which
occurred in Feb.~1992 nearly simultaneously with the detection of this object
by the GRO $\gamma$-ray observatory. The reference magnitude is
$R_0=16.84$.

\noindent
{\bf OJ\,287 (0851+202)} is one of the best studied OVVs of BL\,Lac type.
The Rosemary Hill data reported by WSFL have a 5\,mag range of variability.
Sillanp\"a\"a et al.\ (\cite{STHK85}) who compared the 1972 and 1983
outbursts found a similar morphology. The historical lightcurve, combined
with more recent data, was collected by Sillanp\"a\"a et
al.~(\cite{SHVS88}); it starts 1894 and exhibits 7 outbursts with a
possible periodicity and a period of $\sim$\,11,5\,yrs. Our HQM lightcurve
is clearly undersampled in most parts; in winter 1992, however, the
increasing wing of a flare (see Fig.~4) could be recorded with sufficient
resolution.  Making use of the photometric sequence of Smith et al.\
(\cite{SBHE85}), we were able to determine the reference magnitude of our
lightcurve, $R_0=15.13$.

\noindent
{\bf W\,Com (1219+285)}. The Rosemary Hill data reported for this BL\,Lac
object by WSFL show 2\,mag outbursts occurring with a relatively high
frequency. The HQM lightcurve is well sampled only in February 1991 (see
Fig.~5); we did not detect any dramatic event.

\begin{figure}

\vspace*{0.5cm}

\begin{picture}(7.8 ,2 )(-1,0)
\put(0,0){\setlength{\unitlength}{0.2cm}%
\begin{picture}(38,  10.000)(285,0)
\put(285,0){\framebox(38,  10.000)[tl]{\begin{picture}(0,0)(0,0)
        \put(0,0){\makebox(0,0)[tr]{$\Delta R$\hspace*{0.2cm}}}
        \put(38,0){\makebox(0,0)[tr]{\large{0735+178}\T{1.2}
                                 \hspace*{0.5cm}}}
        \put(38,-
10.000){\setlength{\unitlength}{1cm}\begin{picture}(0,0)(0,0)
        \end{picture}}
    \end{picture}}}

\thicklines
\put(285,0){\setlength{\unitlength}{2.5cm}\begin{picture}(0,0)(0,-0.4)
   \put(0,0){\setlength{\unitlength}{1cm}\begin{picture}(0,0)(0,0)
        \put(0,0){\line(1,0){0.3}}
        \end{picture}}
   \end{picture}}

\put(323,0){\setlength{\unitlength}{2.5cm}\begin{picture}(0,0)(0,-0.4)
   \put(0,0){\setlength{\unitlength}{1cm}\begin{picture}(0,0)(0,0)
        \put(0,0){\line(-1,0){0.3}}
        \end{picture}}
   \end{picture}}

\thinlines
\put(285,0){\setlength{\unitlength}{2.5cm}\begin{picture}(0,0)(0,-0.4)
   \multiput(0,0)(0,0.1){4}{\setlength{\unitlength}{1cm}%
\begin{picture}(0,0)(0,0)
        \put(0,0){\line(1,0){0.12}}
        \end{picture}}
   \end{picture}}

\put(285,0){\setlength{\unitlength}{2.5cm}\begin{picture}(0,0)(0,-0.4)
   \multiput(0,0)(0,-0.1){4}{\setlength{\unitlength}{1cm}%
\begin{picture}(0,0)(0,0)
        \put(0,0){\line(1,0){0.12}}
        \end{picture}}
   \end{picture}}

\put(323,0){\setlength{\unitlength}{2.5cm}\begin{picture}(0,0)(0,-0.4)
   \multiput(0,0)(0,0.1){4}{\setlength{\unitlength}{1cm}%
\begin{picture}(0,0)(0,0)
        \put(0,0){\line(-1,0){0.12}}
        \end{picture}}
   \end{picture}}

\put(323,0){\setlength{\unitlength}{2.5cm}\begin{picture}(0,0)(0,-0.4)
   \multiput(0,0)(0,-0.1){4}{\setlength{\unitlength}{1cm}%
\begin{picture}(0,0)(0,0)
        \put(0,0){\line(-1,0){0.12}}
        \end{picture}}
   \end{picture}}

\put(285,0){\setlength{\unitlength}{2.5cm}\begin{picture}(0,0)(0,-0.4)
   \put(0,-0.2){\setlength{\unitlength}{1cm}\begin{picture}(0,0)(0,0)
        \put(0,0){\line(1,0){0.12}}
        \put(-0.2,0){\makebox(0,0)[r]{\bf -0.2}}
        \end{picture}}
   \put(0,-0.2){\setlength{\unitlength}{1cm}\begin{picture}(0,0)(0,0)
        \put(0,0){\line(1,0){0.12}}
        \put(-0.2,0){\makebox(0,0)[r]{\bf -0.2}}
        \end{picture}}
   \put(0,-0.2){\setlength{\unitlength}{1cm}\begin{picture}(0,0)(0,0)
        \put(0,0){\line(1,0){0.12}}
        \put(-0.2,0){\makebox(0,0)[r]{\bf -0.2}}
        \end{picture}}
   \put(0,0.2){\setlength{\unitlength}{1cm}\begin{picture}(0,0)(0,0)
        \put(0,0){\line(1,0){0.12}}
        \end{picture}}
   \put(0,-0.2){\setlength{\unitlength}{1cm}\begin{picture}(0,0)(0,0)
        \put(0,0){\line(1,0){0.12}}
        \put(-0.2,0){\makebox(0,0)[r]{\bf -0.2}}
        \end{picture}}
   \put(0,-0.2){\setlength{\unitlength}{1cm}\begin{picture}(0,0)(0,0)
        \put(0,0){\line(1,0){0.12}}
        \put(-0.2,0){\makebox(0,0)[r]{\bf -0.2}}
        \end{picture}}
   \put(0,0.0){\setlength{\unitlength}{1cm}\begin{picture}(0,0)(0,0)
        \put(0,0){\line(1,0){0.12}}
        \put(-0.2,0){\makebox(0,0)[r]{\bf 0.0}}
        \end{picture}}
   \put(0,-0.2){\setlength{\unitlength}{1cm}\begin{picture}(0,0)(0,0)
        \put(0,0){\line(1,0){0.12}}
        \put(-0.2,0){\makebox(0,0)[r]{\bf -0.2}}
        \end{picture}}
   \put(0,0.2){\setlength{\unitlength}{1cm}\begin{picture}(0,0)(0,0)
        \put(0,0){\line(1,0){0.12}}
        \end{picture}}
   \end{picture}}

   \put(302.5,  10.000){\setlength{\unitlength}{1cm}\begin{picture}(0,0)(0,0)
        \put(0,0){\line(0,-1){0.2}}
        \put(0,0.2){\makebox(0,0)[b]{\bf 1991 Feb.~15.0}}
   \end{picture}}

    \multiput(285,0)(1,0){38}%
        {\setlength{\unitlength}{1cm}\begin{picture}(0,0)(0,0)
        \put(0,0){\line(0,1){0.12}}
    \end{picture}}
    \put(290,0){\setlength{\unitlength}{1cm}\begin{picture}(0,0)(0,0)
        \put(0,0){\line(0,1){0.2}}
    \end{picture}}
    \put(290,0){\setlength{\unitlength}{1cm}\begin{picture}(0,0)(0,0)
        \put(0,0){\line(0,1){0.2}}
    \end{picture}}
   \put(300,0){\setlength{\unitlength}{1cm}\begin{picture}(0,0)(0,0)
        \put(0,0){\line(0,1){0.2}}
    \end{picture}}
   \put(300,0){\setlength{\unitlength}{1cm}\begin{picture}(0,0)(0,0)
        \put(0,0){\line(0,1){0.2}}
    \end{picture}}
    \put(310,0){\setlength{\unitlength}{1cm}\begin{picture}(0,0)(0,0)
        \put(0,0){\line(0,1){0.2}}
    \end{picture}}
    \put(310,0){\setlength{\unitlength}{1cm}\begin{picture}(0,0)(0,0)
       \put(0,0){\line(0,1){0.2}}
    \end{picture}}
    \put(320,0){\setlength{\unitlength}{1cm}\begin{picture}(0,0)(0,0)
       \put(0,0){\line(0,1){0.2}}
    \end{picture}}

\punktj{31/01/91}{20:21}{288.348}{  0.095}{0.048}{ 100}{3.76}{
845}{3.5CA}
\punktj{31/01/91}{20:32}{288.356}{ -0.034}{0.023}{ 500}{4.06}{
3833}{3.5CA}
\punktj{01/02/91}{00:16}{288.512}{ -0.050}{0.045}{ 100}{3.13}{
957}{3.5CA}
\punktj{01/02/91}{00:24}{288.517}{ -0.045}{0.019}{ 500}{3.10}{
2035}{3.5CA}
\punktj{01/02/91}{01:29}{288.562}{ -0.066}{0.041}{ 100}{2.26}{
681}{3.5CA}
\punktj{01/02/91}{01:36}{288.567}{ -0.050}{0.020}{ 500}{2.35}{
2180}{3.5CA}
\punktj{01/02/91}{21:24}{289.392}{ -0.197}{0.035}{1203}{3.13}{
2278}{3.5CA}
\punktj{02/02/91}{01:22}{289.557}{ -0.246}{0.038}{1200}{4.51}{
3686}{3.5CA}
\punktj{02/02/91}{19:30}{290.313}{ -0.188}{0.024}{ 500}{5.16}{
614}{3.5CA}
\punktj{02/02/91}{20:36}{290.359}{ -0.185}{0.022}{ 500}{4.90}{
562}{3.5CA}
\punktj{02/02/91}{22:15}{290.427}{ -0.161}{0.043}{ 100}{3.44}{
389}{3.5CA}
\punktj{02/02/91}{22:32}{290.440}{ -0.169}{0.026}{ 300}{3.55}{
584}{3.5CA}
\punktj{03/02/91}{00:25}{290.518}{ -0.157}{0.043}{ 100}{2.40}{
595}{3.5CA}
\punktj{03/02/91}{00:33}{290.523}{ -0.132}{0.021}{ 500}{2.52}{
1068}{3.5CA}
\punktj{04/02/91}{18:56}{292.289}{ -0.024}{0.047}{1201}{3.45}{
1089}{3.5CA}
\punktjj{04/02/91}{22:05}{292.420}{ -0.030}{0.103}{1200}{5.22}{
875}{3.5CA}
\punktj{06/02/91}{21:42}{294.404}{ -0.001}{0.039}{1212}{4.41}{
1261}{3.5CA}
\punktj{06/02/91}{22:59}{294.458}{  0.071}{0.045}{ 110}{3.57}{
344}{3.5CA}
\punktj{06/02/91}{23:04}{294.461}{  0.095}{0.031}{ 229}{3.48}{
427}{3.5CA}
\punktj{06/02/91}{23:37}{294.484}{  0.079}{0.023}{ 300}{3.17}{
497}{3.5CA}
\punktj{07/02/91}{00:47}{294.533}{  0.107}{0.041}{ 100}{2.26}{
341}{3.5CA}
\punktj{07/02/91}{19:27}{295.311}{ -0.039}{0.030}{ 100}{4.67}{
335}{3.5CA}
\punktj{07/02/91}{19:34}{295.316}{ -0.047}{0.021}{ 500}{4.74}{
625}{3.5CA}
\punktjj{07/02/91}{20:33}{295.356}{ -0.131}{0.073}{ 200}{4.22}{
422}{3.5CA}
\punktjj{09/02/91}{00:12}{296.509}{  0.119}{0.084}{ 100}{4.30}{
407}{3.5CA}
\punktj{09/02/91}{00:19}{296.513}{  0.028}{0.051}{ 200}{3.92}{
492}{3.5CA}
\punktj{09/02/91}{00:48}{296.534}{  0.024}{0.016}{ 100}{2.90}{
339}{3.5CA}
\punktj{09/02/91}{00:55}{296.539}{  0.025}{0.030}{ 500}{2.55}{
628}{3.5CA}
\punktjj{09/02/91}{02:24}{296.600}{  0.025}{0.108}{ 150}{1.94}{
381}{3.5CA}
\punktjj{09/02/91}{19:42}{297.321}{  0.249}{0.147}{ 100}{1.69}{
328}{3.5CA}
\punktjj{09/02/91}{20:48}{297.367}{  0.233}{0.059}{ 100}{1.32}{
324}{3.5CA}
\punktj{09/02/91}{20:55}{297.372}{  0.218}{0.008}{ 500}{1.42}{
594}{3.5CA}
\punktj{09/02/91}{22:20}{297.431}{  0.238}{0.010}{ 100}{1.63}{
328}{3.5CA}
\punktj{09/02/91}{22:31}{297.439}{  0.243}{0.021}{ 300}{1.73}{
448}{3.5CA}
\punktjj{10/02/91}{00:15}{297.511}{  0.243}{0.043}{ 150}{1.94}{
386}{3.5CA}
\punktjj{10/02/91}{00:22}{297.516}{  0.213}{0.059}{ 150}{2.14}{
379}{3.5CA}
\punktjj{10/02/91}{19:29}{298.312}{  0.116}{0.055}{ 100}{3.20}{
323}{3.5CA}
\punktj{10/02/91}{19:36}{298.317}{  0.150}{0.029}{ 500}{3.12}{
575}{3.5CA}
\punktj{10/02/91}{19:47}{298.325}{  0.183}{0.025}{ 500}{3.59}{
579}{3.5CA}
\punktj{11/02/91}{19:55}{299.330}{  0.120}{0.034}{ 100}{4.82}{
323}{3.5CA}
\punktj{11/02/91}{20:02}{299.335}{  0.135}{0.021}{ 500}{4.74}{
546}{3.5CA}
\punktjj{11/02/91}{20:31}{299.355}{  0.113}{0.063}{ 100}{3.97}{
317}{3.5CA}
\punktjj{11/02/91}{20:36}{299.359}{  0.115}{0.053}{ 206}{4.43}{
392}{3.5CA}
\punktjj{11/02/91}{22:25}{299.435}{  0.154}{0.037}{ 150}{4.50}{
340}{3.5CA}
\punktj{11/02/91}{22:32}{299.439}{  0.127}{0.024}{ 300}{4.32}{
425}{3.5CA}
\punktjj{14/02/91}{18:58}{302.291}{ -0.047}{0.045}{ 100}{3.96}{
343}{3.5CA}
\punktj{14/02/91}{19:06}{302.296}{ -0.023}{0.024}{ 500}{4.24}{
609}{3.5CA}
\punktj{14/02/91}{20:26}{302.351}{ -0.036}{0.022}{ 100}{3.69}{
324}{3.5CA}
\punktj{14/02/91}{20:29}{302.354}{ -0.013}{0.041}{ 110}{3.67}{
338}{3.5CA}
\punktjj{14/02/91}{20:33}{302.356}{ -0.046}{0.048}{  96}{3.70}{
410}{3.5CA}
\punktj{14/02/91}{20:37}{302.359}{ -0.022}{0.018}{ 300}{3.66}{
470}{3.5CA}
\punktjj{14/02/91}{21:36}{302.400}{ -0.029}{0.048}{ 100}{3.53}{
365}{3.5CA}
\punktj{14/02/91}{21:47}{302.408}{ -0.034}{0.016}{ 500}{3.62}{
589}{3.5CA}
\punktjj{14/02/91}{23:13}{302.467}{ -0.021}{0.043}{ 100}{2.82}{
344}{3.5CA}
\punktj{14/02/91}{23:19}{302.472}{ -0.019}{0.016}{ 300}{3.01}{
463}{3.5CA}
\punktj{15/02/91}{01:49}{302.576}{ -0.020}{0.011}{ 100}{2.05}{
340}{3.5CA}
\punktj{15/02/91}{01:57}{302.582}{ -0.020}{0.010}{ 500}{2.17}{
655}{3.5CA}
\punktj{15/02/91}{20:07}{303.339}{ -0.002}{0.009}{ 100}{1.60}{
331}{3.5CA}
\punktj{15/02/91}{20:14}{303.343}{ -0.003}{0.022}{ 300}{1.94}{
470}{3.5CA}
\punktj{15/02/91}{21:45}{303.407}{  0.001}{0.008}{ 100}{1.59}{
384}{3.5CA}
\punktj{15/02/91}{21:53}{303.412}{  0.006}{0.010}{ 300}{1.66}{
408}{3.5CA}
\punktj{16/02/91}{00:02}{303.502}{ -0.001}{0.039}{ 100}{1.61}{
322}{3.5CA}
\punktj{16/02/91}{00:11}{303.508}{  0.010}{0.011}{ 300}{1.63}{
406}{3.5CA}
\punktj{16/02/91}{01:39}{303.569}{  0.010}{0.039}{ 100}{1.53}{
326}{3.5CA}
\punktj{16/02/91}{01:45}{303.573}{  0.013}{0.009}{ 300}{1.46}{
450}{3.5CA}
\punktj{19/02/91}{19:35}{307.317}{  0.082}{0.034}{1200}{2.62}{
1150}{3.5CA}
\punktj{19/02/91}{20:50}{307.368}{  0.086}{0.046}{1391}{2.69}{
1036}{3.5CA}
\punktj{19/02/91}{22:31}{307.438}{  0.090}{0.057}{1200}{2.61}{
922}{3.5CA}
\punktj{20/02/91}{00:32}{307.522}{  0.088}{0.055}{1201}{2.93}{
1034}{3.5CA}
\punktj{20/02/91}{19:29}{308.312}{  0.089}{0.072}{1200}{2.98}{
1173}{3.5CA}
\punktj{20/02/91}{20:57}{308.373}{  0.092}{0.033}{1200}{2.73}{
1222}{3.5CA}
\punktjj{20/02/91}{22:29}{308.437}{  0.122}{0.077}{1426}{2.68}{
1490}{3.5CA}
\punktj{21/02/91}{19:04}{309.295}{  0.114}{0.035}{1200}{3.37}{
1920}{3.5CA}
\punktj{21/02/91}{20:43}{309.363}{  0.091}{0.053}{1320}{3.15}{
2032}{3.5CA}
\punktj{21/02/91}{22:32}{309.439}{  0.083}{0.056}{1200}{3.23}{
1697}{3.5CA}
\punktj{22/02/91}{19:16}{310.303}{  0.149}{0.056}{2189}{3.07}{
1660}{3.5CA}
\punktj{22/02/91}{20:47}{310.366}{  0.158}{0.032}{1200}{2.99}{
1739}{3.5CA}
\punktj{22/02/91}{22:37}{310.442}{  0.144}{0.031}{1200}{3.16}{
1745}{3.5CA}
\punktj{22/02/91}{22:42}{310.446}{  0.182}{0.013}{ 100}{1.54}{
408}{3.5CA}
\punktj{22/02/91}{22:48}{310.450}{  0.179}{0.020}{ 300}{1.61}{
704}{3.5CA}
\punktj{23/02/91}{21:06}{311.380}{  0.294}{0.011}{ 100}{1.01}{
476}{3.5CA}
\punktj{23/02/91}{21:15}{311.386}{  0.297}{0.010}{ 300}{1.09}{
904}{3.5CA}
\punktj{25/02/91}{01:10}{312.549}{  0.315}{0.039}{ 100}{1.94}{
1691}{3.5CA}
\punktj{25/02/91}{01:17}{312.554}{  0.291}{0.026}{ 300}{1.93}{
3960}{3.5CA}

\end{picture}}

\end{picture}

\vspace*{-0.02cm}

\begin{picture}(7.8 ,2 )(-1,0)
\put(0,0){\setlength{\unitlength}{0.2cm}%
\begin{picture}(38,  10.000)(285,0)
\put(285,0){\framebox(38,  10.000)[tl]{\begin{picture}(0,0)(0,0)
        \put(0,0){\makebox(0,0)[tl]{\hspace*{0.5cm}\large{1219+285}\T{0.4}
                                 }}
        \put(38,-
10.000){\setlength{\unitlength}{1cm}\begin{picture}(0,0)(0,0)
            \put(0,-1){\makebox(0,0)[br]{\bf J.D.\,2,440,000\,+}}
        \end{picture}}
    \end{picture}}}

\thicklines
\put(285,0){\setlength{\unitlength}{2.5cm}\begin{picture}(0,0)(0,-0.4)
   \put(0,0){\setlength{\unitlength}{1cm}\begin{picture}(0,0)(0,0)
        \put(0,0){\line(1,0){0.3}}
        \end{picture}}
   \end{picture}}

\put(323,0){\setlength{\unitlength}{2.5cm}\begin{picture}(0,0)(0,-0.4)
   \put(0,0){\setlength{\unitlength}{1cm}\begin{picture}(0,0)(0,0)
        \put(0,0){\line(-1,0){0.3}}
        \end{picture}}
   \end{picture}}

\thinlines
\put(285,0){\setlength{\unitlength}{2.5cm}\begin{picture}(0,0)(0,-0.4)
   \multiput(0,0)(0,0.1){4}{\setlength{\unitlength}{1cm}%
\begin{picture}(0,0)(0,0)
        \put(0,0){\line(1,0){0.12}}
        \end{picture}}
   \end{picture}}

\put(285,0){\setlength{\unitlength}{2.5cm}\begin{picture}(0,0)(0,-0.4)
   \multiput(0,0)(0,-0.1){4}{\setlength{\unitlength}{1cm}%
\begin{picture}(0,0)(0,0)
        \put(0,0){\line(1,0){0.12}}
        \end{picture}}
   \end{picture}}

\put(323,0){\setlength{\unitlength}{2.5cm}\begin{picture}(0,0)(0,-0.4)
   \multiput(0,0)(0,0.1){4}{\setlength{\unitlength}{1cm}%
\begin{picture}(0,0)(0,0)
        \put(0,0){\line(-1,0){0.12}}
        \end{picture}}
   \end{picture}}

\put(323,0){\setlength{\unitlength}{2.5cm}\begin{picture}(0,0)(0,-0.4)
   \multiput(0,0)(0,-0.1){4}{\setlength{\unitlength}{1cm}%
\begin{picture}(0,0)(0,0)
        \put(0,0){\line(-1,0){0.12}}
        \end{picture}}
   \end{picture}}

\put(285,0){\setlength{\unitlength}{2.5cm}\begin{picture}(0,0)(0,-0.4)
   \put(0,-0.2){\setlength{\unitlength}{1cm}\begin{picture}(0,0)(0,0)
        \put(0,0){\line(1,0){0.12}}
        \put(-0.2,0){\makebox(0,0)[r]{\bf -0.2}}
        \end{picture}}
   \put(0,-0.2){\setlength{\unitlength}{1cm}\begin{picture}(0,0)(0,0)
        \put(0,0){\line(1,0){0.12}}
        \put(-0.2,0){\makebox(0,0)[r]{\bf -0.2}}
        \end{picture}}
   \put(0,-0.2){\setlength{\unitlength}{1cm}\begin{picture}(0,0)(0,0)
        \put(0,0){\line(1,0){0.12}}
        \put(-0.2,0){\makebox(0,0)[r]{\bf -0.2}}
        \end{picture}}
   \put(0,-0.2){\setlength{\unitlength}{1cm}\begin{picture}(0,0)(0,0)
        \put(0,0){\line(1,0){0.12}}
        \put(-0.2,0){\makebox(0,0)[r]{\bf -0.2}}
        \end{picture}}
   \put(0,-0.2){\setlength{\unitlength}{1cm}\begin{picture}(0,0)(0,0)
        \put(0,0){\line(1,0){0.12}}
        \put(-0.2,0){\makebox(0,0)[r]{\bf -0.2}}
        \end{picture}}
   \put(0,0.2){\setlength{\unitlength}{1cm}\begin{picture}(0,0)(0,0)
        \put(0,0){\line(1,0){0.12}}
        \put(-0.2,0){\makebox(0,0)[r]{\bf 0.2}}
        \end{picture}}
   \put(0,0.0){\setlength{\unitlength}{1cm}\begin{picture}(0,0)(0,0)
        \put(0,0){\line(1,0){0.12}}
        \put(-0.2,0){\makebox(0,0)[r]{\bf 0.0}}
        \end{picture}}
   \put(0,-0.2){\setlength{\unitlength}{1cm}\begin{picture}(0,0)(0,0)
        \put(0,0){\line(1,0){0.12}}
        \put(-0.2,0){\makebox(0,0)[r]{\bf -0.2}}
        \end{picture}}
   \put(0,0.2){\setlength{\unitlength}{1cm}\begin{picture}(0,0)(0,0)
        \put(0,0){\line(1,0){0.12}}
        \end{picture}}
   \end{picture}}

   \put(302.5,  10.000){\setlength{\unitlength}{1cm}\begin{picture}(0,0)(0,0)
        \put(0,0){\line(0,-1){0.2}}
   \end{picture}}

    \multiput(285,0)(1,0){38}%
        {\setlength{\unitlength}{1cm}\begin{picture}(0,0)(0,0)
        \put(0,0){\line(0,1){0.12}}
    \end{picture}}
    \put(290,0){\setlength{\unitlength}{1cm}\begin{picture}(0,0)(0,0)
        \put(0,0){\line(0,1){0.2}}
        \put(0,-0.2){\makebox(0,0)[t]{\bf 8290}}
    \end{picture}}
    \put(290,0){\setlength{\unitlength}{1cm}\begin{picture}(0,0)(0,0)
        \put(0,0){\line(0,1){0.2}}
        \put(0,-0.2){\makebox(0,0)[t]{\bf 8290}}
    \end{picture}}
   \put(300,0){\setlength{\unitlength}{1cm}\begin{picture}(0,0)(0,0)
        \put(0,0){\line(0,1){0.2}}
        \put(0,-0.2){\makebox(0,0)[t]{\bf 8300}}
    \end{picture}}
   \put(300,0){\setlength{\unitlength}{1cm}\begin{picture}(0,0)(0,0)
        \put(0,0){\line(0,1){0.2}}
        \put(0,-0.2){\makebox(0,0)[t]{\bf 8300}}
    \end{picture}}
    \put(310,0){\setlength{\unitlength}{1cm}\begin{picture}(0,0)(0,0)
        \put(0,0){\line(0,1){0.2}}
        \put(0,-0.2){\makebox(0,0)[t]{\bf 8310}}
    \end{picture}}
    \put(310,0){\setlength{\unitlength}{1cm}\begin{picture}(0,0)(0,0)
       \put(0,0){\line(0,1){0.2}}
        \put(0,-0.2){\makebox(0,0)[t]{\bf 8310}}
    \end{picture}}
    \put(320,0){\setlength{\unitlength}{1cm}\begin{picture}(0,0)(0,0)
       \put(0,0){\line(0,1){0.2}}
        \put(0,-0.2){\makebox(0,0)[t]{\bf 8320}}
    \end{picture}}

\punktl{01/02/91}{01:44}{288.573}{ -0.159}{0.037}{ 100}{2.51}{
715}{1.2CA}
\punktl{01/02/91}{01:50}{288.577}{ -0.259}{0.020}{ 100}{2.60}{
658}{1.2CA}
\punktl{01/02/91}{01:59}{288.583}{ -0.304}{0.014}{ 500}{2.61}{
2108}{1.2CA}
\punktll{01/02/91}{23:47}{289.491}{ -0.360}{0.100}{ 600}{3.76}{
2850}{1.2CA}
\punktl{02/02/91}{05:07}{289.713}{ -0.173}{0.014}{ 600}{3.83}{
4277}{1.2CA}
\punktll{02/02/91}{05:36}{289.733}{ -0.187}{0.115}{ 600}{3.73}{
3796}{1.2CA}
\punktl{05/02/91}{01:15}{292.553}{ -0.005}{0.044}{ 600}{3.60}{
1031}{1.2CA}
\punktl{07/02/91}{00:28}{294.520}{ -0.206}{0.011}{ 100}{2.26}{
326}{1.2CA}
\punktl{07/02/91}{00:32}{294.523}{ -0.214}{0.029}{  64}{2.26}{
306}{1.2CA}
\punktll{07/02/91}{00:39}{294.527}{ -0.107}{0.094}{ 500}{2.40}{
492}{1.2CA}
\punktl{09/02/91}{01:05}{296.545}{ -0.187}{0.036}{ 100}{2.50}{
316}{1.2CA}
\punktl{09/02/91}{01:11}{296.550}{ -0.084}{0.067}{ 200}{2.67}{
404}{1.2CA}
\punktl{09/02/91}{01:17}{296.554}{  0.020}{0.024}{ 300}{2.48}{
481}{1.2CA}
\punktll{09/02/91}{23:41}{297.487}{ -0.128}{0.053}{ 100}{2.34}{
328}{1.2CA}
\punktl{09/02/91}{23:50}{297.494}{ -0.188}{0.013}{ 500}{2.83}{
658}{1.2CA}
\punktl{15/02/91}{00:31}{302.522}{ -0.045}{0.026}{ 100}{4.36}{
310}{1.2CA}
\punktl{15/02/91}{00:57}{302.540}{ -0.054}{0.019}{ 500}{3.56}{
662}{1.2CA}
\punktl{15/02/91}{03:01}{302.626}{ -0.090}{0.011}{ 100}{2.02}{
451}{1.2CA}
\punktl{15/02/91}{03:13}{302.634}{ -0.080}{0.082}{ 500}{2.14}{
526}{1.2CA}
\punktl{15/02/91}{22:02}{303.419}{ -0.107}{0.028}{ 100}{1.97}{
341}{1.2CA}
\punktl{15/02/91}{22:11}{303.425}{ -0.254}{0.076}{ 300}{1.93}{
471}{1.2CA}
\punktl{16/02/91}{00:39}{303.528}{ -0.311}{0.054}{ 300}{1.90}{
385}{1.2CA}
\punktl{16/02/91}{01:53}{303.579}{ -0.343}{0.057}{ 100}{1.69}{
330}{1.2CA}
\punktl{16/02/91}{02:02}{303.585}{ -0.304}{0.056}{ 200}{1.50}{
343}{1.2CA}
\punktl{16/02/91}{04:45}{303.698}{ -0.197}{0.013}{ 100}{1.64}{
329}{1.2CA}
\punktl{16/02/91}{04:55}{303.705}{ -0.224}{0.026}{ 300}{1.55}{
385}{1.2CA}
\punktl{19/02/91}{23:57}{307.498}{ -0.159}{0.060}{ 601}{2.34}{
554}{1.2CA}
\punktl{20/02/91}{01:53}{307.579}{ -0.106}{0.046}{ 600}{2.65}{
546}{1.2CA}
\punktl{20/02/91}{04:19}{307.680}{ -0.154}{0.048}{ 600}{2.73}{
623}{1.2CA}
\punktll{20/02/91}{23:29}{308.479}{ -0.289}{0.108}{ 600}{3.56}{
775}{1.2CA}
\punktl{21/02/91}{04:05}{308.670}{ -0.308}{0.046}{ 600}{2.40}{
688}{1.2CA}
\punktl{21/02/91}{21:39}{309.402}{ -0.133}{0.060}{ 600}{3.04}{
896}{1.2CA}
\punktl{21/02/91}{23:25}{309.476}{ -0.125}{0.063}{ 600}{3.73}{
704}{1.2CA}
\punktl{22/02/91}{21:52}{310.412}{  0.221}{0.048}{ 600}{3.20}{
950}{1.2CA}
\punktl{22/02/91}{23:29}{310.479}{  0.217}{0.043}{ 601}{3.05}{
793}{1.2CA}
\punktl{23/02/91}{03:18}{310.638}{  0.046}{0.079}{ 100}{1.37}{
306}{1.2CA}
\punktl{23/02/91}{03:28}{310.645}{  0.038}{0.076}{ 200}{1.51}{
364}{1.2CA}
\punktl{24/02/91}{21:35}{312.400}{ -0.064}{0.048}{ 600}{2.40}{
1094}{1.2CA}
\punktl{24/02/91}{22:31}{312.438}{ -0.071}{0.048}{ 600}{2.71}{
977}{1.2CA}
\punktl{24/02/91}{23:29}{312.479}{ -0.045}{0.054}{ 601}{3.42}{
901}{1.2CA}
\punktl{25/02/91}{00:53}{312.537}{ -0.066}{0.044}{ 601}{3.11}{
911}{1.2CA}
\punktl{25/02/91}{03:33}{312.649}{ -0.100}{0.047}{ 600}{2.81}{
837}{1.2CA}
\punktll{25/02/91}{22:01}{313.417}{ -0.145}{0.083}{ 600}{2.96}{
1716}{1.2CA}
\punktll{25/02/91}{22:57}{313.457}{ -0.145}{0.081}{ 600}{3.04}{
1608}{1.2CA}
\punktll{25/02/91}{23:42}{313.488}{ -0.191}{0.072}{ 600}{2.82}{
1444}{1.2CA}
\punktll{26/02/91}{01:30}{313.563}{ -0.227}{0.063}{ 600}{2.86}{
1351}{1.2CA}
\punktl{26/02/91}{02:56}{313.623}{ -0.221}{0.050}{ 600}{3.25}{
1282}{1.2CA}
\punktll{26/02/91}{04:50}{313.702}{ -0.175}{0.072}{ 600}{3.65}{
875}{1.2CA}
\punktl{26/02/91}{22:05}{314.421}{  0.162}{0.049}{ 600}{3.32}{
3184}{1.2CA}
\punktl{26/02/91}{23:10}{314.466}{  0.080}{0.046}{ 600}{3.58}{
2528}{1.2CA}
\punktl{27/02/91}{00:22}{314.515}{  0.120}{0.045}{ 600}{3.48}{
2279}{1.2CA}
\punktl{27/02/91}{01:34}{314.566}{  0.110}{0.044}{ 600}{2.97}{
2110}{1.2CA}
\punktll{27/02/91}{02:56}{314.623}{  0.130}{0.121}{ 600}{3.20}{
2271}{1.2CA}
\punktl{27/02/91}{04:14}{314.676}{  0.166}{0.044}{ 600}{3.83}{
2068}{1.2CA}
\punktll{27/02/91}{05:02}{314.710}{  0.188}{0.142}{ 600}{3.97}{
1539}{1.2CA}

\end{picture}}

\end{picture}

\vspace*{1cm}
\caption{Lightcurves of PKS\,0735+178 and W\,Com (1219+285)
         in February 1991}
\end{figure}
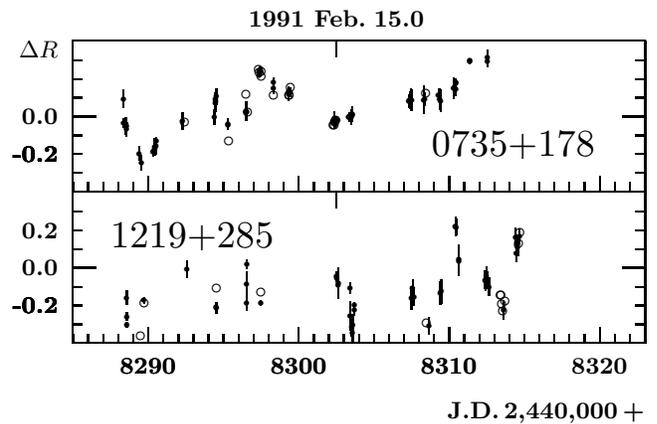

\noindent
{\bf 1E\,1229+645.} There exist to our knowledge no variability studies in
the literature for this BL\,Lac object. Our HQM lightcurve clearly indicates
variations but up to now no OVV behaviour. In about 3$''$ from the quasar
there is another object, unresolved even on good seeing CCD frames.

\noindent
{\bf 3C\,279 (1253$-$055).} The Rosemary Hill lightcurve of this BL\,Lac
object plotted in SNLC
is as undersampled as the HQM curve; however, a flare occurring in 1989 is
represented by more data points. Another flare with the peak value 4\,mag
above the average base level appeared one year before. From 1971 to 1986
the object did not show OVV behaviour. Very impressive is the historical
lightcurve measured by Eachus \& Liller (\cite{EL75}); the object reached
11.27\,$\pm$\,0.07\,mag in 1937 April and thus showed a 6.7\,mag total
range of variability. The most rapid variation was a 2.2\,mag change in
13 days in 1936.

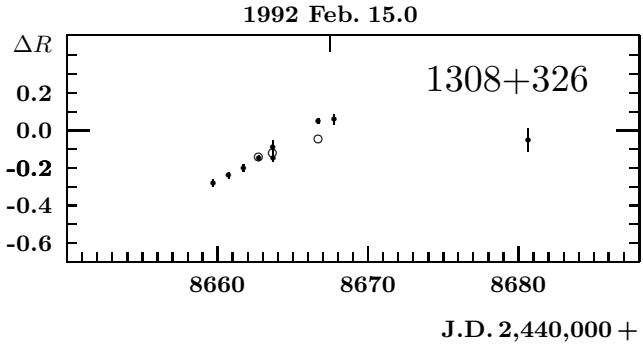
\begin{figure}

\vspace*{0.5cm}

\begin{picture}(7.8 ,3 )(-1,0)
\put(0,0){\setlength{\unitlength}{0.2cm}%
\begin{picture}(38,  15.000)(650,0)
\put(650,0){\framebox(38,  15.000)[tl]{\begin{picture}(0,0)(0,0)
        \put(0,0){\makebox(0,0)[tr]{$\Delta R$\hspace*{0.2cm}}}
        \put(38,0){\makebox(0,0)[tr]{\large{1308+326}\T{0.4}
                                 \hspace*{0.5cm}}}
        \put(38,-
15.000){\setlength{\unitlength}{1cm}\begin{picture}(0,0)(0,0)
            \put(0,-1){\makebox(0,0)[br]{\bf J.D.\,2,440,000\,+}}
        \end{picture}}
    \end{picture}}}

\thicklines
\put(650,0){\setlength{\unitlength}{2.5cm}\begin{picture}(0,0)(0,-0.7)
   \put(0,0){\setlength{\unitlength}{1cm}\begin{picture}(0,0)(0,0)
        \put(0,0){\line(1,0){0.3}}
        \end{picture}}
   \end{picture}}

\put(688,0){\setlength{\unitlength}{2.5cm}\begin{picture}(0,0)(0,-0.7)
   \put(0,0){\setlength{\unitlength}{1cm}\begin{picture}(0,0)(0,0)
        \put(0,0){\line(-1,0){0.3}}
        \end{picture}}
   \end{picture}}

\thinlines
\put(650,0){\setlength{\unitlength}{2.5cm}\begin{picture}(0,0)(0,-0.7)
   \multiput(0,0)(0,0.1){5}{\setlength{\unitlength}{1cm}%
\begin{picture}(0,0)(0,0)
        \put(0,0){\line(1,0){0.12}}
        \end{picture}}
   \end{picture}}

\put(650,0){\setlength{\unitlength}{2.5cm}\begin{picture}(0,0)(0,-0.7)
   \multiput(0,0)(0,-0.1){7}{\setlength{\unitlength}{1cm}%
\begin{picture}(0,0)(0,0)
        \put(0,0){\line(1,0){0.12}}
        \end{picture}}
   \end{picture}}

\put(688,0){\setlength{\unitlength}{2.5cm}\begin{picture}(0,0)(0,-0.7)
   \multiput(0,0)(0,0.1){5}{\setlength{\unitlength}{1cm}%
\begin{picture}(0,0)(0,0)
        \put(0,0){\line(-1,0){0.12}}
        \end{picture}}
   \end{picture}}

\put(688,0){\setlength{\unitlength}{2.5cm}\begin{picture}(0,0)(0,-0.7)
   \multiput(0,0)(0,-0.1){7}{\setlength{\unitlength}{1cm}%
\begin{picture}(0,0)(0,0)
        \put(0,0){\line(-1,0){0.12}}
        \end{picture}}
   \end{picture}}

\put(650,0){\setlength{\unitlength}{2.5cm}\begin{picture}(0,0)(0,-0.7)
   \put(0,-0.4){\setlength{\unitlength}{1cm}\begin{picture}(0,0)(0,0)
        \put(0,0){\line(1,0){0.12}}
        \put(-0.2,0){\makebox(0,0)[r]{\bf -0.4}}
        \end{picture}}
   \put(0,-0.6){\setlength{\unitlength}{1cm}\begin{picture}(0,0)(0,0)
        \put(0,0){\line(1,0){0.12}}
        \put(-0.2,0){\makebox(0,0)[r]{\bf -0.6}}
        \end{picture}}
   \put(0,-0.2){\setlength{\unitlength}{1cm}\begin{picture}(0,0)(0,0)
        \put(0,0){\line(1,0){0.12}}
        \put(-0.2,0){\makebox(0,0)[r]{\bf -0.2}}
        \end{picture}}
   \put(0,-0.2){\setlength{\unitlength}{1cm}\begin{picture}(0,0)(0,0)
        \put(0,0){\line(1,0){0.12}}
        \put(-0.2,0){\makebox(0,0)[r]{\bf -0.2}}
        \end{picture}}
   \put(0,-0.2){\setlength{\unitlength}{1cm}\begin{picture}(0,0)(0,0)
        \put(0,0){\line(1,0){0.12}}
        \put(-0.2,0){\makebox(0,0)[r]{\bf -0.2}}
        \end{picture}}
   \put(0,0.4){\setlength{\unitlength}{1cm}\begin{picture}(0,0)(0,0)
        \put(0,0){\line(1,0){0.12}}
        \end{picture}}
   \put(0,0.0){\setlength{\unitlength}{1cm}\begin{picture}(0,0)(0,0)
        \put(0,0){\line(1,0){0.12}}
        \put(-0.2,0){\makebox(0,0)[r]{\bf 0.0}}
        \end{picture}}
   \put(0,0.2){\setlength{\unitlength}{1cm}\begin{picture}(0,0)(0,0)
        \put(0,0){\line(1,0){0.12}}
        \put(-0.2,0){\makebox(0,0)[r]{\bf 0.2}}
        \end{picture}}
   \put(0,0.2){\setlength{\unitlength}{1cm}\begin{picture}(0,0)(0,0)
        \put(0,0){\line(1,0){0.12}}
        \end{picture}}
   \end{picture}}

   \put(667.5,  15.000){\setlength{\unitlength}{1cm}\begin{picture}(0,0)(0,0)
        \put(0,0){\line(0,-1){0.2}}
        \put(0,0.2){\makebox(0,0)[b]{\bf 1992 Feb.~15.0}}
   \end{picture}}

    \multiput(650,0)(1,0){38}%
        {\setlength{\unitlength}{1cm}\begin{picture}(0,0)(0,0)
        \put(0,0){\line(0,1){0.12}}
    \end{picture}}
    \put(660,0){\setlength{\unitlength}{1cm}\begin{picture}(0,0)(0,0)
        \put(0,0){\line(0,1){0.2}}
        \put(0,-0.2){\makebox(0,0)[t]{\bf 8660}}
    \end{picture}}
    \put(670,0){\setlength{\unitlength}{1cm}\begin{picture}(0,0)(0,0)
        \put(0,0){\line(0,1){0.2}}
        \put(0,-0.2){\makebox(0,0)[t]{\bf 8670}}
    \end{picture}}
   \put(680,0){\setlength{\unitlength}{1cm}\begin{picture}(0,0)(0,0)
        \put(0,0){\line(0,1){0.2}}
        \put(0,-0.2){\makebox(0,0)[t]{\bf 8680}}
    \end{picture}}

\punkto{07/02/92}{04:47}{659.700}{ -0.279}{0.019}{ 500}{1.68}{
464}{2.2CA}
\punkto{08/02/92}{05:33}{660.732}{ -0.239}{0.016}{ 500}{1.35}{
465}{2.2CA}
\punkto{09/02/92}{05:18}{661.721}{ -0.199}{0.018}{ 500}{0.89}{
456}{2.2CA}
\punktoo{10/02/92}{05:17}{662.721}{ -0.142}{0.035}{ 100}{0.91}{
302}{2.2CA}
\punkto{10/02/92}{05:25}{662.726}{ -0.146}{0.012}{ 500}{1.07}{
493}{2.2CA}
\punktoo{11/02/92}{03:42}{663.654}{ -0.118}{0.020}{ 100}{0.91}{
292}{2.2CA}
\punkto{11/02/92}{03:52}{663.662}{ -0.087}{0.033}{ 500}{1.36}{
452}{2.2CA}
\punkto{11/02/92}{04:40}{663.695}{ -0.149}{0.018}{ 500}{0.92}{
460}{2.2CA}
\punktoo{14/02/92}{04:38}{666.693}{ -0.044}{0.015}{ 100}{1.09}{
299}{2.2CA}
\punkto{14/02/92}{04:46}{666.699}{  0.051}{0.013}{ 500}{1.17}{
485}{2.2CA}
\punkto{15/02/92}{05:52}{667.745}{  0.059}{0.025}{ 500}{1.41}{
1438}{2.2CA}
\punkto{28/02/92}{03:21}{680.640}{ -0.052}{0.060}{ 500}{2.66}{
550}{2.2CA}

\end{picture}}

\end{picture}

\vspace*{1cm}
\caption{HQM-lightcurve of B2\,1308+326 in February 1992}
\end{figure}

\noindent
{\bf B2\,1308+326.} The lightcurve shown by WSLF covers the time between
1976 and 1986; it displays a 3.0\,mag total range of variability with several
flares recorded. WSLF note an overall decline in the lightcurve and report
that the BL\,Lac object was at a very faint state in early 1987. Our HQM
lightcurves seems to indicate that the object is brightening again. A
series of data is shown with higher time resolution in Fig.~6. We
have obviously covered the peak of a ``mini-flare''. On our CCD frames there
is a number excess of relatively bright galaxies inside $2\arcmin$.

\noindent
{\bf NRAO\,512 (1638+398)}. The Rosemary Hill lightcurve plotted in SNLC
includes only three data points for the period of our observations; it
shows, however, a series of flares in the early '70s and a 3.2\,mag total
range of variability. The HQM lightcurve shows an isolated data point in
the second half of 1988 which is 1.5\,mag above the average base
brightness; a few 0.5\,mag events are covered by more data points; see also
Fig.~7. Making use of the photometric sequence of Smith et al.\
(\cite{SBHE85}) for 3C\,345, we were able to determine the reference
magnitude of our lightcurve, $R_0=19.31$.

\begin{figure}

\vspace*{0.5cm}

\begin{picture}(7.8 ,2 )(-1,0)
\put(0,0){\setlength{\unitlength}{0.2cm}%
\begin{picture}(38,  10.000)(460,0)
\put(460,0){\framebox(38,  10.000)[tl]{\begin{picture}(0,0)(0,0)
        \put(0,0){\makebox(0,0)[tr]{$\Delta R$\hspace*{0.2cm}}}
        \put(38,0){\makebox(0,0)[tr]{\large{1638+398}\T{0.4}
                                 \hspace*{0.5cm}}}
        \put(38,-
10.000){\setlength{\unitlength}{1cm}\begin{picture}(0,0)(0,0)
        \end{picture}}
    \end{picture}}}

\thicklines
\put(460,0){\setlength{\unitlength}{2.5cm}\begin{picture}(0,0)(0,-0.5)
   \put(0,0){\setlength{\unitlength}{1cm}\begin{picture}(0,0)(0,0)
        \put(0,0){\line(1,0){0.3}}
        \end{picture}}
   \end{picture}}

\put(498,0){\setlength{\unitlength}{2.5cm}\begin{picture}(0,0)(0,-0.5)
   \put(0,0){\setlength{\unitlength}{1cm}\begin{picture}(0,0)(0,0)
        \put(0,0){\line(-1,0){0.3}}
        \end{picture}}
   \end{picture}}

\thinlines
\put(460,0){\setlength{\unitlength}{2.5cm}\begin{picture}(0,0)(0,-0.5)
   \multiput(0,0)(0,0.1){3}{\setlength{\unitlength}{1cm}%
\begin{picture}(0,0)(0,0)
        \put(0,0){\line(1,0){0.12}}
        \end{picture}}
   \end{picture}}

\put(460,0){\setlength{\unitlength}{2.5cm}\begin{picture}(0,0)(0,-0.5)
   \multiput(0,0)(0,-0.1){5}{\setlength{\unitlength}{1cm}%
\begin{picture}(0,0)(0,0)
        \put(0,0){\line(1,0){0.12}}
        \end{picture}}
   \end{picture}}

\put(498,0){\setlength{\unitlength}{2.5cm}\begin{picture}(0,0)(0,-0.5)
   \multiput(0,0)(0,0.1){3}{\setlength{\unitlength}{1cm}%
\begin{picture}(0,0)(0,0)
        \put(0,0){\line(-1,0){0.12}}
        \end{picture}}
   \end{picture}}

\put(498,0){\setlength{\unitlength}{2.5cm}\begin{picture}(0,0)(0,-0.5)
   \multiput(0,0)(0,-0.1){5}{\setlength{\unitlength}{1cm}%
\begin{picture}(0,0)(0,0)
        \put(0,0){\line(-1,0){0.12}}
        \end{picture}}
   \end{picture}}

\put(460,0){\setlength{\unitlength}{2.5cm}\begin{picture}(0,0)(0,-0.5)
   \put(0,-0.2){\setlength{\unitlength}{1cm}\begin{picture}(0,0)(0,0)
        \put(0,0){\line(1,0){0.12}}
        \put(-0.2,0){\makebox(0,0)[r]{\bf -0.2}}
        \end{picture}}
   \put(0,-0.2){\setlength{\unitlength}{1cm}\begin{picture}(0,0)(0,0)
        \put(0,0){\line(1,0){0.12}}
        \put(-0.2,0){\makebox(0,0)[r]{\bf -0.2}}
        \end{picture}}
   \put(0,-0.2){\setlength{\unitlength}{1cm}\begin{picture}(0,0)(0,0)
        \put(0,0){\line(1,0){0.12}}
        \put(-0.2,0){\makebox(0,0)[r]{\bf -0.2}}
        \end{picture}}
   \put(0,-0.2){\setlength{\unitlength}{1cm}\begin{picture}(0,0)(0,0)
        \put(0,0){\line(1,0){0.12}}
        \put(-0.2,0){\makebox(0,0)[r]{\bf -0.2}}
        \end{picture}}
   \put(0,-0.2){\setlength{\unitlength}{1cm}\begin{picture}(0,0)(0,0)
        \put(0,0){\line(1,0){0.12}}
        \put(-0.2,0){\makebox(0,0)[r]{\bf -0.2}}
        \end{picture}}
   \put(0,-0.2){\setlength{\unitlength}{1cm}\begin{picture}(0,0)(0,0)
        \put(0,0){\line(1,0){0.12}}
        \put(-0.2,0){\makebox(0,0)[r]{\bf -0.2}}
        \end{picture}}
   \put(0,0.0){\setlength{\unitlength}{1cm}\begin{picture}(0,0)(0,0)
        \put(0,0){\line(1,0){0.12}}
        \put(-0.2,0){\makebox(0,0)[r]{\bf 0.0}}
        \end{picture}}
   \put(0,-0.4){\setlength{\unitlength}{1cm}\begin{picture}(0,0)(0,0)
        \put(0,0){\line(1,0){0.12}}
        \put(-0.2,0){\makebox(0,0)[r]{\bf -0.4}}
        \end{picture}}
   \put(0,0.2){\setlength{\unitlength}{1cm}\begin{picture}(0,0)(0,0)
        \put(0,0){\line(1,0){0.12}}
        \end{picture}}
   \end{picture}}

   \put(469.5,  10.000){\setlength{\unitlength}{1cm}\begin{picture}(0,0)(0,0)
        \put(0,0){\line(0,-1){0.2}}
        \put(0,0.2){\makebox(0,0)[b]{\bf 1991 Aug.~01.0}}
   \end{picture}}

    \multiput(460,0)(1,0){38}%
        {\setlength{\unitlength}{1cm}\begin{picture}(0,0)(0,0)
        \put(0,0){\line(0,1){0.12}}
    \end{picture}}
    \put(470,0){\setlength{\unitlength}{1cm}\begin{picture}(0,0)(0,0)
        \put(0,0){\line(0,1){0.2}}
    \end{picture}}
    \put(470,0){\setlength{\unitlength}{1cm}\begin{picture}(0,0)(0,0)
        \put(0,0){\line(0,1){0.2}}
    \end{picture}}
   \put(480,0){\setlength{\unitlength}{1cm}\begin{picture}(0,0)(0,0)
        \put(0,0){\line(0,1){0.2}}
    \end{picture}}
   \put(480,0){\setlength{\unitlength}{1cm}\begin{picture}(0,0)(0,0)
        \put(0,0){\line(0,1){0.2}}
    \end{picture}}
    \put(490,0){\setlength{\unitlength}{1cm}\begin{picture}(0,0)(0,0)
        \put(0,0){\line(0,1){0.2}}
    \end{picture}}
    \put(490,0){\setlength{\unitlength}{1cm}\begin{picture}(0,0)(0,0)
       \put(0,0){\line(0,1){0.2}}
    \end{picture}}
    \put(490,0){\setlength{\unitlength}{1cm}\begin{picture}(0,0)(0,0)
       \put(0,0){\line(0,1){0.2}}
    \end{picture}}

\punktkk{29/07/91}{21:34}{467.399}{ -0.389}{0.281}{ 100}{2.80}{
347}{1.2CA}
\punktkk{31/07/91}{21:51}{469.411}{ -0.066}{0.039}{ 100}{1.30}{
327}{1.2CA}
\punktk{31/07/91}{22:00}{469.417}{ -0.039}{0.018}{ 500}{1.37}{
559}{1.2CA}
\punktkk{01/08/91}{21:17}{470.387}{  0.089}{0.030}{ 100}{1.01}{
306}{1.2CA}
\punktk{01/08/91}{21:26}{470.393}{ -0.028}{0.022}{ 500}{1.30}{
459}{1.2CA}
\punktkk{02/08/91}{20:58}{471.374}{  0.032}{0.042}{ 100}{1.37}{
299}{1.2CA}
\punktk{02/08/91}{21:09}{471.382}{  0.001}{0.016}{ 500}{1.34}{
491}{1.2CA}
\punktkk{03/08/91}{22:09}{472.423}{  0.100}{0.055}{ 100}{1.72}{
296}{1.2CA}
\punktk{03/08/91}{22:17}{472.429}{  0.110}{0.014}{ 500}{1.30}{
473}{1.2CA}
\punktkk{04/08/91}{21:52}{473.412}{  0.112}{0.041}{ 100}{1.58}{
293}{1.2CA}
\punktk{04/08/91}{22:01}{473.418}{  0.122}{0.021}{ 500}{1.57}{
480}{1.2CA}
\punktkk{05/08/91}{21:38}{474.402}{ -0.045}{0.074}{ 100}{1.39}{
562}{1.2CA}
\punktk{05/08/91}{21:46}{474.408}{ -0.040}{0.016}{ 500}{1.37}{
452}{1.2CA}
\punktk{05/08/91}{23:44}{474.489}{ -0.037}{0.020}{ 500}{1.39}{
460}{1.2CA}
\punktk{06/08/91}{20:21}{475.348}{ -0.131}{0.026}{ 500}{1.55}{
577}{1.2CA}
\punktkk{07/08/91}{00:59}{475.541}{ -0.147}{0.069}{ 100}{1.41}{
299}{1.2CA}
\punktk{07/08/91}{01:09}{475.548}{ -0.202}{0.026}{ 500}{1.44}{
494}{1.2CA}
\punktkk{08/08/91}{20:29}{477.354}{ -0.290}{0.103}{ 100}{2.05}{
311}{1.2CA}
\punktk{08/08/91}{20:40}{477.361}{ -0.199}{0.036}{ 500}{1.96}{
470}{1.2CA}
\punktkk{09/08/91}{00:57}{477.540}{ -0.178}{0.109}{ 500}{3.04}{
531}{1.2CA}
\punktk{10/08/91}{20:19}{479.347}{ -0.370}{0.086}{ 500}{2.80}{
551}{1.2CA}
\punktkk{11/08/91}{21:02}{480.377}{ -0.351}{0.104}{ 100}{1.86}{
305}{1.2CA}
\punktkk{20/08/91}{00:23}{488.516}{ -0.267}{0.161}{ 500}{2.23}{
708}{1.2CA}
\punktkk{26/08/91}{20:22}{495.349}{ -0.261}{0.116}{ 500}{1.44}{
1366}{1.2CA}

\end{picture}}

\end{picture}

\vspace*{-0.02cm}

\begin{picture}(7.8 ,3 )(-1,0)
\put(0,0){\setlength{\unitlength}{0.2cm}%
\begin{picture}(38,  15.000)(460,0)
\put(460,0){\framebox(38,  15.000)[tl]{\begin{picture}(0,0)(0,0)
        \put(38,0){\makebox(0,0)[tr]{\large{1823+568}\T{0.4}
                                 \hspace*{0.5cm}}}
        \put(38,-
15.000){\setlength{\unitlength}{1cm}\begin{picture}(0,0)(0,0)
            \put(0,-1){\makebox(0,0)[br]{\bf J.D.\,2,440,000\,+}}
        \end{picture}}
    \end{picture}}}

\thicklines
\put(460,0){\setlength{\unitlength}{2.5cm}\begin{picture}(0,0)(0,-0.6)
   \put(0,0){\setlength{\unitlength}{1cm}\begin{picture}(0,0)(0,0)
        \put(0,0){\line(1,0){0.3}}
        \end{picture}}
   \end{picture}}

\put(498,0){\setlength{\unitlength}{2.5cm}\begin{picture}(0,0)(0,-0.6)
   \put(0,0){\setlength{\unitlength}{1cm}\begin{picture}(0,0)(0,0)
        \put(0,0){\line(-1,0){0.3}}
        \end{picture}}
   \end{picture}}

\thinlines
\put(460,0){\setlength{\unitlength}{2.5cm}\begin{picture}(0,0)(0,-0.6)
   \multiput(0,0)(0,0.1){6}{\setlength{\unitlength}{1cm}%
\begin{picture}(0,0)(0,0)
        \put(0,0){\line(1,0){0.12}}
        \end{picture}}
   \end{picture}}

\put(460,0){\setlength{\unitlength}{2.5cm}\begin{picture}(0,0)(0,-0.6)
   \multiput(0,0)(0,-0.1){6}{\setlength{\unitlength}{1cm}%
\begin{picture}(0,0)(0,0)
        \put(0,0){\line(1,0){0.12}}
        \end{picture}}
   \end{picture}}

\put(498,0){\setlength{\unitlength}{2.5cm}\begin{picture}(0,0)(0,-0.6)
   \multiput(0,0)(0,0.1){6}{\setlength{\unitlength}{1cm}%
\begin{picture}(0,0)(0,0)
        \put(0,0){\line(-1,0){0.12}}
        \end{picture}}
   \end{picture}}

\put(498,0){\setlength{\unitlength}{2.5cm}\begin{picture}(0,0)(0,-0.6)
   \multiput(0,0)(0,-0.1){6}{\setlength{\unitlength}{1cm}%
\begin{picture}(0,0)(0,0)
        \put(0,0){\line(-1,0){0.12}}
        \end{picture}}
   \end{picture}}

\put(460,0){\setlength{\unitlength}{2.5cm}\begin{picture}(0,0)(0,-0.6)
   \put(0,-0.4){\setlength{\unitlength}{1cm}\begin{picture}(0,0)(0,0)
        \put(0,0){\line(1,0){0.12}}
        \put(-0.2,0){\makebox(0,0)[r]{\bf -0.4}}
        \end{picture}}
   \put(0,-0.2){\setlength{\unitlength}{1cm}\begin{picture}(0,0)(0,0)
        \put(0,0){\line(1,0){0.12}}
        \put(-0.2,0){\makebox(0,0)[r]{\bf -0.2}}
        \end{picture}}
   \put(0,-0.2){\setlength{\unitlength}{1cm}\begin{picture}(0,0)(0,0)
        \put(0,0){\line(1,0){0.12}}
        \put(-0.2,0){\makebox(0,0)[r]{\bf -0.2}}
        \end{picture}}
   \put(0,-0.2){\setlength{\unitlength}{1cm}\begin{picture}(0,0)(0,0)
        \put(0,0){\line(1,0){0.12}}
        \put(-0.2,0){\makebox(0,0)[r]{\bf -0.2}}
        \end{picture}}
   \put(0,-0.2){\setlength{\unitlength}{1cm}\begin{picture}(0,0)(0,0)
        \put(0,0){\line(1,0){0.12}}
        \put(-0.2,0){\makebox(0,0)[r]{\bf -0.2}}
        \end{picture}}
   \put(0,0.4){\setlength{\unitlength}{1cm}\begin{picture}(0,0)(0,0)
        \put(0,0){\line(1,0){0.12}}
        \put(-0.2,0){\makebox(0,0)[r]{\bf 0.4}}
        \end{picture}}
   \put(0,0.0){\setlength{\unitlength}{1cm}\begin{picture}(0,0)(0,0)
        \put(0,0){\line(1,0){0.12}}
        \put(-0.2,0){\makebox(0,0)[r]{\bf 0.0}}
        \end{picture}}
   \put(0,0.2){\setlength{\unitlength}{1cm}\begin{picture}(0,0)(0,0)
        \put(0,0){\line(1,0){0.12}}
        \put(-0.2,0){\makebox(0,0)[r]{\bf 0.2}}
        \end{picture}}
   \put(0,0.4){\setlength{\unitlength}{1cm}\begin{picture}(0,0)(0,0)
        \put(0,0){\line(1,0){0.12}}
        \end{picture}}
   \end{picture}}

   \put(469.5,  15.000){\setlength{\unitlength}{1cm}\begin{picture}(0,0)(0,0)
        \put(0,0){\line(0,-1){0.2}}
   \end{picture}}

    \multiput(460,0)(1,0){38}%
        {\setlength{\unitlength}{1cm}\begin{picture}(0,0)(0,0)
        \put(0,0){\line(0,1){0.12}}
    \end{picture}}
    \put(470,0){\setlength{\unitlength}{1cm}\begin{picture}(0,0)(0,0)
        \put(0,0){\line(0,1){0.2}}
        \put(0,-0.2){\makebox(0,0)[t]{\bf 8470}}
    \end{picture}}
    \put(470,0){\setlength{\unitlength}{1cm}\begin{picture}(0,0)(0,0)
        \put(0,0){\line(0,1){0.2}}
        \put(0,-0.2){\makebox(0,0)[t]{\bf 8470}}
    \end{picture}}
   \put(480,0){\setlength{\unitlength}{1cm}\begin{picture}(0,0)(0,0)
        \put(0,0){\line(0,1){0.2}}
        \put(0,-0.2){\makebox(0,0)[t]{\bf 8480}}
    \end{picture}}
   \put(480,0){\setlength{\unitlength}{1cm}\begin{picture}(0,0)(0,0)
        \put(0,0){\line(0,1){0.2}}
        \put(0,-0.2){\makebox(0,0)[t]{\bf 8480}}
    \end{picture}}
    \put(490,0){\setlength{\unitlength}{1cm}\begin{picture}(0,0)(0,0)
        \put(0,0){\line(0,1){0.2}}
        \put(0,-0.2){\makebox(0,0)[t]{\bf 8490}}
    \end{picture}}
    \put(490,0){\setlength{\unitlength}{1cm}\begin{picture}(0,0)(0,0)
       \put(0,0){\line(0,1){0.2}}
        \put(0,-0.2){\makebox(0,0)[t]{\bf 8490}}
    \end{picture}}
    \put(490,0){\setlength{\unitlength}{1cm}\begin{picture}(0,0)(0,0)
       \put(0,0){\line(0,1){0.2}}
        \put(0,-0.2){\makebox(0,0)[t]{\bf 8490}}
    \end{picture}}

\punktm{24/07/91}{22:50}{462.452}{ -0.436}{0.034}{ 500}{1.70}{
1500}{1.2CA}
\punktmm{26/07/91}{01:23}{463.558}{ -0.255}{0.133}{ 100}{1.22}{
582}{1.2CA}
\punktm{26/07/91}{01:32}{463.564}{ -0.299}{0.025}{ 500}{1.38}{
1819}{1.2CA}
\punktm{26/07/91}{23:32}{464.481}{  0.121}{0.027}{ 300}{1.33}{
4420}{1.2CA}
\punktm{26/07/91}{23:39}{464.486}{  0.142}{0.027}{ 300}{1.35}{
4268}{1.2CA}
\punktm{28/07/91}{21:19}{466.388}{  0.012}{0.041}{ 300}{1.86}{
1036}{1.2CA}
\punktm{29/07/91}{00:56}{466.539}{ -0.194}{0.031}{ 500}{2.30}{
1489}{1.2CA}
\punktmm{29/07/91}{22:29}{467.437}{ -0.197}{0.130}{ 100}{2.60}{
409}{1.2CA}
\punktm{29/07/91}{22:37}{467.443}{ -0.243}{0.055}{ 500}{2.94}{
886}{1.2CA}
\punktm{30/07/91}{23:17}{468.471}{ -0.081}{0.077}{ 300}{3.34}{
875}{1.2CA}
\punktm{30/07/91}{23:25}{468.476}{ -0.195}{0.078}{ 300}{3.19}{
893}{1.2CA}
\punktm{31/07/91}{22:48}{469.450}{  0.023}{0.010}{ 300}{1.13}{
478}{1.2CA}
\punktm{31/07/91}{22:54}{469.454}{  0.003}{0.065}{ 300}{1.13}{
483}{1.2CA}
\punktm{31/07/91}{23:45}{469.490}{  0.085}{0.013}{ 100}{1.05}{
364}{1.2CA}
\punktm{31/07/91}{23:52}{469.495}{  0.075}{0.063}{ 300}{1.08}{
543}{1.2CA}
\punktm{01/08/91}{22:38}{470.444}{ -0.149}{0.053}{ 500}{1.13}{
591}{1.2CA}
\punktm{01/08/91}{22:44}{470.448}{ -0.143}{0.013}{ 150}{0.95}{
359}{1.2CA}
\punktm{02/08/91}{22:28}{471.437}{ -0.257}{0.024}{ 100}{1.09}{
304}{1.2CA}
\punktm{02/08/91}{22:37}{471.443}{ -0.257}{0.008}{ 500}{1.05}{
527}{1.2CA}
\punktm{03/08/91}{23:02}{472.460}{ -0.360}{0.018}{ 100}{1.22}{
299}{1.2CA}
\punktm{03/08/91}{23:10}{472.466}{ -0.395}{0.062}{ 500}{1.38}{
508}{1.2CA}
\punktm{05/08/91}{23:19}{474.472}{ -0.434}{0.014}{ 100}{1.04}{
291}{1.2CA}
\punktm{05/08/91}{23:28}{474.478}{ -0.437}{0.064}{ 500}{1.11}{
444}{1.2CA}
\punktm{07/08/91}{00:08}{475.506}{ -0.425}{0.048}{ 100}{1.04}{
288}{1.2CA}
\punktm{07/08/91}{00:16}{475.512}{ -0.477}{0.008}{ 500}{1.17}{
473}{1.2CA}
\punktm{08/08/91}{00:18}{476.513}{ -0.380}{0.047}{ 100}{1.63}{
376}{1.2CA}
\punktm{08/08/91}{00:28}{476.520}{ -0.438}{0.062}{ 500}{1.46}{
599}{1.2CA}
\punktmm{08/08/91}{23:26}{477.476}{ -0.270}{0.137}{ 100}{1.86}{
317}{1.2CA}
\punktm{08/08/91}{23:34}{477.482}{ -0.221}{0.059}{ 500}{2.14}{
471}{1.2CA}
\punktmm{09/08/91}{22:09}{478.423}{  0.207}{0.074}{ 100}{3.11}{
379}{1.2CA}
\punktm{09/08/91}{22:22}{478.432}{  0.150}{0.025}{1000}{3.20}{
825}{1.2CA}
\punktm{09/08/91}{23:23}{478.475}{  0.230}{0.044}{1000}{3.85}{
1177}{1.2CA}
\punktm{10/08/91}{00:09}{478.507}{  0.245}{0.027}{1000}{3.59}{
895}{1.2CA}
\punktm{10/08/91}{00:27}{478.519}{  0.263}{0.027}{1000}{3.37}{
914}{1.2CA}
\punktm{10/08/91}{01:33}{478.565}{  0.255}{0.021}{1000}{3.08}{
933}{1.2CA}
\punktm{10/08/91}{01:52}{478.578}{  0.272}{0.027}{1000}{3.50}{
974}{1.2CA}
\punktm{10/08/91}{20:31}{479.355}{  0.286}{0.041}{ 100}{2.62}{
321}{1.2CA}
\punktm{10/08/91}{23:11}{479.466}{  0.237}{0.044}{ 100}{2.91}{
324}{1.2CA}
\punktm{10/08/91}{23:33}{479.482}{  0.294}{0.022}{ 500}{2.63}{
521}{1.2CA}
\punktm{11/08/91}{00:10}{479.508}{  0.327}{0.048}{ 500}{2.67}{
519}{1.2CA}
\punktmm{11/08/91}{19:53}{480.329}{  0.106}{0.093}{ 100}{1.65}{
2167}{1.2CA}
\punktm{11/08/91}{20:13}{480.343}{  0.137}{0.011}{ 500}{1.59}{
870}{1.2CA}
\punktm{11/08/91}{22:52}{480.453}{  0.182}{0.022}{ 100}{1.73}{
330}{1.2CA}
\punktm{11/08/91}{22:59}{480.458}{  0.194}{0.010}{ 500}{1.63}{
556}{1.2CA}
\punktm{12/08/91}{00:34}{480.524}{  0.264}{0.018}{ 100}{1.50}{
323}{1.2CA}
\punktm{12/08/91}{00:42}{480.529}{  0.265}{0.007}{ 500}{1.41}{
597}{1.2CA}
\punktm{12/08/91}{21:01}{481.376}{  0.256}{0.019}{ 100}{1.55}{
328}{1.2CA}
\punktm{12/08/91}{21:11}{481.383}{  0.257}{0.045}{ 500}{1.75}{
491}{1.2CA}
\punktm{12/08/91}{23:37}{481.484}{  0.252}{0.021}{ 100}{1.60}{
352}{1.2CA}
\punktm{12/08/91}{23:47}{481.491}{  0.283}{0.009}{ 500}{1.65}{
534}{1.2CA}
\punktm{19/08/91}{22:40}{488.445}{ -0.288}{0.034}{ 100}{1.11}{
342}{1.2CA}
\punktm{19/08/91}{22:48}{488.450}{ -0.259}{0.019}{ 500}{1.18}{
705}{1.2CA}
\punktm{19/08/91}{23:13}{488.468}{ -0.282}{0.030}{ 100}{1.17}{
333}{1.2CA}
\punktmm{21/08/91}{20:16}{490.345}{ -0.013}{0.150}{ 100}{1.41}{
590}{1.2CA}
\punktm{21/08/91}{20:24}{490.351}{ -0.077}{0.039}{ 500}{1.37}{
1730}{1.2CA}
\punktm{22/08/91}{00:09}{490.507}{ -0.092}{0.126}{ 500}{2.40}{
1757}{1.2CA}
\punktm{22/08/91}{20:03}{491.336}{ -0.253}{0.106}{ 500}{2.75}{
1290}{1.2CA}
\punktmm{23/08/91}{21:35}{492.400}{ -0.247}{0.099}{ 100}{1.78}{
495}{1.2CA}
\punktm{23/08/91}{21:43}{492.405}{ -0.226}{0.059}{ 500}{1.81}{
1500}{1.2CA}
\punktmm{24/08/91}{00:49}{492.534}{ -0.244}{0.108}{ 100}{1.55}{
548}{1.2CA}
\punktm{24/08/91}{00:57}{492.540}{ -0.191}{0.060}{ 500}{1.72}{
1735}{1.2CA}
\punktmm{24/08/91}{21:06}{493.379}{ -0.351}{0.070}{ 100}{1.38}{
506}{1.2CA}
\punktm{24/08/91}{21:13}{493.385}{ -0.293}{0.035}{ 500}{1.41}{
1603}{1.2CA}
\punktmm{25/08/91}{21:37}{494.401}{ -0.272}{0.342}{ 100}{1.60}{
1212}{1.2CA}
\punktm{25/08/91}{21:45}{494.407}{ -0.267}{0.077}{ 500}{1.42}{
3502}{1.2CA}
\punktmm{26/08/91}{22:26}{495.435}{ -0.312}{0.202}{ 500}{2.93}{
2015}{1.2CA}
\punktmm{27/08/91}{20:36}{496.358}{ -0.235}{0.260}{ 100}{2.84}{
421}{1.2CA}
\punktm{27/08/91}{20:46}{496.366}{ -0.247}{0.062}{ 500}{2.78}{
1047}{1.2CA}

\end{picture}}

\end{picture}

\vspace*{1cm}
\caption{Lightcurves of NRAO\,512 (1638+398) and 4C\,56.27 (1823+568)
         in August 1991}
\end{figure}
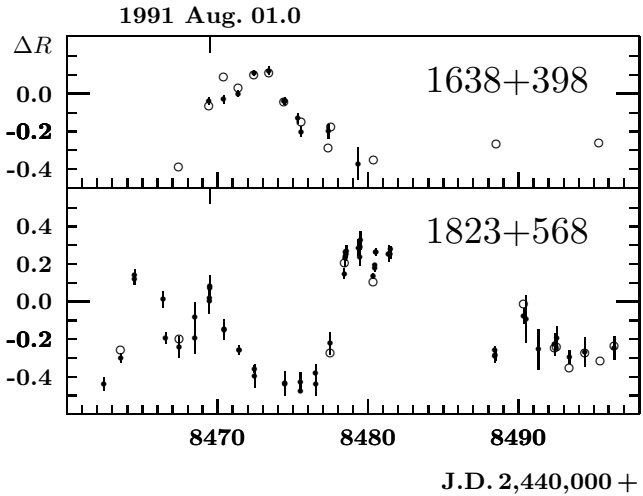

\begin{figure}

\vspace*{0.5cm}

\begin{picture}(7.8 ,2.5 )(-1,0)
\put(0,0){\setlength{\unitlength}{0.2cm}%
\begin{picture}(38,  12.500)(455,0)
\put(455,0){\framebox(38,  12.500)[tl]{\begin{picture}(0,0)(0,0)
        \put(0,0){\makebox(0,0)[tr]{$\Delta R$\hspace*{0.2cm}}}
        \put(0,0){\makebox(0,0)[tl]{\hspace*{0.5cm}\large{2223$-$052}\T{0.4}
                                 }}
        \put(38,-
12.500){\setlength{\unitlength}{1cm}\begin{picture}(0,0)(0,0)
            \put(0,-1){\makebox(0,0)[br]{\bf J.D.\,2,440,000\,+}}
        \end{picture}}
    \end{picture}}}

\put(455,0){\setlength{\unitlength}{2.5cm}\begin{picture}(0,0)(0,-1.4)
   \multiput(0,-0.4)(0,-0.1){10}{\setlength{\unitlength}{1cm}%
\begin{picture}(0,0)(0,0)
        \put(0,0){\line(1,0){0.12}}
        \end{picture}}
   \end{picture}}

\put(493,0){\setlength{\unitlength}{2.5cm}\begin{picture}(0,0)(0,-1.4)
   \multiput(0,-0.4)(0,-0.1){10}{\setlength{\unitlength}{1cm}%
\begin{picture}(0,0)(0,0)
        \put(0,0){\line(-1,0){0.12}}
        \end{picture}}
   \end{picture}}

\put(455,0){\setlength{\unitlength}{2.5cm}\begin{picture}(0,0)(0,-1.4)
   \put(0,-1.2){\setlength{\unitlength}{1cm}\begin{picture}(0,0)(0,0)
        \put(0,0){\line(1,0){0.12}}
        \put(-0.2,0){\makebox(0,0)[r]{\bf -1.2}}
        \end{picture}}
   \put(0,-0.6){\setlength{\unitlength}{1cm}\begin{picture}(0,0)(0,0)
        \put(0,0){\line(1,0){0.12}}
        \end{picture}}
   \put(0,-0.8){\setlength{\unitlength}{1cm}\begin{picture}(0,0)(0,0)
        \put(0,0){\line(1,0){0.12}}
        \put(-0.2,0){\makebox(0,0)[r]{\bf -0.8}}
        \end{picture}}
   \put(0,-1.0){\setlength{\unitlength}{1cm}\begin{picture}(0,0)(0,0)
        \put(0,0){\line(1,0){0.12}}
        \put(-0.2,0){\makebox(0,0)[r]{\bf -1.0}}
        \end{picture}}
        \end{picture}}

   \put(469.5,  12.500){\setlength{\unitlength}{1cm}\begin{picture}(0,0)(0,0)
        \put(0,0){\line(0,-1){0.2}}
        \put(0,0.2){\makebox(0,0)[b]{\bf 1991 Aug.~01.0}}
   \end{picture}}

    \multiput(455,0)(1,0){38}%
        {\setlength{\unitlength}{1cm}\begin{picture}(0,0)(0,0)
        \put(0,0){\line(0,1){0.12}}
    \end{picture}}
    \put(460,0){\setlength{\unitlength}{1cm}\begin{picture}(0,0)(0,0)
        \put(0,0){\line(0,1){0.2}}
        \put(0,-0.2){\makebox(0,0)[t]{\bf 8460}}
    \end{picture}}
    \put(460,0){\setlength{\unitlength}{1cm}\begin{picture}(0,0)(0,0)
        \put(0,0){\line(0,1){0.2}}
        \put(0,-0.2){\makebox(0,0)[t]{\bf 8460}}
    \end{picture}}
   \put(470,0){\setlength{\unitlength}{1cm}\begin{picture}(0,0)(0,0)
        \put(0,0){\line(0,1){0.2}}
        \put(0,-0.2){\makebox(0,0)[t]{\bf 8470}}
    \end{picture}}
   \put(470,0){\setlength{\unitlength}{1cm}\begin{picture}(0,0)(0,0)
        \put(0,0){\line(0,1){0.2}}
        \put(0,-0.2){\makebox(0,0)[t]{\bf 8470}}
    \end{picture}}
    \put(480,0){\setlength{\unitlength}{1cm}\begin{picture}(0,0)(0,0)
        \put(0,0){\line(0,1){0.2}}
        \put(0,-0.2){\makebox(0,0)[t]{\bf 8480}}
    \end{picture}}
    \put(490,0){\setlength{\unitlength}{1cm}\begin{picture}(0,0)(0,0)
       \put(0,0){\line(0,1){0.2}}
        \put(0,-0.2){\makebox(0,0)[t]{\bf 8490}}
    \end{picture}}
    \put(480,0){\setlength{\unitlength}{1cm}\begin{picture}(0,0)(0,0)
       \put(0,0){\line(0,1){0.2}}
        \put(0,-0.2){\makebox(0,0)[t]{\bf 8480}}
    \end{picture}}

\punkti{25/07/91}{01:12}{462.551}{ -1.359}{0.043}{ 500}{1.56}{
2348}{1.2CA}
\punktii{26/07/91}{01:53}{463.579}{ -1.330}{0.139}{ 100}{1.39}{
1048}{1.2CA}
\punkti{01/08/91}{01:16}{469.553}{ -1.233}{0.029}{ 100}{1.20}{
521}{1.2CA}
\punkti{01/08/91}{01:25}{469.559}{ -1.273}{0.059}{ 500}{1.29}{
1579}{1.2CA}
\punkti{02/08/91}{00:23}{470.517}{ -1.245}{0.059}{ 500}{1.60}{
972}{1.2CA}
\punkti{02/08/91}{23:23}{471.475}{ -1.282}{0.030}{ 100}{1.54}{
361}{1.2CA}
\punkti{02/08/91}{23:31}{471.480}{ -1.252}{0.021}{ 500}{1.75}{
776}{1.2CA}
\punkti{04/08/91}{01:34}{472.566}{ -1.138}{0.026}{ 100}{1.47}{
324}{1.2CA}
\punkti{04/08/91}{01:43}{472.572}{ -1.177}{0.012}{ 500}{1.28}{
627}{1.2CA}
\punkti{05/08/91}{01:58}{473.582}{ -1.094}{0.023}{ 100}{1.35}{
334}{1.2CA}
\punkti{05/08/91}{02:08}{473.589}{ -1.127}{0.010}{ 500}{1.32}{
661}{1.2CA}
\punktii{06/08/91}{00:28}{474.520}{ -1.044}{0.121}{ 100}{1.05}{
295}{1.2CA}
\punkti{06/08/91}{00:47}{474.533}{ -1.039}{0.007}{ 500}{1.07}{
471}{1.2CA}
\punkti{07/08/91}{02:28}{475.603}{ -1.039}{0.013}{ 100}{1.06}{
300}{1.2CA}
\punkti{07/08/91}{02:36}{475.609}{ -1.029}{0.008}{ 500}{1.23}{
506}{1.2CA}
\punkti{08/08/91}{01:09}{476.548}{ -0.941}{0.037}{ 100}{2.27}{
336}{1.2CA}
\punkti{08/08/91}{01:17}{476.554}{ -0.970}{0.027}{ 500}{3.12}{
532}{1.2CA}
\punktii{11/08/91}{00:55}{479.539}{ -0.617}{0.134}{ 100}{3.76}{
313}{1.2CA}
\punkti{12/08/91}{01:43}{480.572}{ -0.713}{0.020}{ 100}{1.46}{
324}{1.2CA}
\punkti{12/08/91}{01:50}{480.577}{ -0.717}{0.009}{ 500}{1.55}{
590}{1.2CA}
\punkti{13/08/91}{01:23}{481.558}{ -0.672}{0.019}{ 500}{1.75}{
526}{1.2CA}
\punkti{14/08/91}{03:53}{482.662}{ -0.731}{0.050}{ 300}{1.85}{
448}{1.2CA}
\punktii{16/08/91}{03:33}{484.648}{ -0.774}{0.104}{ 300}{2.05}{
412}{1.2CA}
\punkti{17/08/91}{01:01}{485.543}{ -0.865}{0.026}{ 300}{1.63}{
403}{1.2CA}
\punktii{18/08/91}{04:33}{486.690}{ -0.794}{0.145}{ 300}{2.18}{
1666}{1.2CA}
\punkti{18/08/91}{23:16}{487.470}{ -0.666}{0.057}{ 300}{2.28}{
493}{1.2CA}
\punkti{20/08/91}{01:14}{488.552}{ -0.451}{0.031}{ 100}{1.33}{
301}{1.2CA}
\punkti{20/08/91}{01:22}{488.557}{ -0.425}{0.017}{ 500}{1.20}{
493}{1.2CA}

\end{picture}}

\end{picture}

\vspace*{1cm}
\caption{Lightcurve of 3C\,446 (2223$-$052) in August 1991}
\end{figure}
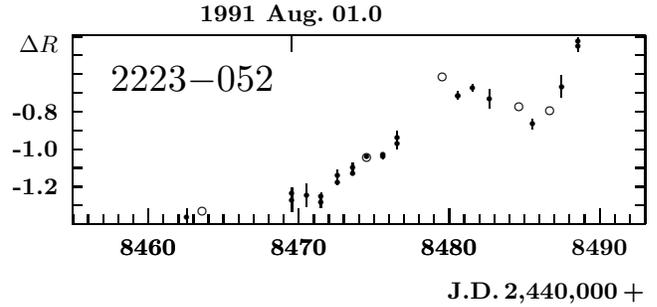

\noindent
{\bf 3C\,345 (1641+399).} The Rosemary Hill lightcurve plotted in SNLC is
not well sampled since 1988. All other literature data and the HQM
lightcurve have been discussed in detail in Schramm et al.\
(\cite{SBCW93}). The lightcurve shown in Fig.~1 includes some more recent
data points, partly obtained with a 0.7\,m telescope at the
Landessternwarte Heidelberg by J.~Heidt. The reference magnitude is
$R_0=15.67$. The new measurements indicate that the object has reached
again a state nearly as faint as
prior to the 1991/92 outburst. This differs from the
predicted lightcurve of Schramm et al.\ but does not contradict the model
of a ``lighthouse effect''; the new data probably
indicate that the opening of the
jet sets in at a shorter distance from the core (for more details see
Borgeest \& Schramm 1993b, Camenzind 1993 and Sillanp\"a\"a \& Valtaoja 1993).

\noindent
{\bf W1\,1749+701.} There is only one publication concerning variability of
this BL\,Lac object: Four measurements obtained by Arp et al.\
(\cite{ASWR76}) in 1975 indicate a 0.5\,mag range of variability. The HQM
lightcurve has a total range of 1.0\,mag and shows rapid brightness changes
of a few 0.1\,mag.

\noindent
{\bf 4C\,56.27 (1823+568).} There exist to our knowledge no variability
studies in the literature for this BL\,Lac object. The
HQM data have a total range of 1.2\,mag.
The lightcurve obtained in August 1991 is shown with higher resolution in
Fig~7; some rapid brightness changes $\ga$\,0.5\,mag occurring within a
few days are obvious.

\noindent
{\bf 3C\,446 (2223$-$052).} The Rosemary Hill lightcurve as shown in WSLF
covers the time from 1971 to 1985; data collected between 1964 and 1974
are plotted in BRZ. This BL\,Lac object exhibits strong, spikelike flares
and slow, large-amplitude variations. Since 1984 the object showed an
almost monotonic decline. The HQM lightcurve exhibits very rapid, large
amplitude variations; one part, collected in August 1991 is shown in Fig.~8
with higher resolution. The most rapid variation occurred in early summer
1989 (Fig.~1) where the object faded down by 1.50\,mag within 12 days.
Making use of the photometric sequence of Smith et al.\
(\cite{SBHE85}), we were able to determine the reference
magnitude of our lightcurve, $R_0=19.31$.

\bigskip

\acknowledgements{We are indebted to
          the MPIA and the Calar Alto staff members for excellent support
          and to D.~Mehlert, L.~Nieser, M.~Schaaf, T.~Schramm
          and H.J.~Witt for their help during observations. We thank
          J.~Heidt and S.J.~Wagner for making available some recent
          measurements of 3C\,345 and
          P.~King as well as the referee A.~Sillanp\"a\"a for critically
          reading the manuscript. This work has
          been supported by the Deutsche Forschungsgemeinschaft
          under Bo$\,$904/1, Re$\,$439/5 and Schr\,292/6.}

\end{document}